\documentclass[a4paper,onecolumn,preprintnumbers,superscriptaddress,nofootinbib,aps,prd,floatfix,10pt]{revtex4}

\usepackage{graphicx,footmisc,newtxtext}
\usepackage{hyperref}
\hypersetup{colorlinks=true, citecolor=blue, urlcolor=blue, linkcolor=blue}

\usepackage{amsmath,amssymb,mathrsfs,slashed,braket,dsfont}
\usepackage{multirow,booktabs}
\usepackage{feynmp}
\usepackage[utf8]{inputenc}

\makeatletter
\@ifundefined{pdfoutput}{}{\DeclareGraphicsRule{*}{mps}{*}{}}
\makeatother

\graphicspath{{./figs_pub/}}

\AtBeginDocument{
  \heavyrulewidth=.08em
  \lightrulewidth=.05em
  \cmidrulewidth=.03em
  \belowrulesep=.65ex
  \belowbottomsep=0pt
  \aboverulesep=.4ex
  \abovetopsep=0pt
  \cmidrulesep=\doublerulesep
  \cmidrulekern=.5em
  \defaultaddspace= .5em
  \setlength{\tabcolsep}{0.5em}
}

\newcommand{\met}{\slashchar{E}_T}
\newcommand{\tnaive}{\hat{T}}
\newcommand{\Dfba}{\mbox{$\raisebox{2mm}{\boldmath ${}^\leftrightarrow$}\hspace{-4mm} D^a$}}

\newcommand{\qqquad}{\qquad \qquad}
\newcommand{\qqqquad}{\qquad \qquad \qquad}

\newcommand{\lag}{\mathscr{L}}

\newcommand{\ope}{\mathcal{O}}
\newcommand{\opt}{\mathscr{O}}
\newcommand{\opp}{\mathcal{P}}

\newcommand{\mat}{\mathcal{M}}
\newcommand{\one}{\mathds{1}}

\newcommand{\pb}{\ensuremath \mathrm{pb}}
\newcommand{\fb}{\ensuremath \mathrm{fb}}

\newcommand{\ifb}{\ensuremath \mathrm{fb}^{-1}}
\newcommand{\iab}{\ensuremath \mathrm{ab}^{-1}}
\newcommand{\mev}{{\ensuremath \mathrm{MeV}}}
\newcommand{\gev}{{\ensuremath \mathrm{GeV}}}
\newcommand{\tev}{{\ensuremath \mathrm{TeV}}}
\newcommand{\GeV}{{\ensuremath \mathrm{GeV}}}
\newcommand{\br}{{\ensuremath \mathrm{BR}}}

\newcommand{\ie}{\textsl{i.e. }}
\newcommand{\eg}{\textsl{e.g. }}

\DeclareMathOperator{\tr}{tr}
\DeclareMathOperator{\Tr}{Tr}

\def\slashchar#1{\setbox0=\hbox{$#1$}           
   \dimen0=\wd0                                 
   \setbox1=\hbox{/} \dimen1=\wd1               
   \ifdim\dimen0>\dimen1                        
      \rlap{\hbox to \dimen0{\hfil/\hfil}}      
      #1                                        
   \else                                        
      \rlap{\hbox to \dimen1{\hfil$#1$\hfil}}   
      /                                         
   \fi}

\numberwithin{equation}{section}

\begin{document}

\title{Higgs Physics: It ain't over till it's over}

\begin{abstract}
We review the theoretical underpinning of the Higgs mechanism of
electroweak symmetry breaking and the experimental status of Higgs
measurements from a pedagogical perspective. The possibilities and
motivations for new physics in the symmetry breaking sector are
discussed along with current measurements. A focus is on the
implications of measurements in the Higgs sector for theoretical
insights into extensions of the Standard Model. We also discuss of
future prospects for Higgs physics and new analysis techniques.
\end{abstract}

\author{Sally Dawson}
\email{dawson@bnl.gov}
\affiliation{Department of Physics, Brookhaven National Laboratory, Upton, N.Y., 11973 U.S.A.\\[0.1cm]}
\author{Christoph Englert} \email{christoph.englert@glasgow.ac.uk}
\affiliation{SUPA, School of Physics and Astronomy, University of Glasgow, Glasgow G12 8QQ, UK\\[0.1cm]}
\author{Tilman Plehn}
\email{plehn@uni-heidelberg.de}
\affiliation{Institut f\"ur Theoretische Physik, Universit\"at Heidelberg, Germany\\[0.5cm]}

\maketitle
\newpage
\tableofcontents
\begin{fmffile}{feynman}
\newpage

\section*{Introduction}

The Higgs discovery in 2012 can be considered the most unusual, yet
arguably most important discovery in the history of particle
physics. ``Most unusual'' because it was entirely expected, which
stands in stark contrast to other paradigm-shifting discoveries such
as $J/\psi$ during the November revolution in 1974. ``Most important''
as it marks the ultimate test of perturbative quantum field theory as
a theoretical framework, validating spontaneous symmetry breaking as
the only remaining realization of gauge symmetry after linear
realizations such as QCD and QED had been tested with tremendous
success.

The Higgs discovery completed the Standard Model (SM) of particle
physics, which after subsequent measurements not only in Higgs related
channels at the Large Hadron Collider (LHC) stands firmly as Occam's
Razor of fundamental particle interactions. The Standard Model is relatively
simple quantum field theory, the minimal symmetry and particle content
required by measurements seems to be perfectly sufficient to explain a
range of phenomena over a many orders of magnitude. Yet we know that
the Standard Model cannot be the final answer to fundamental
interactions in nature. Typically one mentions gravity as the prime
example of a fundamental force that is not included in the Standard Model and it
is fair to say that the basics of its fundamental quantum mechanical
nature are yet to understood. But there are phenomena more closely
related to our usual thinking of particle physics that are not
addressed in the Standard Model, such as dark matter, which could well fit into
our well-established framework of perturbative quantum field
theory. Besides this obvious negligence of phenomenological facts,
there theoretical arguments against the Standard Model, which have served as
motivations to construct various avenues of physics beyond the Standard Model. One
of the most famous ones is probably the hierarchy problem that states
that fundamental scalars imply relevant mass parameters from a
renormalization group point of view. As ``traditional'' gauge symmetry
principles are not sufficient to mend the instability of the Higgs
mass against radiative corrections in the presence of large new
physics scales, qualitatively different extensions of the Standard Model need to
be formulated. These include various scenarios of Supersymmetry and
strong interactions, where the latter seeks to explain the electroweak
scale as a dimensional transmutation effect similar to QCD.

Irrespective of their theoretical appeal or our prejudice as
theoretical physicists, we have effectively entered a era of particle
physics that is marked by its return to a measurement-driven
character. Although this seems unusual in the context of particle
physics where model-building has crucially influenced experimental
measurement and design strategies over the past four decades, it is a
common denominator of all physical sciences. While we still look for
concrete UV scenarios at the LHC, more model-independent approaches
have gained attention and efforts in creating adapted (data-driven)
analysis approaches have been intensified.

It is the purpose of this review to provide a pedagogical snapshot of Higgs
physics six years into the LHC's Higgs phenomenology program. We
review production and decay processes as well as their perturbative
history that led to the Higgs boson discovery in 2012. We discuss
well-motivated extensions beyond the Standard Model, how they have
been searched for at the LHC, as well as their current
phenomenological status. We also discuss how these measurements affect the
current discussion on the direction that the field might take in the
future after the LHC.

\section{Theoretical basis}
\label{sec:basis}

\begin{figure}[t]
\includegraphics[width=0.35\textwidth]{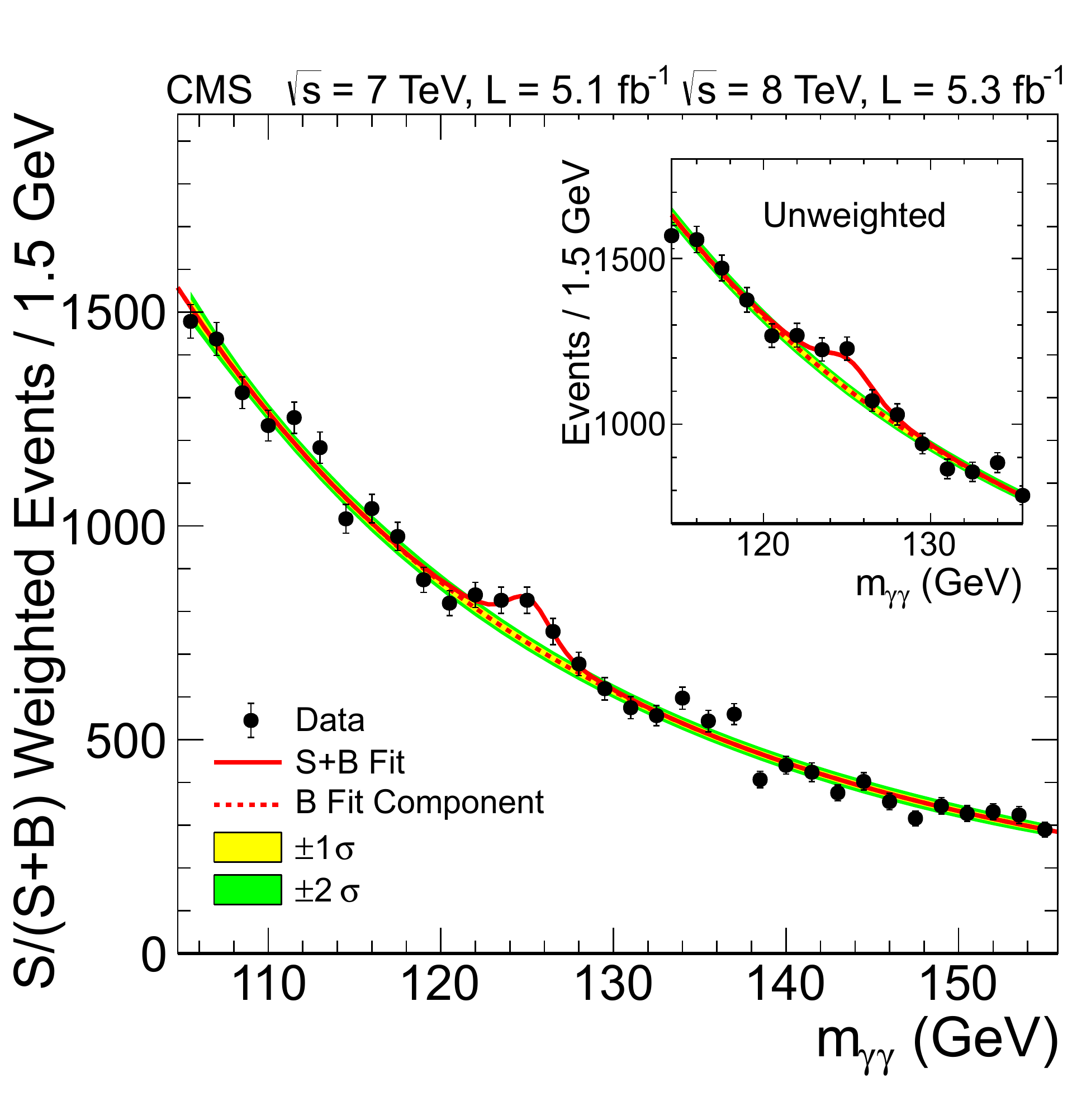}
\hspace*{0.1\textwidth}
\includegraphics[width=0.35\textwidth]{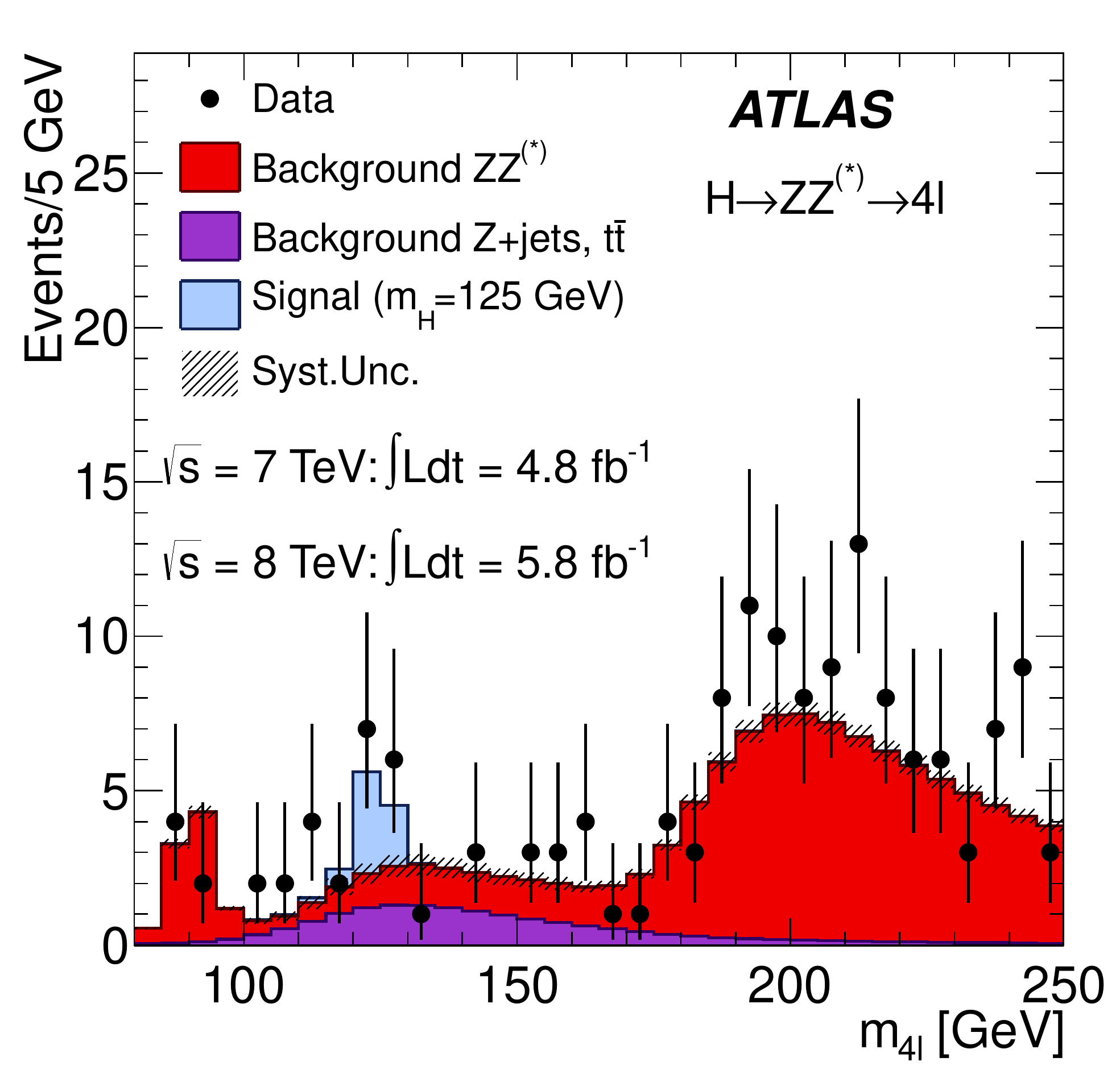}
\caption{2012 Higgs discovery in $\gamma\gamma$ (left) and leptonic $ZZ$
  (right) channels. Figures from from
  Ref.~\cite{CMS:2012nga,Aad:2012tfa}.}
\label{fig:ggdisc}
\end{figure}

Higgs physics is one of the few instances in fundamental physics where
a theoretical idea or concept was discovered experimentally exactly
the way it was
predicted~\cite{Higgs:1964pj,Higgs:1964ia,Englert:1964et,Guralnik:1964eu,Kibble:1967sv,Higgs:1966ev}~\footnote{If you feel strongly that we are missing a reference, please send a friendly email to niceguy\@@englert.com, and we will include it. If you just want us to cite one more of your papers, please email to devnull\@@ihateyoutilman.com.}. In
that sense, the discovery of the Higgs boson in 2012 marks the coming
of age of one of the largest and longest-lasting projects in
fundamental physics. The Higgs boson was discovered in
2012~\cite{CMS:2012nga,Aad:2012tfa} through the predicted resonant
decays to $\gamma \gamma$ and $ZZ$ as seen in Fig.~\ref{fig:ggdisc}.
This discovery is now established and there can be no doubt that a
particle with properties close to those predicted decades ago has been
discovered with the mass is measured rather
accurately~\cite{Aad:2015zhl} to be
\begin{align}
M_h=125.09\pm 0.21 \text{(stat)} \pm 0.11 \text{(syst)}~\text{GeV}\,.
\end{align}

In the first part of this review we discuss the basic structure of the
Higgs mechanism in the Standard Model, discuss extended models and the
formalism of effective field theory as a preferred theory framework
for LHC studies.  Next, we comment on the current experimental status,
organized by LHC signatures. We  touch on possible future
developments in Higgs physics at the LHC, ILC, and future high energy hadron machines.  Finally,
we close with a discussion of new analysis techniques for Higgs physics. 

\subsection{Historical introduction}
\label{sec:basics_history}

In some sense, Higgs physics started with Fermi's theory in
1934~\cite{Fermi:1934hr}. While this theory offers a valid description
of beta decay, muon decay, and many other low-energy phenomena, it
turns out to have a fundamental problem: the interaction strength of
the four-fermion interaction has an inverse mass dimension. This leads
to structural problems when we want to unify it with the proto-typical
fundamental quantum field theory of quantum electrodynamics (QED).
The main difference between these two theories is that the photon is a
massless vector boson, while the agent of the weak interaction has a
mass that can be linked to the inverse Fermi coupling.  This mass
shields the interactions at large distances and transforms a
Coulomb-like potential into a Yukawa potential.  The fundamental
theoretical question therefore is: how can we break the gauge symmetry
describing the weak interaction such that at high energies it can be
unified with QED while remaining compatible with Fermi's theory at low
energies?

The Brout-Englert-Higgs
trick~\cite{Anderson:1963pc,Englert:1964et,Higgs:1964pj,Higgs:1964ia},
or spontaneous symmetry breaking, is to not break the weak gauge
symmetry explicitly or at the level of the Lagrangian. Instead, it
realizes it non-linearly through the presence of an additional scalar
field~\cite{Fujikawa:1972fe,Becchi:1974xu,Becchi:1974md,Becchi:1975nq,Tyutin:1975qk} (see also~\cite{tHooft:1995wad}).
In this way electroweak symmetry can be compatible with a large
non-vanishing background expectation value of the scalar field. A
phenomenological consequence of this approach that is commonly
referred to as the Higgs mechanism is the appearance of longitudinal
degrees of freedom for the involved gauge bosons, which behave like
derivatives of scalar fields in the deep ultraviolet regime of
momenta. The ingenious idea of realizing electroweak symmetry this
way therefore effectively restores the standard power counting
arguments of perturbative quantum field theory~\cite{Weinberg:1959nj} that are at the core of
ultraviolet renormalizability~\cite{tHooft:1972tcz}, which again is the theoretical
underpinning of the theoretical progress that was necessary to finally
discover the Higgs boson. Strictly speaking, \textsl{spontaneous symmetry
breaking} is a slight abuse of language as the gauge symmetry is not
broken and the vacuum is invariant under local gauge transformations.
The semi-classical treatment of non-linearly realized local symmetries
and spontaneously broken global symmetries at the level of Lagrangians
is similar, while the consequences for the quantum dynamics are vastly
different. Given the semi-classical analogy, we can therefore pretend
that we are dealing with identical approaches while implicitly relying
on the fact that the quantum theory will make allowances for our
ignorance.

The fundamental problem of breaking the weak $SU(2)_L$ symmetry is not
introducing an appropriate breaking term in the Lagrangian and
justifying its existence. The problem is that a massless gauge boson
has two degrees of freedom, while a massive gauge boson requires three
degrees of freedom, and the third or longitudinal degree of freedom
has to come from somewhere. The solution is linked to Goldstone's
theorem~\cite{Goldstone:1961eq}, which in the particular case of a
weak $SU(2)_L$ symmetry naively implies that three broken generators
give rise to three massless degrees of freedom in the low energy
spectrum of the theory. When we gauge the symmetry, these degrees of
freedom combine with the two massless degrees of freedom of a massless
gauge boson to form massive $W$-bosons. In addition, a quantum
description predicts excitations around the vacuum expectation value,
the new Higgs boson. In general, the Goldstone modes and the Higgs
mode will therefore not form a common multiplet under a given
symmetry. They define an effective field theory with a non-linear
realization of the weak symmetry, as described in
Sec.~\ref{sec:basis_eft}.

In the ultraviolet extension of Fermi's
theory~\cite{Glashow:1961tr,Salam:1964ry,Weinberg:1967tq} the three Goldstone modes are
part of an $SU(2)_L$ doublet. Such a complex doublet has four degrees
of freedom, and the fourth direction is identified with the vacuum
expectation value (VEV) and the additional scalar field, $h$~\cite{Higgs:1964pj},
\begin{align}
       \phi = 
             \frac{1}{\sqrt{2}}
             \begin{pmatrix}
                    \phi^1+i\phi^2 \\ v+h+i\phi^0 
             \end{pmatrix} \; .
\label{eq:def_phi}
\end{align}
The building block $\phi^\dagger \phi$ is invariant
under all symmetries of the Standard Model.  The Higgs VEV $v$ can be
traced to a non-trivial minimum of the potential
\begin{alignat}{5}
-\lag \supset V_\text{SM}=\mu^2 (\phi^\dagger \phi) +\lambda (\phi^\dagger \phi)^2 +\text{const} \; ,
\label{eq:higgs_pot_sm}
\end{alignat}
with $\mu^2 <0$.  If we limit ourself to this potential, we can relate
the two parameters $\mu$ and $\lambda$ to two (pseudo)observables, the Higgs
VEV $v$ and the physical Higgs boson mass
\begin{align}
v^2 = - \frac{\mu^2}{\lambda}
\qquad \text{and} \qquad  
M_h^2 = 2 \lambda v^2 = - 2\mu^2 \; .
\label{eq:higgs_sm}
\end{align}
The triple and quartic self-couplings of the physical Higgs bosons are
predicted to be 
\begin{align}
\lag \supset
- \frac{M_h^2}{2v} h^3
- \frac{M_h^2}{8v^2} h^4 \; ,
\label{eq:higgs_sm2}
 \end{align}
allowing for a consistency test of the renormalizable Higgs sector of
the Standard Model.

The fact that the Higgs potential in Eq.\eqref{eq:higgs_pot_sm} is
truncated at dimension four reflects our theoretical bias that a
fundamental theory should not feature couplings with inverse mass
dimensions. Forgetting this bias for a moment, we can add higher
powers of the gauge-invariant building block $\phi^\dagger \phi$ to
the Higgs potential,
\begin{align}
-\lag \supset V=\mu^2 (\phi^\dagger \phi) + \lambda (\phi^\dagger \phi)^2 + \frac{f_{\phi,3}}{\Lambda^2} (\phi^\dagger \phi)^3 
 +\ope \left( \dfrac{1}{\Lambda^4} \right) \; ,
\label{eq:higgs_pot_d6}
\end{align}
where we assume $\Lambda \gg v$ for convergence and consistency
reasons.  The Higgs VEV and mass then become,
\begin{align}
v^2 =
 -\dfrac{\mu^2}{\lambda}
 \left[ 1 + 
 \dfrac{3 f_{\phi,3} \mu^2}{4 \lambda^2 \Lambda^2}
 +\ope \left( \dfrac{1}{\Lambda^4} \right) \right]
\qquad \text{and} \qquad 
M_h^2 = 
  2 \lambda v^2 \left[ 1
 + \frac{3 f_{\phi,3} v^2}{2 \Lambda^2\lambda} 
 +\ope \left( \dfrac{1}{\Lambda^4} \right) \right] \;,
\end{align}
which breaks the correlations of Eqs.~\eqref{eq:higgs_sm} and \eqref{eq:higgs_sm2}.
Such higher-dimensional terms including the $SU(2)_L$ doublet $\phi$
are the basis of a linearly realized effective theory, that will be
discussed in detail in Sec.~\ref{sec:basis_eft}. Extended Higgs
potentials, which after integrating out heavier additional Higgs
states lead to higher-dimensional operators, are at the heart of many of the 
theoretical constructions discussed in
Sec.~\ref{sec:basis_weak}.

\subsection{Weinberg-Salam Model}
\label{sec:basis_sm}

In the Standard Model, the Higgs mechanism serves two independent
purposes. First, it introduces masses for the weak gauge bosons in an,
as we know now, renormalizable and perturbative gauge theory. Second,
it allows us to write fermion masses for all fermions where the
left-handed spinors are part of a weak doublet, while the right-handed
spinors are weak singlets. Per se, these two aspects are not
related. Moreover, this kind of mass generation does not include the
neutrinos, which are missing a light right-handed component in the SM context.

\subsubsection{Gauge sector}
\label{sec:basis_sm_gauge}

The Higgs mechanism of the previous section is embedded in the
Weinberg-Salam model of electroweak
interactions~\cite{Weinberg:1967tq,Glashow:1961tr,Goldstone:1962es,Salam:1964ry}.
The theory is an $SU(2)_L \times U(1)_Y$ gauge theory containing three
$SU(2)_L$ gauge bosons, $W_\mu^a$, and one $U(1)_Y$ gauge boson,
$B_\mu$. Their kinetic energy terms read
\begin{align}
\lag_\text{KE} =-{1\over 4}W_{\mu\nu}^a W^{\mu\nu a}
-{1\over 4} B_{\mu\nu} B^{\mu\nu}\, ,
\end{align}
where the index $a=1,2,3$ is summed over and,
\begin{align}
W_{\mu\nu}^a=& \partial_\nu W_\mu^a-\partial _\mu W_\nu^a
+g \epsilon^{abc}W_\mu^b W_\nu^c \, ,
\notag \\
B_{\mu\nu}=&\partial_\nu B_\mu-\partial_\mu B_\nu\quad .
\end{align}
The $SU(2)_L$ and $U(1)_Y$ coupling constants are $g$ and $g^\prime$, respectively. 

The scalar contribution to the Lagrangian is,
\begin{align}
\lag_s=(D^\mu \phi)^\dagger (D_\mu \phi)-V(\phi)\, ,
\label{eq:scalepot}
\end{align}
where $V(\phi)$ is given in Eq.\eqref{eq:higgs_pot_sm} and
\begin{align}
D_\mu=\partial_\mu +i {g\over 2}\sigma^a  W^a_\mu+i{g^\prime\over 2}
B_\mu Y\, .
\end{align}
where $\sigma^a$ are the Pauli matrices. This covariant derivative
describes the Higgs interactions with the massive gauge bosons, and $Y=1$ is the $U(1)_Y$ hypercharge of the Higgs boson. 

In unitary gauge there are no Goldstone bosons and only the physical
Higgs scalar remains in the spectrum after the spontaneous symmetry
breaking in the Higgs potential of Eq.\eqref{eq:higgs_pot_sm} has
occurred.  The spontaneous symmetry breaking results in two charged
gauge fields, $W^\pm$, and two neutral gauge bosons, $Z$ and $\gamma$.
\begin{align}
W^{\pm}_\mu&=
{1\over \sqrt{2}}(W_\mu^1 \mp i W_\mu^2)\notag \\
Z^\mu&= {-g^\prime B_\mu+ g W_\mu^3\over \sqrt{g^2+g^{\prime~2}}}\equiv -\sin\theta_W B_\mu+\cos\theta_W W_\mu^3
\notag \\
A^\mu&= {g B_\mu+ g^{\prime} W_\mu^3\over \sqrt{g^2+g^{\prime~2}}}\equiv \cos\theta_W B_\mu+\sin\theta_WW_\mu^3 \; .
\label{eq:masseig}
\end{align}
The massless neutral mode is the photon, mediating the electromagnetic
interaction. The mixing pattern of Eq.\eqref{eq:masseig} defines the weak mixing 
angle,
\begin{align}
\sin\theta_W = {g^\prime \over \sqrt{g^2+g^{\prime2}}}\;.
\end{align}
Since the massless photon must couple with electromagnetic strength,
$e$, the coupling constants are related to the weak mixing angle $\theta_W$,
\begin{align}
e=g \sin\theta_W  \equiv  g s_W
= g^\prime \cos\theta_W\equiv g^\prime c_W \; .
\label{eq:coupsdef}
\end{align}
The gauge bosons obtain masses from the Higgs mechanism,
\begin{align}
M_W^2 = {1\over 4} g^2 v^2, \qquad \qquad
M_Z^2 = {1\over 4} (g^2 + g^{\prime2})v^2,\qquad\qquad
M_A =  0\, .
\label{eq:gaugemasses}
\end{align}
In a gauge other than unitary gauge we have to account for the Goldstone bosons in the
particle spectrum. A complete set of Feynman rules is conveniently
given in Ref.~\cite{Romao:2012pq}.

\subsubsection{Fermion sector}
\label{sec:basis_sm_fermion}

The $SU(2)_L$ and $U(1)_Y$ charge assignments of the first generation
of fermions are given in Table~\ref{tab:ferm}.  The quantum numbers of
the $2^{nd}$ and $3^{rd}$ generations are identical to those of the first
generation and the hypercharge satisfies the relationship, $Q=(\sigma^3
+Y)/2$.

\begin{table}[b!]
\begin{center}
\renewcommand\arraystretch{1.2}
\begin{tabular}{l|ccr}
\toprule
Field & $SU(3)_C$ & $SU(2)_L$& $U(1)_Y$ \\
\midrule
$Q_L=\left(\begin{array}{c}
u_L\\ d_L \end{array}\right)$
   &    $3$          & $2$&  $~\dfrac{1}{3}$ \\
$u_R$ & $3$ & $1$& $\dfrac{4}{3}$ \\[3mm]
$ d_R$ & $ 3$ & $1$&  $~-\dfrac{2}{3}$ \\
\midrule
$L_L=\left(\begin{array}{c}
\nu_L\\ e_L \end{array}\right)
$  & $1$             & $2$& $~-1$ \\
$e_R$ & $1$             & $1$& $~-2$ \\ 
\midrule
$\phi= \left(\begin{array}{c}
\phi^+\\ \phi^0 \end{array}\right)
$  & $1$             & $2$& $1$ \\
\bottomrule
\end{tabular}
\caption{Quantum numbers of the  SM fermions.\label{tab:ferm}}
\end{center}
\end{table}

Direct fermion mass terms are forbidden by the $SU(2)_L$ assignments
of Table~\ref{tab:ferm}.  The Higgs boson can generate fermion masses
through the couplings
\begin{align}
\lag_M = -y_d {\overline Q}_L \phi\, d_R 
-y_u {\overline Q}_L {\tilde{\phi}} \,u_R -
y_e{\overline L}_L\phi\, e_R+ \text{h.c.}
\label{eq:fermyuk}
\end{align}
where ${\tilde{\phi}} \equiv  \phi^c=  i \sigma^2 \phi^*$ is the charge-conjugated Higgs doublet.
After the Higgs obtains a VEV, 
all of the fermions except the neutrinos obtain masses and the Higgs-fermion 
couplings are proportional to the fermion masses,
\begin{align}
y_f = {\sqrt{2} m_f \over v} \; .
\label{eq:def_yukawas}
\end{align}

For the multi-family case, the Yukawa couplings, $ y_d$ and $y_u$,
become $N_F \times N_F$ matrices, where $N_F$ is the number of
families.  Since the fermion mass matrices and Yukawa matrices are
proportional to each other, the interactions of the Higgs boson with
the fermion mass eigenstates are flavor diagonal and the Higgs boson
does not mediate flavor changing interactions.  This is an important
prediction of the Standard Model and the discovery of flavor changing Higgs decays
would be a clear signal of physics beyond the Standard
Model~\cite{Bjorken:1977vt,Barr:1990vd} .

The numerical value of the  Higgs VEV can be extracted from 
the charged current result for $\mu$ decay,
$\mu\rightarrow e {\overline \nu}_e \nu_\mu$. It is measured
very accurately to be $G_F=1.1663787(6)\times 10^{-5}~\gev^{-2}$.
Since the momentum carried by the $W$ boson is of order $m_\mu$ it
can be neglected in comparison with $M_W$ and we make the identification,
\begin{align}
{G_F\over \sqrt{2}}={g^2\over 8 M_W^2}={1\over 2 v^2}
\qquad \text{or} \qquad 
v= \frac{1}{(\sqrt{2} G_F)^{1/2}} = 246 \, \gev \; .
\end{align}
This leaves the Higgs mass as the only free parameter in the scalar potential.

\subsubsection{Custodial symmetry}
\label{sec:basis_sm_cust}

One of the most important features of the minimalistic Higgs mechanism
in the Standard Model is that all of the couplings of the Higgs boson
to fermions and gauge bosons are completely determined in terms of the
gauge coupling constants,  the Higgs mass and fermion masses, and the VEV, $v$.
Moreover, at tree level, the different gauge
boson masses are related as
\begin{align}
\rho\equiv {M_W^2\over M_Z^2 \cos^2\theta_W}=1\, .
\label{eq:treero}
\end{align}
The relation of the eigenvalues of the $(W^a,B)_\mu$ mixing matrix are aligned with the characteristic rotation 
angle of the associated basis transformation $(W^3,B)_\mu\to (Z,A)_\mu$.
This relation follows from an accidental symmetry of the scalar
potential that  can be seen by rewriting the Higgs field as a
bi-doublet,
\begin{align}
\Phi=\left(
\begin{matrix}
\phi^{0*}&\phi^+\\
-\phi^{+*}& \phi^0
\end{matrix}
\right)\, ,
\label{eq:bidoub}
\end{align}
as $\phi^c$ transforms identically as $\phi$ (the doublet representation of SU(2) is symplectic).
The Higgs potential can now  be written as,
\begin{align}
V_\text{SM}=-\mu^2 \Tr\left(\Phi^\dagger\Phi\right)-\lambda \Tr\left(\Phi^\dagger\Phi\Phi^\dagger\Phi\right)\; .
\label{eq:newpo}
\end{align}
and features a global $SU(2)_L\times SU(2)_R$ symmetry which is broken
to a global diagonal $SU(2)_V$ symmetry when $\Phi$ gets a VEV.  This is
the custodial symmetry. It is broken by the
gauging of hypercharge ($g^\prime$) and by the mass difference between
the top and bottom quarks~\cite{Degrassi:1990tu},
\begin{align}
\label{eq:rhoparameter}
\rho=1
+ {3g^2\over 64\pi^2 }{m_t^2-m_b^2\over M_W^2}
-{11\over 192 \pi^2}g^\prime\phantom{}^2 \; \log \frac{M_h^2}{M_Z^2} 
+ \cdots
\end{align}
The leading one-loop corrections to $\rho=1$ vanish when
$g^\prime\rightarrow 0$ and $m_t\rightarrow m_b$. The power dependence
on the top mass and the logarithmic dependence on the Higgs mass will
be reflected in the sensitivity of electroweak precision measurements
to these parameters.

\begin{figure}[t]
\includegraphics[width=0.4\textwidth]{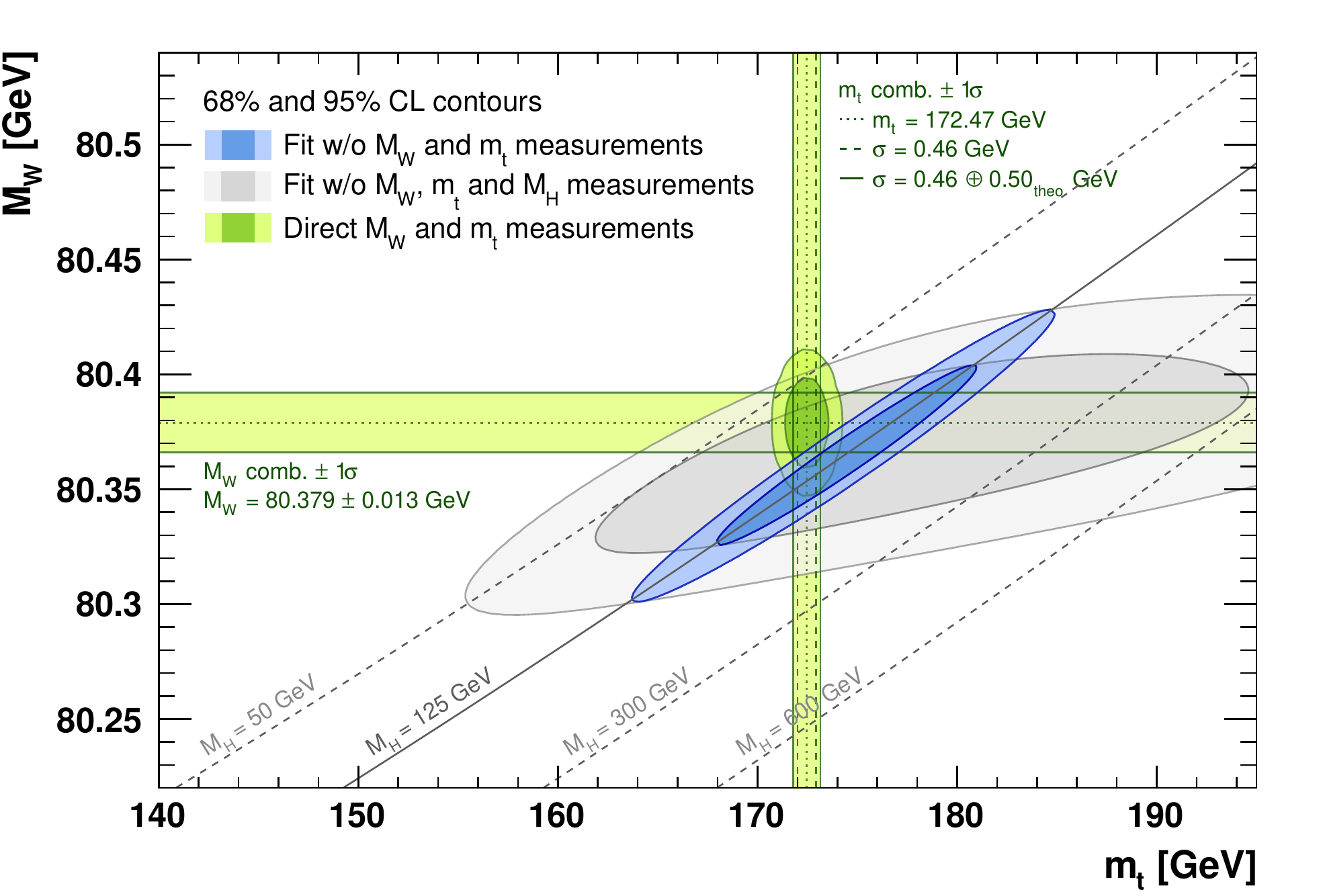}
\hspace*{0.1\textwidth}
\includegraphics[width=0.4\textwidth]{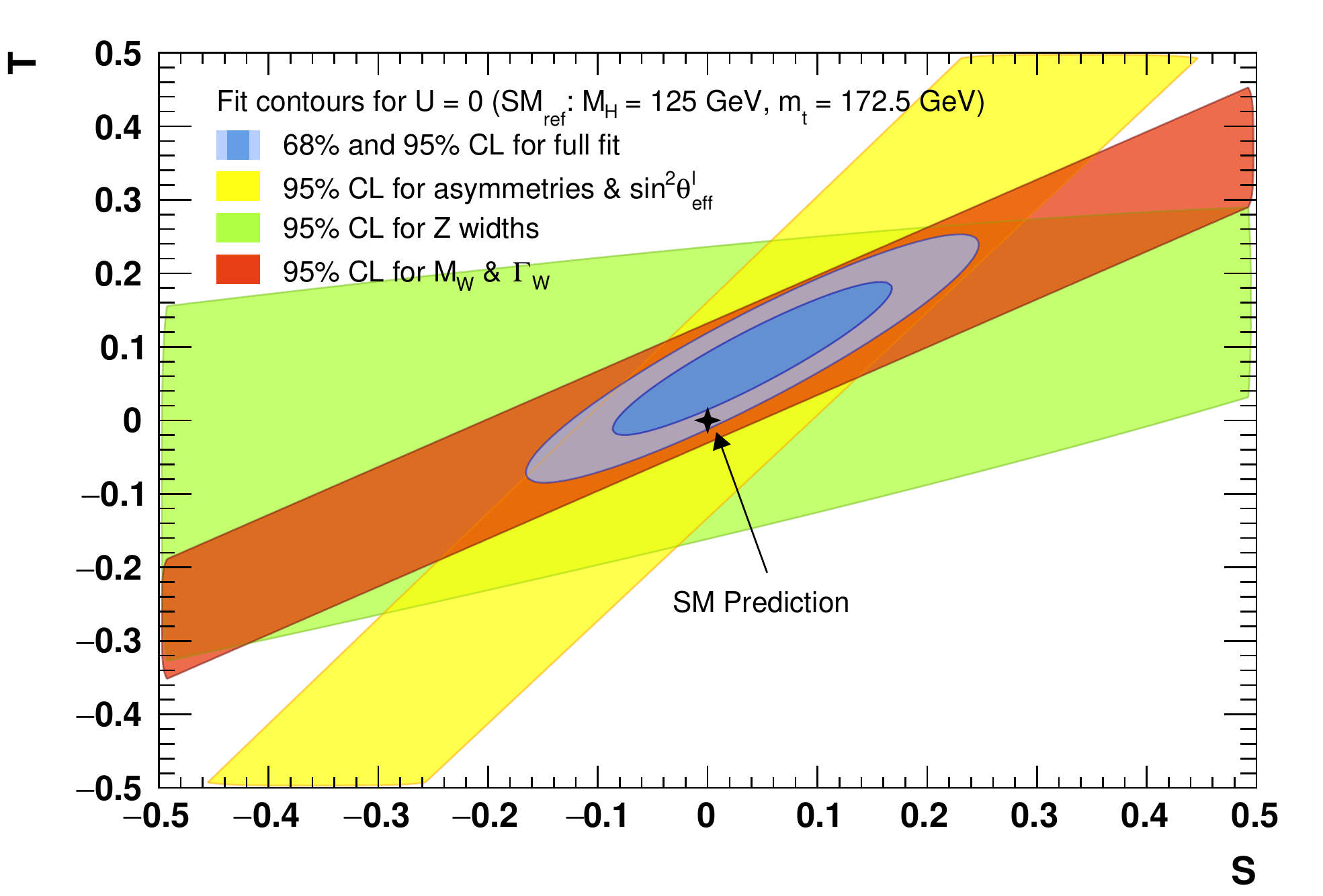}
\caption{Left: experimental limits on $M_W$ and $m_t$ from precision
  electroweak measurements including (blue) and excluding (grey) the
  $M_h$ measurement. The direct measurements of $M_W$ and $m_t$ are
  excluded from the fit. The straight bands are the direct
  measurements of $M_W$ and $m_t$.  Right: Experimental limits on the
  oblique parameters. Figures from Ref.~\cite{Haller:2018nnx}.}
\label{fg:ewprec}
\end{figure}

The full set of electroweak data from LEP~\cite{ALEPH:2005ab}, the
Tevatron and the LHC, including Higgs data, can be used to test the
self consistency of the theory, as demonstrated in 
Fig.~\ref{fg:ewprec}.  The most restrictive data
points are the measurements of the $Zbb$ coupling and the
$W$ boson mass.  When the experimental values of $M_W$, $m_t$, and
$M_h$ are omitted, the fit is in good agreement with the directly
measured values of the masses.  Note that the fit excludes a large
($\sim 100's$ of GeV) value of $M_h$ and so even before the Higgs
boson was discovered, we knew that if there were no new physics
contributions to the predictions for electroweak quantities such as
$M_W$, the Higgs boson could not be too heavy.

The experimental closeness of $\rho$ to $1$ has strong implications
for new physics models.  Extensions of the Standard Model with
modified Higgs sectors are significantly restricted by the requirement
of consistency with the electroweak measurements.  A simple way to
examine whether a theory with an extended Higgs sector is consistent
with electroweak experiments is to use the oblique parameters, $S,
\alpha T=\rho,$ and
$U$~\cite{Peskin:1990zt,Peskin:1991sw,Altarelli:1990zd}.  Using the
oblique parameters to obtain limits on Beyond the Standard Model
physics assumes that the dominant contributions resulting from the
expanded theory are to the gauge boson 2-point functions.  Current
experimental limits on $S$ and $T$ are shown in Fig.~\ref{fg:ewprec},
featuring the typical correlation along the
diagonal~\cite{Bardin:1999yd,Haller:2018nnx,deBlas:2016ojx}.

\subsection{Higgs Decays}
\label{sec:basis_decs}

The Weinberg-Salam model predicts all Higgs production and decay
rates, once the Higgs boson mass is known.  The main feature is that
the Higgs will prefer to decay to the heaviest particles allowed, with longitudinal gauge
boson polarizations playing a particular role.  As
such, it is a testable theory, and non-standard values for Higgs rates
would be indications of new physics.  This has inspired an immense
theoretical effort to obtain predictions for Higgs processes to the
highest possible order in perturbation theory, along with resummation
of large logarithms.  Numerical values including the most up-to-date
theoretical calculations have been tabulated by the LHC Higgs cross
section working group~\cite{Dittmaier:2011ti} (for the most recent update see~\cite{deFlorian:2016spz}) and we review the status
of the Standard Model predictions here.

Expressions for the SM Higgs decay widths at leading order can be
found in Ref.~\cite{Gunion:1989we}, and the QCD and electroweak
corrected rates, with references to the original literature, are given
in Refs.~\cite{Djouadi:2005gi,Spira:2016ztx}.  The NLO corrected decay
rates can be found using the public codes
\textsc{Hdecay}~\cite{Djouadi:1997yw,Djouadi:2018xqq} and \textsc{Prophecy4f}~\cite{Bredenstein:2007ec}.

\subsubsection{Fermions}
\label{sec:basis_decs_f}

The Higgs couplings to fermions are proportional to the fermion mass
and the lowest order width for the Higgs decay to fermions of mass
$m_f$ is,
\begin{align}
\label{eq:decbb}
\Gamma(h\rightarrow f {\overline f} )={G_F m_f^2 N_{ci}\over 
4\sqrt{2}\pi} \; M_h \beta_f^3
\, ,
\end{align}
where $\beta_f\equiv ({1-4m_f^2/M_h^2})^{1/2}$ and $N_{ci}=1 (3)$ for
charged leptons (quarks), related to their color charge.  The leading fermion decay channel is $h
\rightarrow b {\overline{b}}$, which receives large QCD corrections.
A significant portion of the QCD corrections can be accounted for by
expressing the decay width in terms of a running quark mass,
$m_f(\mu)$ evaluated at the scale $\mu=M_h$, indicating that the large
effects are triggered by large logarithms.  The QCD corrected decay
width can be approximated as~\cite{Drees:1990dq,Braaten:1980yq},
\begin{align}
\Gamma(h\rightarrow q {\overline q})=
{3G_F  m_q^2(M_h^2)\over 4 \sqrt{2} \pi} \;
M_h \beta_q^3 \; \left(1+5.67{\alpha_s(M_h^2)\over \pi}+\cdots
\right)   \; ,
\end{align}
where $\alpha_s(M_h^2)$ is defined in the ${\overline{\text{MS}}}$
scheme with $5$ flavors.  The electroweak radiative corrections to $h
\rightarrow f {\overline f}$ amount to the expected few-percent
correction~\cite{Dabelstein:1991ky,Kniehl:1993ay}.

\subsubsection{Weak bosons}
\label{sec:basis_decs_v}

The Higgs boson can also decay to weak boson pairs.  At tree level,
the decays $h\rightarrow WW^*$ and $h\rightarrow ZZ^*$ are possible,
if we allow for the gauge bosons to be off-shell. Over most of phase
space we can assume that one of the two gauge bosons decays on its
mass shell, while the other is pushed into its sizeable Breit-Wigner
tails. The decay width for such a decay, $h\rightarrow Z
Z^*\rightarrow f_1(p_1) f_2(p_2) Z(p_3)$, is,
\begin{align}
\Gamma=\int_0^{(M_h-M_Z)^2} dm_{12}^2 \int dm_{23}^2\,{| A|^2\over 256\pi^3 M_h^3}\; ,
\end{align}
where $m_{ij}=(p_i+p_j)^2$ and
$m_{12}^2+m_{23}^2+m_{13}^2=M_h^2+M_Z^2$. The usual Kaellen function
is $\lambda = m_{12}^4-2m_{12}^2(M_h^2+M_Z^2)+(M_h^2-M_Z^2)^2$, and
the integration boundaries for $m_{23}^2$ are given by
$(M_h^2+M_Z^2-m_{12}^2\pm \sqrt{\lambda})/2$.  The amplitude-squared
is,
\begin{align}
| A (h\rightarrow Z f {\overline {f}})|^2=
32\,(g_{Lf}^{\, 2}+g_{Rf}^{2})\,G_F^2\,
M_Z^4 \; 
{2 M_Z^2m_{12}^2 -m_{13}^2 m_{12}^2 
-M_h^2 M_Z^2+m_{13}^2M_Z^2
 +m_{13}^2\,M_h^2- m_{13}^4\over
 (m_{12}^2-M_Z^2)^2+\Gamma_Z^2M_Z^2} \; ,
   \label{eq:treeamp}
\end{align}
with $g_{Lf}=T_{3}^f-Q_fs_W^2$, $g_{Rf}=-Q_fs_W^2$, and $T_3^f=\pm 1/2$.
Integrating over $dm_{23}^2$ we find the differential decay rate
\begin{align}
{d\Gamma\over dm_{12}^2}(h\rightarrow Z f {\overline {f}})=
(g_{Lf}^{\, 2}+g_{Rf}^{2})\,G_F^2\,\sqrt{\lambda}{M_Z^4\over 48 \pi^3 M_h^3} \;
{12 M_Z^2 m_{12}^2+\lambda
  \over
 (m_{12}^2-M_Z^2)^2+\Gamma_Z^2M_Z^2}
\; .
\end{align}
The result for $h\rightarrow W f {\overline {f^\prime}}$ can be found
by making the appropriate redefinitions of the fermion-gauge boson
couplings.  Performing the $m_{12}^2$ integral over the invariant mass
of the two fermions and summing over the final state fermions gives
us~\cite{Rizzo:1980gz,Keung:1984hn},
\begin{align}
\Gamma(h\rightarrow WW^*)&={3g^4 M_h\over 512 \pi^3} \; F\left({M_W\over M_h}\right)
\notag \\
\Gamma(h\rightarrow ZZ^*)&={g^4 M_h\over 2048  c_W^4 \pi^3}\left(7-{40\over 3}s_W^2
+{160\over 9}s_W^4\right) \; F\left({M_Z\over M_h}\right)\, ,
\end{align}
where
\begin{align}
F(x)=-  |1-x^2 | \left(
{47\over 2}x^2-{13\over 2} +{1\over x^2}\right)
+3(1-6x^2+4x^4) |\log  x| 
+{3(1-8x^2+20x^4)\over \sqrt{4x^2-1}} 
\cos^{-1}\left({3x^2-1\over 2 x^3}\right)\, . 
\end{align}
The NLO QCD and electroweak corrections to the fully off-shell decays,
$h\rightarrow \text{4 fermions}$, are
implemented in the public code
\textsc{Prophecy4f}~\cite{Bredenstein:2007ec}. Going beyond the total
decay rate and instead studying the $m_{12}$-distribution is a
powerful tool in studying the Lorentz structure of the $VVh$ coupling.

\subsubsection{Gluons}
\label{sec:basis_decs_g}

The decay of the Higgs boson to gluons, just like the production
process of the Higgs in gluon fusion, only arises through fermion
loops in the Standard Model and is sensitive to colored particles that interact
with the Higgs~\cite{Georgi:1977gs,Ellis:1979jy,Rizzo:1980gz},
\begin{align} 
\label{eq:decgg}
\Gamma(h\rightarrow gg)={ G_F \alpha_s^2 M_h^3
\over 64 \sqrt{2}\pi^3} \;
\left | \sum_q F_{1/2}(\tau_q)\right|^2\,  ,
\end{align}
where $\tau_q\equiv 4 m_q^2/M_h^2$ and the loop 
function is defined to be,
\begin{align}
F_{1/2}(\tau_q) \equiv -2\tau_q\left[1+(1-\tau_q)f(\tau_q)\right]
\, .
\label{eq:etadef}
\end{align}
It includes one power of the Yukawa coupling, expressed in terms of
the mass, and the re-scaled scalar one-loop three-point function
\begin{align}
f(\tau_q)=
\begin{cases}
\left[\sin^{-1} \sqrt{1/\tau_q} \right]^2,&\hbox{if~}
\tau_q\ge 1
\\
-\dfrac{1}{4}\left[\log \dfrac{1+\sqrt{1-\tau_q}}{1-\sqrt{1-\tau_q}}
-i\pi\right]^2,
&\hbox{ if~}\tau_q<1,
\end{cases}
\label{fundef}
\end{align}
In the limit in which the (bottom) quark mass is much less than the
Higgs boson mass the loop function becomes
\begin{align}
F_{1/2}(\tau_b) \rightarrow {2 m_b^2\over M_h^2} \; \log^2 \biggl({4m_b^2\over M_h^2}\biggr)
\; .
\label{eq:lighth}
\end{align}
On the other hand, for a heavy top quark, $\tau_q\rightarrow\infty$,
it approaches a constant,
\begin{align}
F_{1/2}(\tau_t) \rightarrow -{4\over 3} \; .
\label{eq:f12}
\end{align}
These two limits make it clear that the top quark loop is the dominant
contribution.  QCD corrections to the decay $h\rightarrow gg$ are
known to N$^3$LO~\cite{Baikov:2006ch}.
Another
theoretical aspect we can read off Eq.\eqref{eq:f12} is that the top
contribution to the effective Higgs-gluon coupling does not vanish for
large top masses. The reason for this non-decoupling
behavior~\cite{Appelquist:1974tg} is that the Yukawa coupling in the
numerator exactly cancels the linear kinetic decoupling of fermions from the loop
function $f(\tau_t)$. We can turn this feature around and use it to
probe the existence of heavy quarks which get their mass through a
Yukawa coupling to the Higgs field. This argument can be extended to
heavy quark masses until the corresponding Yukawa coupling breaks the
perturbative expansion~\cite{Kribs:2007nz,Denner:2011vt,Eberhardt:2012gv}.

\subsubsection{Photons}
\label{sec:basis_decs_a}

\begin{figure}[t]
\includegraphics[width=0.4\textwidth]{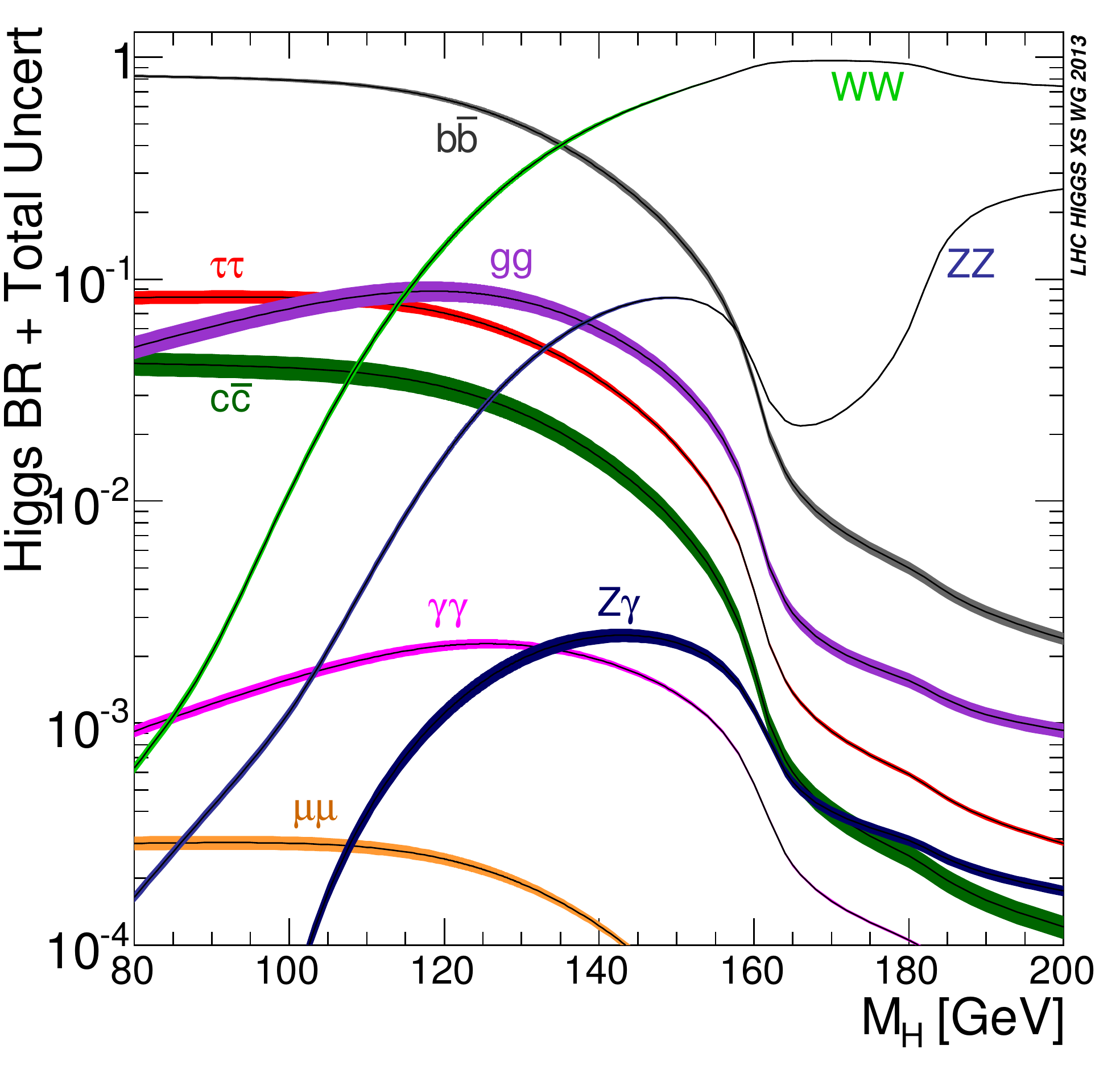}
\hspace*{0.1\textwidth}
\includegraphics[width=0.4\textwidth]{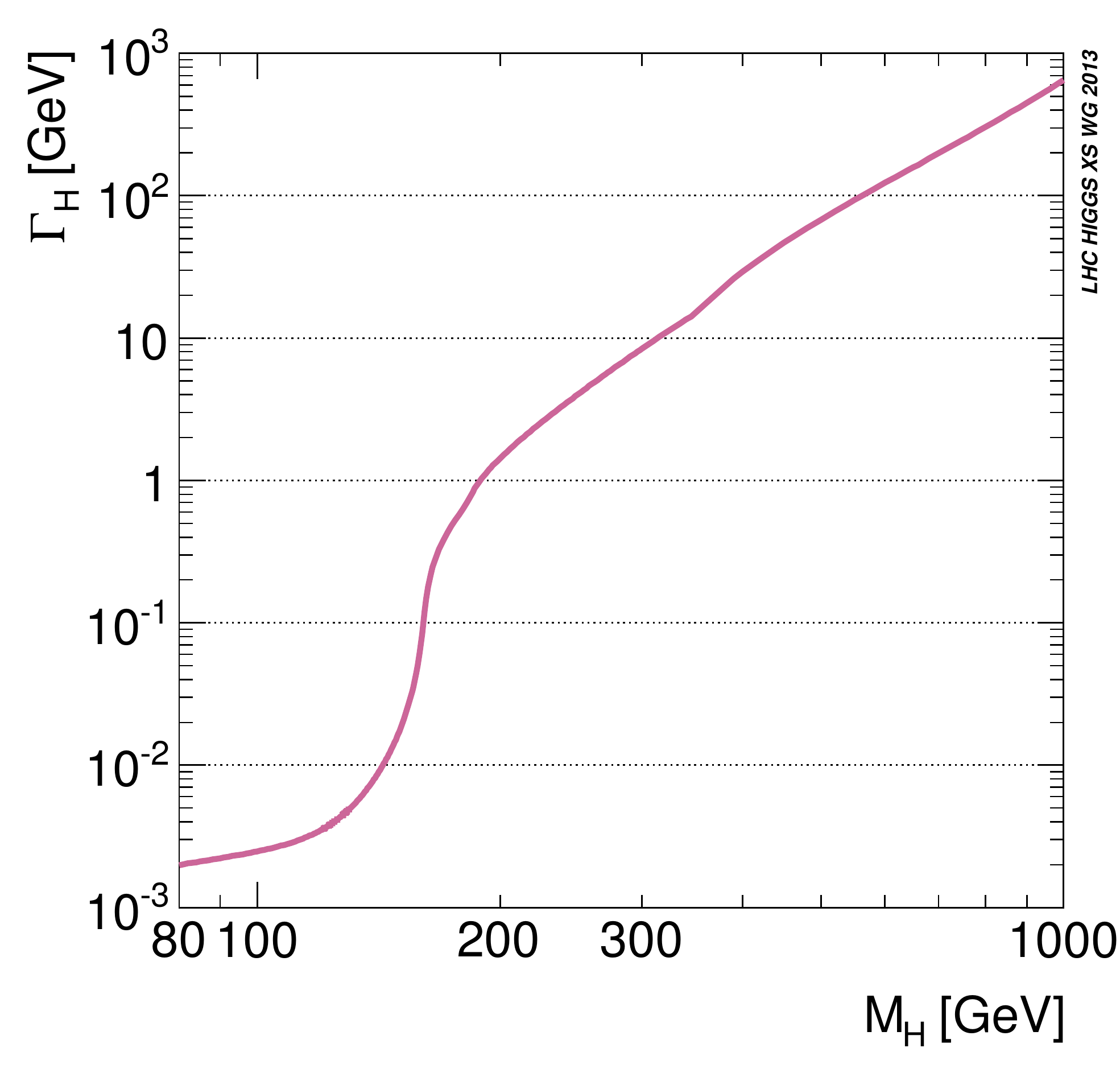}
\caption{ Left: SM Higgs branching ratios.  The widths of the curves are estimates
of the theoretical uncertainty.  Right: total width for a
  SM-like Higgs boson of arbitrary mass.  In this figure, $H$
  is the SM Higgs boson that is denoted by $h$ in this review. Figure
  from Ref.~\cite{Denner:2011mq}.}
\label{fig:hbrs}
\end{figure} 

Finally, the decay $h\rightarrow \gamma \gamma$ arises from fermion
loops and $W$-loops and was a discovery channel for the Higgs boson at
the LHC, despite its small branching ratio. The reason is that the LHC
experiments are built to very efficiently reconstruct a narrow
$\gamma\gamma$ mass peak over a large continuum background.  At lowest
order the width is~\cite{Gunion:1989we,Ellis:1975ap}
\begin{align}
\label{eq:hdecaa}
\Gamma(h\rightarrow \gamma\gamma)={\alpha^2 G_F\over 128\sqrt{2} \pi^3}
M_h^3\left| 
  \sum_f N_{c,f} Q_f^2 F_{1/2} (\tau_f) 
+ F_1(\tau_W)
+ \sum_S N_{c,S} \frac{g_{hSS}}{m_S^2} F_0(\tau_S)
\right|^2\, ,
\end{align}
where the sum is over charged fermions, vector bosons, and a possible BSM
scalar contribution. We use $\tau_{W,S}=4 M_{W,S}^2/M_h^2$, $N_{c,f}=1
(3)$ for leptons (quarks), and $Q_i$ for the electric charge in units
of $e$.  The scalar loop function $F_{1/2}(\tau_q)$ is given in
Eq.\eqref{eq:etadef}, and
\begin{align}
F_1(\tau_W) &= 2+3\tau_W \left[ 1+(2-\tau_W)f(\tau_W) \right] \notag\;, \\
F_0(\tau_S) &= \tau_S \left[ 1 - \tau_S f(\tau_S) \right] \; .
\label{fdef}
\end{align}
These functions $F_j$
differ from the expressions in Ref.~\cite{Plehn:2015dqa} by a global
factor $-2$. In the (unphysical) limit $\tau_W\rightarrow\infty$ we
find $F_1\rightarrow 7$, so the top quark and $W$ contributions have
opposite signs. In the limit $\tau_S \to \infty$ we find
$F_0 \to -1/3$. The decay $h\rightarrow \gamma\gamma$ is sensitive to
the sign of the top quark Yukawa coupling through the interference of
the $W$ and $t$ loops.
Similarly, the rate for $h\rightarrow Z\gamma$ receives contributions
from both fermions and the $W$ boson~\cite{Cahn:1978nz,Bergstrom:1985hp}.  The analytic formula is given
in Ref~\cite{Gunion:1989we} and the $Z\gamma$ width is quite small.

\begin{table}[b!]
\begin{tabular}{cccc|cccc|cc}
\toprule
$XX$ & $\br(h\rightarrow XX)$ & $\delta_\text{QCD}$ & $\delta_\text{ew}$ &
$XX$ & $\br(h\rightarrow XX)$ & $\delta_\text{QCD}$ & $\delta_\text{ew}$ &
$XX$ & $\br(h\rightarrow XX)$ \\
\midrule
$b{\overline b}$ & $.5824$& $0.2\%$ & $0.5\%$ &
$gg$& $8.187\times 10^{-2}$& $3\%$   & $1\%$ &
$(l^+l^-)(l^+l^-)$&$2.745\times 10^{-4}$ 
\\
$\tau^+\tau^-$ & $6.272\times10^{-2}$& & $0.5\%$ &
$\gamma \gamma $&$2.27\times 10^{-3}$& $<1\%$ & $<1\%$ &
$(l^+l^-)(\nu {\overline {\nu}})$&$2.338\times10^{-8}$ 
\\
$\mu^+\mu^-$ & $ 2.176 \times 10^{-4}$& & $0.5\%$ 
&
$WW$ & $2.15\times 10^{-1}$& $<0.5\%$ & $0.5\%$ &
$(\nu {\overline {\nu}})(\nu {\overline {\nu}})$&$1.044\times 10^{-3}$  
\\
$c {\overline c}$ & $2.891\times 10^{-2}$ & $0.2\%$ & $0.5\%$ &
$ZZ$& $2.619\times 10^{-2}$& $<0.5\%$ & $<0.5\%$ &
$(l^+l^-)(q {\overline q}$)&$3.668\times 10^{-3}$
\\
&&&&
$Z\gamma$ & $1.533\times 10^{-3}$& $5\%$ & $<1\%$ &
$(q {\overline q})(q {\overline q})$&$1.089\times 10^{-1}$
\\
\bottomrule
\end{tabular}
\caption{Standard Model branching ratios for $M_h=125$~GeV.  In the
  right column we show the final states from the $WW$ and $ZZ$
  decays. For leptons we sum over $l=e, \mu, \tau$, while for quarks
  we sum over $q=u,d,c,s,b$. The $\delta$ values are relative
  theoretical uncertainties due to missing higher orders. Details on
  the uncertainties for each branching ratio are given in
  Ref.~\cite{deFlorian:2016spz}.}
\label{tab:smbr}
\end{table} 

The Higgs branching ratios are shown in Fig.~\ref{fig:hbrs} for a SM
Higgs boson of arbitrary mass. The width of the curves is an estimate
of the theoretical uncertainties on the branching ratios.  The
branching ratios assume SM couplings and no new decay channels and
include all known radiative corrections~\cite{Dittmaier:2011ti,deFlorian:2016spz}.  Also
shown in Fig.~\ref{fig:hbrs} is the Higgs total decay width as a
function of Higgs mass.  For $M_h=125~\GeV$, the total width is very
small, $\Gamma_h=4~\mev \ll M_h$. The reason is that the
leading coupling entering the decays is the small bottom Yukawa
coupling $Y_b = \sqrt{2} m_b/v$.  Detailed predictions for the
Standard Model branching ratios are given in Table~\ref{tab:smbr}.
The sizes of the uncertainties on the Higgs branching ratios from
various sources has been estimated in Ref.~\cite{deFlorian:2016spz}.

\subsubsection{Exotic  and rare decays}
\label{sec:exoticrare}

Little is known about the interplay of the Higgs with the light
fermionic degrees of freedom in the Standard Model that directly reflects the mass
hierarchy of the three generations of fermions. The generation of the
observed fermion mass hierarchy is a long-standing open question of
particle physics and, contrary to the modeling of a light Higgs boson
and electroweak symmetry breaking, only a small number of proposals
exist to address the issue. Arguably the most famous among them is the
Froggatt-Nielsen mechanism~\cite{Froggatt:1978nt}, although the original Froggatt-Nielsen
model with the Higgs as a flavon is not phenomenologically viable.

It is possible that our naive SM picture of reflecting the mass
hierarchy of the light fermions in their Yukawa couplings is not quite
correct. The Higgs interactions could be a source of non-minimal
flavor violation. If we allow such interactions, then approaches
reminiscent of the Froggatt-Nielsen approach can explain the fermion
mass hierarchy by making the Yukawa couplings
Higgs-dependent~\cite{Giudice:2008uua}, which also breaks the SM
relation $m\propto v$ which holds irrespective of the particle
species. This scaling has been established only approximately so far,
and measurements at the LHC are limited to the heavy particles of the
Standard Model.

Unearthing the Higgs interactions with the light first and second
generation of quarks is notoriously difficult. The hierarchy of
Yukawas expected in the Standard Model indirectly generalizes to small branching
probabilities of the Higgs into light fermions which systematically
limits direct searches, see Eq.\eqref{eq:decbb}. The largest effect in
the quark sector that qualifies as rare can be expected from $h\to
c\bar c$, whose branching ratio in the Standard Model is
2.9\%~\cite{Denner:2011mq}. Studying the interactions of the
Higgs with the lightest quark flavors $u,d$ is phenomenologically not
possible as there is no ways to isolate these partons from the
overwhelming QCD background. An observation of Higgs boson decays to
charm quarks is also limited by large hadronic backgrounds at the LHC,
as well as the strong trigger criteria required to facilitate the data
taking in busy final states. However there are various possibilities
linked to heavier quarks forming distinct meson final states.

One possibility is relying on distinctive signatures that are
associated with the $c\bar c$ decay itself. For instance if an
additional photon or $Z$ boson is considered, $h\to c\bar c \gamma/Z$,
the charm quarks can recombine to form a $J/\psi$ vector meson which
decays to muons about 6\% of the time. Phenomenologically, vector
boson-associated meson production leads to a highly distinctive and
clean signature of two leptons recoiling against a photon, thus
providing good a priori prospects to isolate this signature from the
relatively small backgrounds that one expects here.

\begin{figure}[!b]
\includegraphics[width=3.5cm]{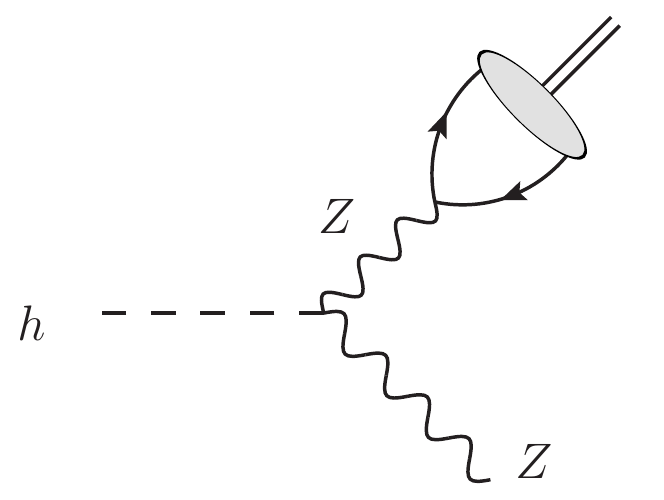}
\includegraphics[width=3.5cm]{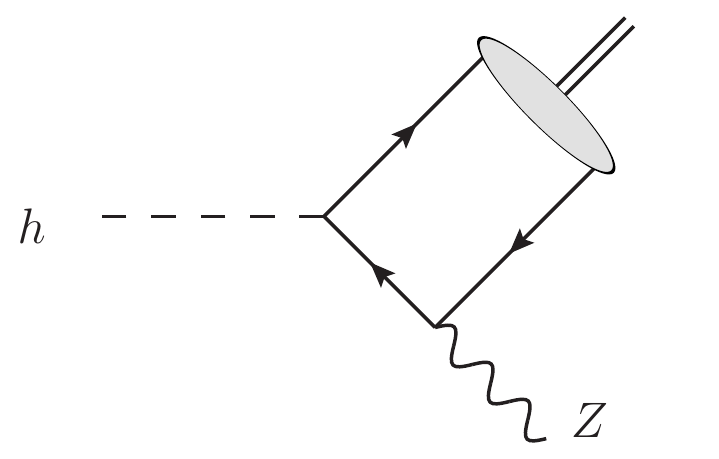}
\includegraphics[width=3.5cm]{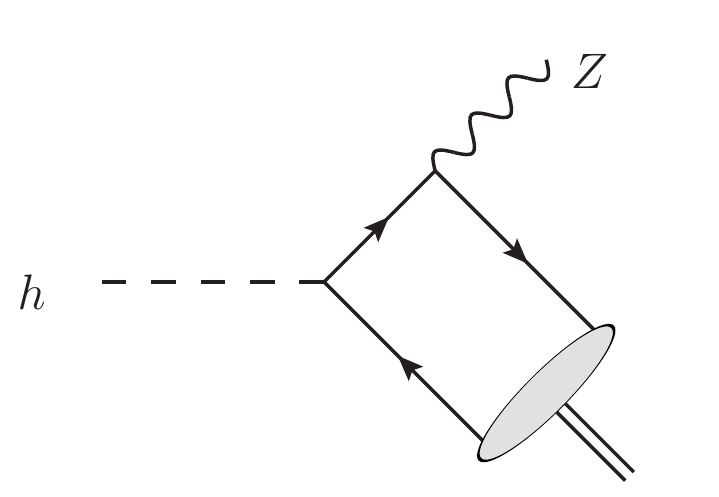}
\includegraphics[width=3.5cm]{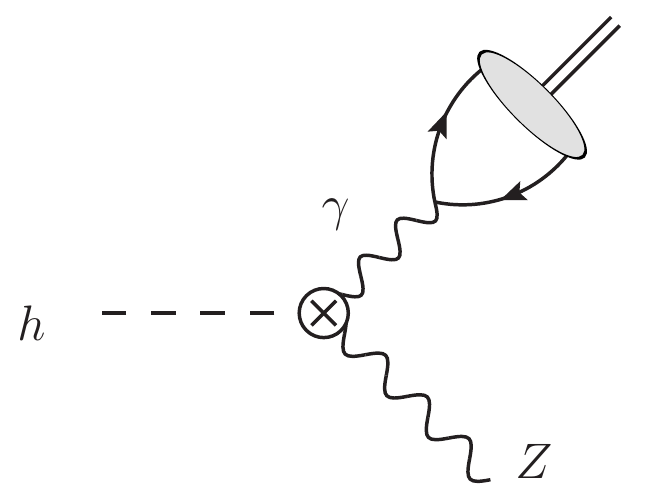}
\caption{Feynman diagrams contributing to the decay of the Higgs boson
  into a meson and a $Z$ boson. The crossed vertex refers to potential
  new Higgs interactions as well as one-loop SM contributions, while
  the first two diagrams probe the Higgs interactions with the quarks
  inside the meson. Taken from
  Ref.~\cite{Alte:2016yuw}.}
\label{fig:exdecay}
\end{figure}

Interfering processes leading to an identical final state are $h\to
(\gamma/Z)^\ast \gamma,(\gamma/Z)^\ast \to c\bar c$, which also are
enhanced close to the meson
thresholds~\cite{Bodwin:2013gca,Bodwin:2014bpa}, as well as prompt
$h\to\mu^+\mu^-\gamma/Z$, see Fig.~\ref{fig:exdecay}. In the Standard Model these
contributions interfere
destructively~\cite{Koenig:2015pha,Alte:2016yuw} which can be used to
place limits on new physics.

\begin{figure}[!t]
\includegraphics[height=5cm]{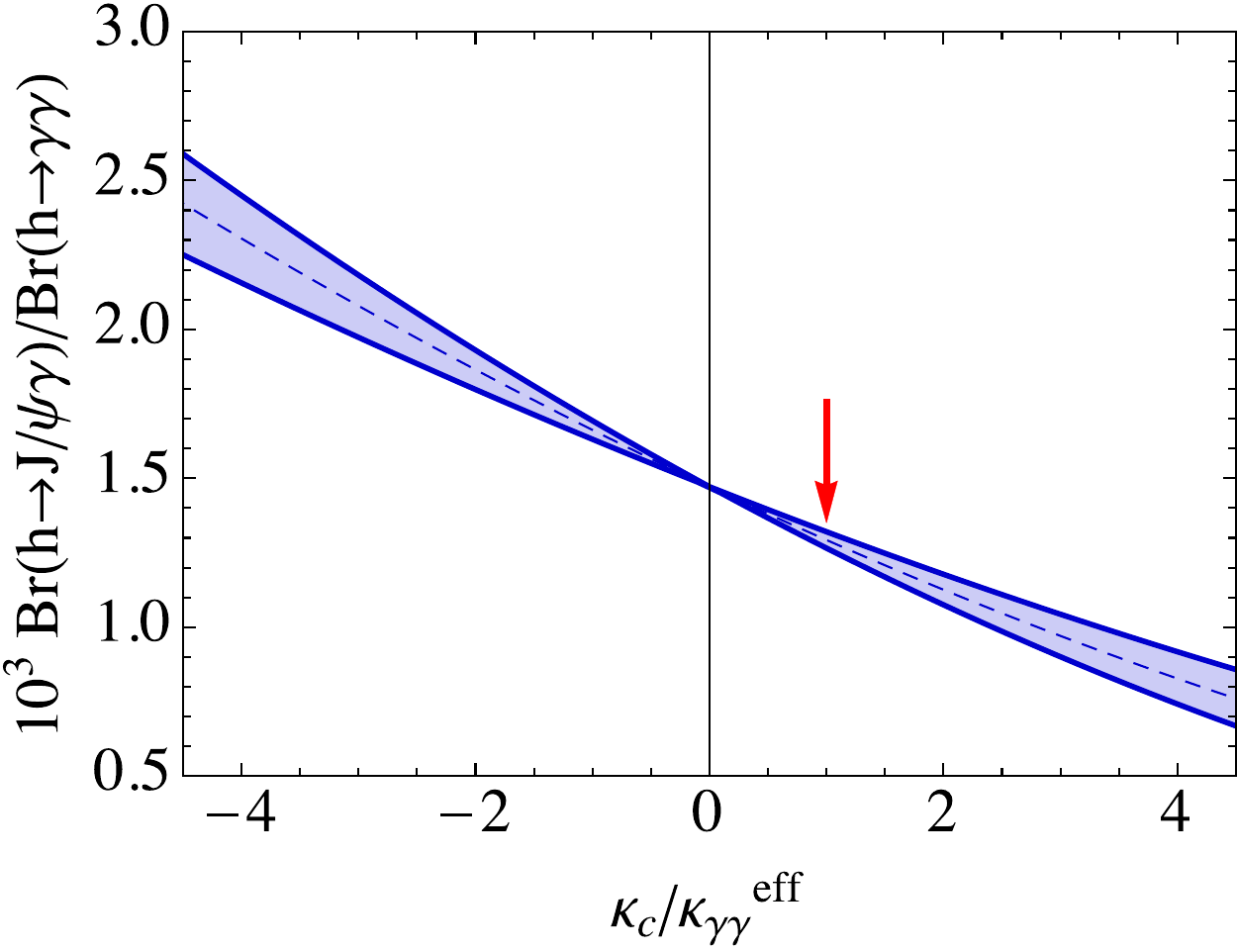}
\hspace*{2cm}
\includegraphics[height=5cm]{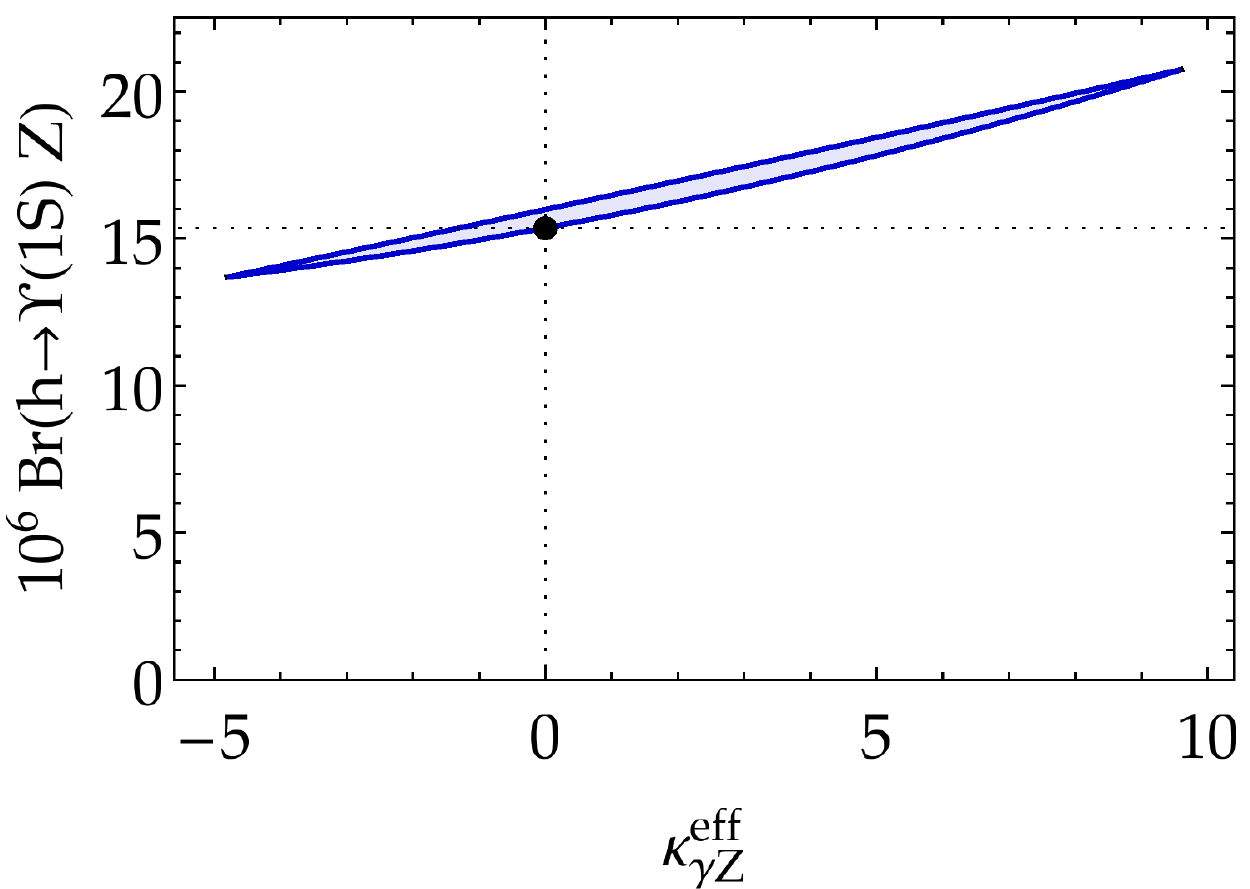}
\caption{Left: Decay of the Higgs boson to a $J/\psi$ meson in
  association with a photon as a function of a charm coupling
  rescaling $\kappa_c$ in units of new contributions from modified
  $h\gamma Z$ interactions. The branching ratio is normalized to the
  $h\to \gamma \gamma$ branching ratio, and therefore not sensitive to
  the total Higgs decay width. The red arrow denotes the Standard Model and the
  width of the band indicates the theoretical uncertainty. Figure
  taken from Ref.~\cite{Koenig:2015pha}.  Right: Decay of the Higgs
  boson to a 1S Upsilon meson in association with a $Z$
  boson. $\kappa_{\gamma Z}^{\text{eff}}$ summarizes new contributions
  to the $h\gamma Z$ interactions, and can be tightly constrained
  using this decay. The black dot refers to the Standard Model, the blue region
  includes a measurement in the $h\to \gamma Z$ channel using CMS
  constraints based on Ref.~\cite{Chatrchyan:2013vaa}. Figure from
  Ref.~\cite{Alte:2016yuw}.}
\label{eq:updecay} 
\end{figure}

To discuss a specific and illustrative example, we consider the decay
rates of a Higgs boson into a pseudo-scalar meson $P$ and a $Z$
boson~\cite{Alte:2016yuw}
\begin{align}
\Gamma(h\to PZ) = {M_h^3\over 4\pi v^4} \left[\left(1-{M_Z^2\over M_h^2}-{m_P^2\over M_h^2}\right)^2-{4M_Zm_P\over M_h^2}\right]^{3/2} |F^{PZ}|^2\;,
\end{align}
where the form factor $F^{PZ}$ is dominated by contributions that are
related to local hadronic matrix elements of the form
\begin{align}
F^{PZ} = \sum_q f_P^q {\sigma^3\over 2},\quad \langle P(k) | \bar q \gamma^\mu\gamma_5 q | 0 \rangle = -i f_P^q k^\mu \,,
\end{align}
\ie the probability to excite a pseudo-scalar meson from the
strongly-interacting QCD vacuum (see also
Sec.~\ref{sec:basis_strong}). $\sigma^3$ stands for the axial-vector
couplings of the $Z$ boson to the corresponding quark. Contributions
that are directly related to the charm quark coupling are suppressed
$m_c^2/M^2_h$. While limits can be placed on the charm quark coupling,
see Fig.~\ref{eq:updecay}, and its parity structure (through angular
correlations of the vector meson decay
leptons~\cite{Bhattacharya:2014rra}), modifications of the $h\gamma Z$
interactions are typically more relevant.  For the pseudo-scalar
mesons,~Ref.~\cite{Alte:2016yuw} finds
\begin{align}
F^{\pi^0Z} \simeq  46.1\,, \quad F^{\eta Z} \simeq  27.7\,, \quad F^{\eta' Z} \simeq  27.7\,.
\end{align}
The phenomenologically most relevant decays at the LHC are the Higgs
decays involving heavy $b$ and $c$-flavored vector mesons and a
photon due to their final states and rates. The expected branching
ratios in the Standard Model for the decays $h\to V\gamma$
are~\cite{Koenig:2015pha}
\begin{align}
\text{BR}(h\to J/\psi \,\gamma) & = (2.95\pm 0.07_{f_J/\psi} \pm 0.06_\text{direct} \pm 0.14_{h\to\gamma\gamma} )\times 10^{-6}\,,\\
\text{BR}(h\to \Upsilon(1S)\, \gamma) & = (4.61\pm 0.06_{f_{\Upsilon(1S)}} \phantom{}^{+1.75}_{-1.21}\phantom{}_{\text{direct}} \pm 0.22_{h\to\gamma\gamma} )\times 10^{-9} \,,\\
\text{BR}(h\to \Upsilon(2S)\, \gamma) & = (2.34\pm 0.04_{f_{\Upsilon(2S)}} \phantom{}^{+0.75}_{-0.99}\phantom{}_{\text{direct}} \pm 0.11_{h\to\gamma\gamma} ) \times 10^{-9}\,,\\
\text{BR}(h\to \Upsilon(3S)\, \gamma) & = (2.13\pm 0.04_{f_{\Upsilon(3S)}} \phantom{}^{+0.75}_{-1.12}\phantom{}_{\text{direct}} \pm 0.10_{h\to\gamma\gamma} ) \times 10^{-9}\,,
\end{align}
where uncertainties related to decay constants $f_i$ and the theoretical
uncertainty related to direct coupling of the Higgs to the meson and
the dominant uncertainty from $h\to\gamma \gamma$ are
included. Although the associated branching ratios are typically of
the order of $\sim 10^{-6}$, a measurement of decays of the Higgs to
mesons can provide a strong hint for the presence of new physics. For
instance the decay of $h\to \Upsilon(1S) Z$ is highly sensitive to the
presence of new $h\gamma Z$ interactions as can be seen from
Fig.~\ref{eq:updecay}.

Most of the difficulties that arise when considering meson-associated
decays of the Higgs boson are avoided when turning to rare leptonic
Higgs decays as these are only impacted by QCD on the production
side. A phenomenologically clear avenue to a rare leptonic Higgs decay
is therefore $h\to \mu^+\mu^-$. In the Standard Model, this decay has a branching
ratio of~\cite{Heinemeyer:2013tqa}
\begin{align}
\text{BR}(h\to \mu^+\mu^-) = (0.022 \pm 0.001)\%
\end{align}
Although the Higgs decay to muons is rare, muons are well-understood
objects at colliders and the major limiting factor of analyses is the
low signal yield of a resonance signal at $m_{\mu^+\mu^-}\simeq
125~\text{GeV}$.

A phenomenologically striking signature would be a lepton flavor
violating Higgs decay~\cite{Dorsner:2015mja}.  Such a situation can
appear when the fermion mass generation is not aligned with the Yukawa
couplings, for example in general two-Higgs doublet models. Seesaw
mechanisms typically imply lepton flavor-violating Higgs couplings at
the loop level since the mixing of neutrinos as part of a $SU(2)_L$
doublet will induce a lepton flavor misalignment of the Higgs
couplings. For charged leptons, however, the associated branching
ratios are very small, $\lesssim 10^{-15}$~\cite{Arganda:2004bz}.

\subsection{Implications for high scales}
\label{sec:basis_implications}

The biggest impact of the Higgs discovery on theoretical physics is
that it structurally completes the Standard Model as a perturbative,
renormalizable gauge theory. This means that we can use
renormalization group evolution to connect very different energy
regimes, from the QCD scale to Grand Unification or even the Planck
scale.  On the other hand, fundamental scalars in quantum field theory
face two challenges, which we can explain in physics terms when we use
an effective field theory picture: first, as discussed in
Sec.~\ref{sec:basis_decs_g} since  the Higgs Yukawa couplings to
fermions are proportional to the fermion mass, they do not fulfill the conditions necessary for the decoupling theorem to be valid, one of the fundamental
assumptions we like to make in an effective field theory. As a side
remark, SM-like Yukawa couplings larger than the top Yukawa
immediately lead to nearby Landau poles where the Higgs self coupling blows up.
  Second, the hierarchy
problem can be viewed as a quadratic dependence of the one-loop Higgs
mass corrections on the exact value of the matching scale.  This also
contradicts our usual assumption that heavy physics does not affect
predictions far below the matching scale. Both of those problems can,
in practice, be ignored, because they do not invalidate our picture of
the Standard Model as structurally complete. On the other hand, they
might be able to lead us in our search for more fundamental theories
at higher scales, and one of the main motivations of particle
physicists is still to understand the most fundamental structures
behind physics.

Before discussing renormalization group effects, we discuss a simple
self-consistency criterion on the Higgs
sector~\cite{LlewellynSmith:1973yud,Cornwall:1974km,Lee:1977yc,Lee:1977eg,Chanowitz:1978uj,Chanowitz:1978mv}.
It is based on the fact that the Higgs field and the Higgs potential
in Eq.\eqref{eq:higgs_pot_sm} also includes three Goldstone modes.  In
the high-energy limit, the Goldstone modes can be consistently
identified with the longitudinal degrees of freedom of the weak
bosons.  For the charged Goldstone modes, defined as $w^\pm = (\phi_1
\pm i \phi_2)/\sqrt{2}$ (cf Eq.\eqref{eq:def_phi}), we can compute the
scattering amplitude for the process
\begin{align}
w^\pm w^\pm \to w^\pm w^\pm \; ,
\end{align}
approximating the scattering of physical, longitudinal gauge bosons at high energies.
In the absence of a physical Higgs boson this scalar scattering
process is perturbatively not unitary. A Higgs exchange in the $s$-channel unitarizes
the amplitude, but only if the Higgs field is light enough to
contribute below the cut-off scale defined by perturbative unitarity
violation.  This translates into an upper limit on the Higgs mass,
assuming that all Higgs and Goldstone interactions have SM strengths,
\begin{align}
 M_h^2 \lesssim 4\pi v^2 = (870~\gev)^2 \; .
\label{eq:unitarity_higgs}
\end{align}
While the exact numerical value of this limit depends on the detailed
interpretation, this relation has historically lead us to believe that
the LHC had to either discover the Higgs boson or find some
spectacular new physics effect. Given that the observed Higgs has
approximately SM properties, longitudinal gauge boson scattering is
approximately unitary, leaving little room for effects from strongly
interacting gauge bosons at large energies. Similar unitarity arguments
can be made based on Higgs-vector boson and fermion-vector boson
scattering, but with a subleading effect on the Higgs mass limit.  Unitarity arguments
place strong restrictions on the parameters of BSM theories.  

\begin{figure}[t]
\includegraphics[width=0.4\textwidth]{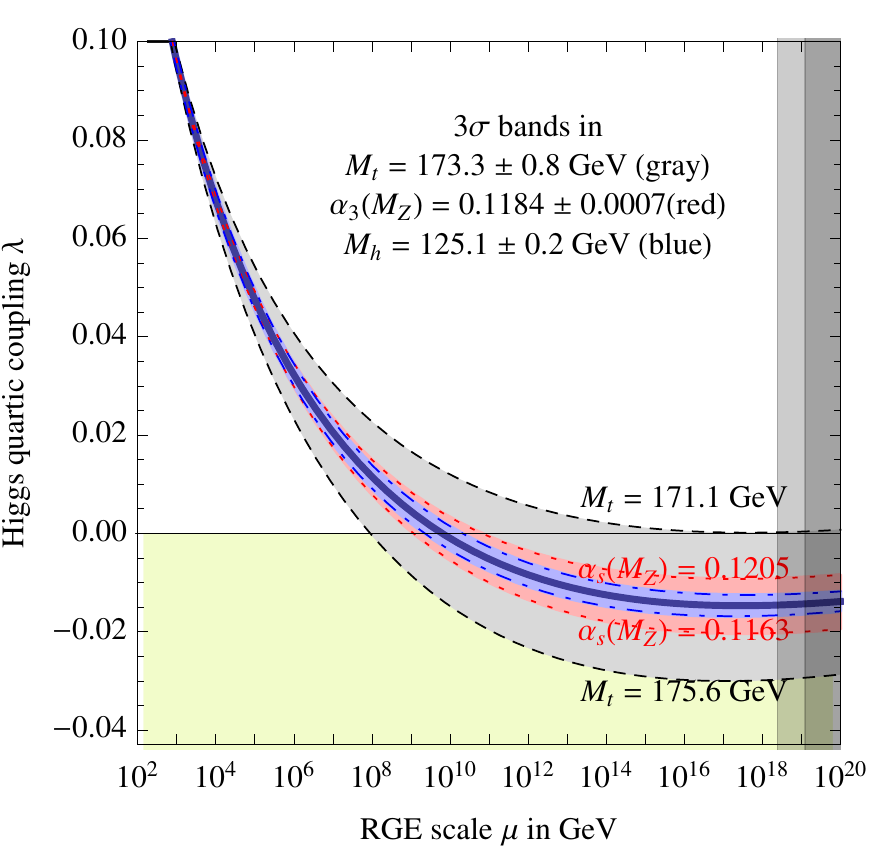}
\hspace{0.1\textwidth}
\includegraphics[width=0.42\textwidth]{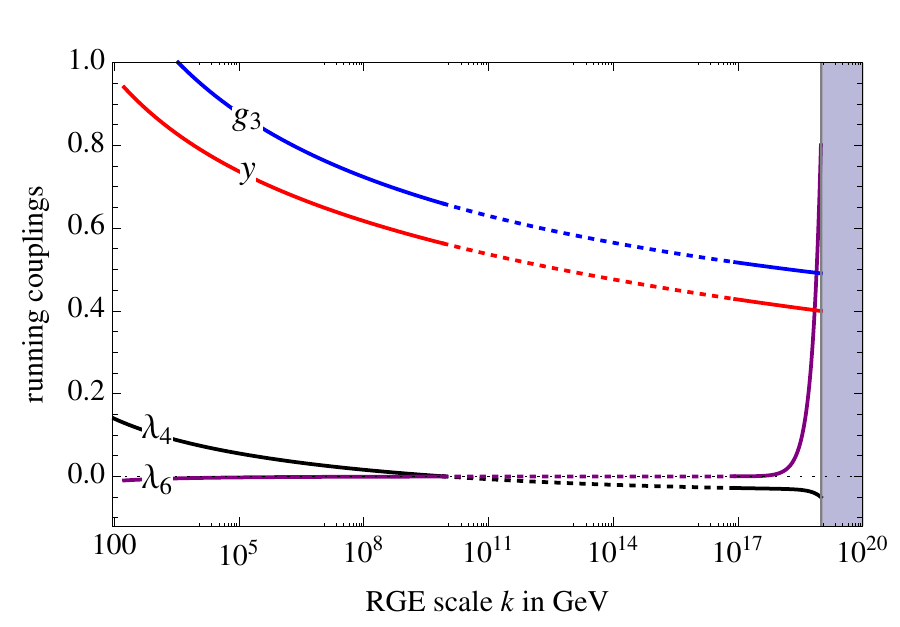}
\caption{Left: running Higgs quartic coupling including uncertainties
  on the input parameters. Figure from from
  Ref.~\cite{Buttazzo:2013uya}. Right: effect from a dimensional-6
  operator, including a metastable intermediate phase.  $k=\mu$ is the
  renormalization scale.  Figure from Ref.~\cite{Eichhorn:2015kea}.}
\label{fig:potlam}
\end{figure}

Using the renormalization group we can evolve the Higgs potential of
the Standard Model to high energy scales. To good approximation, this
evolution is driven by the self coupling $\lambda$, which to leading
order runs as
\begin{alignat}{5}
 \frac{d\,\lambda}{d\,\log{\mu^2}}=\frac{1}{16\pi^2}
 \left[12\lambda^2+6\lambda y_t^2 - 3y_t^4
 -\frac{3}{2}\lambda \left( 3g^2+g'^2 \right)
 +\frac{3}{16} \left( 2g^4+(g^2+g'^2)^2 \right) \right]
\; ,
\label{eq:lambda_rge}
\end{alignat}
in terms of the top Yukawa $y_t$ and the gauge couplings $g$ and $g'$.
There are two distinct features in this running. First, a large Higgs
self-coupling will rapidly become stronger towards larger energy
scales and hit a Landau pole, driven by the first term on the right
side of Eq.\eqref{eq:lambda_rge}. This behavior is also referred to as
the triviality bound~\cite{Maiani:1977cg,Dashen:1983ts} and can be
turned into an upper limit on the Higgs
mass~\cite{Lindner:1985uk}. This limit is well above the observed
Higgs mass of 125~GeV even when we require Landau poles to be absent
all the way to the Planck scale.  For smaller Higgs self-couplings,
the negative top contribution drives the self-coupling negative at
large
energies~\cite{Krive:1976sg,Krasnikov:1978pu,Lindner:1988ww,Sher:1988mj,Holland:2004sd}.
In the absence of higher-dimensional operators of the kind shown in
Eq.\eqref{eq:higgs_pot_d6} the running self-coupling can ruin vacuum
stability around $\Lambda\sim 10^{10}$~GeV, as shown in the left panel
of Fig.~\ref{fig:potlam}. If we include higher-dimensional operators,
parameterized as
\begin{align}
V\sim {\lambda_4(\mu^2)\over 4} h^4+{\lambda_6(\mu^2)\over 8 \mu^2}h^6\; , 
\end{align}
where $\lambda_6$ is identical to $f_{\phi,3}$ of
Eq.\eqref{eq:higgs_pot_d6} and $\mu\sim\Lambda$, we find that the
scale of new physics curing the stability problem (the point where
$\lambda_4$ goes negative) can be a few orders of magnitude above this
instability scale shown on the left of Fig.~\ref{fig:potlam}. However,
in the right panel of Fig.~\ref{fig:potlam} we see that even allowing
for a metastability phase the new states should appear well below the
GUT or Planck scale.

Renormalization group analyses become especially interesting when they
feature infrared or ultraviolet
fixed-points~\cite{Schrempp:1996fb}. This does not only include
individual parameters, but also pseudo-fixed-point in ratios of
parameters~\cite{Pendleton:1980as,Hill:1980sq}, which effectively
removes one parameter from the quantitative renormalization group
analysis. In our case, the running of the top Yukawa coupling offers
little information, but an interesting feature occurs in the running
of the ratio of the Higgs self-coupling to the top Yukawa
coupling~\cite{Wetterich:1981ir}. The ratio $R = \lambda/y_t^2$ runs
like
\begin{align}
\frac{dR}{d\log \mu^2} =
\frac{3\lambda}{32\pi^2R} \;
\left( 8R^2 + R - 2
\right) \; ,
\end{align}
leading to an approximate infrared fixed-point in this ratio. The
numerical value turns out to be very close to the observed values of
the Higgs and top masses~\cite{Shaposhnikov:2009pv}. The observed values of
the top and Higgs masses will generically be predicted by a wide range
of parameter choices at some high-scale, implying a limited
sensitivity of the relation between those two parameters to the
details of the model in the ultraviolet. This still leaves open the
question  of why the electroweak scale is where it is.

Finally, baryogenesis is one of the great mysteries of particle
physics. It can be related to the three Sakharov
conditions~\cite{Sakharov:1967dj}, namely baryon number violation, $C$
and $CP$ violation, and deviation from thermal equilibrium. None of
the three conditions are convincingly fulfilled by the Standard
Model. On the other hand, the second and third conditions can be
linked to modified Higgs potentials. Concerning the third condition,
it has been known for a long time that a renormalizable SM Higgs
potential with a Higgs mass below roughly 60~GeV could lead to a
strong first-order electroweak phase transition, leading to a
deviation from thermal equilibrium around that time. Given the
observed Higgs mass, there are two ways to turn the phase transition
from second to first order~\cite{Sher:1988mj}: either we introduce
additional scalar particles with masses around the Higgs or top mass
or we modify the Higgs
potential~\cite{Grojean:2004xa,Reichert:2017puo}.

\begin{figure}[t]
\centering
\includegraphics[width=0.35\textwidth]{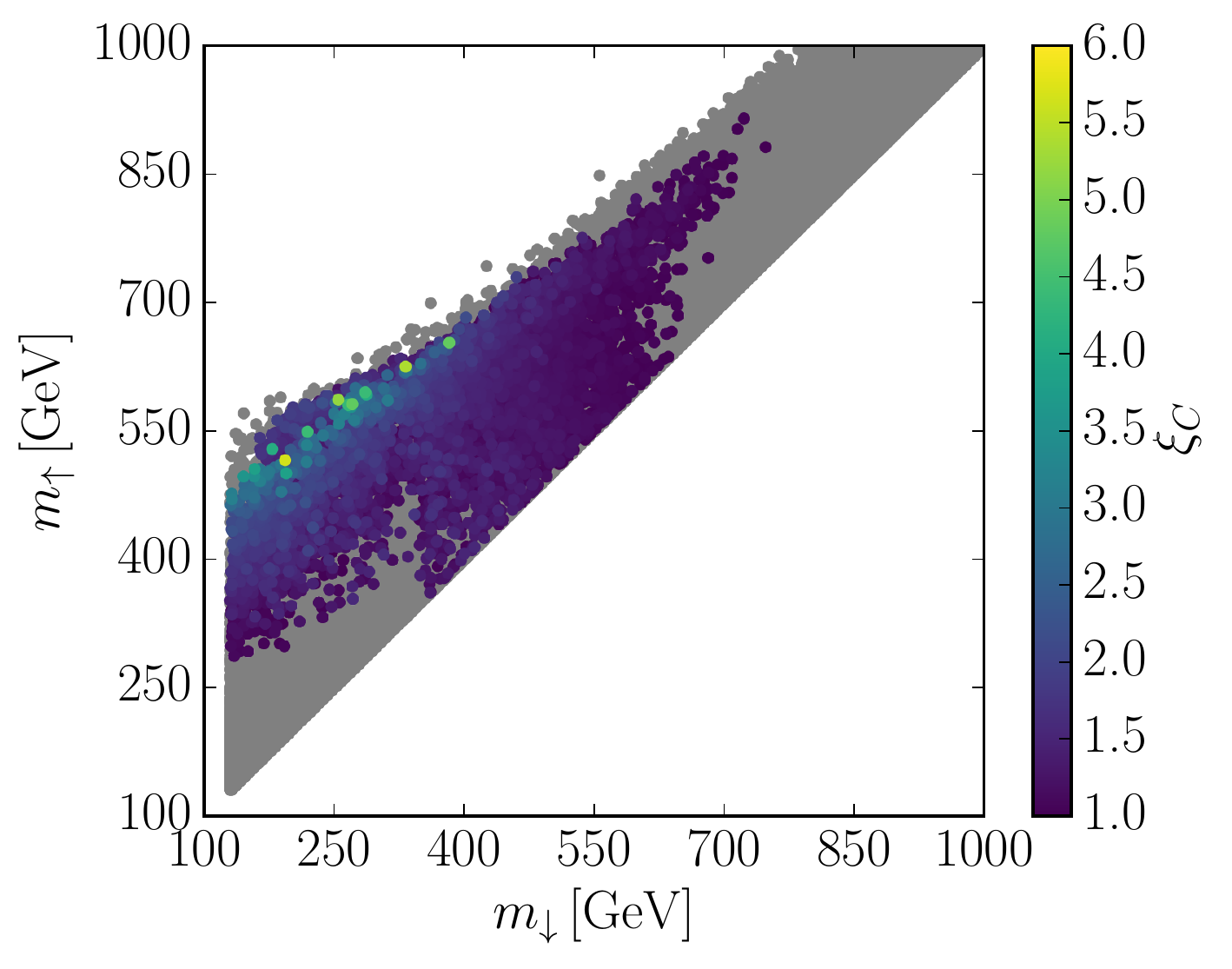}
\hspace*{0.1\textwidth}
\includegraphics[width=0.45\textwidth]{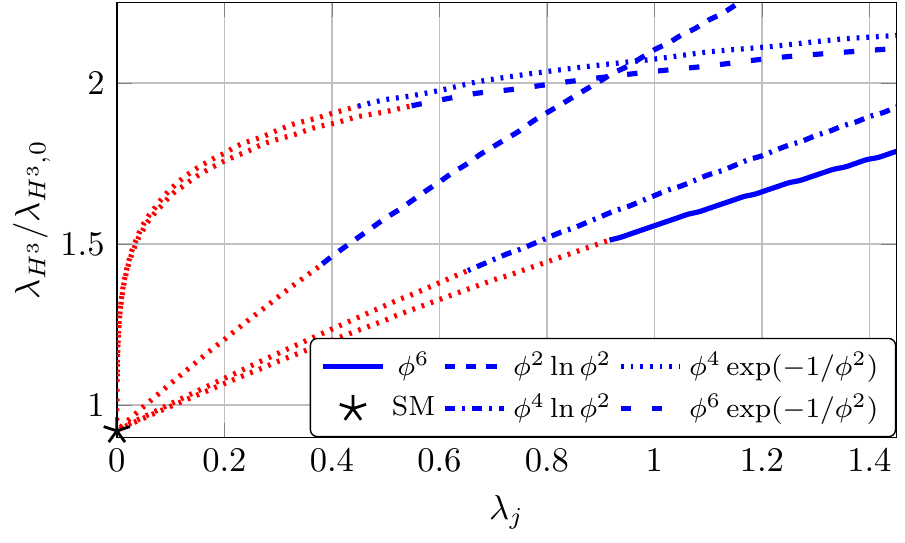}
\caption{Left: $\xi_c = v_c/T_c$ as a function of the masses of the
  lighter vs heavier additional Higgs particle in the $CP$-violating
  2HDM. The grey points pass all constraints. The right amount of
  $CP$-violation is not required. Figure from
  Ref.~\cite{Basler:2017uxn}.  Right: dependence of the physical
  triple Higgs coupling on the parameters of extended Higgs
  potentials. The change in color indicates a strong first-order phase
  transition correlated with a sufficiently large
  self-coupling. Figure adapted from Ref.~\cite{Reichert:2017puo}.}
\label{fig:phase_transition}
\end{figure}

Following the first approach, the singlet extension discussed in
Sec.~\ref{sec:basic_weak_singlet} accommodates a strong first-order
electroweak phase transition for specific choices of input parameters,
typically $m_H < 250$~GeV~\cite{Profumo:2014opa,Kurup:2017dzf,Chen:2017qcz,Chiang:2017nmu}.  This
possibility requires $\kappa\ne 0$ in Eq.\eqref{eq:singlet-potential}
and places significant restrictions on the parameters of the
potential. This first order electroweak phase transition does not
occur in the $Z_2$-symmetric singlet model.  A simple model which
allows for $CP$-violation as well as a first-order phase transition is
the complex 2HDM~\cite{Basler:2017uxn}. Unlike for the 2HDM discussed
in Sec~\ref{sec:basis_weak_doublet}, the VEVs now have to be
complex. In this case, the three neutral Higgs particles $H_{1,2,3}$
are not $CP$-eigenstates. In the type-I setup of the fermion sector all
fermions couple to the same Higgs doublet. We can for example assume
that the lightest of the three Higgs scalars is the observed SM-like
Higgs boson. In that case the two heavier masses are free parameters
of the model. In the left panel of Fig.~\ref{fig:phase_transition} we
show the critical parameter as a function of the two heavy BSM Higgs
masses, where $\xi_c = v_c/T_c > 1$ implies a strong first-order phase
transition. Lighter masses are in general preferred. The complex 2HDM
has the advantage that in addition it can provide the necessary
$CP$-violation for electroweak baryogenesis. For the points shown in
Fig.~\ref{fig:phase_transition} this second condition is easiest, but
not exclusively, fulfilled along the diagonal.

Using the second approach, the classic result is that a sufficiently
large Wilson coefficient of the effective operator $(\phi^\dagger
\phi)^3$ also leads to a strong first-order phase
transition~\cite{Grojean:2004xa}. This can be viewed as the main
motivation for testing the Higgs self-coupling for order-one
deviations.  This relation between the triple Higgs coupling,
accessible at the LHC, and the conditions for baryogenesis exists more
generally. We can test it by expanding the SM Higgs potential in
different ways,
\begin{align} 
 \Delta V_6  &= \lambda_6 \, \frac{\phi^6}{\Lambda^2}\,, \notag\\
 \Delta V_{\ln,2}  &= -\lambda_{\ln,2} \, \frac{\phi^2\Lambda^2}{100} \; \ln \frac{\phi^2}{2\Lambda^2}\,, 
 &
 \Delta V_{\ln,4}  &= \lambda_{\ln,4} \, \frac{\phi^4}{10} \; \ln \frac{\phi^2}{2\Lambda^2}  \,, \notag\\
 \Delta V_{\exp,4}  &= \lambda_{\exp,4} \phi^4  \exp\left(-\frac{2\Lambda^2}{\phi^2}+23\right)\,, 
 &
 \Delta V_{\exp,6}  &= \lambda_{\exp,6} \frac{\phi^6}{\Lambda^2}\exp\left(-\frac{2\Lambda^2}{\phi^2} +26 \right)\,.
\label{eq:pots}
\end{align}
The scale $\Lambda$ describes some kind of new physics.  Neither the
logarithmic nor the exponential potentials can be expanded around
$\phi=0$.  From a more general viewpoint, the set of power law,
logarithmic and exponential potential functions does not only reflect
the physics structures arising from local vertex expansions, one-loop
determinants or semi-classical approximations. It also includes the set
of functions to be expected if the effective potential permits a
potentially resurgent trans-series expansion~\cite{Dunne:2012ae}.  In
the right panel of Fig.~\ref{fig:phase_transition} we show the nature
of the electroweak phase transition as a function of the underlying
model parameters, directly correlated with the size of the physical
Higgs self-coupling. For all cases we observe that a strong
first-order phase transition is only possible for Higgs self-couplings
significantly above the SM value. Specifically, the HL-LHC should be
able to conclusively test these parameter regions and confirm or rule
out the possibility of electroweak baryogenesis in a very
model-independent manner.

\subsection{Extended Higgs sectors}
\label{sec:basis_weak}

Perturbative extensions of the Standard Model are strongly restricted
by experimental constraints, cosmological considerations, and
theoretical limitations. New particles can be scalars, fermions, or
gauge bosons. From a Higgs physics perspective, new fermions mostly
enter the loop-induced Higgs decays described in
Sec.~\ref{sec:basis_decs} and the gluon fusion production rate
described in Sec.~\ref{sec:exp_gf}.  The requirement that these rates
have close to the observed rates forbids a $4th$ chiral
generation~\cite{Kribs:2007nz,Denner:2011vt}. Extended gauge sectors are a key ingredient of many
strongly interacting extensions described in
Sec.~\ref{sec:basis_strong}.  Similarly, extended scalar potentials are a generic
extension of the Standard Model and have direct impacts on Higgs
physics. They can, but do not have to, be related to a broader
ultraviolet completion of the Standard Model. 

For a scalar extension of the
Standard Model, we classify the new scalar fields by their
representation of $SU(2)_L$.  This representation is directly linked
to the main constraint on many such extensions, namely electroweak
precision constraints.  From Sec.~\ref{sec:basis_implications} we know
that the constraints typically require a global custodial symmetry at least at
tree level. For additional scalar or Higgs fields with weak isospins
$T_i$, hypercharges $Y_i$, and vacuum expectation values $v_i$ we
find~\cite{Langacker:1980js,Chen:2006pb,Cheng:1973nv}
\begin{align}
\frac{M^2_W}{M^2_Z c_W^2} 
= \frac{\sum_i\left[ T_i(T_i+1) - \cfrac{1}{4}Y_i^2\right]v^2_i}{\cfrac{1}{2}\sum_i Y_i^2\,v^2_i} 
\stackrel{\text{doublets}}{=} 
  \frac{\sum_i\left[ \dfrac{3}{4} - \cfrac{1}{4} \right]v^2_i}{\cfrac{1}{2}\sum_i v^2_i} 
= 1 \; ,
 \label{eq:rho-tree}
\end{align}
where $T_i=1/2$ for a doublet and $0$ for a singlet.  This means that
aside from additional scalar singlets, models with additional Higgs
doublets also respect custodial symmetry at tree level.  Higher
representations, like triplets, typically force us to live with very strong
constraints~\cite{Barger:2009me}.

\subsubsection{Additional singlet}
\label{sec:basic_weak_singlet}

The addition of a real $SU(2)_L$ scalar singlet, $S$, is the simplest
extension of the minimal Higgs sector, with very few free parameters,
allowing for definite predictions of Higgs signals.  The most general
scalar interactions
are~\cite{Silveira:1985rk,OConnell:2006rsp,Bowen:2007ia,Pruna:2013bma,Espinosa:2011ax,Dawson:2015haa}
\begin{alignat}{5}
\label{eq:singlet-potential}
V(\phi,S) = 
  \mu^2\,(\phi^\dagger\,\phi) 
+ \lambda\,(\phi^{\dagger}\phi)^2 +a_1 S
+ \mu^2_2\,S^2 
+ \kappa S^3 
+ \lambda_2\,S^4 
+ \lambda_3\,(\phi^{\dagger}\,\phi)S^2 \; .
\end{alignat}

Based on this form of the potential, the VEV of $S$ can always be set
to $0$ by shifts in the input parameters.  After spontaneous symmetry
breaking, the neutral component of $\phi$ and $S$ mix to form the
physical states $h^0$ and $H^0$.  Explicit expressions for the masses,
$m_h$ and $m_H$, and the mixing angle $\alpha$  that diagonalizes the scalar 
mass matrix, in terms of the
parameters of Eq.\eqref{eq:singlet-potential} can be found in the
appendices of Ref.~\cite{Chen:2014ask} or
Ref.~\cite{Lopez-Val:2013yba}.  We usually assume the lighter mass
eigenstate is the observed SM-like particle, $h^0 \approx H_{125}$.
This implies that all LHC production rates are suppressed by
$\cos^2\alpha\equiv c_\alpha^2$, while the branching ratios are unchanged. This kind of
model allows for a particularly easy interpretation of the global
Higgs coupling analyses described in Sec.~\ref{sec:exp_global}.  If
the new scalar is light, $m_h \lesssim 63$~GeV, and the heavier
scalar, $H^0$, is the observed Higgs boson, the model predicts
Higgs-to-Higgs decays with the partial width
\begin{align}
\Gamma(H^0 \to h^0h^0) = \frac{|\lambda_{Hhh}|^2}{32\pi m_{H}}\,\sqrt{1-\frac{4 m^2_{h}}{m^2_{H}}} \; .
\label{eq:singlet-width} 
\end{align}
in terms of the physical triple-Higgs coupling $\lambda_{Hhh}$, which
can be extracted from the potential~\cite{Espinosa:2011ax}.

In the $Z_2$-symmetric version of the theory there is a symmetry under
$S\rightarrow -S$, and it is in general not possible to set $\langle S
\rangle = 0$.  Whenever there are  two non-zero VEVs, as is the case in this example, we define the
ratio of the two VEVs as $\tan\beta\equiv t_\beta = v/\langle S
\rangle$.  The theory is completely described by the parameters,
\begin{align}
m_h,\, m_H\, c_\alpha, t_\beta \, .
\end{align}

\begin{figure}[t]
\centering
\includegraphics[width=0.45\textwidth]{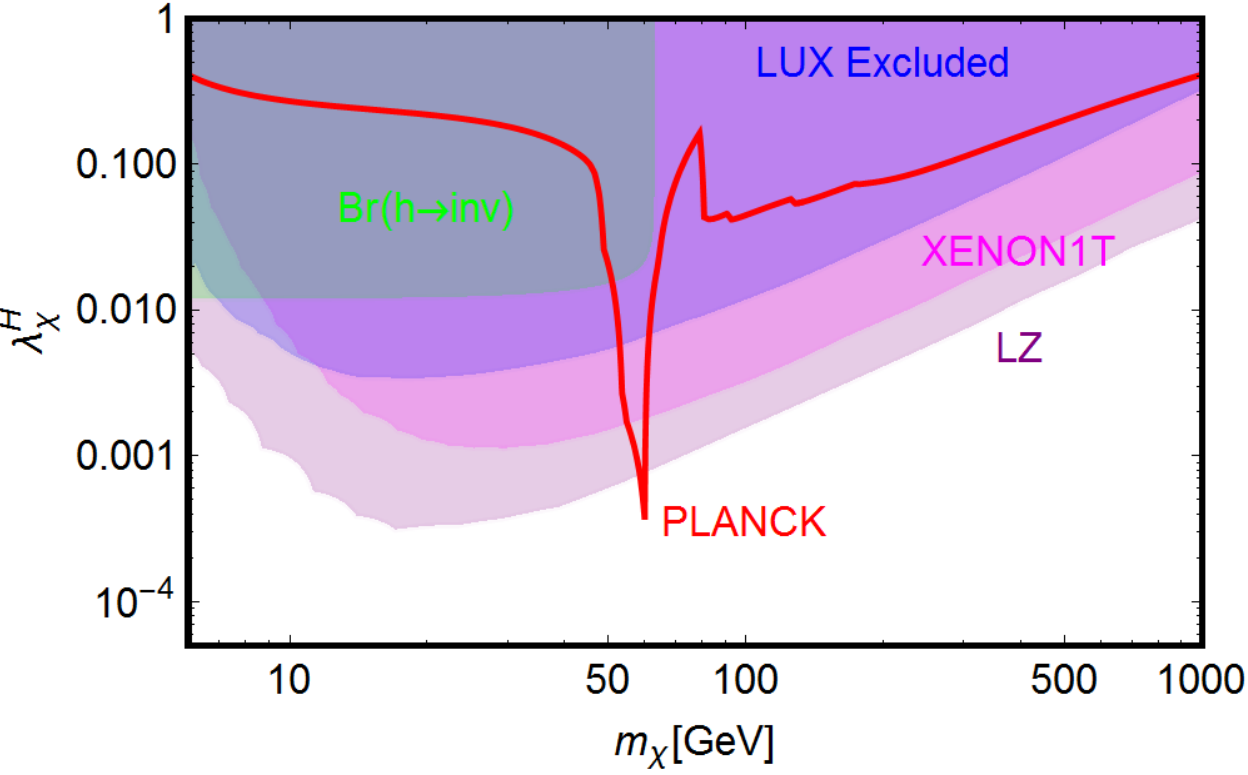}
\caption{Current and future constraints on the Higgs portal model with
  a scalar  dark matter candidate, where
  $m_X$ denotes the mass of the scalar dark matter agent and
  $\lambda_X^H$ the portal coupling. Figure from
  Ref.~\cite{Arcadi:2017kky}.  }
\label{fig:higgs_portal}
\end{figure}

A Higgs portal dark matter
model~\cite{Silveira:1985rk,McDonald:1993ex,Burgess:2000yq} arises in
the $Z_2$-symmetric scenario when we adjust the parameters such that
$\langle S \rangle$ is
0~\cite{Binoth:1996au,Schabinger:2005ei,Patt:2006fw,Lebedev:2011iq}. The
entire dark matter model is then described by three parameters: the
singlet mass term $\mu^2_2$, the portal coupling $\lambda_3$, and the
singlet self-interactions $\lambda_2$. For most observables, the
singlet self-interaction is phenomenologically irrelevant over a wide
range of model parameters.  The portal coupling determines, for
example, the dark matter annihilation rate or the scattering rate with
nucleons. The interaction relevant for direct detection is mediated by
the SM Higgs boson, coupling to virtual top quark loops generated by
the non-relativistic gluon content of the
nucleus~\cite{Shifman:1978zn}. Away from the pole condition $m_H = 2
m_h$ and for reasonably light dark matter masses, this rules out the
model as an explanation of the observed relic density. Some current
and future constraints including the relic density constraints and
direct detection experiments are shown in Fig.~\ref{fig:higgs_portal}.

For a light enough dark matter state, the Higgs-to-Higgs decays shown
in Eq.\eqref{eq:singlet-width} lead to invisible Higgs decays~\cite{Shrock:1982kd} at the
LHC, as described in Sec.~\ref{sec:exp_wbf}. In
Fig.~\ref{fig:higgs_portal} we see that invisible Higgs decays can
probe the parameter regions which lead the to correct relic density,
provided that the dark matter is lighter than the SM Higgs
boson. Unfortunately, this parameter region is, for the simple singlet
extension, essentially ruled out by direct detection experiments.  In
the opposite case, where the dark matter scalar is heavier than
$m_h/2$ we will not be able to extract the small production rate from
the backgrounds. If the new singlet scalar becomes even heavier, the
Higgs-to-Higgs decays shown in Eq.\eqref{eq:singlet-width} correspond
to a decay of the heavy new state and makes an important resonance
contribution to $gg\rightarrow h^0 h^0$, as discussed in
Sec.~\ref{sec:fcc_hh}.

\subsubsection{Additional doublet}
\label{sec:basis_weak_doublet}

The two Higgs doublet model (2HDM)~\cite{Haber:1978jt,Branco:2011iw} adds a second
$SU(2)_L$ doublet with weak hypercharge one. This way it allows for
$CP$-violation and a complex vacuum structure~\cite{Gunion:2005ja}, can
be linked to neutrino masses~\cite{Ma:2006km}, electroweak
baryogenesis~\cite{Cline:1995dg,Fromme:2006cm}, or dark matter~\cite{Gong:2012ri}.
The general $CP$-invariant renormalizable potential reads
\begin{alignat}{5}
 V(\phi_1,\phi_2) 
&= m^2_{11}\,\phi_1^\dagger\phi_1
 + m^2_{22}\,\phi_2^\dagger\phi_2
 - \left[ m^2_{12}\,\phi_1^\dagger\phi_2 + \text{h.c.} \right] \notag \\
&+ \frac{\lambda_1}{2} \, (\phi_1^\dagger\phi_1)^2
 + \frac{\lambda_2}{2} \, (\phi_2^\dagger\phi_2)^2
 + \lambda_3 \, (\phi_1^\dagger\phi_1)\,(\phi_2^\dagger\phi_2) 
 + \lambda_4 \, |\phi_1^\dagger\,\phi_2|^2 \notag \\
&+ \left[ \frac{\lambda_5}{2} \, (\phi_1^\dagger\phi_2)^2 
        + \lambda_6 \, (\phi_1^\dagger\phi_1) \, (\phi_1^\dagger\phi_2)
        + \lambda_7 \, (\phi_2^\dagger\phi_2)\,(\phi_1^\dagger\phi_2) + \text{h.c.} 
   \right] \; .
\label{eq:2hdmpotential}
\end{alignat}
Hermiticity of the potential requires $m^2_{11},m^2_{22}$ and
$\lambda_{1,2,3,4}$ to be real.  The parameters $m^2_{12}$ and
$\lambda_{5,6,7}$ can be complex,  in which case all neutral Higgs
fields will mix and define physical fields that  are no longer
$CP$-eigenstates. We only illustrate the $CP$-even case, for a comprehensive discussion
of the $CP$-violating case see~\cite{Pilaftsis:1999qt}.

If we consider the 2HDM to be a valid ultraviolet completion of the Standard Model, 
the vacuum must be  stable. This can be ensured by
requiring~\cite{Gunion:2002zf,Kanemura:1999xf}
\begin{align} 
\lambda_1 > 0\, , \qqquad 
\lambda_2 > 0\, , \qqqquad 
\sqrt{\lambda_1\lambda_2} + \lambda_3 
    + \text{min}\left( 0, \lambda_4 - |\lambda_5| \right) > 0 \; .
\label{eq:vacuum}
\end{align}
Electroweak symmetry breaking, as described in Sec.~\ref{sec:basis_sm}, now
requires $v_1^2 + v_2^2 = v^2 = (246~\gev)^2$.  The remaining free
parameter is  $\tan\beta\equiv t_\beta =
v_2/v_1$.  The physical spectrum consists of two neutral $CP$-even
scalars $h^0,H^0$, one neutral $CP$-odd scalar $A^0$, and a charged
scalar $H^\pm$.  The two mass eigenstates $h^0$ and $H^0$ arise
from a rotation by the angle $\alpha$, 
\begin{align}
h^0 =& \sqrt{2}\left[-\text{Re}(\phi_1^0-v_1)\sin\alpha+\text{Re}(\phi_2^0-v_2)\cos\alpha\right] \notag \\
H^0 =&\sqrt{2}\left[\text{Re}(\phi_1^0-v_1)\cos\alpha+\text{Re}(\phi_2^0-v_2)\sin\alpha\right]\, ,
\end{align}
and the $CP$-odd and charged Higgs
masses can be easily extracted from the potential,
\begin{align}
m_{H^\pm}^2 &= 
  \frac{m_{12}^2}{s_\beta c_\beta} 
- \frac{v^2}{2}\,\left[\lambda_4 + \lambda_5 
                       + \frac{\lambda_6}{t_\beta} + \lambda_7 t_\beta \right] 
\notag \\
m_{A^0}^2 &=
  \frac{m_{12}^2}{s_\beta c_\beta}
- \frac{v^2}{2}\,\left[2\lambda_5 + \frac{\lambda_6}{t_\beta} + \lambda_7\,t_\beta \right] \; ,
\label{eq:2hdm-masses}
\end{align}
and $c_\beta\equiv \cos\beta, s_\beta\equiv \sin\beta$. 
The range of possible spectra includes compressed
masses ($m_{h^0} \simeq m_{H^0}$), twisted masses ($m_{A^0} <
m_{h^0,H^0}$), a single mass hierarchy ($m_{h^0} \ll
m_{H^0,A^0,H^\pm}$), or multiple mass hierarchies ($m_{h^0} \ll
m_{H^0} \ll m_{A^0,H^\pm}$), each of which have quite different phenomenology\cite{Craig:2015jba}. The limit $m_{h^0} \ll
m_{H^0,A^0,H^\pm}$ is termed the decoupling limit, and in this limit the couplings
of $h^0$ approach those of the SM Higgs boson~\cite{Gunion:2002zf,Carena:2013ooa,Dev:2014yca}.

Since custodial symmetry strongly constrains extended Higgs sectors,
we can assign the Higgs masses in the 2HDM accordingly.  Barring
fine-tuning in the mixing angles, two scenarios generally respect
custodial symmetry. First, a compressed mass spectrum with only
moderately split Higgs masses directly follows the pattern of
Eq.\eqref{eq:rho-tree}. For degenerate masses the contributions to the
electroweak precision parameters at tree level are $\Delta S = \Delta
T = 0$.  Second, for a light SM-like Higgs with mass-degenerate heavy
companions $H^0, A^0, H^\pm$ all Higgs states fall into the singlet
and triplet representations of the custodial symmetry group
$SU(2)_{V}$,
\begin{alignat}{7}
\phi_1 \supset& 
\begin{pmatrix} H^+ \\ A^0 \\ H^- \end{pmatrix} 
\oplus H^0 \quad  \text{or} \quad
\phi_1 \supset 
\begin{pmatrix} H^+ \\ H^0 \\ H^- \end{pmatrix}
 \oplus A^0
\qqqquad 
\phi_2\supset& 
\begin{pmatrix} \omega^+ \\ \omega^0 \\ \omega^- \end{pmatrix} 
\oplus \frac{v+ h^0}{\sqrt{2}} 
 \; ,
\label{eq:spectrum-custodial}
\end{alignat}
typically with a rich vacuum structure, see~\cite{Battye:2011jj,Pilaftsis:2011ed}. We can integrate out all heavy Higgs fields simultaneously and
retrieve an effective field theory in terms of $\phi_2$ only. This 
gives in the limit of heavy additional Higgs states~\cite{He:2001tp}
\begin{alignat}{5}
\Delta S = -\cfrac{1}{12\pi}\, \log \cfrac{m^2_{H^\pm}}{m^2_{A^0}}
 \qqqquad
\Delta T = \frac{\sqrt{2} G_F}{16\pi^2 \alpha_{EW}}\,\left(m^2_{H^{\pm}}-m^2_{A^0}\right)  \; .
\label{eq:stlimits}
\end{alignat}

The gauge sector of all 2HDM setups is especially simple. The 
couplings of the light Higgs to the weak bosons always scale like
\begin{align}
\frac{g_{VVh^0}}{ g_V^\text{SM}} = \sin(\beta-\alpha)\equiv s_{\beta - \alpha} \; ,
\label{eq:vvh_2hdm}
\end{align}
where we define $g_V^\text{SM} \equiv g_{VVh}^\text{SM}$ as
systematically introduced in Sec.~\ref{sec:basis_couplings}, and
possible pre-factors drop out of the ratio.  The decoupling limit is
given by $s_{\beta -\alpha} \to 1$. On the side of the Yukawa
couplings, the 2HDM setup offers a lot of freedom.  In the Standard
Model, for example, a global symmetry group $U(3)^3$ leaves the CKM
matrix invariant and guarantees the absence of tree-level FCNCs.  The symmetry
is broken by large Yukawa couplings.  This protection  against FCNCs does not hold in the
presence of a second Higgs doublet~\cite{Hall:1981bc}.  Natural flavor conservation
imposes a global flavor-blind symmetry $\phi_{1,2} \to \mp \phi_{1,2}$
and requires that any fermion family can  couple to only one Higgs
doublet. This simple Higgs flavor symmetry satisfies the
Glashow-Weinberg theorem~\cite{Glashow:1976nt}, forbids the mass term
$m^2_{12}$ and the self interactions $\lambda_{6,7}$, and defines four
canonical setups for the couplings of $\phi_1$ and $\phi_2$ to fermions: 
\begin{itemize}
\item type-I, where all fermions only couple to $\phi_2$;
\item type-II, where up-type (down-type) fermions couple
  exclusively to $\phi_2$ ($\phi_1$);
\item lepton--specific, with a type-I quark sector and a type-II
  lepton sector;
\item flipped, with a type-II quark sector and a type-I lepton
  sector.
\end{itemize}
If we extend the flavor symmetry to the Yukawa sector as a
Peccei-Quinn (PQ)  symmetry, it removes all sources of tree-level FCNC
interactions. The mass splitting between the $H^0$ and $A^0$ is
controlled by the size of the PQ-breaking terms $m^2_{12}$ and
$\lambda_{6,7}$.  Yukawa structures respecting natural flavor
conservation are a special case of a class of models in which
tree-level FCNCs are largely suppressed via minimal flavor
violation~\cite{DAmbrosio:2002vsn,Dery:2013aba}. In the 2HDM
realization~\cite{Pich:2009sp} the fermions couple to both Higgs
doublets with aligned Yukawa matrices, \ie linked to one another by
linear shifts $\epsilon_f$,
\begin{align}
y_{u,1} = \epsilon_u \; y_{u,2} \qqquad  
y_{d,2} = \epsilon_d \; y_{d,1} \qqquad  
y_{l,2} = \epsilon_{\tau} \; y_{l,1} \; ,
\label{eq:alignment}
\end{align}
with $y_{f,i} = \sqrt{2} m_f/v_i$ as defined in
Eq.\eqref{eq:def_yukawas}.  This way the fermion masses and Yukawa
matrices are diagonalized simultaneously. We can parameterize the
shifts in terms of angles $\gamma_{b,\tau}$~\cite{Davidson:2005cw}
modifying the bottom and tau Yukawas
\begin{align}
\frac{m_{b,\tau}}{v}
= {y_{b,\tau}\over \sqrt{2} v}\braket{\phi_1}\cos\gamma_{b,\tau}
+{ y_{b,\tau}\over \sqrt{2} v}\,\braket{\phi_2}\,\sin\gamma_{b,\tau} 
= \frac{y_{b,\tau}}{\sqrt{2}} \,\cos(\beta-\gamma_{b,\tau}) \; .
\label{eq:alignment2}
\end{align}
The neutral Higgs couplings to fermions in the 2HDM in terms of the SM
Higgs couplings are given in Tab.~\ref{tab:2hdmcouplings}.  For the
light SM-like Higgs couplings to fermions there exist two sources of
coupling modifications: first, the fermionic 2HDM mixing structure
includes a $CP$-odd gauge boson $A^0$. This leads to non-trivial
ratios of functions of $\alpha$ and $\beta$. Second, in the aligned
model where $\cos(\alpha-\beta)\rightarrow 0$, the couplings to the
top quark, bottom quark, and tau can be varied independently.  Note
that the aligned limit can be realized without the decoupling of
$H^0,A^0$ and $H^\pm$.

\begin{table}[b!]
\begin{center} 
\begin{tabular}{l|ccc} 
\toprule
 & $h^0$ & $H^0$ & $A^0$ \\ \midrule
$\dfrac{g_{VV\phi}}{g^\text{SM}_V}$ & $s_{\beta-\alpha}$  & $c_{\beta-\alpha}$ & 0 \\[3mm]
$\dfrac{y_t}{y^\text{SM}_t}$ & $\dfrac{c_\alpha}{s_\beta}$ &  $\dfrac{s_\alpha}{s_\beta}$ & $\dfrac{1}{t_\beta}$ \\[3mm]
$\dfrac{y_b}{y^\text{SM}_b}$ & $-\dfrac{\sin(\alpha-\gamma_b)}{\cos(\beta-\gamma_b)}$ & $\dfrac{\cos(\alpha-\gamma_b)}{\cos(\beta-\gamma_b)}$ & $\tan(\beta-\gamma_b)$ \\[3mm]
$\dfrac{y_\tau}{y^\text{SM}_\tau}$ & $-\dfrac{\sin(\alpha-\gamma_\tau)}{\cos(\beta-\gamma_\tau)}$ & $\dfrac{\cos(\alpha-\gamma_\tau)}{\cos(\beta-\gamma_\tau)}$ & $\tan(\beta-\gamma_\tau)$ \\
\bottomrule
\end{tabular}
\end{center}
\caption{Neutral Higgs couplings in the 2HDM, all normalized to the SM
  Higgs couplings. We use $\phi$ to represent $h^0,H^0,A^0$. The
  angles $\gamma_b$ and $\gamma_\tau$ are defined in
  Eq.\eqref{eq:alignment2}. See also
  Refs.~\cite{Eriksson:2009ws,Lopez-Val:2013yba}.}
\label{tab:2hdmcouplings}
\end{table}

When we embed the extended Higgs sector in a perturbative ultraviolet
completion of the Standard Model, we need to ensure that the Landau poles of all
Higgs self-couplings lie outside the validity range of our model. For
weakly interacting models, a stronger constraint arises from
perturbative unitarity~\cite{Lee:1977eg}. The leading effects at high
energies constrain the 
combinations~\cite{Casalbuoni:1986hy,Casalbuoni:1987cz,Kanemura:1993hm,Akeroyd:2000wc,Gunion:2002zf,Ginzburg:2005dt,Osland:2008aw,Pruna:2013bma} of parameters in the scalar potential given in
Eq.\eqref{eq:2hdmpotential},
\begin{alignat}{5} 
a_{\pm} &= 
\frac{1}{16\pi}\,\left[3(\lambda_1+\lambda_2)\, \pm \sqrt{9(\lambda_1-\lambda_2)^2 + 4(2\lambda_3+\lambda_4)^2} \right] 
\qqquad 
&f_1 &= f_2 = \frac{1}{8\pi}\,\left(\lambda_3 + \lambda_4\right) \notag \\
b_{\pm} &= 
\frac{1}{16\pi}\,\left[(\lambda_1+\lambda_2)\, \pm \sqrt{(\lambda_1-\lambda_2)^2 + 4\lambda_4^2} \right] 
&f_+ &= \frac{1}{8\pi}\,\left(\lambda_3 + 2\lambda_4 + 3\lambda_5\right) \notag \\
c_{\pm} &= 
\frac{1}{16\pi}\,\left[(\lambda_1+\lambda_2)\, \pm \sqrt{(\lambda_1-\lambda_2)^2 + 4\lambda_5^2} \right] 
&f_- &= \frac{1}{8\pi}\,\left(\lambda_3 + \lambda_5\right) \notag \\
e_1 &= \frac{1}{8\pi}\left(\lambda_3 + 2\lambda_4 - 3\lambda_5\right) 
&p_1 &= \frac{1}{8\pi}\left(\lambda_3 - \lambda_4\right) \notag \\
e_2 &= \frac{1}{8\pi}\,\left(\lambda_3 -\lambda_5\right)
\label{eq:2hdm-unitarity1}
\end{alignat}
 to remain smaller than unity. Assuming the usual 2HDM pattern
imposes a set of sum rules~\cite{LlewellynSmith:1973yud,Cornwall:1973tb,Cornwall:1974km,Weldon:1984wt,Gunion:1990kf}
\begin{alignat}{5}
g_{VV h^0}^2 + g_{VV H^0}^2 &= \left( g_V^\text{SM} \right)^2
\qqqquad 
&y^2_{ff h^0} + y^2_{ff H^0} + y^2_{ff A^0} &= \left( y_f^\text{SM} \right)^2
 \notag \\
g^2_{h^0A^0Z^0} +  g^2_{H^0A^0Z^0} &= \frac{g^2}{4c_W^2} 
\qqqquad 
&g^2_{\phi ZZ} + 4 M_Z^2 g^2_{\phi A^0Z^0} &= \frac{g^2 M_Z^2}{c_W^2} \notag \\ 
y_{ff h^0}\,g_{VV h^0} + y_{ff H^0}\,g_{VV H^0} &= y_f^\text{SM} \,g_{V}^\text{SM}\, ,
\label{eq:sumrules}
\end{alignat}
with $\phi = h^0,H^0$.  These tree--level sum rules imply that the
Higgs couplings to weak bosons can be at most as strong as in the
Standard Model. Moreover, all vertices containing at least one gauge
boson and exactly one heavy Higgs field are proportional to
$c_{\beta-\alpha}$.  Finally, the $A^0$ Yukawa coupling comes with an
additional factor $i \gamma_5$.  According to the sum rule, at least
one of the $CP$-even Yukawa couplings can therefore lie above
$y^\text{SM}_f$.

If, in addition, we require that all 2HDM Yukawas remain perturbative
at the weak scale, $y_f/\sqrt{2} < \sqrt{4 \pi}$, we find $t_\beta >
0.28$ for all natural flavor-conserving models, $t_\beta < 140 $ for
type-II and flipped models, and $t_\beta < 350$ in the
lepton--specific case~\cite{Chen:2013kt}.

Interesting variations of the 2HDM are dark portal or inert doublet
models~\cite{Deshpande:1977rw,Ma:2006km,Majumdar:2006nt,Barbieri:2006dq,LopezHonorez:2006gr}. They
require a $Z_2$-symmetry in Eq.\eqref{eq:2hdmpotential}, under which
one doublet transforms as odd.  For the 2HDM potential given in
Eq.\eqref{eq:2hdmpotential} this means that $\lambda_{6,7} = 0$ and
$m_{12} = 0$.  The new, $Z_2$-odd doublet does not participate in
electroweak symmetry breaking. This means that it can only interact
with SM particles when it appears twice in the vertex. For the
fermions such an interaction would not be renormalizable, but for the
weak bosons it is induced by the covariant interactions in the kinetic
term.  The corresponding scalar or pseudo-scalar dark matter agent
therefore couples to the SM particles through the usual Higgs portal
discussed in Sec.~\ref{sec:basic_weak_singlet}, complemented by a
possible direct annihilation to $WW$ or $ZZ$ pairs.

\subsubsection{Additional triplet}
\label{sec:basis_weak_triplet}

Naive triplet extensions of the SM Higgs sector tend to be in violent
conflict with electroweak precision data already at tree level,
Eq.\eqref{eq:rho-tree}. There are three ways to deal with this: (i)
live with very strong model constraints; (ii) carefully align
different triplet Higgs fields, like in the Georgi-Machacek
model~\cite{Georgi:1985nv}, at the price of additional
fine-tuning~\cite{Gunion:1990dt}; or (iii) combine exotic
representations~\cite{Hisano:2013sn}.  Theoretical motivations for
triplet models are provided by left-right
symmetries~\cite{Mohapatra:1974gc,Mohapatra:1974hk,Senjanovic:1975rk}
or models for neutrino mass
generation~\cite{Konetschny:1977bn,Magg:1980ut,Cheng:1980qt}.
Interesting implications have also been highlighted in the context of
non-minimal SUSY extensions~\cite{Delgado:2013zfa}.

There are two possible hypercharge assignments for an $SU(2)_L$ triplet:
\begin{itemize}
\item
$Y=0$, real scalar
\begin{align}
\xi=\left(\begin{matrix}
\xi^+\\
\xi^0+v_\xi\\
\xi^-
\end{matrix}
\right)
\end{align}
\item
$Y=1$, complex scalar 
\begin{align}
\chi=\left(\begin{matrix}
\chi^{++}\\
\chi^+\\
\chi^0=v_\chi+(h_\chi+ia_\chi)/\sqrt{2}
\end{matrix}
\right)\, .
\end{align}
\end{itemize}
The interactions of the triplets with the $SU(2)_L\times U(1)$ gauge kinetic terms
generates contributions to the gauge boson masses,
\begin{align}
M_W^2 =&{g^2\over 4} \left( v_\phi^2+4v_\chi^2+4v_\xi^2 \right)\notag \\
M_Z^2 =& {g^2\over 4 c_W^2} \left( v_\phi^2+8 v_\chi^2 \right) \; ,
\end{align}
where $v_\phi$ is the doublet VEV. 
The addition of triplet scalars contributes to custodial symmetry breaking,
\begin{align}
\rho = {M_W^2\over c_W^2 M_Z^2}={v^2+4(v_\xi^2+v_\chi^2)\over v^2+8 v_\chi^2}\, .
\label{eq:rhotrip}
\end{align}
From this formula, we see that choosing $v_\xi = v_\chi$ protects the
model from constraints arising due to electroweak precision data at
tree level.  The Georgi-Machacek
model~\cite{Georgi:1985nv,Hartling:2014aga,Englert:2013zpa} realizes
this mechanism with the help of a real scalar triplet and a complex
scalar triplet, in addition to a doublet which is still needed for
fermion mass generation, and predicts $\rho=1$ at tree level.  The
physical spectrum includes two $CP$-even scalars $h^0, H^0$, one
$CP$-odd scalar $A^0$, one singly charged scalar $H^\pm$, and one
doubly-charged state $H^{\pm \pm}$. This way we define three mixing
angles in the scalar sector, the usual angle $\alpha$ in the $CP$-even
scalar sector, plus the angle $\beta_c$ in the charged scalar sector
and the angle $\beta_n$ in the $CP$-odd scalar sector.  The heavy
`states can be combined into a heavy $SU(2)_L$ triplet, minimizing the
tree-level breaking of the custodial symmetry,
\begin{align}
  \Xi=\left(\begin{matrix} \chi_0^{\ast} & \xi^+ & \chi^{++} \\
      -\chi^{+\ast} & \xi^0 & \chi^+ \\
      \chi^{++\ast} & -\xi^{+\ast} & \chi^0 \\
    \end{matrix}\right)\, .
    \label{eq:georg}
\end{align}
In terms of the common VEV $\langle \chi^0\rangle =\langle
\xi^0\rangle \equiv v_\Xi$, the $W$ and $Z$ masses are both given by
the same combination $v_\phi^2+8v_\Xi^2 = v^2 =
(246~\gev)^2$. Custodial symmetry $SU(2)_V$ then acts in the triplet
representation on Eq.\eqref{eq:georg} similar to
Eq.\eqref{eq:bidoub}.

A structural difference between the doublet and triplet extensions is
that the triplet models predict tree-level $H^\pm W^\mp Z$
couplings~\cite{Han:2005ru}. The smoking gun at the LHC for complex
triplets is the doubly charged Higgs, produced either in a Drell-Yan
process or in weak boson fusion through a $H^{++}W^-W^-$
coupling. These exotic couplings are directly related to the triplet
character of the vacuum expectation value and they scale with a
characteristic angle $s_\theta^2 = 8 v^2_\Xi /v^2$.

\begin{table}[b!]
\begin{center} 
\begin{tabular}{l|l|ll|ll|ll|ll} 
\toprule
extension & model & \multicolumn{4}{|c}{$g_{hVV}$} & \multicolumn{4}{|c}{$g_{hff}$} \\
&& \multicolumn{2}{c}{universal} & \multicolumn{2}{c|}{non-universal} & \multicolumn{2}{c}{universal} & \multicolumn{2}{c}{non-universal} \\ \midrule
\multirow{2}{1.8cm}{singlet}
& inert ($v_S = 0$) &&&&&&&& \\ 
& EWSB ($v_S \neq 0$) & $\alpha$ & $\Delta_V< 0$ &&& $\alpha$ & $\Delta_f< 0$ & &  \\ \midrule
\multirow{5}{1.8cm}{doublet} 
& inert ($v_d = 0$) &&&&&&&& \\
& type-I & $\alpha-\beta$ & $\Delta_V < 0$ & $\mathcal{O}(y_f,\lambda_H)$ & $\Delta_V \gtrless 0$ \
         & $\alpha-\beta$ & $\Delta_f\gtrless 0$ & $\mathcal{O}(y_f,\lambda_H)$ & $\Delta_f\gtrless 0$  \\
& type-II-IV & $\alpha-\beta$ & $\Delta_V< 0$&   $\mathcal{O}(y_f,\lambda_H)$ & $\Delta_V\gtrless 0$ \           &&&  $\alpha,\beta$, $\mathcal{O}(y_f,\lambda_H)$ & $\Delta_f\gtrless 0$ \\ 
&aligned, MFV & $\alpha-\beta$ & $\Delta_V< 0$ &  $\mathcal{O}(y_f,\lambda_H)$ & $\Delta_V\gtrless 0$ 
              & $y_f$ & $\Delta_f\gtrless 0$ & $y_f \mathcal{O}(y_f,\lambda_H)$ & $\Delta_f\gtrless 0$  \\ \midrule
triplet & complex scalar  &  & & $\alpha,\beta_n,\beta_c$ & 
$\Delta_V\gtrless 0$ 
  & $\beta_n$ & $\Delta_f \gtrless 0$&  $\mathcal{O}(Y_f,\lambda_H)$& 
$\Delta_f\gtrless 0$ 
\\ \bottomrule
\end{tabular}
\end{center}
\caption{Interaction patterns for a light Higgs boson to weak gauge
  and fermions, allowing for universal or non-universal departures
  from the Standard Model interactions.  We indicate the relevant
  model parameters and a possible enhancement or suppression;
  $\mathcal{O}(y_f,\lambda_H)$ stands for fermion--mediated or
  Higgs--mediated loop contributions. $v_d=0$ implies that there are
  only weak portal couplings. Table from
  Ref.~\cite{Lopez-Val:2013yba}.}
\label{tab:couplings_structure}
\end{table}

In the spirit of precision Higgs physics, the key question for all
extended Higgs sectors is how they affect the SM-like Higgs
couplings. We parameterize the coupling deviations for the SM-like
Higgs from the Standard Model predictions as~\cite{Lafaye:2009vr}
\begin{align}
\Delta_x = \frac{g_{hxx}}{g_{Hxx}^\text{SM}} -1 
         \equiv \kappa_x -1 \; .
\end{align}
We start with the simple coupling structures to gauge bosons given in
Table~\ref{tab:couplings_structure}. For the singlet and doublet
extensions, the universal coupling modifications are simply a rotation
by an angle $\alpha$ or $\alpha - \beta$. The implication is that
extended Higgs sectors can only reduce the couplings to gauge
bosons. In addition, non-universal fermion--mediated or
Higgs--mediated loop contributions can change the couplings to gauge
bosons in either direction.

The SM-like fermionic coupling variations are much more model-dependent. While the singlet model is fully described by one mixing
angle, the aligned 2HDM model  allows for essentially free
coupling modifications with either sign.

Finally, Higgs interactions with  photons and gluons are generated by
loops already in the Standard Model. For the gluon case this requires colored
states, so we can immediately apply the modified quark Yukawa patterns
of an extended Higgs sector. The photon coupling depends on the three
heavy Yukawa couplings and on $\Delta_W$, but also receives corrections due
to new charged scalars in the Higgs sector.  The effect of additional
states is relatively enhanced as it spoils to the destructive
interference between the leading top and $W$ contributions in the
Standard Model~\cite{Kribs:2007nz}.

\subsubsection{(N)MSSM}
\label{sec:basis_weak_susy}

Ultraviolet extensions of the Standard Model predicting an extended
Higgs sector include supersymmetric
models~\cite{Gunion:1984yn,Gunion:1986nh}. The reason is that in the
Standard Model, we use $\phi$ and $i\sigma^2 \phi^\ast$ to give mass
to the up-type and down-type fermions, while in a supersymmetric
Lagrangian we are not allowed to use conjugate super-fields. The way
out is to give mass to the fermions with two Higgs doublets, $\phi_u$
and $\phi_d$.  As in any 2HDM setup, the two VEVs combined give the
weak boson masses, implying $v_u^2 + v_d^2 = v^2$ or $v_u/v_d = \tan
\beta$.

\begin{figure}[t]
\includegraphics[width=0.45\textwidth]{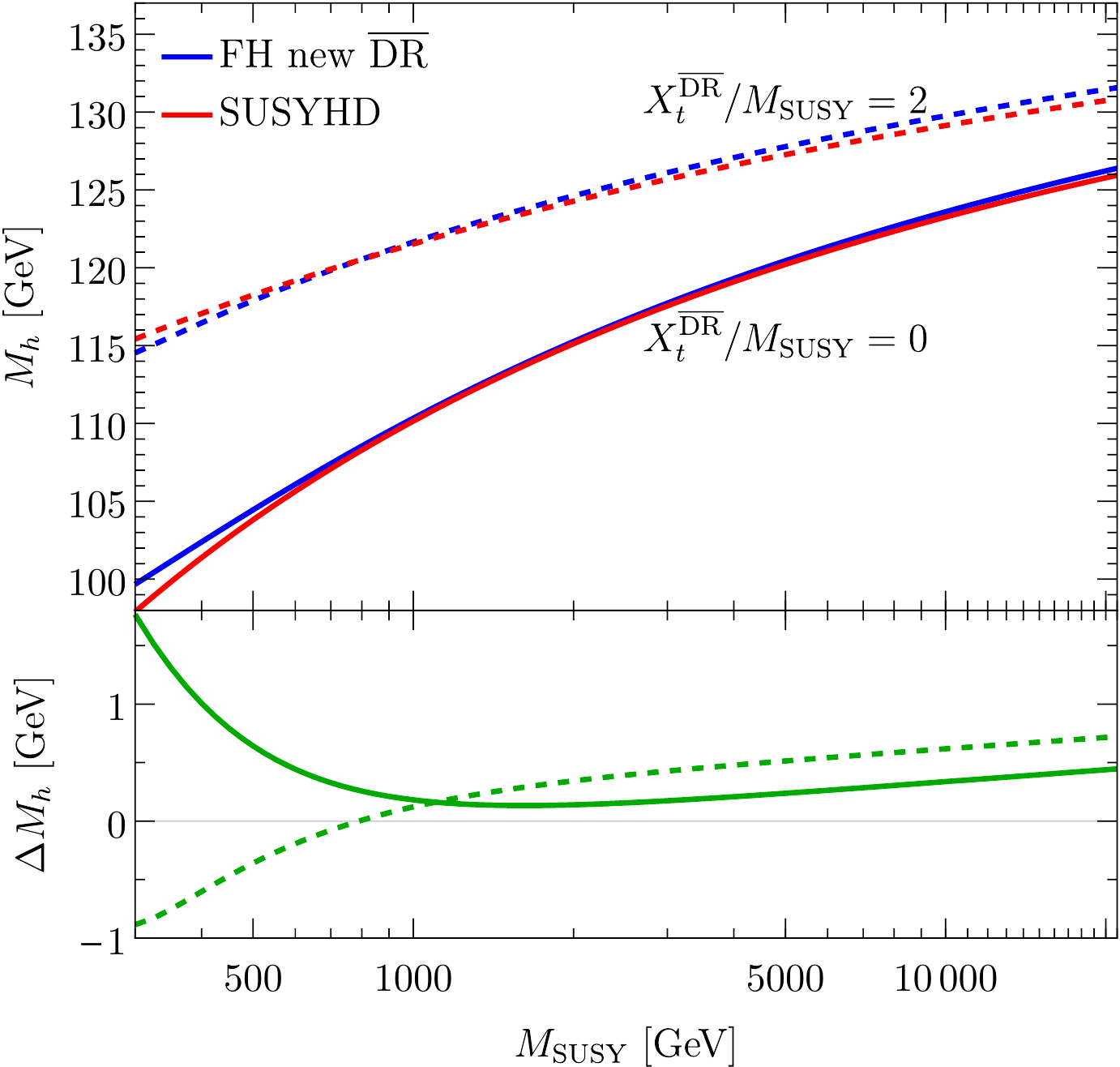}
\hspace*{0.05\textwidth}
\includegraphics[width=0.45\textwidth]{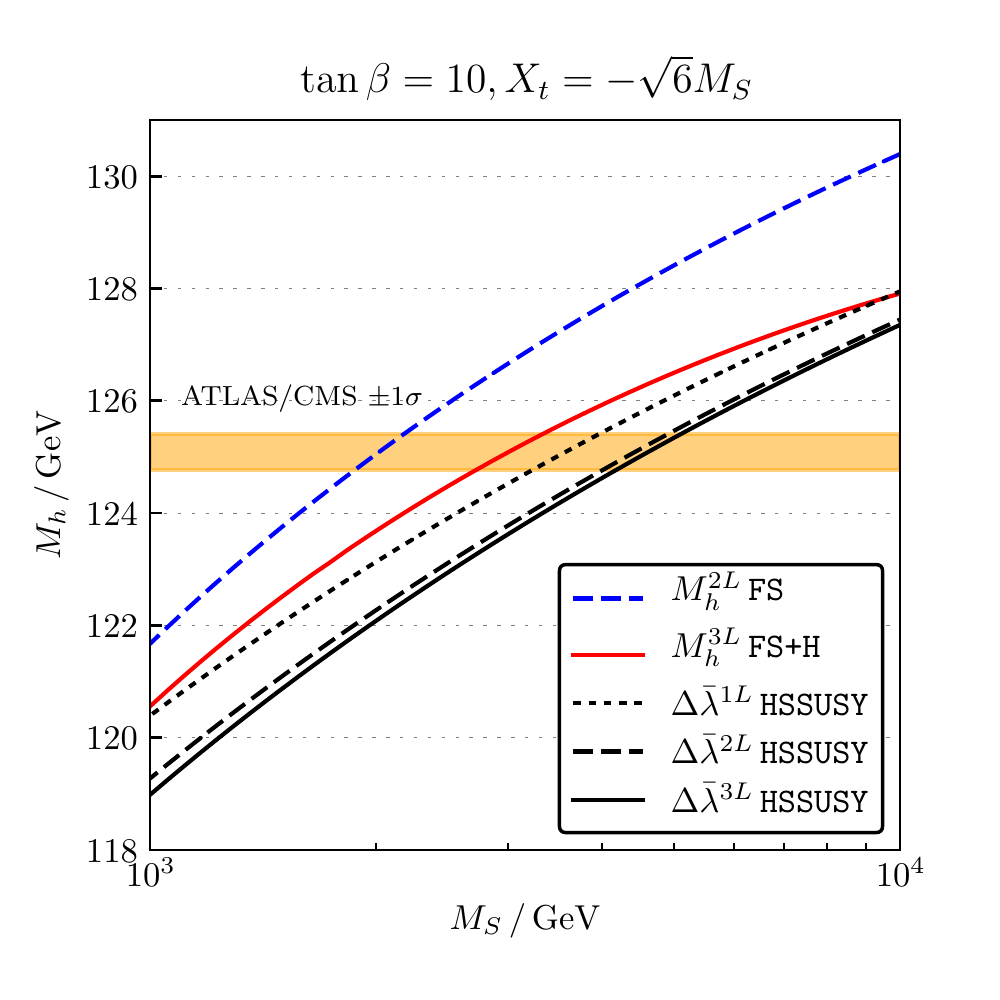}
\caption{Dependence of the SM-like Higgs mass on the supersymmetric
  mass scale from two perturbative approaches at two loops
  (left)~\cite{Bahl:2017aev} and three loops
  (right)~\cite{Harlander:2018yhj}. In the right panel the $\Delta
  \bar{\lambda}_\text{EFT}$ refer to different matching coefficients
  to the 3-loop fixed-order predictions. The experimental
  uncertainties on the Higgs mass are indicated by the yellow band.}
\label{fig:mssm_mh}
\end{figure}

There are three kinds of contributions to the minimal supersymmetric
Higgs potential. First, $F$-terms from the SUSY-conserving scalar
potential include scalar quark interactions proportional to Yukawa
couplings, as well as Higgs mass terms proportional to a new parameter
$|\mu|^2$.  Second, gauge--coupling mediated SUSY-conserving $D$ terms
lead to four-scalar interactions proportional to SM gauge couplings.
The sign of the $D$ terms in the Lagrangian is indeed predicted to be
negative.  Finally, Higgs masses appear as soft breaking parameters.
The $D$-term gauge couplings determine the Higgs potential of the
MSSM~\cite{Chung:2003fi,Bhattacharyya:2017ksj}, cf. Eq.\eqref{eq:2hdmpotential},
\begin{align}
\lambda_1 = \lambda_2 = \cfrac{g^2 s_W}{4 c_W}\;,
\qqquad 
\lambda_3 = \cfrac{g^2 (c_W^2-s_W^2)}{4c_W^2}\;,
\qqquad 
\lambda_4 = -\cfrac{g^2}{2}\;,
\qqquad  
\lambda_{5,6,7} = 0 \; .
\label{eq:susylimit}
\end{align}
The main feature of the MSSM setup is that these self-couplings are
limited from above, implying an upper limit on the lighter
of the two $CP$-even Higgs states.

At tree level the entire Higgs sector is determined by two parameters,
often chosen as $m_{A^0}$ and $\tan\beta$. Typically, the MSSM Higgs
sector shows a mass hierarchy $m_{h^0} \ll m_{H^0,A^0,H^\pm}$, in line
with the generic hierarchical benchmark.  Custodial symmetry and
tree-level FCNC suppression are guaranteed. The lighter $CP$-even mass
is then predicted in terms of the MSSM parameters, and the observed
value of $m_h = 125$~GeV requires large quantum
corrections~\cite{Haber:1996fp,Degrassi:2002fi,Hahn:2009zz}. Given the
LHC constraints for example on the top squarks, the corresponding
calculation at the required precision is plagued by sizable logarithms
$\log m_{\tilde{t}}/m_h$. We show the comparison between
logarithms-enhanced fixed-order prediction and the EFT-based
prediction~\cite{Vega:2015fna} in the left panel of
Fig.~\ref{fig:mssm_mh}, indicating an agreement at the 1~GeV level
even for heavy supersymmetric states. In the right panel of
Fig.~\ref{fig:mssm_mh} we show the results from an EFT-enhanced 4-loop
prediction, compared to the fixed order 2-loop (FS) and 3-loop (FS+H)
results. For large mixing in the supersymmetric stop sector, the EFT
corrections are typically large and negative, and there exists a
sizeable dependence on the matching to the fixed-order 3-loop results.

In supersymmetric models with $R$-parity, the lightest superpartner
serves as a dark matter candidate. In many cases, this is the lightest
neutralino, $\chi$, a Majorana fermion coupling to the Higgs bosons through
gaugino-higgsino mixing. The Higgs sector in the MSSM provides three
dark matter mediators, the light Higgs boson as well as the $CP$-even
and $CP$-odd heavy Higgs bosons. In all three cases, the observed
relic density combined with a standard thermal production process
requires an on-shell condition $m_{h,H,A} \approx 2 m_{\tilde{\chi}}$,
similar to the scalar Higgs portal discussed in
Sec.~\ref{sec:basic_weak_singlet}. In contrast to the scalar case
shown in Fig.~\ref{fig:higgs_portal} the SM-Higgs portal with
Majorana fermion dark matter will be fully accessible by the next
generation of direct detection experiments.

Adding singlets to the supersymmetric 2HDM setup hardly changes its
main LHC features in the Higgs sector, but it can completely change
the neutralino dark matter
phenomenology~\cite{Ellwanger:2009dp,Maniatis:2009re}.  The main
virtue of the NMSSM is that it generates the non-supersymmetric higgsino
mass term through a singlet contribution $\lambda S \hat{H}_u\,\hat{H}_d +
k/3\,\hat{S}^3$ in the Higgs potential and the corresponding singlet
VEV.  The particle spectrum now includes three neutral $CP$-even
states and two neutral $CP$-odd scalars.  A SM-like Higgs boson can be
realized for large $\lambda$ and small $\tan\beta$ values. In this
case the two lightest mass eigenstates are typically close in mass.

If, in the spirit of the NMSSM,  we assume that the additional singlet
develops a VEV, we find a light Higgs state $h^0 =
c_\theta\,(c_\alpha\,H_2^0 - s_\alpha\,H_1^0 ) + s_\theta\,S^0$
with couplings
\begin{align}
\frac{g_{VVh^0}}{ g_V^\text{SM}} = c_\theta\, s_{\beta-\alpha}
\qqqquad 
\frac{y_{ffh^0}}{y_f^\text{SM}} = c_\theta\,\cfrac{c_\alpha}{s_\beta}  \; ,
\label{eq:param7}
\end{align}
where we assume a type-I 2HDM structure.  The 2HDM interaction pattern
is simply rescaled.

A major difference between the MSSM  and the NMSSM appears in the dark matter sector. On
the one hand, the additional  fermionic gauge singlet (singlino) present in the NMSSM
 gives us more freedom to adjust
the parameters of the neutralino sector. On the other hand, the
additional $CP$-even and $CP$-odd singlet field can be light with
sizeable couplings to the singlino and reduced couplings to the
Standard Model. This turns them into promising dark matter mediators
for dark matter neutralinos as light as a few GeV and not accessible
to nuclear recoil direct detection experiments. 

\begin{figure}[t]
\centering
\includegraphics[width=0.35\textwidth]{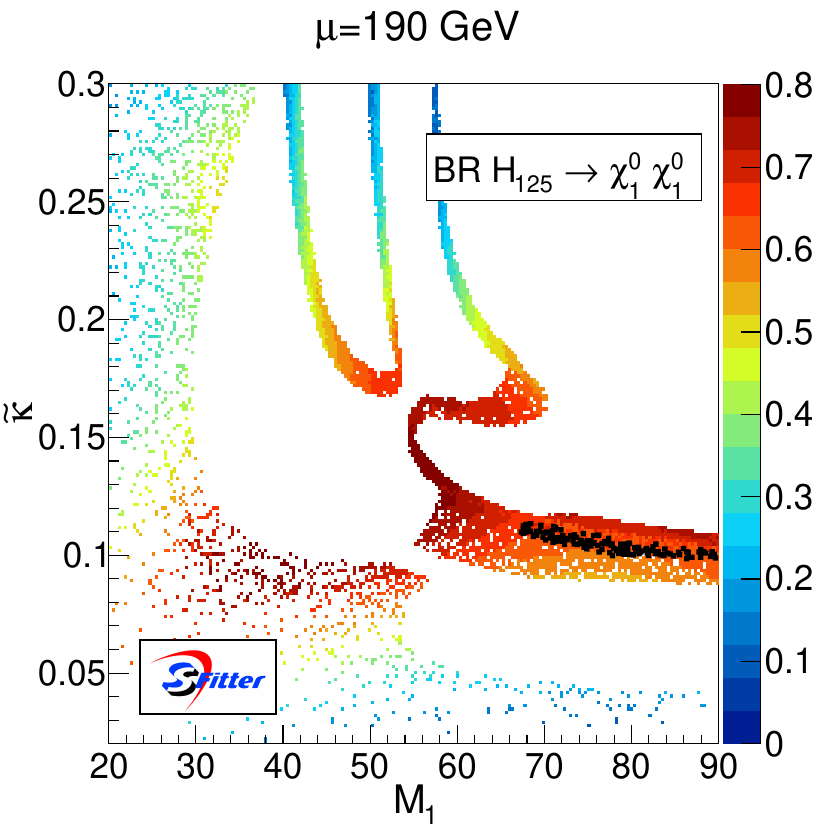}
\caption{Invisible branching ratio of the SM-like Higgs in the NMSSM,
  requiring the observed relic density and before taking into account
  the limits from invisible Higgs searches.  Figure from
  Ref.~\cite{Butter:2015fqa}.}
\label{fig:nmssm_inv}
\end{figure}

Light NMSSM neutralinos are one of the few actual models which
motivate invisible Higgs searches. Even including all other
experimental constraints, the invisible branching ratio of the SM-like
Higgs can vastly exceed the current experimental limits. In
Fig.~\ref{fig:nmssm_inv} we show the predicted invisible branching
ratio for a slice in the MSSM parameter space with fixed Higgsino
mass. The parameter $\tilde \kappa$ is  given by the singlino mass
parameter normalized by $2 \mu$ and $M_1$ is the usual bino mass. All
shown points predict the correct relic density, and the black dots
explain the Fermi galactic center excess. The predicted invisible
branching ratios can be as large as 70\%, strongly correlated with
signal rates in indirect dark matter detection.

\subsubsection{Strong couplings}
\label{sec:basis_strong}

Strong interactions are typically considered as viable contenders for
explaining a hierarchy between the electroweak scale and any other UV
scale which resolves the Standard Model into a more fundamental
theory.  We do not yet know if the Higgs is a fundamental particle or
a composite particle resulting from some strong dynamics, provided the
strong dynamics predicts a very narrow lightest particle.  Theories of
strong electroweak symmetry breaking take inspiration from the
precedent of QCD as one well-understood theory of a strong
interactions with approximate global symmetry~\cite{Scherer:2002tk}.
Simply ``copying'' QCD at higher scales, referred to as
Technicolor~\cite{Eichten:1979ah,Farhi:1980xs}, has become
increasingly difficult after the Higgs discovery, as obtaining a light
SM Higgs-like particle from a strongly-interacting sector with a
significant mass gap between the high scale physics and the weak scale
does not follow the QCD
paradigm~\cite{Donoghue:1988ed,Gasser:1983yg}. Insights from first
principle lattice calculations~\cite{Aoki:2013ldr} are far from
straightforward, in particular because the symmetry and matter content
of realistic candidate theories of electroweak type interactions is
much bigger than that of QCD.

An approach that takes the pion mass splitting as a key inspiration
realizes the light SM-like Higgs boson as a composite pseudo-Nambu
Goldstone field. The pion is governed by strong QCD effects in the
presence of small quark masses, and falls outside the perturbative
realm. The {\emph{weak}} gauging of QED, however, opens up the possibility of
using perturbative techniques within a strongly-interacting picture,
leading to perturbative predictions which are remarkably close to the
actual experimental findings.  Concretely, this utilizes the
weak gauging of QED into the 
\begin{align}
SU(2)_L\times SU(2)_R \to SU(2)_V
\end{align}
symmetry of 2-flavor QCD in the chiral limit, with symmetry breaking 
induced by the strong dynamics of QCD. 

As the axial symmetries associated with the currents
\begin{align}
J^{a,\mu}_{5}(x)
=  J^{a,\mu}_{R}(x) - J^{a,\mu}_{L}(x) 
=  \bar \Psi_R(x) \gamma^\mu  {\sigma^a\over 2}  \Psi_R(x) - \bar \Psi_L(x) \gamma^\mu  {\sigma^a\over 2}  \Psi_L(x) 
= \bar \Psi(x) \gamma^\mu \gamma_5 {\sigma^a\over 2}  \Psi(x) \; ,
\label{eq:axcur}
\end{align}
are broken, there are 3 associated massless Nambu Goldstone
fields. The associated broken symmetry transformations gives these
fields their pseudo-scalar quantum numbers. As the fundamental quark
masses can be chosen to be zero, we still have $\partial_\mu
J^{a,\mu}_5=0$.

The currents of Eq.\eqref{eq:axcur} have the right quantum numbers to
excite the pseudo-scalar state $|p,b\rangle$ with momentum $p_\mu$
from the vacuum
\begin{align}
\label{eq:axcurpion}
\langle 0 | J^{a,\mu}_{5} | p,b\rangle = i p^\mu f \delta^{ab}\,,
\end{align}
with the characteristic pion decay constant $f$ (which can be related
to the quark condensate $\langle 0 | \bar \Psi \Psi | 0 \rangle $ in
the strongly interacting vacuum). Axial current conservation then
implies $p^2=0$, \ie the {\emph{partially-conserved}} axial current
hypothesis.  Using Eq.\eqref{eq:axcurpion} we can express general
Green's functions, \eg
\begin{align}
\label{eq:pionpole}
-i\int\hbox{d}^4 x\, e^{ipx} \left\langle \text{T} \left\{ J^{a,\mu}_{5}(0) J^{b,\nu}_{5}(x) \right\} \right\rangle \stackrel{p^2\to 0}{\simeq} {p^\mu p^\nu \over p^2 + i\epsilon} f^2  \delta^{ab} \, .
\end{align} 
Under the assumption that the low energy behavior of the Greens functions is dominated by the pseudoscalar mesons (\emph{pion pole-dominance}), we can introduce
an interpolating field 
\begin{align}
J^{a,\mu}_{5}(x)= -f \partial^\mu \pi^a(x)
\end{align}
when transitioning to the strongly interacting picture. The detailed
investigation (including the strange quarks) of
Ref.~\cite{Gasser:1984gg} shows that exponentials of these
interpolating fields
\begin{align}
\label{eq:nonlsigma}
U=\exp\left( \frac{i \sigma^a \pi^a}{f} \right)
\end{align}
provide an appropriate way of capturing the dynamics in an effective
theory.  This can be traced to the particular linear and non-linear
symmetry transformation properties of
$U$~\cite{Coleman:1969sm,Callan:1969sn} under the $SU(2)_L\times
SU(2)_R\to SU(2)_V$ transformations. A suitable low energy effective
theory for the pions is then given by
\begin{alignat}{5}
{\cal{L}} &= {f^2\over 4} \tr \left[ \partial_\mu U \partial^\mu U^\dagger \right]+...\\
& = \frac{1}{2} \partial_\mu \pi^a \partial^\mu \pi^a + {\cal{O}}(\pi^4)\,,
\end{alignat}
where the ellipses refer to higher order terms in the low momentum expansion.

The presence of gauge interactions external to the strong sector and
associated with only a subgroup $U(1)_\text{QED} \subset SU(2)_V$
means that the true vacuum of the theory can be misaligned from the
the tree-level vacuum that conserves $SU(2)_V$: gauging a subgroup
means explicitly breaking the global symmetry. Treating the gauging as
as perturbative effect we can compute the effective effective one-loop
pion potential from photon
contributions~\cite{Contino:2010rs,Panico:2015jxa}, yielding
\begin{align}
V(\pi^0,\pi^+,\pi^-)\sim   C_{LR} {\sin^2(\pi/f_\pi) \over  \pi^2}\,  \pi^+\pi^- \,.
\end{align}
with $\pi^2=\sum_i (\pi^i)^2=\pi^+\pi^- +({\pi^0})^2$. The dimension 4 constant
\begin{align}
C_{LR}={3\over 16\pi^2} \int \hbox{d} p^2  \, p^2 \, \Pi_{LR}(p^2)
\end{align}
is related to an $SU(2)_L\times SU(2)_R$ correlator related to the
currents of Eq.\eqref{eq:axcur}
\begin{align}
i\int \hbox{d}^4x \,e^{ipx}\left \langle \hbox{T}\left\{ J^{3,\mu}_L(x) J^{3,\nu}_R \right\} \right\rangle = \Pi_{LR}(p^2) \left[ p^2g^{\mu\nu}-p^\mu p^\nu \right]\;.
\label{eq:pion}
\end{align}
It is straightforward to express $\Pi_{LR}$ in terms of the axial
current correlators $\Pi_{AA}$ (cf. Eq.\eqref{eq:axcur}) as well as
analogously introduced vector current correlators $\Pi_{VV}$, and
analyze its low energy behavior with Eq.\eqref{eq:pionpole},
\begin{align}
\Pi_{LR}(p^2) = \Pi_{AA}(p^2)- \Pi_{VV}(p^2) \stackrel{p^2\to 0}{\simeq}\frac{f^2}{p^2}
\end{align}
as the vector currents are not broken in the QCD vacuum.  $\Pi_{LR}$
is strictly positive for vector-like confining gauge
theories~\cite{Witten:1983ut} and the high energy behavior can be
estimated in QCD using the Weinberg sum rules~\cite{Weinberg:1967kj}
and the vector meson dominance approximation~\cite{Sakurai:1960ju},
which means that the first vector mesons dominate the phenomenology
expressed by the Weinberg sum rules. The Weinberg sum rules imply that
the high-energy behavior of $C_{LR}$ is well-behaved and the integral
converges as $\Pi_{LR}(p^2)\sim p^{-4}$ for large $p^2$.

A positive $C_{LR}$ means that the pion potential is minimized for
$\left\langle \pi^\pm\right\rangle =0$ and the vacuum is dynamically
aligned in the gauge symmetry-conserving direction; the photon stays
massless. The charged pion, however, obtains a non-zero radiative
mass~\cite{Das:1967it} turning it into a pseudo-Nambu Goldstone Boson
(pNGB)
\begin{align}
m^2_{\pi^\pm}-m^2_{\pi^0} \simeq {3\alpha_{EW}\over 4\pi} {m_\rho^2 m_{a_1}^2\over m_{a_1}^2-m_{\rho}^2} \log {m_{a_1}^2 \over m_{\rho}^2} \simeq 5.8~\text{MeV}\,,
\end{align}
where $\rho,a_1$ are the first (and dominant) vector and axial vector
resonances with masses of 770 MeV and 1260 MeV, respectively. In this
formulation the neutral pion stays massless as chiral symmetry is only
broken by the non-zero quark masses that are neglected here. Even
when these effects are included, the mass splitting between the charged
and uncharged pions is dominated by the QED radiative effects, in
good agreement with the experimentally observed splitting of 4.6 MeV.

\begin{figure}[t]
\includegraphics[width=0.4\textwidth]{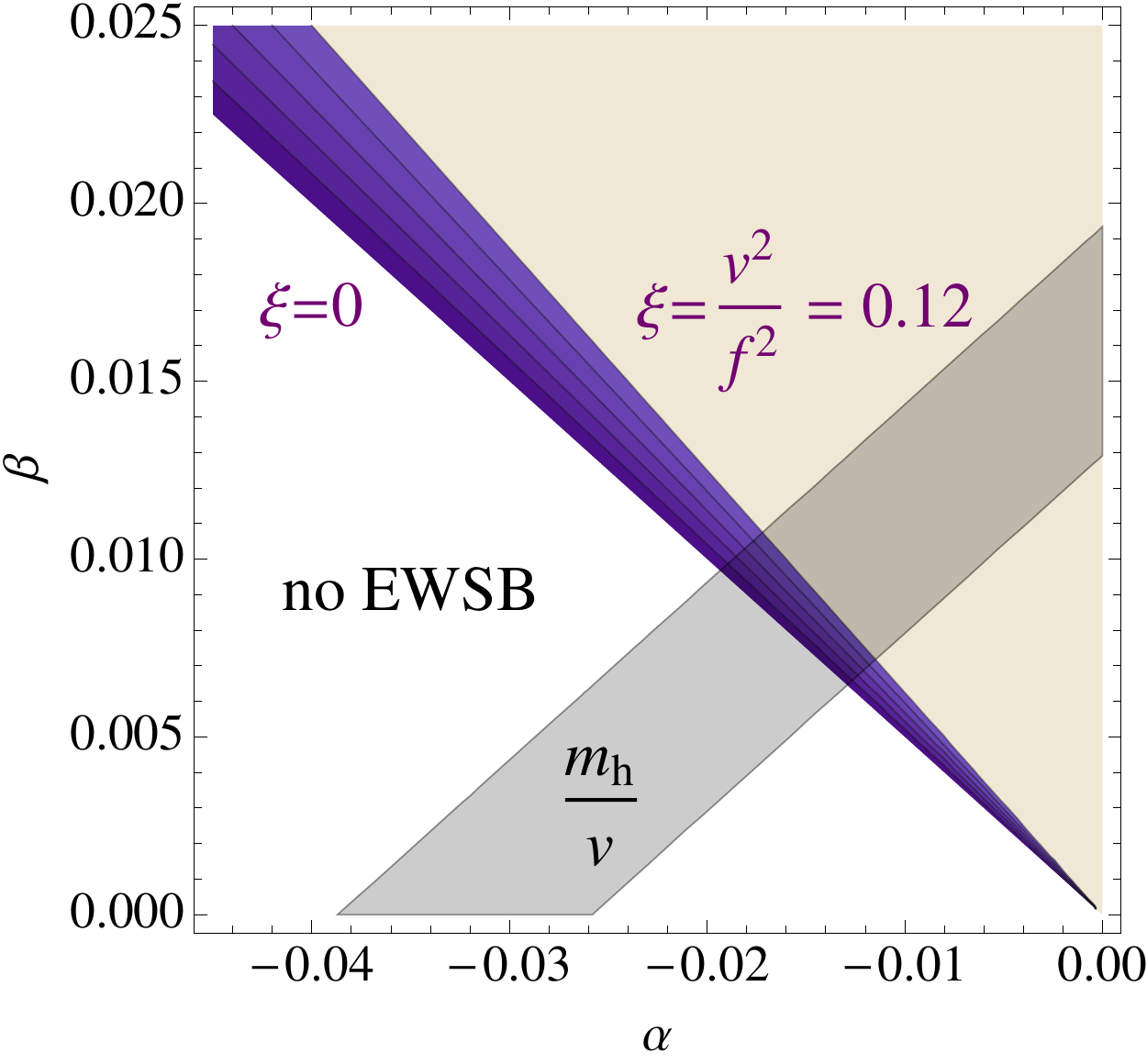}
\hspace*{0.1\textwidth}
\includegraphics[width=0.35\textwidth]{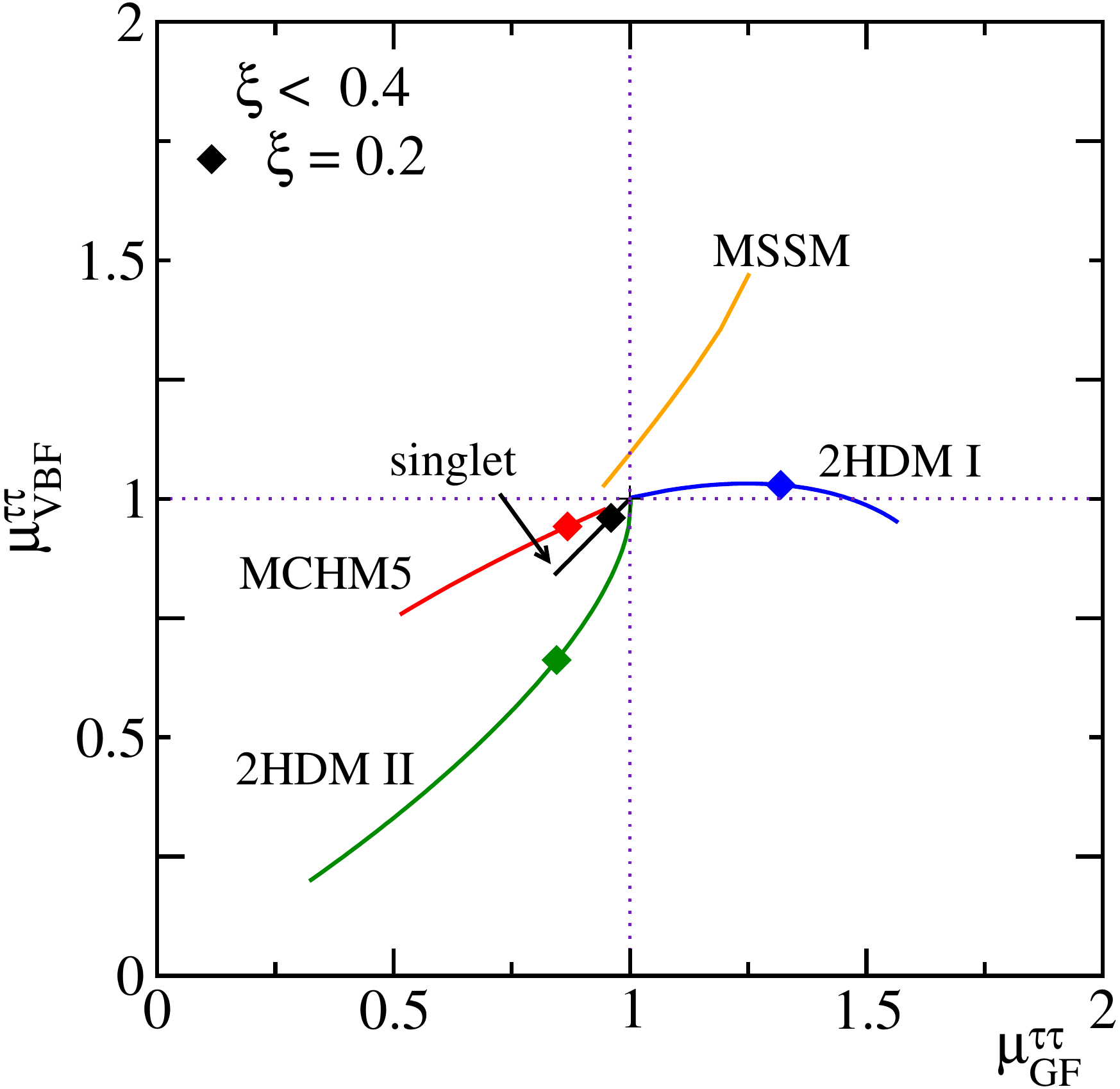}
\caption{Left: constraints on the low energy constants by a
  representative LHC Higgs measurement $\xi\lesssim 0.12$. UV models
  need to fall into the intersection of the purple and grey
  band. Figure from Ref.~\cite{DelDebbio:2017ini}. The limit of
  $\xi\lesssim 0.12$ results from a re-interpretation of the combined
  ATLAS/CMS measurements of~\cite{Khachatryan:2016vau}.  Right:
  expected effect of different models, including MCHM5, on the LHC
  signal strengths with a Higgs decay to taus. Figure from
  Ref.~\cite{Cranmer:2013hia}.}
\label{fig:letcomp}
\end{figure}

\begin{table}[!b]
\begin{tabular}{c|ccccc}
\toprule
Model & $hVV$ & $hhVV$ & $hf\bar f$ & $hhf\bar f$ & $hhh$ \\
\midrule
MCHM4 & $\sqrt{1-\xi} $& $1-2\xi $&$ \sqrt{1-\xi} $&$ \frac{\xi m_f}{v^2} $& $\sqrt{1-\xi}$  \\[2mm]
MCHM5 &$\sqrt{1-\xi} $ & $1-2\xi $ &  $\dfrac{1-2\xi}{\sqrt{1-\xi}} $   & $\frac{4 \xi m_f}{v^2} $ & $\dfrac{1-2\xi}{\sqrt{1-\xi}} $ \\
\bottomrule
\end{tabular}
\caption{Coupling modifiers for different minimal composite Higgs
  models, which are used as LHC benchmark scenarios.}
\label{tab:mchms}
\end{table}

Composite Higgs models aim at adopting the above strategy to the electroweak case.
This is not straightforward: we have to properly break electroweak
symmetry rather than only introducing parametrically small masses for
the Higgs boson as a pNGB; no massless states should remain in the spectrum. A key ingredient of the solution is partial
compositeness~\cite{Kaplan:1991dc,Contino:2006nn} of heavy fermions,
which explains the observed mass hierarchy of fermions in terms of
mixing with baryons of the strongly interacting sector through (extended) hypercolor interactions. 
Partial compositeness introduces new, vacuum-misaligning one-loop terms to the
effective potential, which then takes a
schematic form~\cite{Contino:2003ve,Agashe:2004rs,Contino:2010rs}
\begin{align}
V(h)= 4 f^4\beta \left( \sin^2 \frac{h}{f} -\xi \right)^2\,.
\end{align}
Here $f$ is the analogue of the above pion decay constant
for the electroweak symmetry breaking case and the parameter $\xi$ that
measures the radiatively-generated electroweak scale in units of $f$
\begin{align}
\xi = {v^2\over f^2} = {\alpha + 2\beta\over 4\beta}\,.
\end{align}
The dimensionless coefficients $\alpha,\beta$ are related to
combinations of two-point and four-point correlator functions of the
microscopic theory similar to Eq.\eqref{eq:pion},
see~\cite{DeGrand:2016htl,Golterman:2015zwa,Golterman:2017vdj,DelDebbio:2017ini}. Calculating
these functions using lattice simulations is possible in principle,
however, for realistic theories with an enlarged matter and symmetry
content~\cite{Ferretti:2016upr} this is unfortunately beyond the
current state of the art although significant progress has already
been achieved~\cite{Ayyar:2017qdf,Ayyar:2017uqh,Ayyar:2018zuk}.

\begin{figure}[!t]
\includegraphics[width=0.45\textwidth]{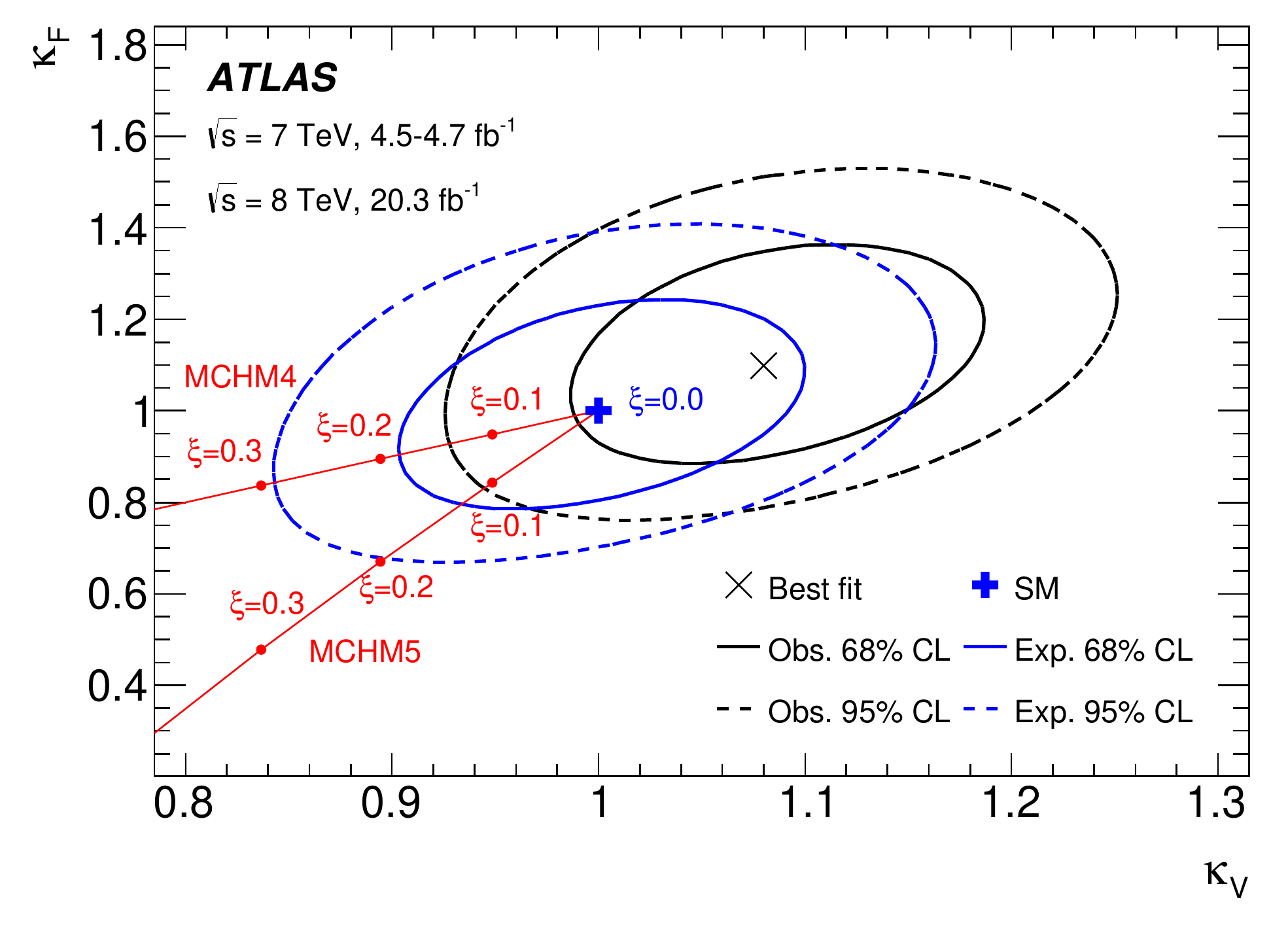}
\hspace*{0.1\textwidth}
\includegraphics[width=0.4\textwidth]{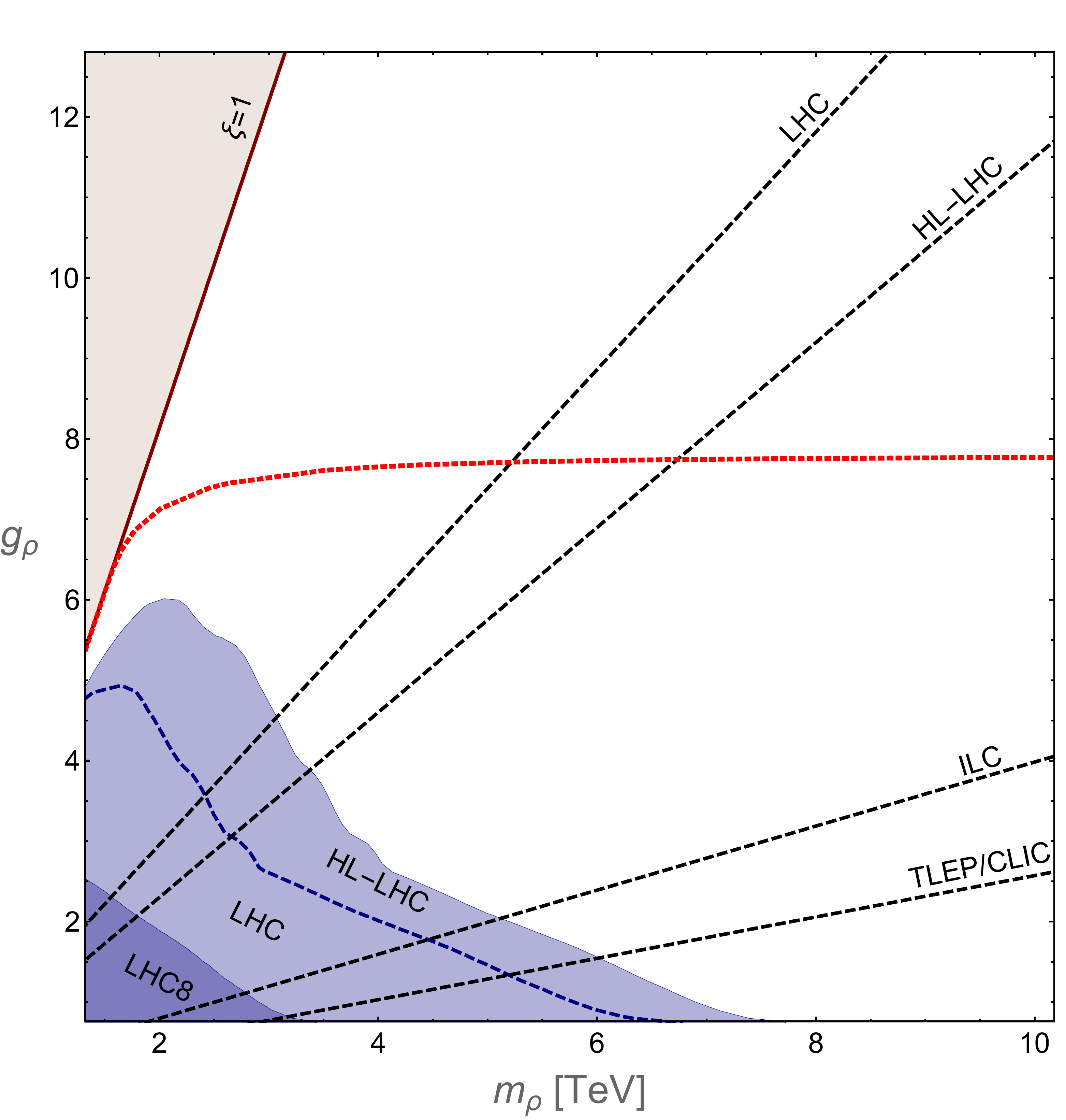}
\caption{Left: constraints on minimal composite Higgs models' coupling
  modifiers as reported by ATLAS. Figure from
  Ref.~\cite{Aad:2015pla}. Right: projected constraints on minimal
  composite Higgs models. Figure from Ref.~\cite{Thamm:2015zwa}.}
\label{fig:atlasmeasmchm}
\end{figure}

Although concrete UV completions of composite Higgs scenarios are
currently unclear (this also includes the validity of the QCD-specific
assumptions made above for the pion example), using the effective low
energy constants $\alpha,\beta$ and their relation to measurable
quantities like the Higgs mass and the Higgs couplings, constraints
can formulated which can inform future lattice investigations. In
particular, the condition for electroweak symmetry breaking requires
$\alpha+2\beta >0$ and the Higgs mass measured in units of the
electroweak vacuum expectation value correlates $\xi$ and $\beta$
\begin{align}
0.26\simeq {M_h^2\over v^2} = {V''(\langle h \rangle )\over v^2} = 32 \beta \xi (1-\xi)\,.
\end{align} 
The parameter $\xi$ parameterizes the differences in Higgs
phenomenology compared to the Standard Model. Scenarios, which are typically
adopted as benchmarks of a pNGB origin of electroweak symmetry
breaking are models based on a $SO(5)\to SO(4)$ non-linear sigma
model~\cite{Agashe:2005dk, Barbieri:2007bh,Bellazzini:2014yua},
analogous to Eq.\eqref{eq:nonlsigma}, with fermions either embedded
in the four-dimensional spinorial or five-dimensional fundamental
representation of $SO(5)$. These models are referred to as minimal
composite Higgs models MCHM4 and MCHM5. They imply different Higgs
interactions with SM fermions, as tabulated in
Tab.~\ref{tab:mchms}~\cite{Giudice:2007fh}.  A $\sim 10\%$ constraint
on $\xi$ can therefore be used to contrast predictions from specific
UV models, Fig.~\ref{fig:letcomp}. MCHM5 is a particularly
well-motivated phenomenological candidate as it includes enough
symmetry to satisfy LEP constraints~\cite{Contino:2006qr} and a range
of viable parameter choices consistent with LHC data can be
formulated~\cite{Gillioz:2012se}. In the right panel of
Fig.~\ref{fig:letcomp} we show how this specific model can be
identified based on LHC rate modifications.  In
Fig.~\ref{fig:atlasmeasmchm} we show an ATLAS interpretation of single
Higgs measurements in these models~\cite{Aad:2015pla}.

Other scenarios of strong interactions that approach the light nature
of the Higgs boson from a different angle are theories of collective
symmetry breaking, known as Little Higgs
theories~\cite{ArkaniHamed:2002qy,Csaki:2002qg}, where the Higgs is
interpreted as a pNGB that typically (but not necessarily) arises
through breaking of a product group of symmetries. These models have
been reviewed in other places in
detail~\cite{Schmaltz:2005ky,Perelstein:2005ka}. Other versions of
pNGB type Higgs physics include Twin Higgs models~\cite{Chacko:2005pe}
as well as models motivated from holography~\cite{Contino:2003ve}.
Scalar sectors in these kinds of models typically share
phenomenological similarities with the 2HDM or the Georgi-Machacek
model, as discussed in Sec.~\ref{sec:basis_weak_triplet}, because they
form representations of $SU(2)_R\times SU(2)_L$ by
construction. Current constraints from a range of existing searches
are loose, in particular when the triplets do not contribute to
electroweak symmetry breaking~\cite{DelDebbio:2017ini}.

Minimal models like MCHM4 and MCHM5 have currently no known
(non-supersymmetric) UV completion. Extended
theories~\cite{Ferretti:2013kya,Ferretti:2014qta,Cacciapaglia:2015yra,Ferretti:2016upr}
typically include other exotic states on top of top and bottom quark
partners~\cite{Pomarol:2012qf,DeSimone:2012fs,Matsedonskyi:2014mna},
such hyper-pions~\cite{Belyaev:2016ftv}, axion-like
states~\cite{Belyaev:2016ftv}, (doubly) charged Higgs
bosons~\cite{DelDebbio:2017ini} and
vectors~\cite{Marzocca:2012zn,Thamm:2015zwa}.  Searches for the heavy
particle in these models typically provide complementary information
from measurement of Higgs couplings. In a model with heavy vector
resonances of mass, $m_\rho$ and coupling $g_\rho$, the parameter
$\xi\sim g_\rho^2 v^2/m_\rho^2$ and typically limits are shown in the
right panel of Fig.~\ref{fig:atlasmeasmchm}.

\subsection{Characterizing the Higgs boson}
\label{sec:basic_char}

Understanding the properties of the SM-like Higgs boson and the
symmetry structure of the Higgs Lagrangian is the first task when we
want to interpret LHC Higgs measurements in terms of a quantum field
theory.

\subsubsection{Mass and Lifetime}
\label{sec:basic_char_mass}

The mass of the Higgs boson was the only unknown parameter of the
Standard Model prior to its discovery. Electroweak symmetry breaking
puts the Higgs mass at the center of the reliability of the
electroweak series expansion. For instance, the parameter that
measures custodial isospin is logarithmically dependent on the Higgs
mass. This together with the constraints on the $S$ parameter allowed
tight constraints to be put on the mass of the Higgs $M_h\lesssim
140~\text{GeV}$, which forced these logarithms to be small, therefore
requiring electroweak physics to follow the standard perturbative QFT
paradigm. More concrete evidence of the Higgs mass being the ``order
parameter'' of the convergence of the electroweak series expansion can
derived from the Higgs width~\cite{Goria:2011wa}: for Higgs masses approaching 
the $WW$ threshold $h\to b\bar b$ is about a factor 2
smaller than $h\to 4$~fermions. As the decay widths are related to
imaginary parts of the Higgs self-energy, the importance of 4 body
decays compared to 2 body decays can be related to 3 loop diagrams
being equally important as one loop contributions for such Higgs
masses, which signalizes a slow convergence.

\begin{figure}[t]
\includegraphics[width=0.75\textwidth]{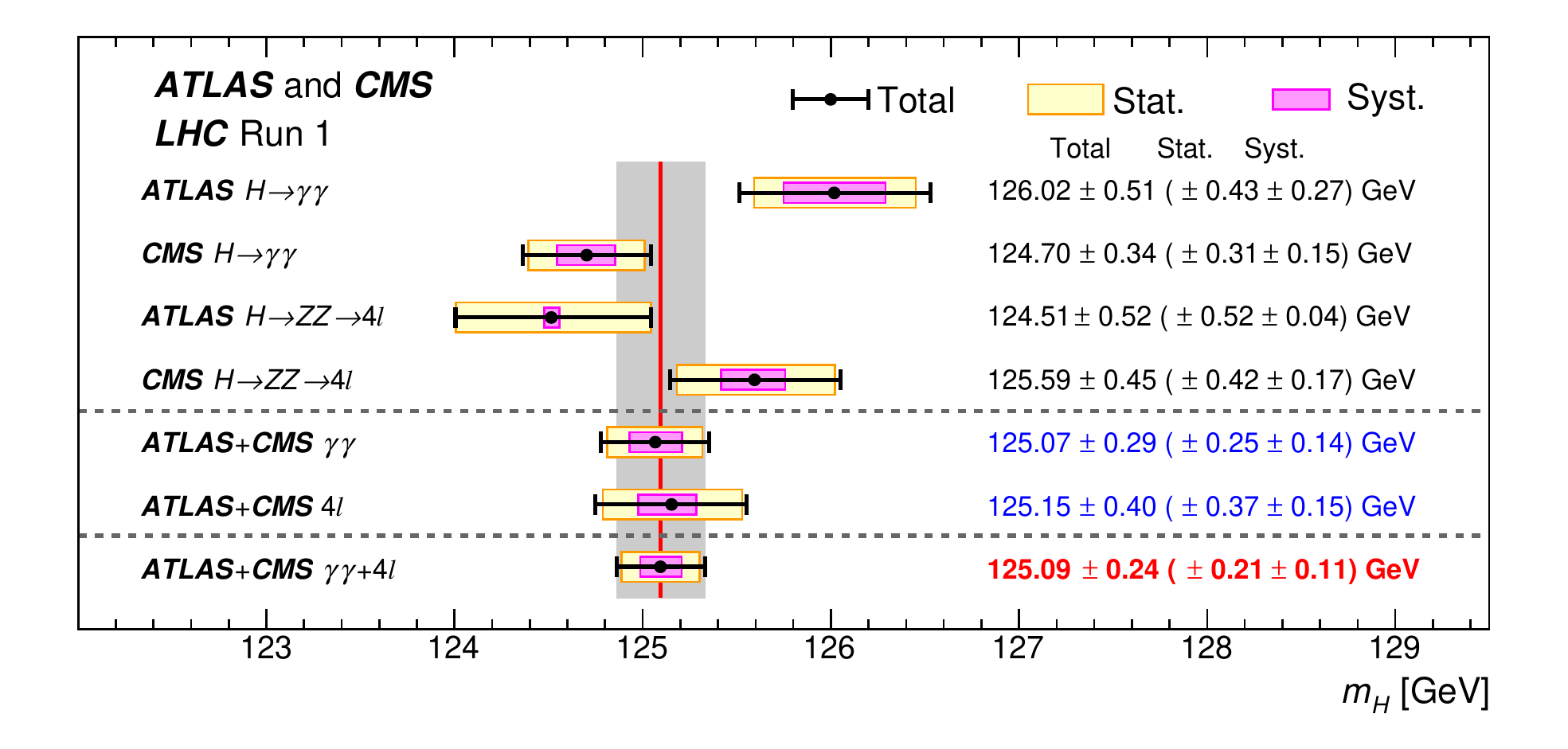}
\caption{Measurement of the Higgs mass across
  the $ZZ$ and $\gamma\gamma$ channels using the combined run 1
  datasets of ATLAS and CMS. Figure from Ref.~\cite{Aad:2015zhl}.}
\label{fig:hmassmeas} 
\end{figure}

Because the Higgs mass  reach was the limiting factor in LEP searches, a
SM-like Higgs is excluded by LEP for $M_h < 114.4$~GeV at 95\%
C.L.~\cite{Barate:2003sz}.  At the LHC the Higgs mass crucially
impacts signal yields. Hence, studying the observed signal rates as a
function of the Higgs boson mass can be used to constrain the Higgs
mass, as done in a combined fit of ATLAS and CMS
data~\cite{Aad:2015zhl} for 7~TeV and 8~TeV data. The combination of
ATLAS and CMS measurements in the $h\to \gamma \gamma$ and $h\to ZZ\to
4l$ channels gives the currently best measurement on the Higgs mass in the
Standard Model, Fig.~\ref{fig:hmassmeas},
\begin{align}
M_h=125.09 \pm0.21~\text{(stat.)} \pm 0.11~\text{(syst.)}~\text{GeV}\,.
\end{align}
The Higgs width or lifetime is a crucial parameter for BSM
searches. For instance, it can be related to dark matter through
scalar extensions as described above. With a SM value of $\Gamma_h
\sim 4~\text{MeV}$ for the 125~GeV Higgs boson, the width is much
smaller than the experimental resolution of about $1~\text{GeV}$ and
line-shape analyses similar to the $Z$ boson are not immediately
promising at the LHC. A range of proposals have been advertised to
circumvent this limitation through on-shell interference
effects~\cite{Dixon:2013haa,Dixon:2003yb,Martin:2012xc,Martin:2013ula,Coradeschi:2015tna,Campbell:2017rke}. They
rely on improved experimental systematics or on correlating different
regions of phase
space~\cite{Caola:2013yja,Campbell:2013wga,Campbell:2014gua,Englert:2014aca}.
It is worth mentioning that most of the Monte Carlo tools relevant for
Higgs and electroweak physics at the LHC rely on $\Gamma_h/M_h \ll
1$~\cite{Nowakowski:1993iu,Argyres:1995ym,Papavassiliou:1997fn,Papavassiliou:1997pb,Denner:2006ic,Passarino:2010qk}. The
naive Breit-Wigner hypothesis that is typically employed in these
simulations cannot be supported when the width becomes
large~\cite{Passarino:2013bha,Goria:2011wa}.

The basic idea for extracting the Higgs width from off-shell Higgs
production employs properties of the longitudinal
$Z$-polarizations~\cite{Kauer:2012hd,Caola:2013yja,Kauer:2013qba}. The
process 
\begin{align}
gg \rightarrow ZZ^{(*)} \to 4 l
\end{align}
contains contributions from the Higgs resonance, but also
contributions independent of the resonance. From the Breit-Wigner
propagator, $1/[(m_{4l}^2-M_h^2)^2 + M_h^2 \Gamma_h^2]$,  we know that
the Higgs amplitude squared and the interference with the continuum
background on and above the resonance give very different information
on the Higgs width.  The analysis idea is that by measuring the $gg
\rightarrow 4l$ rate above and on the resonance we can extract
information about the Higgs width.  
The advantage of the decay $h \to ZZ^*$ is that
 at $8$~TeV approximately
$15\%$ of the total cross section has $m_{4l}>140~\gev$, offering
enough off-shell Higgs events for a measurement.  ATLAS and CMS have
used this technique to place limits on the Higgs
width~\cite{Khachatryan:2014iha,Aad:2015xua},
\begin{align}
\Gamma_h < (4-5)\Gamma_h^\text{SM} \; .
\end{align}
However, the Higgs coupling to gluons is loop-induced and can have a
non-trivial dependence on all masses
involved~\cite{Gainer:2014hha,Englert:2014ffa,Cacciapaglia:2014rla,Ghezzi:2014qpa,Azatov:2014jga,Buschmann:2014sia,Logan:2014ppa},
and so the extraction the Higgs width requires us to make (often implicit)
assumptions that all interactions are those of the Standard Model.

\subsubsection{Spin}
\label{sec:basic_char_spin}

An obvious first step after the Higgs boson discovery was to establish
the new scalar's spin, or constrain the spin-associated interaction terms
of an effective Lagrangian. The effective interactions that were used
to constrain such couplings or to contrast with the SM Higgs in simple
hypothesis tests need to be understood as straw-man proposals. They are
not motivated by actual models or theories, and they typically cannot
be understood in terms of renormalizable theories. Interaction terms
are typically considered for spin $j=0,1,2$ for Higgs
\textsl{imposters} $h, A, Y_\mu,G_{\mu\nu}$. We focus on the $CP$-even
case as an example, for which the following interaction terms can be
considered~\cite{Englert:2010ud,Artoisenet:2013puc,Maltoni:2013sma,Gainer:2014hha}).

\noindent Spin 0:
\begin{align}
{\cal{L}}_{j=0} &=
  g_1^{(0)} \,h V_\mu V^\mu 
  -\frac{g_2^{(0)}}{4} \; h \, V_{\mu\nu}V^{\mu\nu}
  -\frac{g_3^{(0)}}{4} \; A \, V_{\mu\nu}\widetilde V^{\mu\nu} 
  -\frac{g_4^{(0)}}{4} \; h \, G_{\mu\nu}G^{\mu\nu}
  -\frac{g_5^{(0)}}{4} \; A \, G_{\mu\nu}\widetilde G^{\mu\nu} \; ,
\label{eq:spinzero}
\end{align}

\noindent Spin 1:
\begin{align}
{\cal{L}}_{j=1} &= 
    i g_1^{(1)} (W^+_{\mu\nu}W^{- \mu} - W^-_{\mu\nu}W^{+\mu}) \; Y^{(e)\nu} 
   + i g_2^{(1)} W^+_\mu W^-_\nu \; Y^{(e)\mu\nu}  
\notag \\
  &+ g_3^{(1)} \epsilon^{\mu\nu\rho\sigma}(W^+_\mu  \overleftrightarrow{\partial}_\rho W^-_\nu)Y^{(e)}_\sigma   
   + ig_4^{(1)} W^+_{\sigma\mu}W^{- \mu\nu} Y^{(e)\sigma}_{\nu} 
\notag \\
  &- g_5^{(1)} W^+_\mu W^-_\nu(\partial^\mu Y^{(o)\nu} + \partial^\nu Y^{(o)\mu}) 
   + ig_6^{(1)} W^+_\mu W^-_\nu \widetilde{Y}^{(o){\mu \nu}}  
   + ig_7^{(1)} W^+_{\sigma\mu} W^{-\mu\nu} \widetilde{Y}^{(o)\sigma}_\nu
\notag \\
  &+ g_8^{(1)} \epsilon^{\mu\nu\rho\sigma} Y^{(e)}_\mu  Z_\nu (\partial_\rho Z_\sigma)  
   + g_9^{(1)} Y^{(o)}_\mu  (\partial_\nu Z^\mu )Z^\nu \;.
\label{eq:spinone}
\end{align}

\noindent Spin 2:
\begin{alignat}{5}
{\cal{L}}_{j=2} =
  -g_1^{(2)} \; G_{\mu\nu}T^{\mu\nu}_V
  -g_2^{(2)} \; G_{\mu\nu}T^{\mu\nu}_G
  -g_3^{(2)} \; G_{\mu\nu}T^{\mu\nu}_f \; ,
  \label{eq:spintwo}
\end{alignat}
where the energy momentum-tensor parts of the fermions, gluons and
vector bosons can be found in~\cite{Han:1998sg,Hagiwara:2008jb}. All
ad-hoc parameters $g_{i}^{(j)}$ are free couplings associated with the
(higher-dimensional) interactions.

\begin{figure}[t]
\centering
\includegraphics[width=0.4\textwidth]{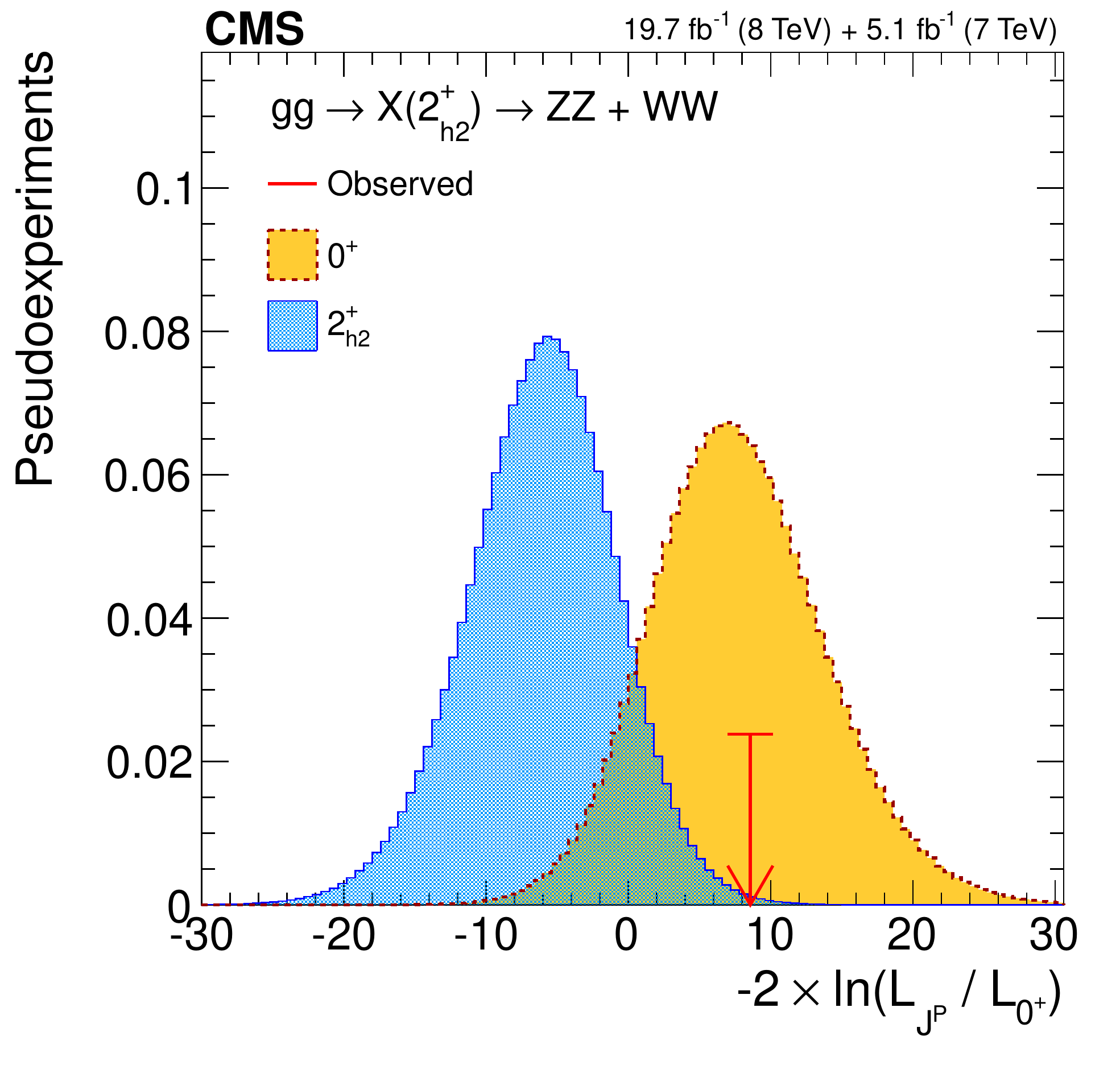}
\hspace*{0.1\textwidth}
\includegraphics[width=0.40\textwidth]{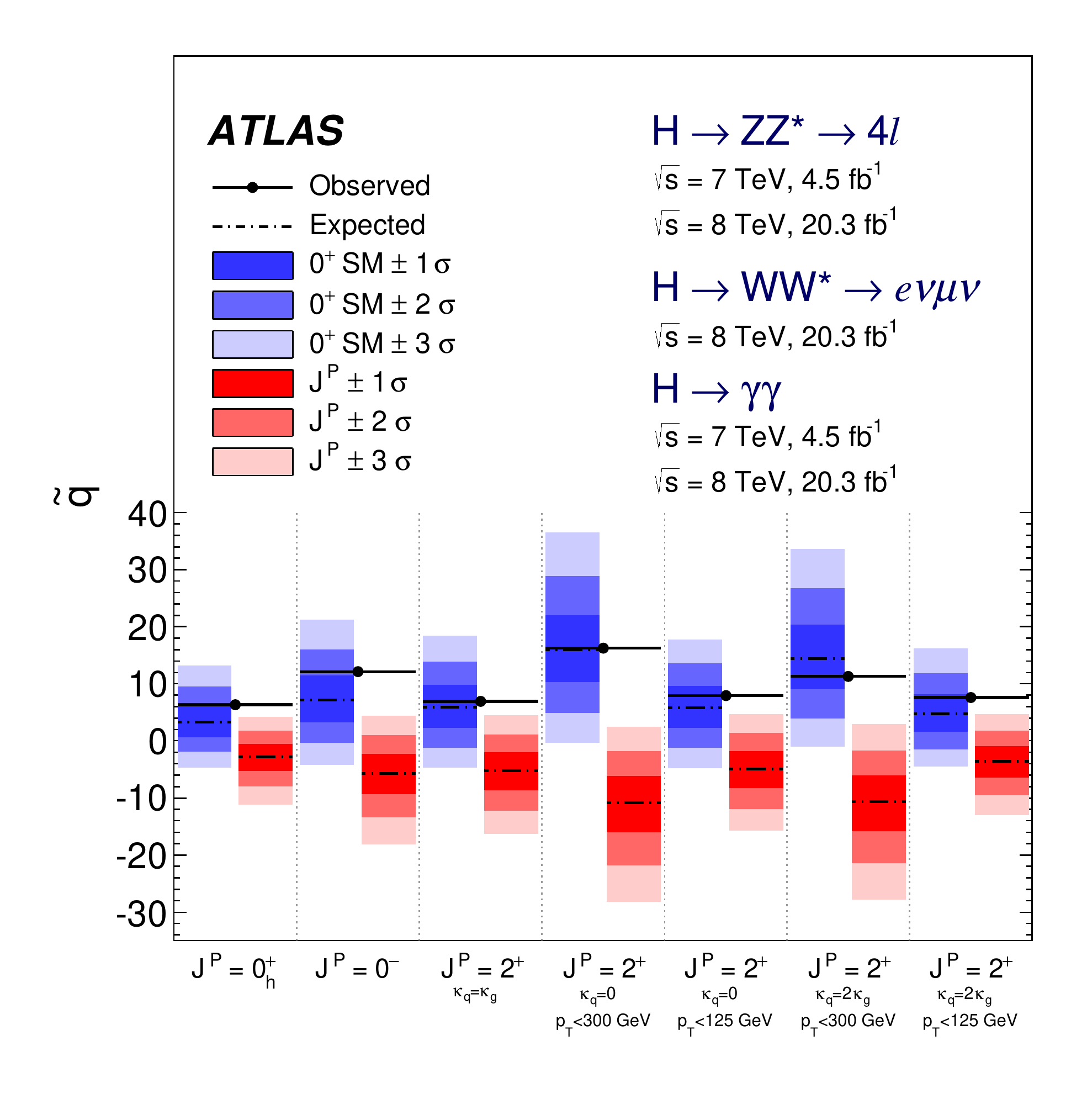}
\caption{Left: comparison of $CP$-even vs. $CP$-odd scalar
  hypotheses. Figure from Ref.~\cite{Chatrchyan:2012jja}. Right:
  comparisons of various alternate hypothesis against the SM Higgs
  boson. Figure from Ref.~\cite{Aad:2015mxa}.}
\label{fig:higgs_spin_res}
\end{figure}

\begin{figure}[b!]
\centering
\includegraphics[width=0.45\textwidth]{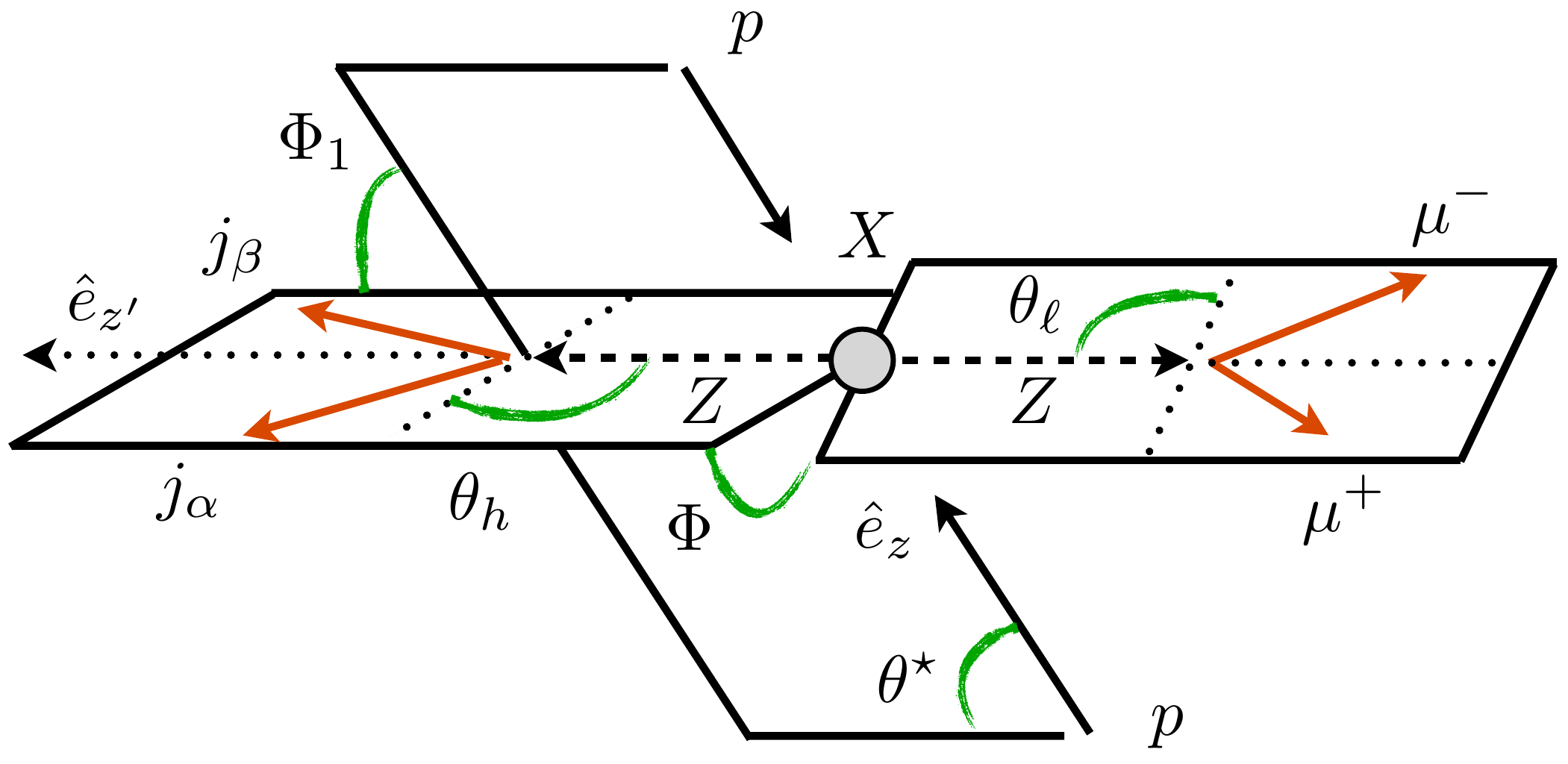}
\caption{Angles sensitive to the spin and $CP$ characterization
  program. Figure from Ref.~\cite{Englert:2010ud}.}
\label{fig:higgs_spin}
\end{figure}

There are various ways to constrain the above interaction terms,
including the classic study of angular decay distribution shapes in
various
channels~\cite{DellAquila:1985mtb,DellAquila:1985jin,Collins:1977iv,Nelson:1986ki,Miller:2001bi,Choi:2002jk,DeRujula:2010ys,Gao:2010qx,Choi:2012yg,Bolognesi:2012mm,Englert:2017bme}. They
are inspired by flavor physics, where they are used to determine the
quantum numbers of different mesons also decaying to four
leptons~\cite{Trueman:1978kh,Cabibbo:1965zzb}.  We study the angular
correlations in the decays
\begin{align}
X\to WW,ZZ \; ,
\end{align}
where $X$ denotes the Higgs or its imposter. It is easiest to
illustrate for the clean $X \to ZZ$ case.  We start by defining
\begin{align}
{\vec{p}}_{Z_h} = {\vec{p}}_{\alpha} + {\vec{p}}_{\beta}
\qqquad
{\vec{p}}_{Z_l} = {\vec{p}}_{+} + {\vec{p}}_{-}
\qqquad 
{\vec{p}}_{X}= {\vec{p}}_{Z_h} + {\vec{p}}_{Z_l} \; .
\end{align}
In addition, we denote the normalized unit vector along the beam axis measured in the $X$ rest frame as $\hat e_z$, and the unit
vector along the $ZZ$ decay axis in the $X$ rest frame as $\hat e_{z'}$. The $CP$- and spin-sensitive angles of Fig.~\ref{fig:higgs_spin} are then defined as 
\begin{alignat}{8}
\cos\theta_h &=   { {\vec{p}}_\alpha \cdot {\vec{p}}_{X}  \over  \sqrt{  {\vec{p}}^{\,2}_\alpha\, {\vec{p}}^{\,2}_{X}    }} \Bigg|_{Z_h} 
&\qquad 
\cos\theta_l &=   { {\vec{p}}_- \cdot {\vec{p}}_{X}  \over \sqrt{  {\vec{p}}^{\,2}_-\, {\vec{p}}^{\,2}_{X}    }} \Bigg|_{Z_l} 
&\qquad 
\cos\theta^\star &=  { {\vec{p}}_{Z_l} \cdot {\hat e_{z}}  \over \sqrt{  {\vec{p}}^{\,2}_{Z_l} }} \Bigg|_{X}
\notag \\
\cos\Phi_1 &=  { (\hat e_z\times \hat e_{z'})\cdot ({\vec{p}}_\alpha \times {\vec{p}}_\beta) \over 
 \sqrt{( {\vec{p}}_\alpha \times {\vec{p}}_\beta )^2} } \Bigg|_X 
&\qquad 
\cos\Phi &= { ( {\vec{p}}_\alpha \times {\vec{p}}_\beta ) \cdot ( {\vec{p}}_- \times {\vec{p}}_{+} ) \over
\sqrt{( {\vec{p}}_\alpha \times {\vec{p}}_\beta )^{\,2}\,  ( {\vec{p}}_- \times {\vec{p}}_{+} )^2 }}  \Bigg|_X \; .
\end{alignat}
The subscripts indicate the reference system in which the angles are
evaluated. Note that the Collins-Soper angle $\theta^\star$ can also
be evaluated for decays $X \to \gamma \gamma$ and can be included to a
combined analysis of spin properties.  These angles were used to
constrain the $CP$ and spin character of the 125~GeV resonance for
various alternate hypotheses~\cite{Khachatryan:2014kca,Aad:2015mxa}
early on in the LHC Higgs program. One CMS example is shown in
Fig.~\ref{fig:higgs_spin_res}, together with an overview by ATLAS,
comparing different $CP$ hypotheses for the most interesting scalar
case.

As an alternative to the angular analysis, we can use integrated cross
sections and kinematic
features~\cite{Englert:2010ud,Ellis:2012xd,Ellis:2012jv,Ellis:2013ywa,Englert:2012xt,Englert:2013opa}. The
latter contrasts SM Higgs interactions with alternative interactions
described by higher-dimensional operators. This means that in addition
to angular correlations we can include observables reflecting a
possible momentum dependence of these interactions, as discussed in
more detail in Sec.~\ref{sec:basic_char_cp}.  As the total cross
section is  sensitive to this momentum dependence, it already
provides hints on the spin properties of the associated
couplings. However, these kind of measurements should be understood as
a hypothesis test rather than a determination of the Higgs quantum
numbers.

\subsubsection{$CP$-properties}
\label{sec:basic_char_cp}

The three discrete symmetries consistent with Lorentz invariance and a
hermitian Hamiltonian are charge conjugation
($C$), parity ($P$), and time reversal ($T$). They are symmetry
properties of the underlying Lagrangian and act on a complex scalar
field like the Higgs field as (see e.g.~\cite{Peskin:1995ev} for a pedagogical introduction)
\begin{align}
 C \, \phi(t, \vec x) \, C^{-1} = \eta_C \; \phi^*(t, \vec x) \qquad 
 P \, \phi(t,\vec x) \, P^{-1}  = \eta_P \; \phi(t, -\vec x) \qquad
 T \, \phi(t, \vec x) \, T^{-1} = \phi(-t, \vec x),
\end{align}
The phases $\eta_j$ define the symmetry properties of the field
$\phi$.  $C$ and $P$ are unitary transformations, while $T$ is
anti-unitary~\cite{Wigner:102713}, implying that a phase can be rotated away. Switching to
momentum--spin space the transformations read
\begin{align}
  C \ket{\phi (p,s)} = \ket{\phi^* (p, s)} \qquad 
  P \ket{\phi (p,s)} = \eta_\phi \ket{\phi(-p, s)} \qquad 
  T \ket{\phi (p,s)} = \bra{\phi (-p, -s)} \; ,
\label{eq:dev_cpt}
\end{align}
where $\eta_\phi$ is the intrinsic parity of the field. It is
convenient to omit the exchange of initial and final states and define
the `naive time
reversal'~\cite{DeRujula:1978bz,Korner:1980np,Hagiwara:1984hi,Brandenburg:1995nv,Hagiwara:2007sz,Brehmer:2017lrt}
\begin{align}
  \tnaive \ket{\phi (p, s)} = \ket{\phi (-p, -s)} \; .
\label{eq:def_that}
\end{align}
Specific observables can be defined such that they directly probe the
transformation properties of the underlying transition amplitude.  In
general, they are functions of 4-momenta, spins, flavors, and charges
of the initial and final states. A $U$-even or $U$-odd observable is
defined as $O ( U \ket{i} \to U \ket{f} ) = \pm \, O \, ( \ket{i} \to
\ket{f} )$ for the four transformations defined above. The problem
with these observables is that they do not access the properties of
the underlying Lagrangian. For this purpose, we can define genuine
$U$-odd observables through a vanishing expectation value in a
$U$-symmetric Lagrangian~\cite{Brehmer:2017lrt}.  If the probability distribution of the
initial states is $U$-symmetric, the second definition is slightly
weaker and we find
\begin{align}
O ( U \ket{i} \to U \ket{f} ) = - \,  O \, ( \ket{i} \to \ket{f} ) \quad \text{(odd)} 
\;\;
\Longrightarrow
\; \;
\braket{O}_{\lag =U \lag U^{-1}} = 0 \quad \text{(genuine odd)} \; .
\label{eq:odd_relation}
\end{align}
Note that this approach to symmetry properties of the underlying
Lagrangian shifts the focus away from the corresponding phases of the
Higgs field. Instead, they will allow us to study the symmetry
properties of (parts of) the Higgs Lagrangian.

For LHC analyses it is useful to relate $CP$ transformation properties
to the $\tnaive$ transformation~\cite{Han:2009ra,Christensen:2010pf}.
If we require
\begin{itemize}
\item the theory is $CPT$-invariant~\cite{Luders:1954zz,Lueders:1992dq};
\item the phase space and the initial state distributions are invariant under $\tnaive$;
\item and there are no re-scattering effects,
\end{itemize}
a finite expectation value of a genuine $\tnaive$-odd observable $O$
indicates a $CP$-violating theory~\cite{Atwood:2000tu}. This implies that
whenever we cannot construct genuine $CP$ observables, we can analyze
genuine $\tnaive$ observables instead.

\begin{table}[t!]
\centering
  \begin{tabular}{lccccr}
  \toprule 
  Observable & Theory & Re-scattering  & Symmetry argument && Prediction \\
  \midrule
  \multirow{4}{*}{$CP$-odd, $\tnaive$-odd} & \multirow{2}{*}{$CP$-symmetric} & no & $CP$ and $\tnaive$: symmetric $\sigma_\text{int}$, odd $O$ & $\Rightarrow$& $\langle O \rangle=0$ \\ 
  & & yes &  $CP$: symmetric $\sigma_\text{int}$, odd $O$ & $\Rightarrow$& $\langle O \rangle=0$\\ 
  \cmidrule{2-6}
  & \multirow{2}{*}{$CP$-violating} & no & \multicolumn{3}{r}{can have $\langle O \rangle \neq 0$} \\ 
  & & yes & \multicolumn{3}{r}{can have $\langle O \rangle \neq 0$}  \\ 
  \midrule
  \multirow{4}{*}{$CP$-odd, $\tnaive$-even} & \multirow{2}{*}{$CP$-symmetric} & no &  $CP$: symmetric $\sigma_\text{int}$, odd $O$ & $\Rightarrow$& $\langle O \rangle=0$\\ 
  & & yes &  $CP$: symmetric $\sigma_\text{int}$, odd $O$ & $\Rightarrow$& $\langle O \rangle=0$\\ 
  \cmidrule{2-6}
  & \multirow{2}{*}{$CP$-violating} & no &  $\tnaive$: anti-symmetric $\sigma_\text{int}$, even $O$& $\Rightarrow$& $\langle O \rangle=0$\\ 
  & & yes & \multicolumn{3}{r}{can have $\langle O \rangle \neq 0$} \\ 
  \bottomrule
  \end{tabular}
  \caption{Predictions for $CP$-odd observables $O$ based on the
    theory's symmetries and the observable's transformation properties
    under $\tnaive$. In all cases we assume that the initial state or
    its probability distribution is symmetric under both $CP$ and
    $\tnaive$. }
\label{tab:tn-odd}
\end{table}

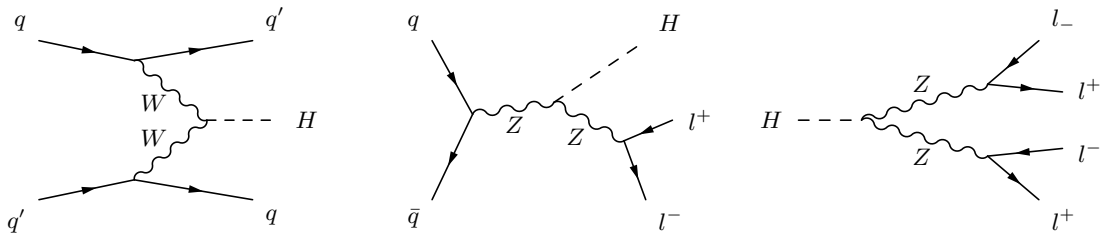
\begin{figure}[b!]
  \begin{center}
    \begin{fmfgraph*}(100,60)
      \fmfset{arrow_len}{2mm}
      \fmfleft{i2,i1}
      \fmfright{o3,o2,o1}
      \fmf{fermion,width=0.6,lab.side=left,tension=2}{i1,v1}
      \fmf{fermion,width=0.6,lab.side=left,tension=1}{v1,o1}
      \fmf{photon,width=0.6,lab.side=right,label=$W$,label.dist=2,tension=1}{v1,v3}
      \fmf{dashes,width=0.6,lab.side=left,tension=2}{v3,o2}
      \fmf{photon,width=0.6,lab.side=right,label=$W$,label.dist=2,tension=1}{v3,v2}
      \fmf{fermion,width=0.6,lab.side=left,tension=2}{i2,v2}
      \fmf{fermion,width=0.6,lab.side=left,tension=1}{v2,o3}
      \fmflabel{$q$}{i1}
      \fmflabel{$q'$}{i2}
      \fmflabel{$q'$}{o1}
      \fmflabel{$q$}{o3}
      \fmflabel{$H$}{o2}
    \end{fmfgraph*}
    \hspace*{15mm}
    \begin{fmfgraph*}(100,60)
      \fmfset{arrow_len}{2mm}
      \fmfleft{i2,i1}
      \fmfright{o3,o2,o1}
      \fmf{fermion,width=0.6,lab.side=left,tension=2}{i1,v1}
      \fmf{fermion,width=0.6,lab.side=left,tension=2}{v1,i2}
      \fmf{photon,width=0.6,lab.side=right,label=$Z$,label.dist=3,tension=2}{v1,v2}
      \fmf{dashes,width=0.6,lab.side=left,tension=1}{v2,o1}
      \fmf{photon,width=0.6,lab.side=right,label=$Z$,label.dist=3,tension=1}{v2,v3}
      \fmf{fermion,width=0.6,lab.side=left,tension=1}{o2,v3}
      \fmf{fermion,width=0.6,lab.side=left,tension=1}{v3,o3}
      \fmflabel{$q$}{i1}
      \fmflabel{$\bar{q}$}{i2}
      \fmflabel{$l^+$}{o2}
      \fmflabel{$l^-$}{o3}
      \fmflabel{$H$}{o1}
    \end{fmfgraph*}
    \hspace*{15mm}
    \begin{fmfgraph*}(100,60)
      \fmfset{arrow_len}{2mm}
      \fmfleft{i1}
      \fmfright{o4,o3,o2,o1}
      \fmf{dashes,width=0.6,lab.side=left,tension=4}{i1,v1}
      \fmf{photon,width=0.6,lab.side=left,label=$Z$,label.dist=4,tension=1}{v1,v2}
      \fmf{fermion,width=0.6,lab.side=left,tension=1}{o1,v2}
      \fmf{fermion,width=0.6,lab.side=left,tension=1}{v2,o2}
      \fmf{photon,width=0.6,lab.side=left,label=$Z$,label.dist=4,tension=1}{v3,v1}
      \fmf{fermion,width=0.6,lab.side=left,tension=1}{o3,v3}
      \fmf{fermion,width=0.6,lab.side=left,tension=1}{v3,o4}
      \fmflabel{$H$}{i1}
      \fmflabel{$l_-$}{o1}
      \fmflabel{$l^+$}{o2}
      \fmflabel{$l^-$}{o3}
      \fmflabel{$l^+$}{o4}
    \end{fmfgraph*}
  \end{center}
  \caption{Feynman diagrams describing three processes useful for
    analyzing Higgs CP properties: WBF Higgs production, associated
    $Zh$ production, and $h \to 4 l$ decays.}
  \label{fig:feyn_char}
\end{figure}

A shortcoming of hadron colliders is that the spins, the light-quark
flavors, and the quark charges in the initial or final states are
typically not measurable. This means that all observables have to be
constructed as functions of the 4-momenta. In many Higgs production
and decay processes we can reconstruct four independent external
4-momenta and combine them into ten scalar products.  Four scalar
products correspond to the masses of the external particles and the
remaining six describe the kinematics of the hard interaction. Scalar
products are $P$-even and by definition also $\tnaive$-even. It can be
shown that four of the scalar product are $C$-even and two of them are
$C$-odd.  In addition to the scalar products, there exists exactly one
$P$-odd and $\tnaive$-odd observable, constructed from four
independent 4-momenta and the Levi-Civita
tensor~\cite{Plehn:2001nj,Hankele:2006ma}.

In terms of $CP$, this means that for a wide class of Higgs processes
we can define three $CP$-odd observables, two $\tnaive$-even and one
$\tnaive$-odd. For processes where the initial state is at least
approximately $CP$-even and $\tnaive$-even, like $q\bar{q}$
scattering, these two types of $CP$-odd observables are:
\begin{enumerate}
\item $CP$-odd and $\tnaive$-odd: the observable is also genuine
  $CP$-odd and genuine $\tnaive$-odd. In a $CP$-symmetric theory its
  expectation value vanishes, implying that a non-zero expectation
  value requires $CP$ violation, regardless of re-scattering.  The
  different cases are illustrated in the upper half of
  Table~\ref{tab:tn-odd}.
\item $CP$-odd and $\tnaive$-even: the observable is also genuine
  $CP$-odd, so in a $CP$-symmetric theory its expectation value
  vanishes. In the lower half of Table~\ref{tab:tn-odd} we show the
  different scenarios: if the theory is $CP$-violating, the
  corresponding expectation values do not vanish.  If we ignore
  re-scattering, the theory also appears $\tnaive$-violating, but the
  expectation value of the $\tnaive$-even observable combined with an
  anti-symmetric amplitude will still vanish.  However, in the
  presence of re-scattering or another complex phase, this unwanted
  condition from the $\tnaive$ symmetry vanishes, and the expectation
  for $\braket{O}$ matches the symmetry of the theory.
\end{enumerate}

\begin{figure}[t]
\centering
\includegraphics[width=0.45\textwidth]{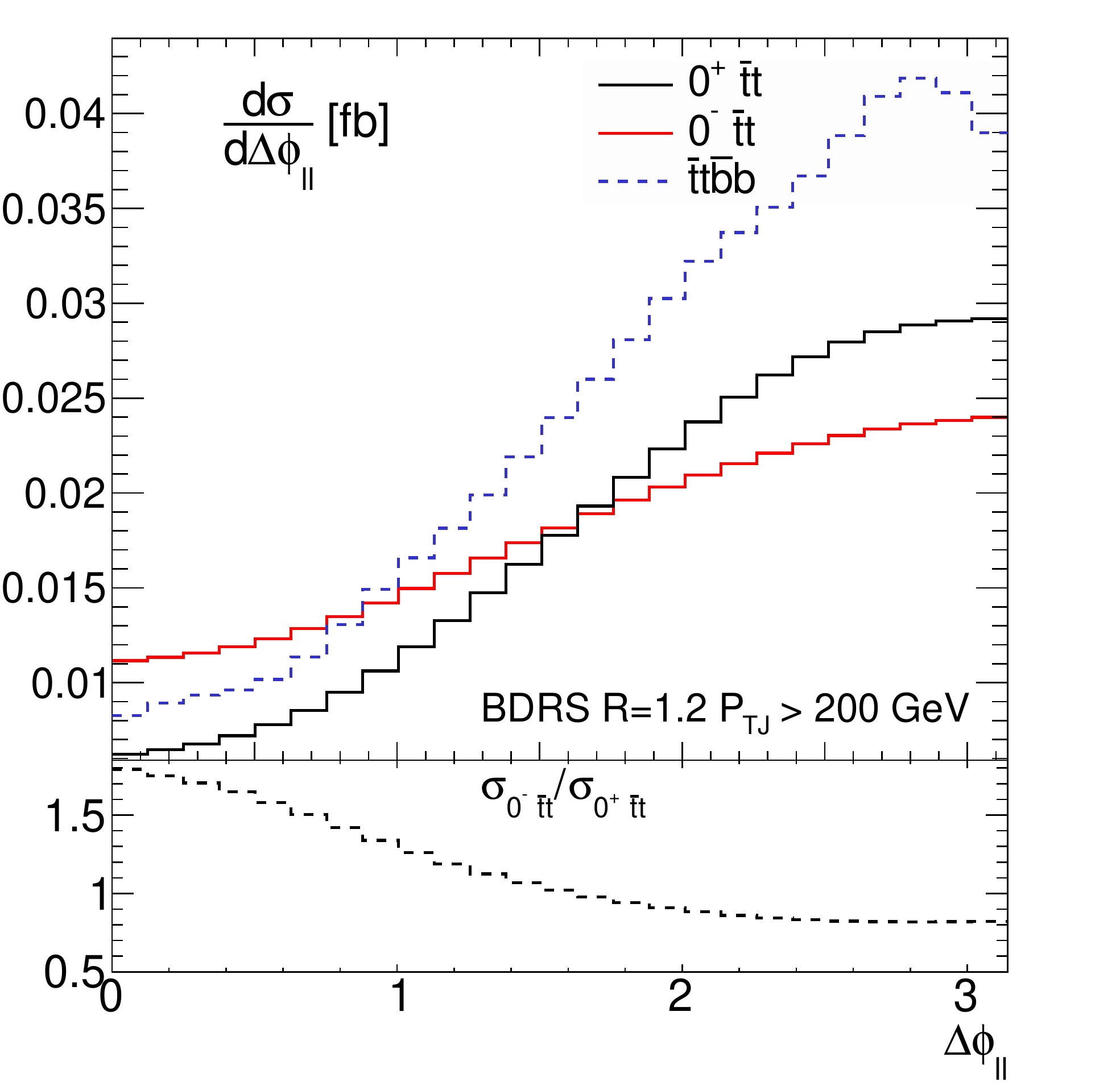}
\includegraphics[width=0.45\textwidth]{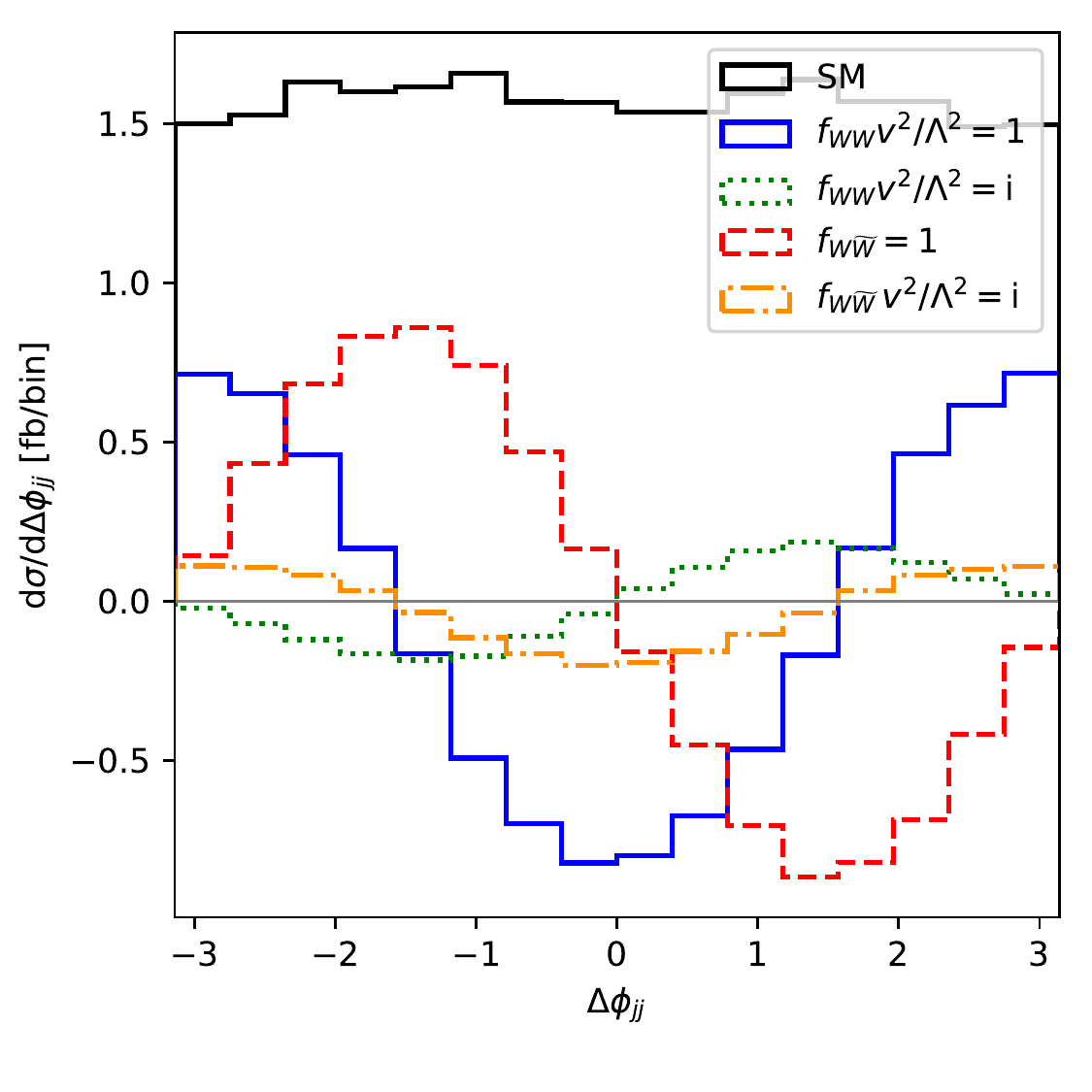}
\caption{Left: azimuthal angle between the top decay leptons for
  $t\bar{t}h$ production for $CP$-even and $CP$-odd couplings, figure
  from Ref.~\cite{Buckley:2015vsa}.  Right: signed azimuthal angle between
  the tagging jets in WBF Higgs production for $CP$-even and $CP$-odd
  couplings with real and (hypothetically) imaginary Wilson
  coefficients, figure from Ref.~\cite{Brehmer:2017lrt}.}
\label{fig:higgs_cp}
\end{figure}

At the LHC, we typically study the $C$, $P$, and $T$ symmetries of the
Higgs interactions described by a given Lagrangian. To this end, the
Higgs-fermion and Higgs-gauge Lagrangians can be studied
separately. In the  case of the Higgs interactions with fermions, like the
top quark or the tau lepton, we can easily extend the usual $CP$-even
structure given in Eq.\eqref{eq:fermyuk} to include a $P$-odd and
$CP$-odd operator~\cite{Buckley:2015vsa},
\begin{align}
-y_u \overline{Q}_L \tilde{\phi} u_R 
\longrightarrow 
-y_u \overline{Q}_L \tilde{\phi} ( c_\alpha \one + i s_\alpha \gamma_5 ) u_R 
\end{align}
The mixing angle $\alpha$ ensures that the combined size of the Yukawa
coupling reproduces the observed $CP$-insensitive Higgs production
rate.  This modification of the dimension-4 Lagrangian is not
suppressed explicitly, but a suppression by some new physics scale
will typically be induced by the mixing angle. The angular correlation is 
especially well-suited for the determination of the $CP$-properties of
the top Yukawa coupling is the azimuthal angle between the two
top-decay leptons in the boosted regime. The different signal and
background distributions are shown in the left panel of
Fig.~\ref{fig:higgs_cp}. Similarly arguments can be made for using the
$\tau$ polarization of the decay $H \to \tau \tau$ to determine the
$CP$-properties of the SM-like Higgs~\cite{Berge:2008wi,Berge:2011ij}.

The situation is different for the Higgs-gauge Lagrangian, where
$CP$-violation only occurs at dimension six.  The general effective
theory framework describing Higgs physics at the LHC is introduced in
Sec.~\ref{sec:basis_eft}.  In the effective Higgs-gauge Lagrangian,
$CP$-odd couplings are described by the operators
\begin{align}
\ope_{B\tilde{B}}  
&= -\frac{g'^2}{4} \, (\phi^\dagger \phi) \, \widetilde{B}_{\mu\nu} \, B^{\mu\nu} 
\equiv -\frac{g'^2}{4} \, (\phi^\dagger \phi) \, \epsilon_{\mu \nu \rho \sigma} B^{\rho\sigma} \, B^{\mu\nu} \notag \\
\ope_{W\widetilde{W}}  
&= -\frac{g^2}{4} \, (\phi^\dagger \phi) \, \widetilde{W}^k_{\mu\nu} \, W^{\mu\nu\, k} 
\equiv -\frac{g^2}{4} \, (\phi^\dagger \phi) \, \epsilon_{\mu \nu \rho \sigma} W^{\rho\sigma\, k} \, W^{\mu\nu\, k} \; .
\label{eq:op_cpodd}
\end{align}
With the Levi-Civita tensor, these operators are $C$-conserving,
$P$-violating and $CP$-violating.  The hard process which will allow
us to determine the $CP$-properties of the gauge-Higgs interactions
needs to include the $WWh$ or $ZZh$ coupling. As illustrated in
Fig.~\ref{fig:feyn_char}, this points to WBF Higgs production,
associated $Vh$ production, or $h \to 4 l$ decays.  The leading
observable to test $CP$-violation is the one $P$-odd and $\tnaive$-odd
observable constructed with the help of the Levi-Civita tensor. For
the three different hard processes it corresponds to the signed
azimuthal angle between the WBF tagging jets, the azimuthal angle
between the leptons from the $Z$-decay, or the decay plane
correlations in the Higgs decays. In the right panel of
Fig.~\ref{fig:higgs_cp} we show the distribution of the signed
azimuthal angle between the WBF tagging jets for different Wilson
coefficients defined in Eq.\eqref{eq:op_cpodd}. The interference with
the SM rate is indeed anti-symmetric for the $CP$-odd
coupling. Imaginary Wilson-like coefficients are not part of a
consistent EFT prescription, but they can be used to illustrate the
effect of re-scattering.

\begin{figure}[t]
\centering
\includegraphics[width=0.70\textwidth]{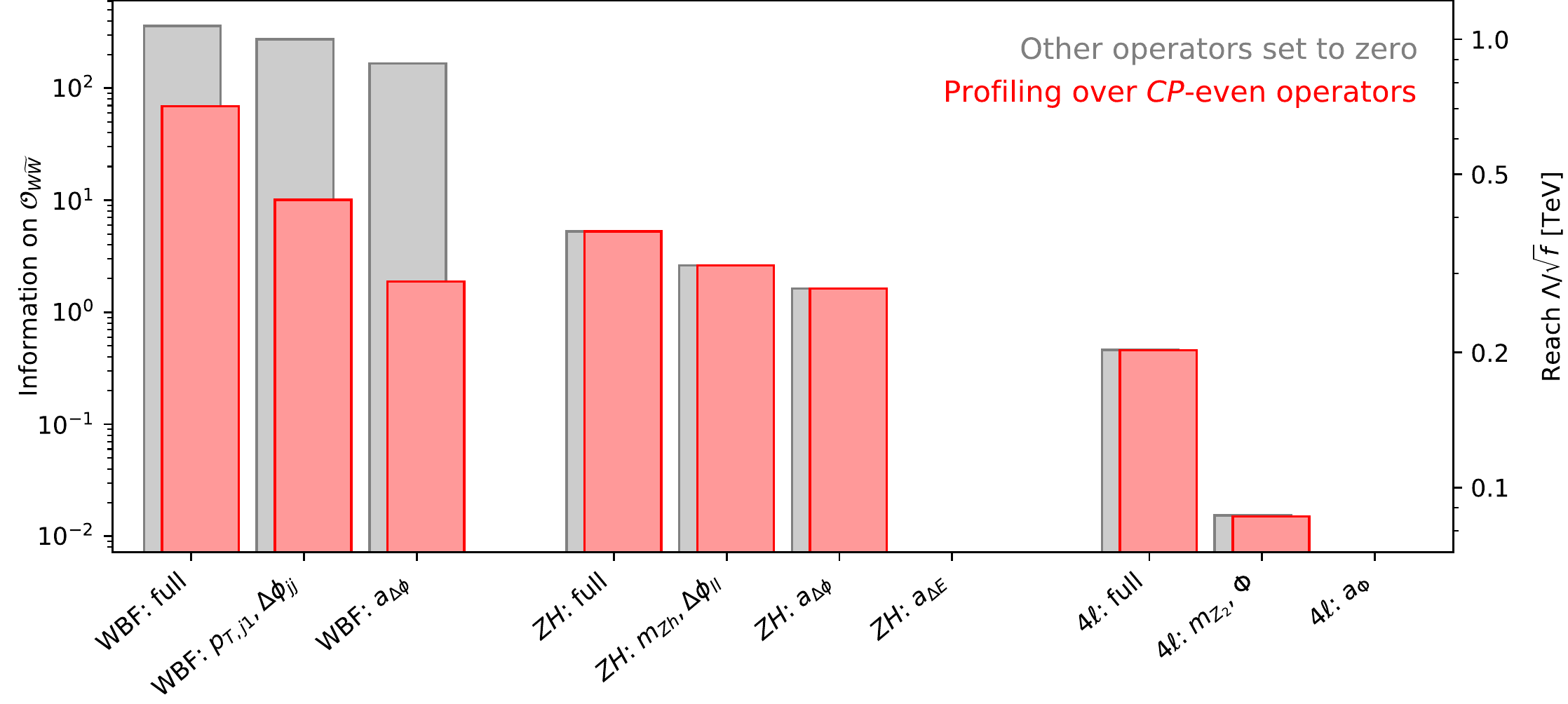}
\caption{Comparison of WBF Higgs production, $Zh$ production, and $h
  \to 4l$ decays, in terms of their reach for $CP$-violating Higgs
  couplings.  The reach is defined in terms of the effective
  interaction ${f/\Lambda^2} \ope_{W\widetilde{W}}$ defined in
  Eq. \ref{eq:op_cpodd}.  The reach is computed from matching an order
  one new physics effect at new physics scale $\tilde{\Lambda}$:
  $1/\tilde{\Lambda}^2=f/\Lambda^2$.}  Figure from
Ref.~\cite{Brehmer:2017lrt}.
\label{fig:cp_comp}
\end{figure}

When we use appropriate kinematic distributions to search for effects
from the specific $CP$-odd operators given in Eq.\eqref{eq:op_cpodd},
the performance of angular correlations can be enhanced by including
momenta in the $CP$-observables. The reason is that the dimension-6
operators have a new Lorentz structure, which comes with powers of
4-momenta in the Feynman rules. In Fig.~\ref{fig:cp_comp} we show a
comparison of the reach of the three leading signatures for
$CP$-violation in the Higgs-gauge sector. The corresponding Feynman
diagrams are shown in Fig.~\ref{fig:feyn_char}. The reach in terms of
the new physics scale $\Lambda$ is shown on the right axis; it is
computed using the Fisher information and information
geometry~\cite{Brehmer:2016nyr,Brehmer:2017lrt}. For each channel we
compute the information available from the full phase space after
basic acceptance cuts, from a combination of momentum-sensitive and
$CP$-sensitive observables, and from the (optimal) $CP$-sensitive
observable. For WBF Higgs production the latter is defined as an
asymmetry in the signed azimuthal angle between the tagging jets,
\begin{align}
a_{\Delta \Phi_{jj}}
= \frac{d\sigma(\Delta \phi_{jj})-d\sigma(-\Delta \phi_{jj})}
       {d\sigma(\Delta \phi_{jj})+d\sigma(-\Delta \phi_{jj})} \; .
\end{align}
For the other two channels, we use the appropriate combination of
final state momenta instead.  While WBF Higgs production and $Zh$
production indicate a similar reach, the Higgs decay channels are
clearly less powerful.

Once we have convinced ourselves through these dedicated measurements
that $CP$ is not a good symmetry of the gauge-Higgs sector, the
operators in Eq.\eqref{eq:op_cpodd} need to be included in global
Higgs analyses described in Sec.~\ref{sec:basis_eft} and
Sec.~\ref{sec:exp_global}.

\subsection{Effective field theory}
\label{sec:basis_eft}

In the absence of any ideas guiding new physics searches in the Higgs
sector, we need a flexible and theoretically sound framework to
describe and interpret precision Higgs studies. The goal of such a data-driven
framework is to parameterize experimental limits on physics beyond the
Standard Model --- always assuming that in the presence of convincing
anomalies we will aim for an interpretation in terms of fundamental
physics models. 

\subsubsection{Coupling modifications}
\label{sec:basis_couplings}

Historically~\cite{Zeppenfeld:2000td,Duhrssen:2004cv,Lafaye:2009vr},
indirect new physics effects on the couplings of the SM-like Higgs
boson are parameterized as simple coupling modifiers
\begin{alignat}{6}
 g_x &= g_x^\text{SM} \; (1 + \Delta_x) \notag \\
 g_{g,\gamma} &
= g_{g,\gamma}^\text{SM} \; (1 + \Delta_{g,\gamma}^\text{SM} + \Delta_{g,\gamma} ) 
\equiv g_{g,\gamma}^\text{SM} \; (1 + \Delta_{g,\gamma}^\text{SM+NP} ) \; .
\label{eq:def_delta}
\end{alignat} 
Analyses of this kind appeared soon after the Higgs
discovery~\cite{Klute:2012pu,Plehn:2012iz,Belanger:2013xza,Giardino:2013bma,Bechtle:2014ewa,Cheung:2014noa}.
Modifications of the tree-level couplings appearing in the Standard
Model loops are encoded into $\Delta_{\gamma,g}^\text{SM}$ while extra
possible new physics contributions are included as
$\Delta_{\gamma,g}$.  In general, the loop-induced couplings have a
non-trivial momentum dependence; for the Higgs couplings measurement
we assume that all three external momenta are fixed, reducing the
coupling to a single number. In terms of a Lagrangian we can write
this hypothesis as~\cite{Corbett:2015ksa}
\begin{alignat}{5}
\lag 
= \lag_\text{SM} 
&+ \Delta_W \; g M_W h \; W^\mu W_\mu
+ \Delta_Z \; \frac{g}{2 c_W} M_Z h \; Z^\mu Z_\mu
- \sum_{\tau,b,t} \Delta_f \; 
\frac{m_f}{v} h \left( \bar{f}_R f_L + \text{h.c.} \right) \notag \\
&+  \Delta_g F_G \; \frac{h}{v} \; G_{\mu\nu}G^{\mu\nu}
+  \Delta_\gamma F_A \; \frac{h}{v} \; A_{\mu\nu}A^{\mu\nu} 
+ \text{invisible decays} \; ,
\label{eq:lag_delta}
\end{alignat} 
where
\begin{align}
F_G =&-{\alpha_s\over 16\pi}\sum_q F_{1/2}(\tau_q)\notag  \\
F_A =&-{\alpha\over 8 \pi}\biggl\{\sum_i N_{ci}e_i^2F_{1/2}(\tau_i)+F_0(\tau_W)\biggr\}\, .
\end{align}
This Lagrangian is based on all SM-like structures at mass dimension
four, but with shifts in the numerical values of the Higgs couplings.
These modified couplings obviously affect all loop-induced Higgs
couplings. In addition, it includes new higher-dimensional operators
coupling the Higgs to photons and gluons.  They arise from potential
new particles in the loop and are normalized to their Standard Model
values $F_G$ and $F_A$. In the limit of heavy top masses these
normalization constants read $F_G^{(\infty)} \to \alpha_s/(12
\pi)$. In principle, we could include a more complete and
gauge-invariant set of these kind of operators, but in all other cases
they would compete with finite tree-level values and are
phenomenologically less relevant.  An alternative notation for the
same framework is~\cite{Duhrssen:2004cv}
\begin{align}
\kappa_x \equiv 1 + \Delta_x \; .
\label{eq:def_kappa}
\end{align}
The same way that the modified couplings approach is only valid for
SM-like couplings, $\Delta_x \ll 1$, we have to ensure $\kappa_x \sim
1$. Moreover, unlike the unique description of
Eq.\eqref{eq:def_delta}, in the $\kappa$-framework there exist
several ways to treat modified loop-induced couplings which have to
specified case by case.  A possible Higgs decay to invisible states
can be described by a wide variety of Lagrangian terms, but the
experimental results are most easily quoted in terms of an invisible
branching ratio.

The immediate benefit of the coupling modifications framework is that
it defines a way to combine many Higgs rate measurements with a wide
variety of initial states and final states. An open question is the
treatment of the Higgs width.

First, we can keep the width as a free parameter. For illustration
purpose, we assume that all Higgs couplings scale universally with
$\Delta_x$.  If we allow for an unobserved contribution to the total
Higgs width the observable LHC rates then scale like
\begin{align}
\dfrac{\sigma \times \text{BR} }{ [\sigma \times \text{BR}]_{\text{SM}} } 
=& \dfrac{g_p^2 g_d^2}{(g_p^2 g_d^2)_\text{SM}}
   \dfrac{\Gamma_h^\text{SM}}{\Gamma_h} 
= (1+\Delta_x)^4 \; \dfrac{\Gamma_h^\text{SM}}{\Gamma_h^\text{obs} + \Gamma_h^\text{unobs}} 
= \dfrac{(1+\Delta_x)^2}{1+\text{BR}_\text{unobs}/\text{BR}_\text{obs}}
\stackrel{\Delta_x \to -1}{\longrightarrow} 0
\label{eq:sigbrfit}
\end{align}
This relation defines a non-trivial scaling of the Higgs rates with
the modified couplings in the presence of additional decay modes.

Alternatively, we can use the sum of the partial width from the
observed Higgs channels as a lower limit on the physical Higgs widths,
\begin{align}
\Gamma_h > \sum_{j \; \text{observed}} \Gamma(h \to jj)\; .
\label{eq:higgs_width}
\end{align}
One practical problem with this approach is that the charm Yukawa
coupling is not small on the scale of the precision of the different
LHC measurements. Given that to date the LHC has no sensitivity to
this parameter in its SM range, we can use a flavor symmetry to
identify $\Delta_t \equiv \Delta_c$. Higgs couplings to the light
$u,d,s$ quarks are strongly constrained by the inclusive Higgs
production rate, because they lead to tree-level production $q\bar{q}
\to h$.  Obviously, so-called invisible Higgs decays are
perfectly measurable and contribute to Eq.\eqref{eq:sigbrfit}.

An upper limit on the Higgs width comes from the unitarity constraints
on $WW \to WW$ scattering. For this process, the Standard Model value
of the $WWH$ coupling saturates the unitarity limit. A new physics
contribution to this coupling should therefore drive it away from the
SM value, $\Delta_W < 0$. Inside the Higgs coupling analysis this
defines an upper limit on the Higgs width.

In practice, many global Higgs analyses promote Eq.\eqref{eq:higgs_width}
to a definition of the Higgs width. This implies that we assume that,
up to small corrections, we have observed all relevant Higgs couplings
with the corresponding production and decay processes.

In spite of its great success during the LHC Run~I,
this framework has three distinct problems:
\begin{enumerate}
\item It is not renormalizable, which means that we cannot include
  electroweak precision predictions;
\item It only describes modifications of total rates and cannot be
  used to interpret kinematic distributions;
\item It does not allow us to combine Higgs measurements with other
  precision measurements.
\end{enumerate}
%

\subsubsection{Linear realization}
\label{sec:basis_eft_linear}

From flavor physics we know how to describe effects from
heavy particles on SM-like observables using effective field
theory~\cite{Buchmuller:1985jz,Grzadkowski:2010es}. In this framework
a Lagrangian is defined by its particle content and its
symmetries. The expansion parameter is the dimensionality of the
individual operators, or inverse powers of a large matching scale.
For example, we can construct a $SU(2)_L\times U(1)_Y$-symmetric Higgs
Lagrangian including the usual Higgs-Goldstone doublet $\phi$ defined
in Eq.\eqref{eq:def_phi}. Truncated to dimension six the
corresponding Lagrangian has the general
form~\cite{Hagiwara:1993ck,Grzadkowski:2010es}
\begin{align}
\lag = \sum_x \frac{f_x}{\Lambda^2} \; \ope_x \; ,
\label{eq:def_f}
\end{align}
where $\Lambda \gg v$ is the scale of the assumed UV complete model. 
It is constructed to be  invariant under the full gauge symmetry of the Standard Model,
but is not fully renormalizable or unitary. Obviously, we can extend this
approach to the full SM Lagrangian, defining the general Standard
Model Effective Field Theory
(SMEFT)~\cite{GonzalezGarcia:1999fq,Contino:2013kra,Englert:2014uua,deBlas:2018tjm}\footnote{For a
  comprehensive review of effective field theories of the Higgs sector
  we recommend Ref.~\cite{Brivio:2017vri}.}. Obviously, the full set
of SMEFT operators is constrained by a huge number of
measurements. Therefore, the question becomes which of these operators
affect Higgs observables at the LHC and which of them are not already
constrained at a level that we can neglect their impact on LHC
measurements. As the Run-II measurements become more precise, it
becomes necessary to include increasing numbers of operators in the fits.  The power of the SMEFT 
is that interdependent constraints arise from precision EW measurements, from di-boson production,
and from Higgs production and decay. The current level of precision requires the inclusion of
not only bosonic operators contributing to Higgs and gauge boson interactions, but also with fermion operators 
modifying the coupling of the $Z$ and $W$ to the fermions when analyzing Higgs properties.

The minimum independent set of dimension-6 operators with the SM
particle content (including the Higgs boson as an $SU(2)_L$ doublet)
and compatible with the SM gauge symmetries as well as baryon number
conservation contains 59 operators, up to flavor and Hermitian
conjugation~\cite{Grzadkowski:2010es}. 
 We start by imposing $C$ and $P$
invariance and employing for the bosonic sector the classical
non-minimal set of dimension-6 operators in the
Hagiwara-Ishihara-Szalapski-Zeppenfeld
conventions~\cite{Hagiwara:1993ck}. In this case, the operators
contributing to the Higgs interactions with gauge bosons are
\begin{alignat}{9}
 \ope_{GG} &= \phi^\dagger \phi \; G_{\mu\nu}^a G^{a\mu\nu}  
& \ope_{WW} &= \phi^{\dagger} \hat{W}_{\mu \nu} \hat{W}^{\mu \nu} \phi  
& \ope_{BB} &= \phi^{\dagger} \hat{B}_{\mu \nu} \hat{B}^{\mu \nu} \phi 
\notag \\
 \ope_{BW} &=  \phi^{\dagger} \hat{B}_{\mu \nu} \hat{W}^{\mu \nu} \phi 
& \ope_W &= (D_{\mu} \phi)^{\dagger}  \hat{W}^{\mu \nu}  (D_{\nu} \phi) 
& \ope_B &=  (D_{\mu} \phi)^{\dagger}  \hat{B}^{\mu \nu}  (D_{\nu} \phi)
\notag \\
 \ope_{\phi,1} &=  \left ( D_\mu \phi \right)^\dagger \phi\  \phi^\dagger
                  \left ( D^\mu \phi \right ) \qquad 
& \ope_{\phi,2} &= \frac{1}{2} \partial^\mu\left ( \phi^\dagger \phi \right)
                            \partial_\mu\left ( \phi^\dagger \phi \right) \qquad
& \ope_{\phi,4} &= \left ( D_\mu \phi \right)^\dagger \left(D^\mu\phi \right)
                 \left(\phi^\dagger\phi \right ) 
\notag \\
 \ope_{\phi,3} &= ( \phi^\dagger \phi )^3
\; .
\label{eq:eff}  
\end{alignat}
The Higgs doublet covariant derivative is $D_\mu\phi=
\left(\partial_\mu+ i g' B_\mu/2 + i g \sigma^a W^a_\mu/2 \right)\phi
$, the hatted field strengths are $\hat{B}_{\mu \nu} = i g' B_{\mu
  \nu}/2$ and $\hat{W}_{\mu\nu} = i g\sigma^a W^a_{\mu\nu}/2$.  The
$SU(2)_L$ and $U(1)_Y$ gauge couplings are $g$ and $g^\prime$. 
The operator  $\ope_{\phi,3}$ contributes to the Higgs self-coupling and does not
contribute to  single Higgs production at tree level, while  $\ope_{\phi,4}$  can be eliminated using the equations of motion.

In the gauge-Higgs sector, the  basis operators important for Higgs fits are
\begin{alignat}{5}
\lag_\text{eff}^{hVV} = &
- \frac{\alpha_s }{8 \pi} \frac{f_{GG}}{\Lambda^2} \ope_{GG}  
+ \frac{f_{BB}}{\Lambda^2} \ope_{BB} 
+ \frac{f_{WW}}{\Lambda^2} \ope_{WW} 
+ \frac{f_{BW}}{\Lambda^2} \ope_{BW} 
+ \frac{f_B}{\Lambda^2} \ope_B 
+ \frac{f_W}{\Lambda^2} \ope_W 
+ \frac{f_{\phi,1}}{\Lambda^2} \ope_{\phi,1} 
+ \frac{f_{\phi,2}}{\Lambda^2} \ope_{\phi,2} \; .
\label{eq:ourleff}
\end{alignat}
They give rise to a variety of dimension-5 Higgs interactions with SM
gauge boson pairs~\cite{Hagiwara:1993qt},
\begin{alignat}{5}
\lag^{hVV} 
&= g_{hgg} \; h G^a_{\mu\nu} G^{a\mu\nu} 
+  g_{h \gamma \gamma} \; h A_{\mu \nu} A^{\mu \nu}
+ g^{(1)}_{h Z \gamma} \; A_{\mu \nu} Z^{\mu} \partial^{\nu} h 
+  g^{(2)}_{h Z \gamma} \; h A_{\mu \nu} Z^{\mu \nu} \notag \\
&+ g^{(1)}_{h Z Z}  \; Z_{\mu \nu} Z^{\mu} \partial^{\nu} h 
+  g^{(2)}_{h Z Z}  \; h Z_{\mu \nu} Z^{\mu \nu} 
+  g^{(3)}_{h Z Z}  \; h Z_\mu Z^\mu \notag \\
&+ g^{(1)}_{h W W}  \; \left (W^+_{\mu \nu} W^{- \, \mu} \partial^{\nu} h 
                            +\text{h.c.} \right) 
+  g^{(2)}_{h W W}  \; h W^+_{\mu \nu} W^{- \, \mu \nu} 
+  g^{(3)}_{h W W}  \; h W^+_{\mu} W^{- \, \mu} \; ,
\label{eq:lhvv}
\end{alignat}
with the usual field strengths. In this form we see that there are two
classes of higher-dimensional operators, those leading to effects
scaling with $v/\Lambda$ and those leading to effects scaling with
$E/\Lambda$. While the former can be constrained by total rate
measurements, the latter describe modifications of kinematic
distributions, like for example transverse momenta.

The effective couplings are related to
the Wilson coefficients from Eq.\eqref{eq:ourleff} through
\begin{alignat}{5}
g_{hgg} &=
         -\frac{\alpha_s}{8 \pi} \frac{f_{GG} v}{\Lambda^2} 
& g^{(1)}_{h Z \gamma} &= \frac{g^2 v}{2 \Lambda^2} \; \frac{s_W (f_W - f_B) }{2 c_w} \notag \\
g_{h \gamma \gamma} &= - \frac{g^2 v s_W^2}{ 2\Lambda^2} \; \frac{f_{BB} + f_{WW} - f_{BW}}{2} 
& g^{(2)}_{h Z \gamma} &= \frac{g^2 v}{2 \Lambda^2} \; \frac{s_W}{c_W} \; \frac{2 s_W^2 f_{BB} - 2 c_W^2 f_{WW} + (c_W^2 - s_W^2) f_{BW}}{2} 
\label{eq:eft_couplings1} 
\end{alignat}
and
\begin{alignat}{5}
g^{(1)}_{h Z Z} &= \frac{g^2 v}{ 2\Lambda^2} \; \frac{c_W^2 f_W + s_W^2 f_B}{2 c_W^2} 
&g^{(1)}_{h W W} &= \frac{g^2 v}{2\Lambda^2} \; \frac{f_W}{2} \notag \\
g^{(2)}_{h Z Z} &= - \frac{g^2 v}{2\Lambda^2} \; \frac{s_w^4 f_{BB} +c_W^4 f_{WW} + s_W^2 c_W^2 f_{BW}}{2 c_W^2} 
&g^{(2)}_{h W W} &= - \frac{g^2 v }{2\Lambda^2} \; f_{WW} \notag \\
g^{(3)}_{h Z Z} &= M_Z^2 (\sqrt{2} G_F)^{1/2} \left(1+\frac{v^2}{4\Lambda^2} (f_{\phi,1} - 2 f_{\phi,2}) \right) \qqquad 
&g^{(3)}_{h W W} &= M_W^2(\sqrt{2} G_F)^{1/2} \left(1-\frac{v^2}{4 \Lambda^2} (f_{\phi,1} + 2 f_{\phi,2} ) \right) \; .
\label{eq:eft_couplings2} 
\end{alignat}

Gauge boson couplings from spontaneous symmetry breaking, just as
Yukawa couplings, violate the Appelquist--Carazzone decoupling
theorem~\cite{Appelquist:1974tg}. This means that the Higgs couplings to photons
and gluons are suppressed by $1/v$.  New physics generally gives rise
to dimension-6 operators suppressed by $1/\Lambda^2$, leading to Higgs
coupling strengths to photons and gluons scaling like $v/\Lambda^2$.

Some of the above Higgs operators are strongly constrained by the
electroweak precision data discussed in Sec.~\ref{sec:basis_sm}.  For
example, two gauge-Higgs operators contribute to the oblique
parameters as
\begin{align}
\Delta T = \frac{1}{\alpha} \; \frac{f_{\phi,1} v^2}{2 \Lambda^2} \qqqquad 
\Delta S = \frac{e^2}{\alpha} \; \frac{f_{BW} v^2}{\Lambda^2} \; .
\end{align}
Even a generous limit of the order $\Delta T \lesssim 0.1$ can be
translated into a new physics scale $\Lambda \gtrsim 6$~TeV for
$f_{\phi,1} \approx 1$, considerably smaller than what we can expect
from the Higgs analysis.   
In specific UV complete models, the $f_i$ are generated with
some specific scaling.  Assuming $f_i\sim1$ corresponds to the assumption
that the high scale physics is weakly interacting.  In Ref.~\cite{Arzt:1994gp}, the EFT operators are classified as operators that can be
generated at tree level (and hence $f_i\sim$1), and operators that are generated
at loop level ($f_i\sim g^2$).  Alternatively, in models where the effective theory
results from high-scale strong interactions, such as the SILH model~\cite{Giudice:2007fh}, the scaling
of the $f_i$ depends on the strong coupling and heavy resonance scale.

There are many ways to expand this set of operators to a  complete basis.  
We
follow Ref.~\cite{Corbett:2012dm}.
Other
basis choices include the general Warsaw basis~\cite{Grzadkowski:2010es} or the
model-inspired SILH set~\cite{Giudice:2007fh}. Translations between actual
bases are simple linear transformations~\cite{Falkowski:2015wza}.

For the Higgs interactions with fermion
pairs~\cite{Grzadkowski:2010es} the problem is a lack of appropriate
observables in the LHC Higgs measurements.  This allows us to limit
the analysis to the flavor-diagonal Yukawa structures
\begin{align}
\ope_{e\phi,33}=(\phi^\dagger\phi)(\bar L_3 \phi e_{R,3}) 
\qquad 
\ope_{u\phi,33}=(\phi^\dagger\phi)(\bar Q_3 \tilde \phi u_{R,3})
\qquad 
\ope_{d\phi,33}=(\phi^\dagger\phi)(\bar Q_3 \phi d_{R,3}) \; ,
\label{eq:hffop}
\end{align}
with $\tilde \phi=i\sigma^2\phi^*$. The corresponding effective
Lagrangian reads
\begin{alignat}{5}
\lag_\text{eff}^{hff} = &
& \frac{f_\tau m_\tau}{v \Lambda^2} \ope_{e\phi,33} 
+ \frac{f_b m_b}{v \Lambda^2} \ope_{d\phi,33} 
+ \frac{f_t m_t}{v \Lambda^2} \ope_{u\phi,33} 
+ \frac{f_{\phi,2}}{\Lambda^2} \ope_{\phi,2} \; .
\label{eq:ourleff2}
\end{alignat}
The operator $\ope_{\phi,2}$ leads to a finite renormalization of the
Higgs field and hence a universal shift of all Higgs couplings to
Standard Model fields.  In analogy to the Higgs-gluon coupling we
scale the fermionic Wilson coefficients $f_x$ by a factor $m/v$ to
reflect the chiral nature of the Higgs coupling operators.  In terms of
the physical Higgs field we find
\begin{align}
 \lag_\text{eff}^{hff} = g_f h \bar f_{L} f_{R} + \text{h.c.} 
\qquad \text{with} \quad 
g_f  =  - \frac{m_f}{v} \left( 1-\frac{v^2}{2\Lambda^2}f_{\phi,2} - \frac{v^2}{\sqrt{2}\Lambda^2}  f_f \right) \; ,
\label{eq:lhff}
\end{align}
where for $f = \tau,b,t$ we define the physical masses and fermions in
the mass basis.

Comparing Eq.\eqref{eq:lhvv} and Eq.\eqref{eq:lhff} to
Eq.\eqref{eq:def_delta} we see that the SMEFT framework for the 
Higgs sector is closely related to the coupling modifiers. All
modified Higgs couplings $\Delta_j$ are present in the SMEFT
description; the only caveat is that the SMEFT Lagrangian at dimension
six features custodial symmetry, translating into the hard-wired
condition $\Delta_W = \Delta_Z$. This is, if anything, a shortcoming
of the SMEFT approach and a reason to think about including
dimension-8 operators. In addition to the six independent coupling
modifiers $\Delta_j$, the SMEFT Lagrangian features new Lorentz
structures linked to $g_{hVV}^{(1)}$ and $g_{hVV}^{(2)}$ as well as
modifications of the still unmeasured $hZ\gamma$ interaction.

A key feature of the effective Higgs-gauge Lagrangian in
Eq.\eqref{eq:ourleff} is that it also leads to deviations of the
triple-gauge couplings~\cite{Falkowski:2015jaa,Falkowski:2016cxu,Berthier:2016tkq,deBlas:2018tjm,Liu:2018pkg}. This
means that pair production of electroweak bosons at the LHC has to be
combined with the global Higgs EFT analysis.  The one additional
purely bosonic operator describing the process $pp \to VV'$ at the LHC
is~\cite{Butter:2016cvz}
\begin{alignat}{5}
\ope_{WWW} &= \tr ( \hat{W}_{\mu \nu} \hat{W}^{\nu \rho} \hat{W}_\rho^\mu ) 
\label{eq:eff2}  
\end{alignat}
The combined operators from Eq.\eqref{eq:eff} and Eq.\eqref{eq:eff2}
are related to the historic notation for anomalous gauge
couplings~\cite{Gaemers:1978hg},
\begin{align}
\lag_\text{eff}^{WWV} = 
 -i g_{WWV} \left[
 g_1^V \left( W^+_{\mu\nu} W^{- \, \mu} V^{\nu} 
            - W^+_\mu V_\nu W^{- \, \mu\nu} \right)
+ \kappa_V W_\mu^+ W_\nu^- V^{\mu\nu}
+ \frac{\lambda_V}{M_W^2} W^+_{\mu\nu} W^{- \, \nu\rho} V_\rho^{\; \mu}
 \right] \; ,
\label{eq:deltakappalambda}
\end{align}
with $g_{WW\gamma} = e$ , $g_{WWZ} = c_W g $, $g_1^V=1+\Delta g_1^V$,
$\Delta g_1^\gamma=0$, and $\kappa_V=1+\Delta \kappa_V$.
Equations~\eqref{eq:deltakappalambda} and \eqref{eq:deltakappalambda2}
use $\kappa_V$ for the modification of the $W^+W^-V$ vertex since this
is standard notation, while in the remainder of this work $\kappa$
always refers to Higgs couplings.  The gauge-invariant form of the
anomalous gauge couplings at mass dimension six reads (see also~\cite{deBlas:2018tjm} for a detailed comparison)
\begin{alignat}{5}
\Delta \kappa_\gamma &= \frac{g^2 v^2}{8\Lambda^2} (f_W + f_B)  &\qqquad 
\Delta \kappa_Z &=   \frac{g^2 v^2}{8 c_W^2 \Lambda^2} (c_W^2 f_W - s_W^2 f_B )  \notag \\  
\lambda_\gamma &= \lambda_Z = \frac{3 g^2 M_W^2}{2 \Lambda^2} f_{WWW} & \qqquad 
\Delta g_1^Z &= \frac{g^2 v^2}{8 c_W^2 \Lambda^2}f_W 
\; . 
\label{eq:deltakappalambda2}
\end{alignat}
Of the five modified triple gauge couplings only three are
independent when written in terms of an effective gauge theory. This
linear relation reflects the fact the the traditional variables are
nothing by a gauge-dependent re-parametrization of the Wilson
coefficients of the SMEFT. Using anomalous couplings for an analysis
does not improve any shortcomings of a SMEFT analysis, it only removes
weak gauge invariance from the list of assumptions.

Finally, we have to account for the fact that di-boson observables at
the LHC are also affected by fermionic operators of the form
\begin{align}
\ope_{\phi Q,ij}^{(1)} =\phi^\dagger (i\,{\Dfba}_{\!\!\mu} \phi) (\bar Q_{i}\gamma^\mu Q_j) 
\qquad \text{and} \qquad
\ope_{\phi Q,ij}^{(3)} =\phi^\dagger (i\,{\Dfba}_{\!\!\mu} \phi) (\bar Q_i\gamma^\mu \sigma_a Q_j) 
\end{align}
with the $SU(2)$ generators $\sigma_a$. The quark doublet $Q$ can be
replaced by the lepton doublet or, for the singlet case, by
right-handed fermions.  Because of the hermitian structure of the
Lagrangian, these operators do not actually contribute to any Higgs
couplings, but they do contribute to anomalous weak gauge
couplings. To simplify the analysis, we can for instance assume that
the fermion operators are generation universal~\cite{Alves:2018nof},
which allows us to remove the operators coupling to lepton
doublets. In addition, at the expected level of accuracy in the
Higgs-gauge fit we can neglect all operators with a non-trivial flavor
structure.  While anomalous gauge boson couplings to light fermions
are constrained by electroweak precision data~\cite{Franceschini:2017xkh}, they have to be
included in an LHC analysis unless we have a firm reason to expect a
hierarchy of scales between LEP and LHC
limits~\cite{Baglio:2017bfe,Baglio:2018bkm,Ellis:2018gqa,Alves:2018nof}. In that case
an additional set of terms in the effective Lagrangian becomes
\begin{align}
\lag_\text{eff}^{Vff} = 
 \frac{f^{(1)}_{\phi Q}}{\Lambda^2}   \ope^{(1)}_{\phi Q}
  + \frac{f^{(3)}_{\phi Q}}{\Lambda^2}   \ope^{(3)}_{\phi Q}
  + \frac{f^{(1)}_{\phi u}}{\Lambda^2}   \ope^{(1)}_{\phi u}
  + \frac{f^{(1)}_{\phi d}}{\Lambda^2}   \ope^{(1)}_{\phi d}
  + \frac{f^{(1)}_{\phi e}}{\Lambda^2}   \ope^{(1)}_{\phi e} \; , 
\label{eq:eff3}  
\end{align}
where we ignore the contribution to the Fermi constant from 4-lepton
interactions. Note that eventually the Higgs-gauge analysis in the
SMEFT framework also has to be combined with the top sector, for more
information we refer to the \textsc{TopFitter}
analyses~\cite{Buckley:2015nca,Buckley:2015lku}.

\subsubsection{Non-linear realization}
\label{sec:basis_eft_nonlinear}

An alternative effective Lagrangian of the Higgs sector is based on a
non-linear realization of electroweak symmetry breaking
(HEFT)~\cite{Alonso:2012px,Brivio:2013pma,Buchalla:2013eza,Buchalla:2013rka,Gavela:2014vra,Brivio:2014pfa}. Here,
the Higgs is not embedded in a doublet together with the Goldstone
modes, as motivated by the symmetry structure of certain UV
completions. We can nevertheless link the non-linear and linear
operator sets in terms of canonical dimensions. This connection is
usually established through the ratio of scales $(v/f)^2$, where $f$
can be related to the scale of strong dynamics.  A detailed analysis
of the non-linear model reveals a double ordering: first, there is the
chiral expansion, and in addition there is the classification in
powers of $(v/f)^2$. Following
Refs.~\cite{Buchalla:2015qju,Corbett:2015mqf,Brivio:2016fzo}, a subset
of the non-linear Lagrangian can be linked to the $\Delta$-framework
of Eq.\eqref{eq:lag_delta}, with the additional assumption
$\Delta_W=\Delta_Z$~\cite{Klute:2012pu}. As for the linear
representation, an extended set of non-linear operators provides a
natural extension to include kinematic distributions at the LHC and can 
be used for global LHC analyses~\cite{deBlas:2018tjm}.

To illustrate the link between the linear and non-linear effective
Lagrangians we start with the reduced bosonic $CP$-even operator set
of order $(v/f)^2$~\cite{Brivio:2013pma},
\begin{alignat}{7}
\opp_C &= -\frac{v^2}{4}\tr(\mathbf{V}^\mu \mathbf{V}_\mu) \; \mathcal{F}_C(h) \qqquad
&\opp_T &= \frac{v^2}{4} \tr(\mathbf{T}\mathbf{V}_\mu)\tr(\mathbf{T}\mathbf{V}^\mu) \; \mathcal{F}_T(h) \qqquad
&\opp_h &= \frac{1}{2}(\partial_\mu h)(\partial^\mu h) \; \mathcal{F}_h(h)
\notag \\
\opp_W &=-\frac{g^2}{4} W_{\mu \nu}^a W^{a\mu\nu} \; \mathcal{F}_W(h) 
&\opp_B &=-\frac{g'^2}{4}B_{\mu \nu} B^{\mu \nu} \; \mathcal{F}_B(h) 
&\opp_G &= -\frac{g_s^2}{4}G_{\mu\nu}^a G^{a\mu\nu} \; \mathcal{F}_G(h)
\notag \\
\opp_{\square h} &=\frac{1}{v^2}(\partial_\mu \partial^\mu h)^2 \; \mathcal{F}_{\square h}(h)
\label{eq:non-linear1}
\end{alignat}
and
\begin{alignat}{7}
\opp_1 &= \phantom{\frac{1}{2}}gg' B_{\mu \nu} \tr(\mathbf{T} W^{\mu \nu}) \; \mathcal{F}_1(h)
&\opp_2 &= \phantom{\frac{1}{2}}ig'B_{\mu \nu} \tr(\mathbf{T}[\mathbf{V}^\mu,\mathbf{V}^\nu]) \; \mathcal{F}_2(h)  \; ,
&\opp_3 &= ig\tr(W_{\mu \nu} [\mathbf{V}^\mu,\mathbf{V}^\nu]) \; \mathcal{F}_3(h) 
\notag \\
\opp_4 &= ig' B_{\mu \nu} \tr(\mathbf{T}\mathbf{V}^\mu) \partial^\nu \; \mathcal{F}_4 
&\opp_5 &= ig \tr(W_{\mu \nu}\mathbf{V}^\mu) \partial^\nu \; \mathcal{F}_5(h) 
&\opp_6 &= (\tr(\mathbf{V}_\mu\mathbf{V}^\mu))^2 \; \mathcal{F}_6(h) 
\notag \\
\opp_7 &= \tr(\mathbf{V}_\mu\mathbf{V}^\mu) \partial_\nu \partial^\nu \; \mathcal{F}_7(h)  
&\opp_8 &= \tr(\mathbf{V}_\mu\mathbf{V}_\nu) \; \partial^\mu \mathcal{F}_{8}(h) \; \partial^\nu \mathcal{F}_8'(h) 
&\opp_9 &= \tr((D_\mu\mathbf{V}^\mu)^2) \; \mathcal{F}_{9}(h) 
\notag \\
\opp_{10} &= \tr(\mathbf{V}_\nu D_\mu\mathbf{V}^\mu) \partial^\nu\; \mathcal{F}_{10}(h) \; , 
\label{eq:non-linear2}
\end{alignat}
where $\mathbf{V}_\mu\equiv \left(D_\mu U \right) U ^\dagger$ and
$\mathbf{T}\equiv U \sigma^3 U ^\dag$ are the vector and scalar chiral
fields in the adjoint representation of $SU(2)_L$.  The unitary matrix
$U$ contains the Goldstone modes ($\pi^a$) , $U=e^{i (\sigma \cdot \pi)/v}$, and
transforms as a bi-doublet $ U \rightarrow L\, U R^\dagger$ under the
global $SU(2)_{L,R}$ transformations. The covariant derivatives act on
the fields as~\cite{Brivio:2013pma}
\begin{align}
D_\mu  U 
&=\partial_\mu U +\frac{i}{2}gW_{\mu}^a\sigma_a U 
                             - \dfrac{ig'}{2} B_\mu  U \sigma_3 \notag \\
D_\mu \mathbf{V}_\nu 
&=\partial_\mu \mathbf{V}_\nu +i g \left[ W^a_\mu\frac{\sigma_a}{2}, \mathbf{V}_\nu \right] \; ,
\tag{5}
\end{align}
The functions $\mathcal{F}_i(h)$ are power series of the Higgs singlet
field and introduce the anomalous Higgs couplings. For the Higgs
analysis at the LHC we truncate their expansion after their linear
term in $h/v$. A link of single Higgs rates to Higgs pair production,
as provided in the SMEFT approach, is not possible in the non-linear
HEFT setup.

This number of operators can be reduced for an LHC analysis.  First,
operators containing the combination $D_\mu\mathbf{V}^\mu$ are
irrelevant for on-shell gauge bosons or when the fermion masses in the
process are neglected. This removes $\opp_9$ and $\opp_{10}$ for the
Higgs analysis.  While $\opp_2$ and $\opp_3$, together with $\opp_4$
and $\opp_5$, will eventually help us to distinguish linear from
non-linear electroweak symmetry breaking~\cite{Brivio:2013pma}, they
do not affect three-point Higgs couplings. For the same reason we can
also omit $\opp_6$ and $\opp_8$.  For on-shell Higgs amplitudes
$\opp_7$ and $\opp_{\square H}$ lead to coupling
shifts~\cite{Brivio:2014pfa}, \ie their effect can be accounted for by a
re-definition of the remaining non-linear operator coefficients.
Finally, for illustration purpose we neglect $\opp_1$ and $\opp_T$,
because of their contributions to the electroweak oblique parameters.

Again in analogy to the linear ansatz of Eq.\eqref{eq:ourleff2}
we 
add three Yukawa-like non-linear operators of the type
\begin{align}
\opp_t
&= \frac{m_{t}}{\sqrt{2}} \; \overline{Q}_L  U 
   \mathcal{F}_t(h) t_R   +\text{h.c.} \; .
\end{align}
Given this simplified fermion sector we can absorb a combination of
$\opp_C$ and $\opp_h$ through the equations of motion and arrive at a
9-dimensional non-linear operator set.  If we use a particularly simple
form for the Wilson coefficients
\begin{align}
\lag_\text{eff}
= \frac{a_G}{\Lambda^2} \opp_G
 +\frac{a_B}{\Lambda^2} \opp_B
 +\frac{a_W}{\Lambda^2} \opp_W
 +\frac{a_h}{\Lambda^2} \opp_h
 +\frac{a_4}{\Lambda^2} \opp_4
 +\frac{a_5}{\Lambda^2} \opp_5
 +\frac{a_\tau}{\Lambda^2} \opp_\tau
 +\frac{a_b}{\Lambda^2} \opp_b
 +\frac{a_t}{\Lambda^2} \opp_t \; .
\end{align}
we can link the non-linear Lagrangian and the linear Lagrangian
relevant for our Higgs analysis,
\begin{align}
\frac{v^2}{2} f_{BB} & =  a_B\qquad
& \frac{v^2}{2} f_{WW} & =  a_W \qquad
& \frac{v^2}{(4\pi)^2} f_{GG} & = a_G  \notag \\
\frac{v^2}{8} f_B & =  a_4 \qquad
& -\frac{v^2}{4} f_W & =  a_5\qquad
& v^2 f_{\phi,2} & =  c_h  \notag \\
v^2 f_t & =  a_t\qquad
& v^2 f_b & =  a_b \qquad
& v^2 f_\tau & =  a_\tau \; .
\end{align}
While the link between the non-linear effective Lagrangian to the
$\Delta$-framework is well known, there also exists a one-to-one
correspondence between the linear and a non-linear operator set. The
LHC results in the two approaches can be translated into each other
through a simple operator rotation. However, these relations are valid
only when we study the effects of the operators restricted to Higgs
interactions with two SM fields and do not include anomalous gauge
boson couplings. The link between anomalous Higgs couplings and triple
gauge couplings in the SMEFT approach does not have any correspondence
in the HEFT.

\subsubsection{Decoupling the 2HDM}
\label{sec:basis_eft_2hdm}

To illustrate the link between the effective field theory and for
example the extended Higgs sectors introduced in
Sec.~\ref{sec:basis_weak} we show how the 2HDM with a large mass
hierarchy can be matched to the effective theory. Per se it not clear
if this matching should be performed in the unbroken or broken phase,
\ie including terms of the order $v$ or setting $v=0$. This
choice is particularly relevant because the decoupling theorem does not
hold in the presence of a vacuum expectation value, which implies that
terms proportional to $v$ do not have to decouple with the new heavy
particles. For simplicity we assume that all additional Higgs bosons
are heavy,
\begin{align}
m_{H^0,H^{\pm},A^0} \approx m_\text{heavy} \gg v \; . 
\end{align}
To construct the effective theory~\cite{Gunion:2005ja,Randall:2007as}
we start from the relations between the Higgs self-couplings,
masses and mixing angles in the decoupled regime
\begin{alignat}{5}
(\lambda_1 c_\beta^2 - \lambda_2 s_\beta^2)\,v^2 
=& m_\text{heavy}^2 c_{2\beta} + (m^2_{H^0} - m^2_{h^0})\, c_{2\alpha} \notag \\
(\lambda_3+\lambda_4 + \lambda_5)\,v^2 
=& m_\text{heavy}^2 
  + (m^2_{H^0} - m^2_{h^0})\, \dfrac{s_{2\alpha}}{s_{2\beta}} 
  - \frac{\lambda_6}{t_\beta} - \lambda_7 t_\beta \; .
\label{eq:comb2}
\end{alignat}
In terms of the combination
\begin{align}
\hat{\lambda} = 
s_\beta c_\beta \left( \lambda_1 c_\beta^2 - \lambda_2 s_\beta^2 
     - ( \lambda_3 + \lambda_4 + \lambda_5 ) c_{2\beta} \right) 
+ \lambda_6 c_\beta c_{3\beta}
+ \lambda_7 s_\beta s_{3\beta}
\label{eq:comb3}
\end{align}
we can define a convenient decoupling parameter related to the 
scale separation as
\begin{align}
\xi = \hat{\lambda} \; \frac{v^2}{m_\text{heavy}^2} 
\approx  s_{\beta - \alpha} c_{\beta - \alpha} \ll 1 \; .
\end{align}
If the effective field theory description of the 2HDM approaches the
decoupling limit $s_{\beta-\alpha} \to 1$, the expansion parameter
scales like $\xi \approx c_{\beta-\alpha} \ll 1$, requiring $c_\beta
\approx s_\alpha$ The correlation of the mixing angles is a key
property of the hierarchical 2HDM.  For sufficiently large $t_\beta$
this gives for the expansion parameter
\begin{align}
\xi = \dfrac{2 t_\beta}{1 + t_\beta^2} \; .
\label{eq:xi_tgb}
\end{align}
We can now expand all modifications of the light Higgs couplings given
in Tab.~\ref{tab:2hdmcouplings} in terms of $\xi$. We show the
complete list of coupling modifications in Tab. \eqref{tab:eff-coup1}. The
main feature is that the gauge couplings in the 2HDM decouple
significantly faster than the Yukawa couplings, especially the
down-type Yukawa couplings with their $t_\beta$ enhancement.

\begin{table}[b!]
\begin{center} 
\begin{tabular}{l|l} 
\toprule
$1 + \Delta_W$ & $1-\dfrac{\xi^2}{2}$  \\[3mm] \\
$1 + \Delta_Z$ & $1-\dfrac{\xi^2}{2}$  \\[3mm]
$1 + \Delta_t$ & $ 1 + \dfrac{\xi}{t_\beta} - \dfrac{\xi^2}{2} + \mathcal{O}(\xi^3)$ \\[3mm] 
 $1 + \Delta_b$ & $ 1 - \tan(\beta-\gamma_b)\xi - \dfrac{\xi^2}{2} + \mathcal{O}(\xi^3)$ \\[3mm]
  $1 + \Delta_{\tau}$ & $ 1 - \tan(\beta-\gamma_\tau)\xi - \dfrac{\xi^2}{2} + \mathcal{O}(\xi^3)$ \\[3mm] 
  \bottomrule
\end{tabular}
\end{center}
\caption{Light Higgs boson couplings to fermions and gauge bosons in
  the 2HDM with a large mass hierarchy, expanded to
  $\mathcal{O}(\xi^3)$. The flavor sector features a Yukawa alignment
  structure.}
\label{tab:eff-coup1}
\end{table}

In addition to the tree-level mixing effects, the heavy charged Higgs
states also contribute to the effective Higgs-photon interaction. To
compute it, we need to consistently expand the corresponding trilinear
Higgs coupling,
\begin{align}
 \lambda_{h^0H^+H^-} &=
\dfrac{-e}{2\,M_W\,s_w}\,\left[\left(1-\dfrac{\xi^2}{2}\right)\,(m^2_{h^0} + 2m^{\pm\,2}_{H}
-\lambda\,v^2) + \dfrac{\xi}{2}\,(2m^2_{h^0} - \tilde{\lambda}\,v^2)\,(\cot\beta-\tan\beta) \right] + \ope( \xi^3) \; .
\label{eq:triple-hier}
\end{align}
Note that this coupling, like all triple Higgs couplings, does not
decouple. The proper mass dimension of the Wilson coefficients of
$\ope_{BB,WW}$ is ensured by the heavy particles in the loop, as
illustrated in \eg Eq.\eqref{eq:hdecaa}.

\clearpage
\section{Experimental results}

\subsection{Gluon Fusion}
\label{sec:exp_gf}

\begin{figure}[b!]
\centering
\parbox{0.25\textwidth}{\vspace{-5cm}\includegraphics[width=0.25\textwidth]{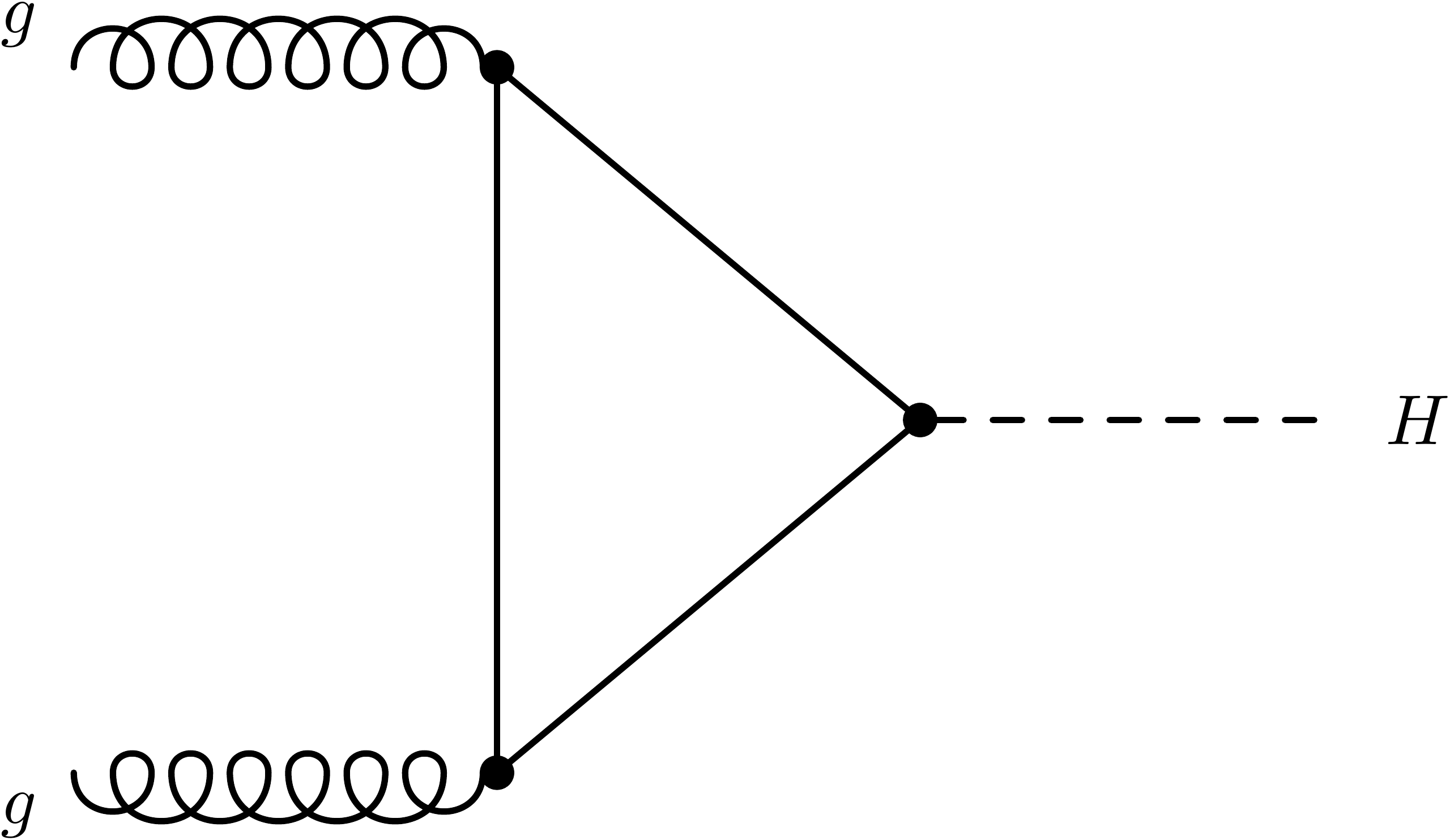}}
\hspace*{0.1\textwidth}
\includegraphics[width=0.50\textwidth]{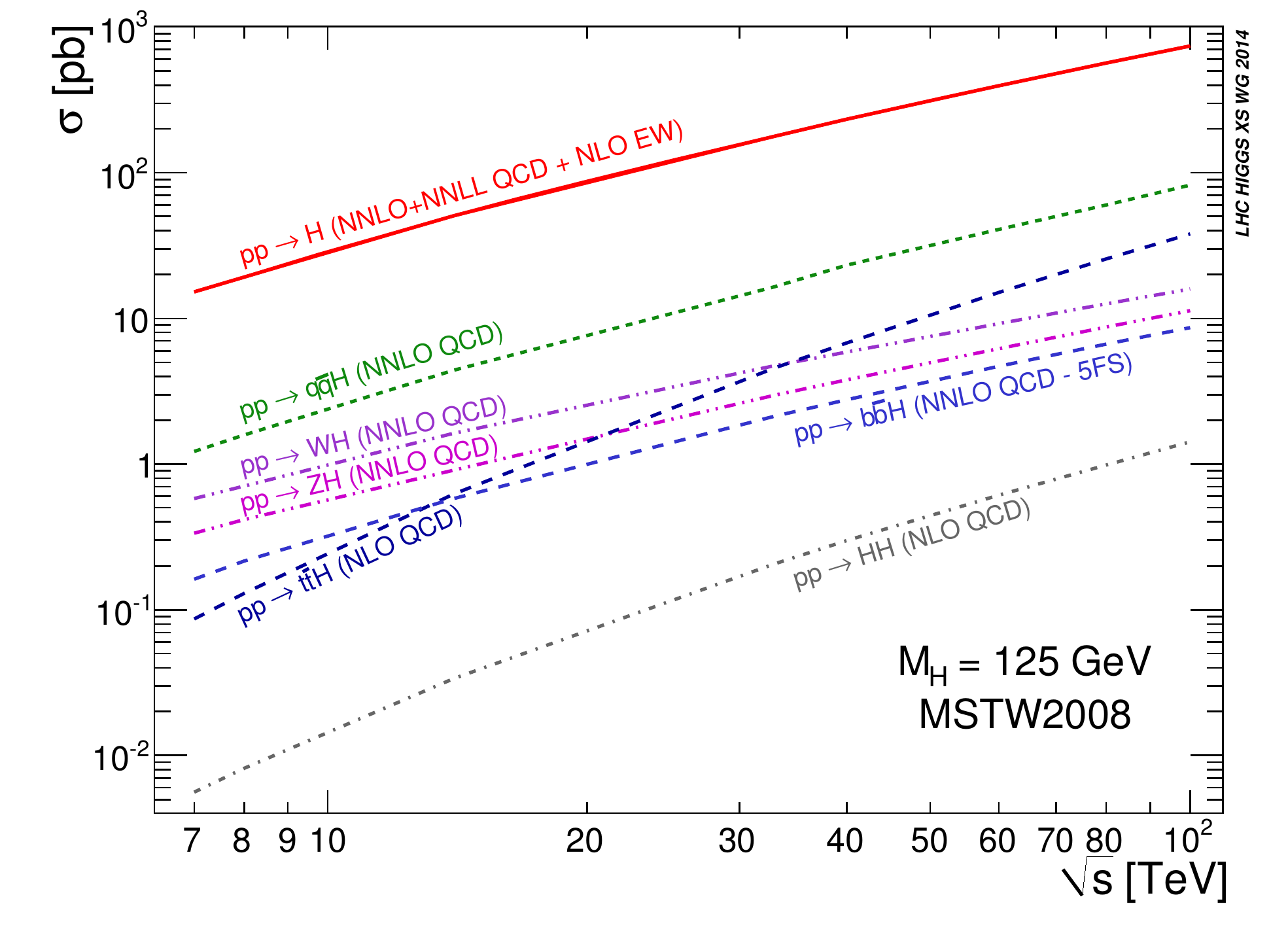}
\caption{Contribution to Higgs boson production from gluon fusion
  (left) and  the total Higgs production cross section (right), taken from
  Ref.~\cite{deFlorian:2016spz}. This figure includes most known higher
  order corrections.}
\label{fig:prodgg}
\end{figure}

The dominant production mechanism for the Higgs boson in hadronic
collisions is the gluon fusion  process shown in Fig.~\ref{fig:prodgg},
\begin{align}
g g \to h + X\,.
\label{eq:def_gf}
\end{align}
The gluon fusion production mode was the driving force behind the
Higgs discovery in 2012, Fig.~\ref{fig:ggdisc}, subsequent Higgs mass
measurements, Fig.~\ref{fig:hmassmeas}, and most measurements of 
Higgs couplings so far. Due to the loop-induced gluon-Higgs coupling
discussed in Sec.~\ref{sec:basis_decs}, gluon fusion  not only enables
high-precision measurements of the total production rate, it also
offers unique opportunities for coupling measurements from its kinematic distributions.

\subsubsection{Motivation and signature}
\label{sec:exp_gf_mot}

Gluon fusion, with the subsequent decays
\begin{align}
pp \to h \to 
\begin{cases} 
 \gamma \gamma \\ 
 ZZ \\ 
 W^+W^- \\
   \tau^+ \tau^- \\
\end{cases}
\end{align}
is one of the most important channels for Higgs studies and as seen in
Fig.~\ref{fig:prodgg}, the gluon fusion rate is significantly larger
than that of the other channels. Due to overwhelming QCD backgrounds
and the lack of additional signatures to tag or trigger on, $h\to
b\bar b$ is not accessible. Given the plethora of Higgs final states
that can be accessed through gluon fusion, this process sits at the
heart of the LHC Higgs phenomenology program.

In particular the observation of Higgs decay to a photon pair and a
leptonically decaying $Z$-boson pair is a tell-tale indication of the
Higgs interactions and electroweak symmetry breaking.  On the one
hand, as the photons do not have a mass, the decay of the Higgs to
photons is loop-suppressed and therefore sensitive to the interplay of
the gauge-Higgs and fermion-Higgs sectors. In the Standard Model there is  a delicate
interference between the charged $W$ and top loops as detailed in
Sec.~\ref{sec:basis_decs_a}. In particular the fermion-Higgs
interactions in new physics models can be vastly different from the
Standard Model, as indicated by the well-motivated two-Higgs doublet models. The
observation of $h\to \gamma\gamma$ at close to the predicted rate is therefore a strong validation of
the Standard Model.

Phenomenologically, searches for $h\rightarrow \gamma \gamma$ amount
to hunting for  bumps on top of a continuously falling di-photon
distribution, as illustrated in Fig.~\ref{fig:ggdisc}. Photons are
experimentally well-understood, including the fact that most jets
consist of photons due to the anomaly-mediated $\pi^0$ decay, and care
has to be taken to understand this background.  In practice, the
combined background is estimated in a data-driven approach through
fitting polynomials. This way, even though the decay $h\to \gamma
\gamma$ only happens at the 2 per-mille level, the impact of
systematic uncertainties is relatively small and the Higgs boson could
be observed at a statistically significant level already early on in
the LHC program.

On the other hand, the decays $h\to ZZ$ and $h \to WW$ highlight the
Higgs couplings to the massive weak bosons and their longitudinal
degrees of freedom. Such an interaction is a direct implication of the
Higgs mechanism itself.  The decay $h\to ZZ$ is typically observed in
fully leptonic final states which are experimentally clean and lead
to a clear resonance peak at $m(4l)=M_h$. Again,
electrons and muons are well-understood objects in the ATLAS
and CMS detectors, and the final state can be reconstructed
without systematic limitations and with small irreducible
backgrounds. Compared to the $h\to \gamma\gamma$ case, $h\to ZZ$ is
statistically limited through the leptonic branching ratios,
which significantly reduces the naive $h\to ZZ$ branching ratio of
2.6\% in the Standard Model. The decays $h\to \tau \tau$ and $h\to WW$ require
special attention to the involved hadronic or missing energy
signatures and are in general more complicated to reconstruct.

\subsubsection{Precision prediction}
\label{sec:exp_gf_pref}

\begin{figure}[t]
\centering
\includegraphics[width=0.44\textwidth]{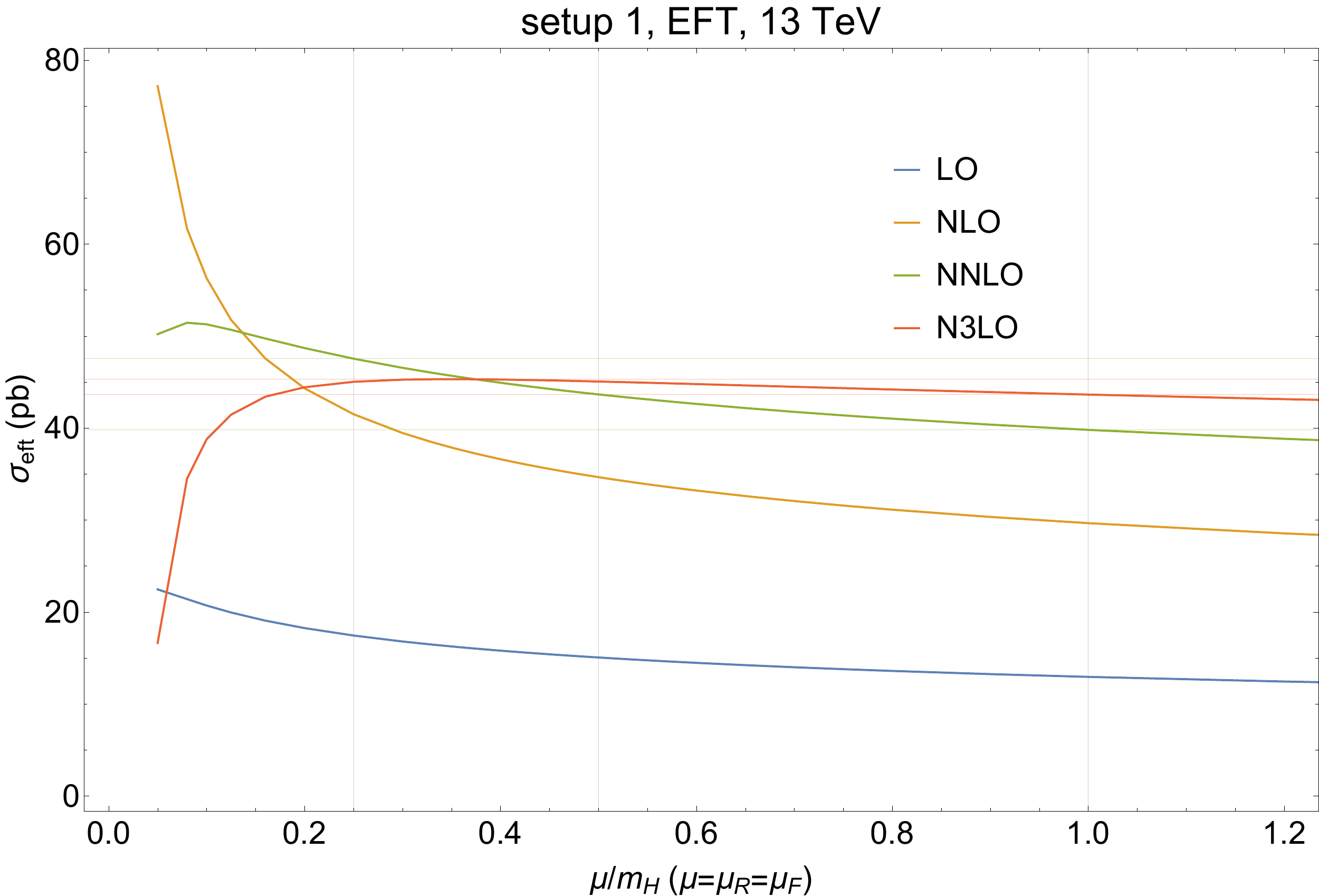}
\caption{Theoretical prediction for Higgs cross
  section~\cite{Anastasiou:2016cez} at various orders in
  $\alpha_s$. $\mu$ is the renormalization and factorization scales,
  which are taken to be equal.}
\label{fig:sig_theory} 
\end{figure}

\begin{table}[b!]
\normalsize\setlength{\tabcolsep}{2pt}
\begin{center}
    \begin{tabular}{rrrrr}
        \toprule
        \multicolumn{1}{c}{$\textrm{E}_{\textrm{CM}}$}&
        \multicolumn{1}{c}{$\sigma$}&
        \multicolumn{1}{c}{$\delta(\textrm{theory})$}&
        \multicolumn{1}{c}{$\delta(\textrm{PDF})$}&
        \multicolumn{1}{c}{$\delta(\alpha_s)$}\\\midrule
        7 TeV & 16.87 pb & ${}^{+0.70\textrm{pb}}_{-1.14\textrm{pb}}\,\left({}^{+4.17\%}_{-6.76\%}\right) $ & $\pm\,0.31\,\textrm{pb}\,(\pm\,1.89\%)$ & ${}^{+0.44\textrm{pb}}_{-0.45\textrm{pb}}\,\left({}^{+2.66\%}_{-2.68\%}\right) $ \\\midrule
8 TeV & 21.45 pb & ${}^{+0.90\textrm{pb}}_{-1.43\textrm{pb}}\,\left({}^{+4.18\%}_{-6.69\%}\right) $ & $\pm\,0.40\,\textrm{pb}\,(\pm\,1.87\%)$ & ${}^{+0.56\textrm{pb}}_{-0.56\textrm{pb}}\,\left({}^{+2.63\%}_{-2.66\%}\right) $ \\\midrule
13 TeV & 48.68 pb & ${}^{+2.07\textrm{pb}}_{-3.16\textrm{pb}}\,\left({}^{+4.26\%}_{-6.48\%}\right) $ & $\pm\,0.89\,\textrm{pb}\,(\pm\,1.85\%)$ & ${}^{+1.25\textrm{pb}}_{-1.26\textrm{pb}}\,\left({}^{+2.59\%}_{-2.62\%}\right) $ \\\midrule
14 TeV & 54.80 pb & ${}^{+2.34\textrm{pb}}_{-3.54\textrm{pb}}\,\left({}^{+4.28\%}_{-6.46\%}\right) $ & $\pm\,1.00\,\textrm{pb}\,(\pm\,1.86\%)$ & ${}^{+1.40\textrm{pb}}_{-1.42\textrm{pb}}\,\left({}^{+2.60\%}_{-2.62\%}\right) $ \\\midrule
28 TeV & 154.63 pb & ${}^{+7.02\textrm{pb}}_{-9.93\textrm{pb}}\,\left({}^{+4.54\%}_{-6.42\%}\right) $ & $\pm\,2.98\,\textrm{pb}\,(\pm\,1.96\%)$ & ${}^{+4.10\textrm{pb}}_{-4.03\textrm{pb}}\,\left({}^{+2.70\%}_{-2.65\%}\right) $ \\\midrule
100 TeV & 808.23 pb & ${}^{+44.53\textrm{pb}}_{-56.95\textrm{pb}}\,\left({}^{+5.51\%}_{-7.05\%}\right) $ & $\pm\,19.98\,\textrm{pb}\,(\pm\,2.51\%)$ & ${}^{+24.89\textrm{pb}}_{-21.71\textrm{pb}}\,\left({}^{+3.12\%}_{-2.72\%}\right) $ \\\bottomrule
    \end{tabular}
  \caption{Gluon fusion rates at N$^3$LO with estimates of theoretical
    uncertainties. Taken from Ref.~\cite{Mistlberger:2018etf}.}
\label{tab:ggrates}
  \end{center}
\end{table}

Since the extraction of Higgs properties often relies on normalizing
the rate to the SM predictions, an extraordinary effort has been made
to obtain the most precise theoretical predictions.

The primary contribution to gluon fusion is through the couplings to
heavy fermions, $g g \rightarrow h$, which is shown in
Fig.~\ref{fig:prodgg}. This process is dominated by the contributions
of the top quark loop and the loop with a bottom quark contributes
roughly $-5\%$ to the SM cross section. The QCD corrections to the
rate are large and known to N$^3$LO order in the heavy-top limit, and they increase the rate
from the LO prediction by more than a factor of 2, as seen in
Fig.~\ref{fig:sig_theory}. The significant reduction in the scale
dependence at higher orders is clear.  Historically, the gluon fusion
rate was computed in the large top mass limit.  In the heavy quark
limit, the lowest order partonic cross section is independent of the
top quark mass and becomes a constant for infinite top quark mass,
\begin{align}
{\hat \sigma}_0(gg\rightarrow h)\sim {\alpha_s^2\over
576 \pi v^2}      \,.
\label{eq:siginf}
\end{align}
Heavy chiral fermions whose masses arise from couplings to the Higgs
field's vacuum expectation value do not decouple at high energy and
the gluon fusion rate essentially counts the number of SM-like chiral
quarks. Hence, the observation of a Higgs production rate consistent
with the SM expectation (\eg in the decay to $Z$ bosons) 
immediately rules out the possibility of a $4^{\text{th}}$ generation
of SM-like chiral fermions~\cite{Kribs:2007nz,Denner:2011vt,Anastasiou:2011qw}.

\begin{figure}[t]
\centering
\includegraphics[height=0.3\textwidth]{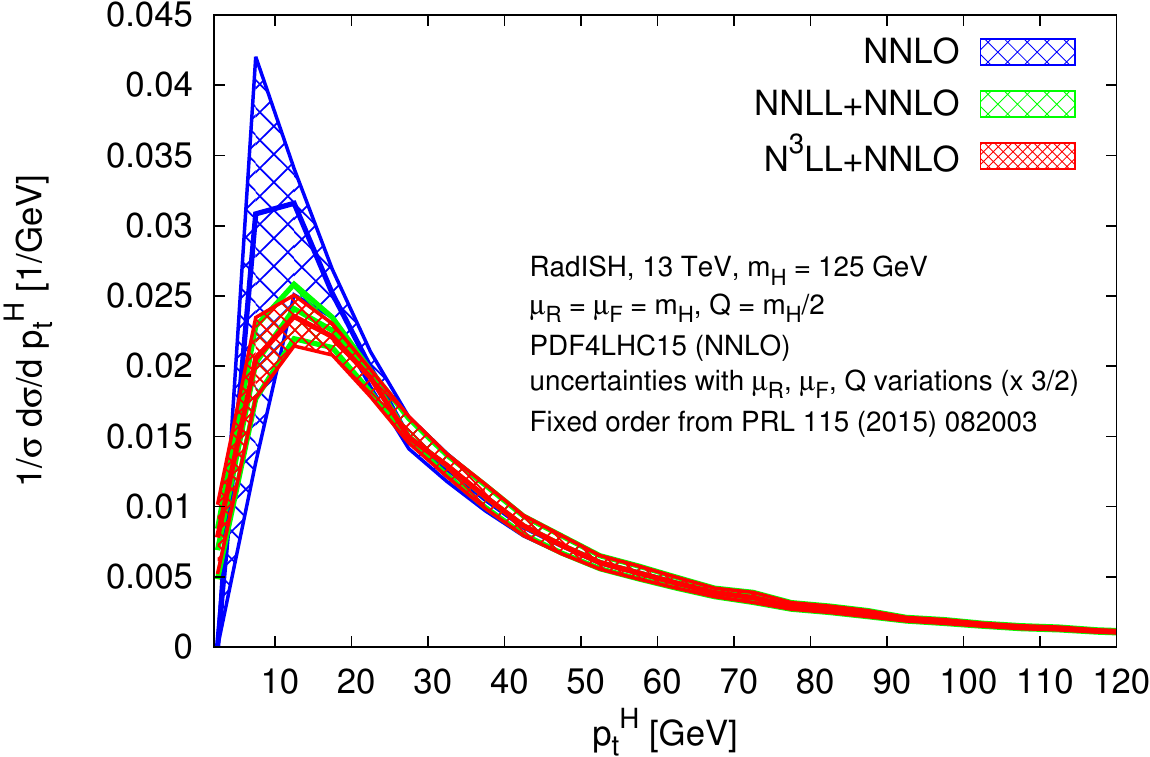}
\hspace*{0.1\textwidth}
\includegraphics[height=0.3\textwidth]{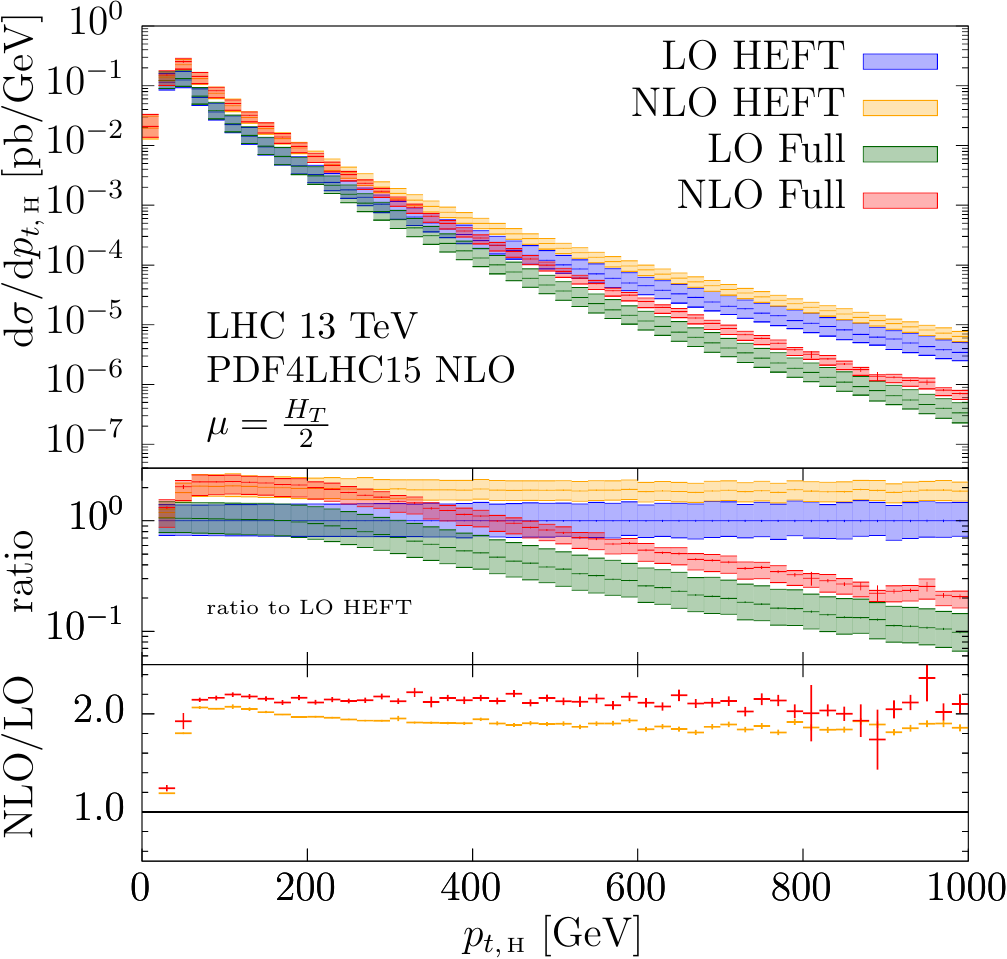}
\caption{Left: Higgs transverse momentum distribution. Figure from
  Ref.~\cite{Bizon:2017rah}. Right: $p_T$ spectrum with top mass
  effects (full) and in the $m_t\rightarrow\infty$ limit
  (HEFT). Figure from Ref.~\cite{Jones:2018hbb}.}
\label{fig:higgspt}
\end{figure} 

The NLO~\cite{Dawson:1990zj,Graudenz:1992pv,Djouadi:1991tka},
NNLO~\cite{Harlander:2002wh,Anastasiou:2002yz,Ravindran:2003um}, and
N$^3$LO~\cite{Anastasiou:2014lda,Anastasiou:2014vaa,Anastasiou:2015ema,Mistlberger:2018etf,Cieri:2018oms}
results are known in the $m_t\rightarrow \infty$ limit, while the NLO
results~\cite{Graudenz:1992pv,Spira:1995rr} are known analytically with the full
top quark mass dependence.  Fig.~\ref{fig:sig_theory} illustrates the
convergence of the QCD expansion and the reduction of scale dependence
at higher orders. 
The effects of a finite top mass can be included to
NNLO~\cite{Harlander:2009mq,Pak:2009dg,Harlander:2009bw} numerically as an expansion in
$1/m_t$, while 
N$^3$LO results are available for $m_t\to\infty$~\cite{Anastasiou:2016cez}. Electroweak corrections are known to
NLO~\cite{Aglietti:2004nj,Actis:2008ug,Actis:2008ts} and mixed
QCD/electroweak
corrections~\cite{Anastasiou:2008tj,Bonetti:2018ukf,Bonetti:2017ovy}
are partially known.  The electroweak corrections increase the rate by
$\sim 5\%$.  The fixed order rates along with estimated uncertainties
have been computed by the LHC Higgs Cross Section Working
group~\cite{deFlorian:2016spz}, with fixed order results shown in
Tab.~\ref{tab:ggrates}. The gluon fusion rate increases significantly
with beam energy, as demonstrated in Fig.~\ref{fig:prodgg}.  This
figure includes most known higher order corrections and the widths of
the curves represent an estimate of the uncertainties. The differential rate is known at NNLO~\cite{Dulat:2017prg,Dulat:2017brz}.

The NNLO rate can be further improved by the resummation of threshold
effects from soft, virtual, and collinear
gluons~\cite{Kramer:1996iq}. These contributions to the total gluon
fusion rate are known at N$^3$LL~\cite{Bonvini:2014joa}. Comparison of a combined 
threshold and high energy resummation
with the fixed order N$^3$LO result shows that the resummation effects
are small at this order~\cite{Bonvini:2016frm,Bonvini:2018ixe}.

As we will discuss in Sec.~\ref{sec:exp_gf_kin}, the kinematics of
Higgs production in gluon fusion can be used to search for new physics
effects.  At LO, the Higgs boson has no $p_T$ and a transverse
momentum spectrum for the Higgs is first generated by the process
$gg\rightarrow g h$, as well as (anti)quark scattering $qg\to qh$.
The Higgs transverse momentum is shown on the left of
Fig.~\ref{fig:higgspt}.  The total rate for Higgs plus jet production
receives $\sim 2\%$ corrections from the contributions of the $b$
quark. Fixed order NLO and
NNLO~\cite{Glosser:2002gm,Boughezal:2013uia,Boughezal:2015aha,Boughezal:2015dra,Dulat:2017prg,Chen:2014gva,deFlorian:1999zd}
radiative corrections to the Higgs plus jet rate are known for
infinite top quark mass and the NLO rate with full top quark mass
dependence is now
available~\cite{Lindert:2018iug,Jones:2018hbb,Neumann:2018bsx}.  In
Fig.~\ref{fig:higgspt} we see how for $p_T >m_t$ the top loop can be
resolved.  The Higgs plus jet processes are part of the NLO (real
emission) contribution to the gluon fusion
process~\cite{Ellis:1987xu}, which break the LO color correlation of
the initial state as required by the color singlet Higgs at LO. The
large radiative corrections to the Higgs plus jet spectrum can
therefore be understood in terms of kinematically un-suppressed QCD
emission of the initial state in the presence of a large effective
color charge of the gluon (the Casimir of the adjoint representation
of $SU(3)$ is $C_A=3$), rather than a breakdown of perturbation
theory.  Integrating over the QCD radiation in the NNLO gluon fusion
calculation yields an NNLO prediction for the Higgs $p_T$ spectrum.
The fully differential Higgs production rate to NNLO is implemented in
the codes \textsc{FEHip}~\cite{Anastasiou:2004xq,Anastasiou:2005qj} and
\textsc{Hnnlo}~\cite{Catani:2007vq,Grazzini:2008tf}. The NNLL
resummation~\cite{Bozzi:2005wk,deFlorian:2011xf,Cao:2009md} of the
small $p_T$ and also the N$^3$LL~\cite{Chen:2016vqn,Bizon:2017rah}
contributions are known and both the fixed order and matched results
are shown in Fig.~\ref{fig:higgspt}. The resummation significantly
reduces the peak of the $p_T$ spectrum relative to the fixed order
prediction.

The backgrounds to the different Higgs decays is very different
depending on how many jets are accompanying the Higgs.  This is
particularly evident in the decay $h\rightarrow W^+W^-$, where in the
$0$-jet bin the dominant background is from continuum $W^+W^-$
production, while the $1$- and $2$- jet bins have large backgrounds
from top quark pair production~\cite{Stewart:2011cf}.  To minimize the
background, the experiments split the signal into different jet
multiplicity bins which introduces a scale, $p_T(\text{cut})$ and
hence large logarithms of the form $\log(M_h/p_T(\text{cut}))$.
Resumming these logarithms and estimating the uncertainty from the
separation of scales introduces a significant theoretical uncertainty.

\subsubsection{LHC analyses}
\label{sec:exp_gf_ana}

\begin{figure}[t]
\centering
\includegraphics[width=0.5\textwidth]{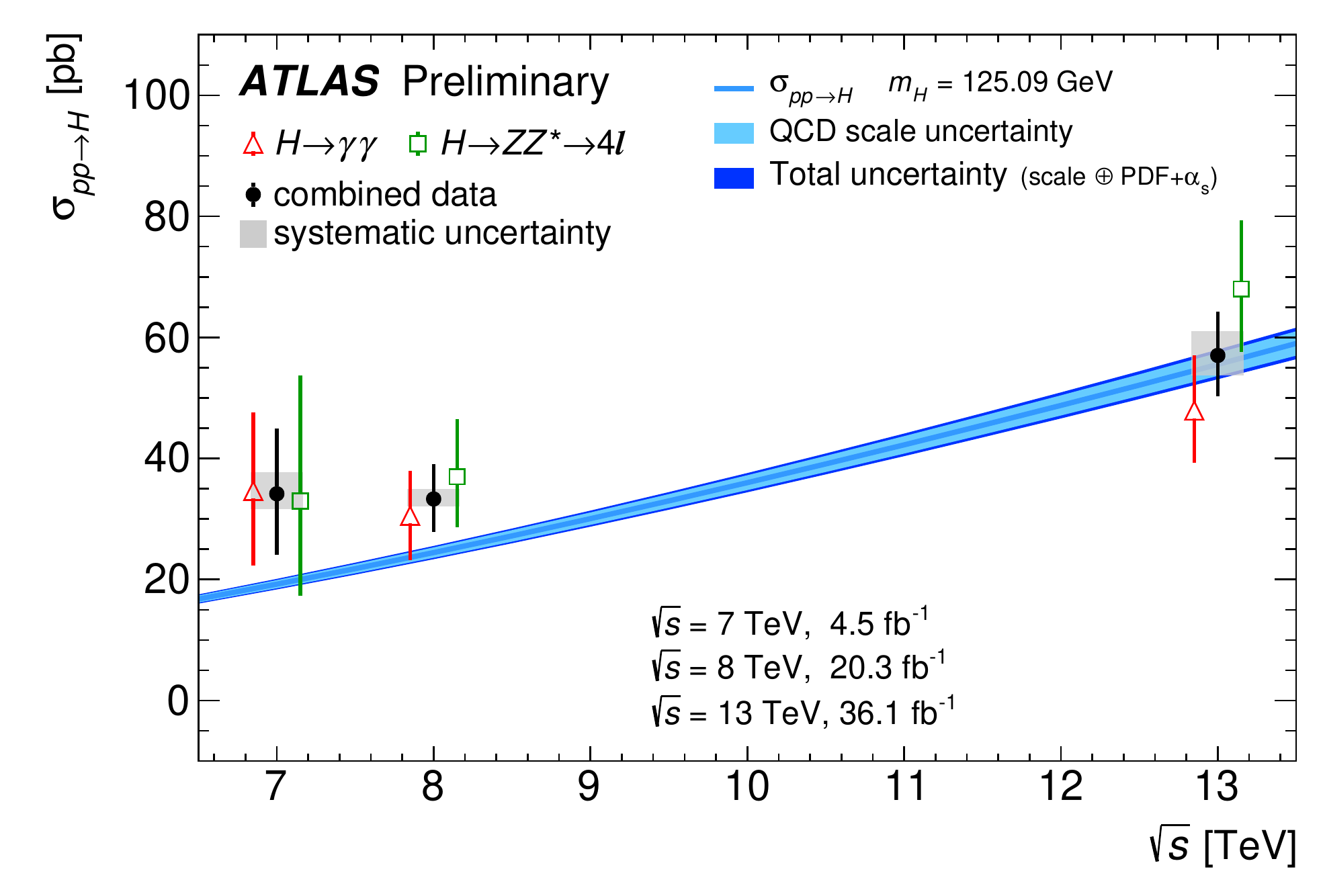}
\caption{Total Higgs cross sections from ATLAS,  Ref.~\cite{ATLAS:2017ovn}.}
\label{fig:ggsigs} 
\end{figure} 

\begin{figure}[t]
\includegraphics[width=0.4\textwidth]{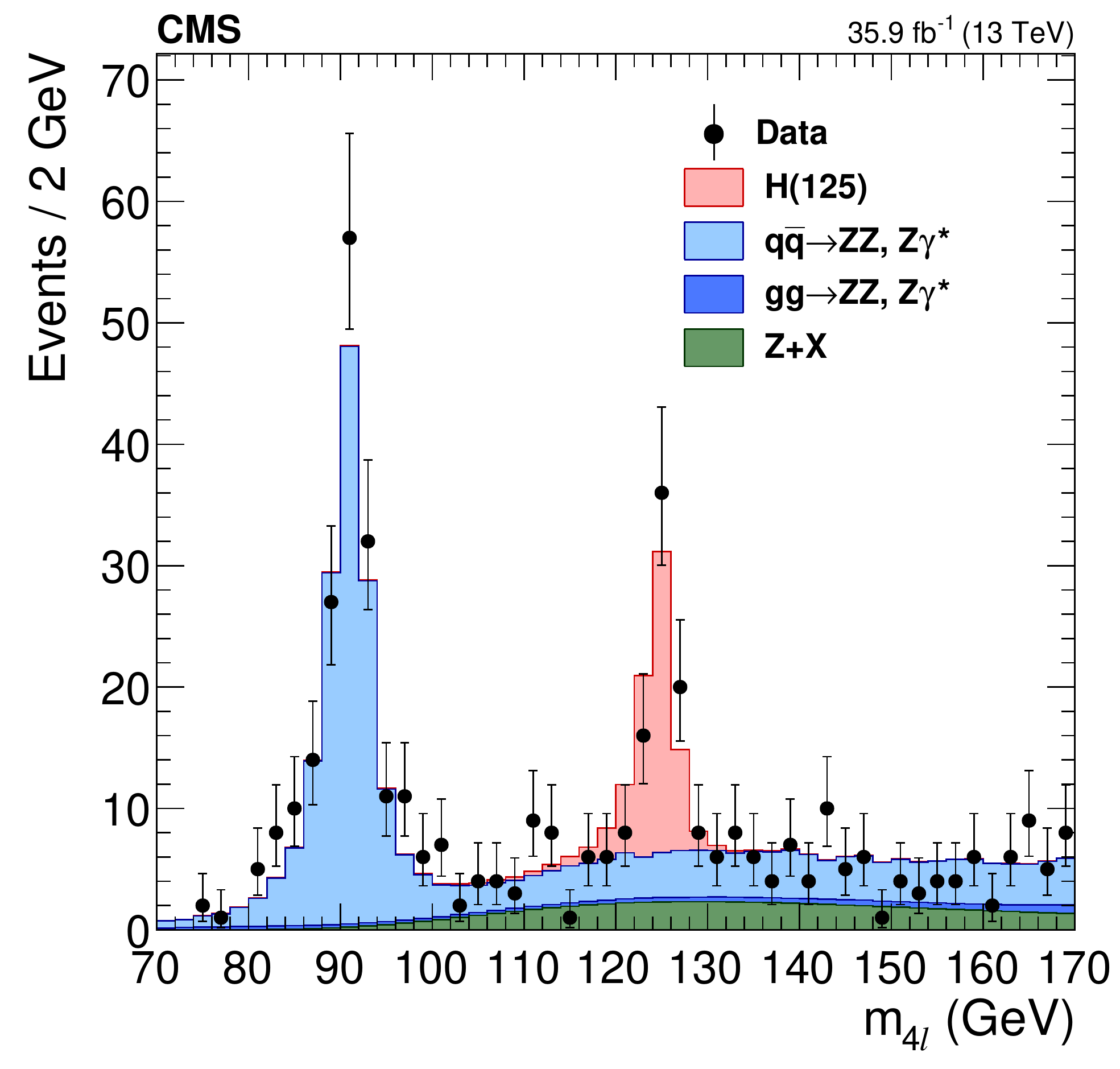}
\hspace{1cm}
\includegraphics[width=0.4\textwidth]{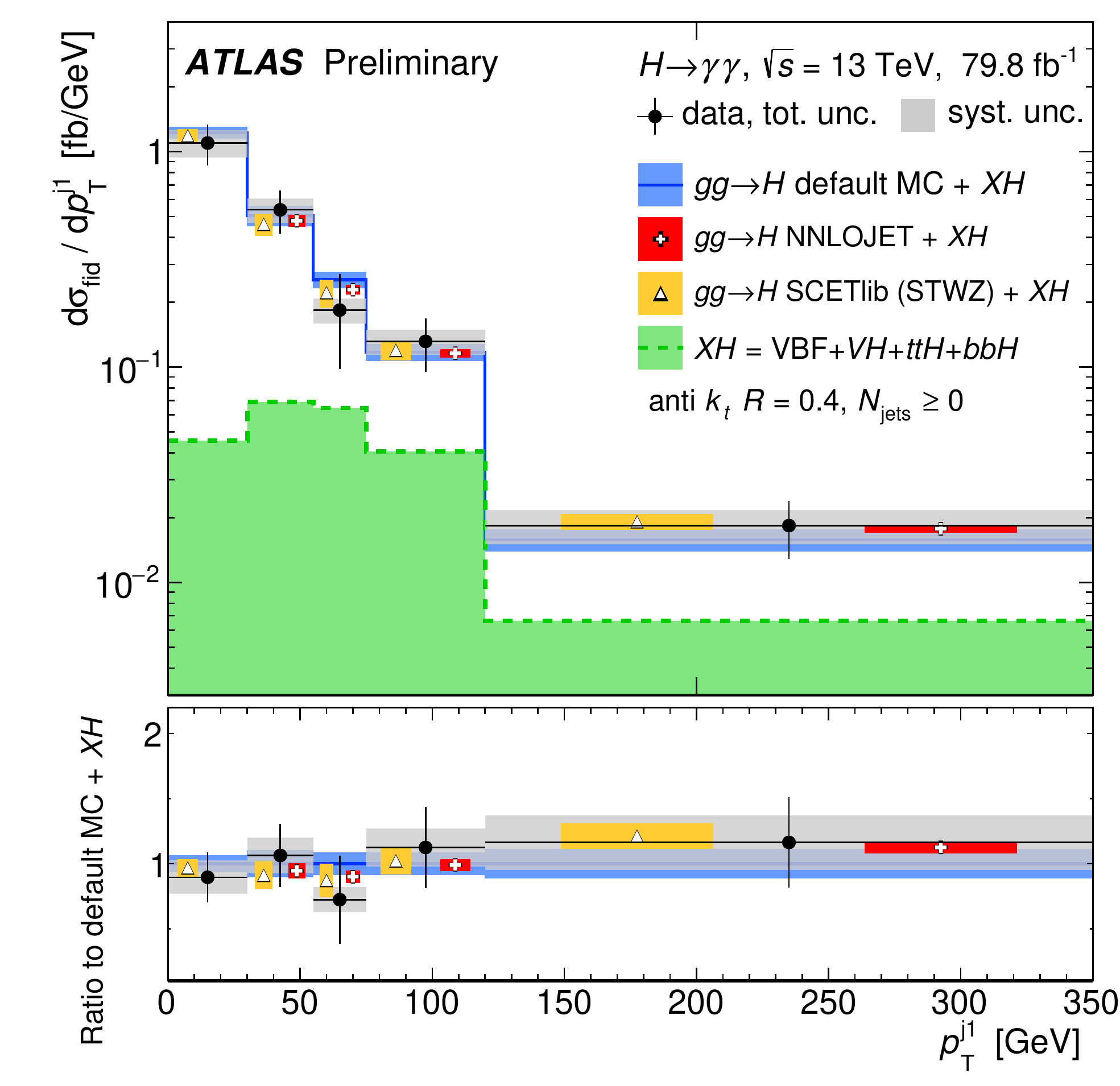}
\caption{Left: $h\rightarrow ZZ\rightarrow$ 4 lepton signal at
  $13~\tev$~\cite{Sirunyan:2017exp}. Right: Jet-transverse momentum
  distribution in gluon fusion Higgs events with the subsequent decay
  $h\to \gamma\gamma$. Taken from Ref.~\cite{ATLAS:2018uso}.}
\label{fig:hzzplot}
 \end{figure}

The measured Higgs total cross section from ATLAS (in the $ZZ$ and
$\gamma \gamma$ channels) is shown in Fig.~\ref{fig:ggsigs} and can be
seen to be in reasonable, but not perfect, agreement with theory. At
13 TeV, the Higgs has been well-established in the $ZZ$ channels,
Fig.~\ref{fig:hzzplot}. Arguably the cleanest signature of gluon
fusion is $h\to \gamma \gamma$, see
\cite{ATLAS:2018uso,Sirunyan:2018kta}, which gives
\begin{align}
\text{ATLAS:}~\quad\sigma_{ggh} \text{BR}(h \to \gamma\gamma) = 98 \pm 11 ~\text{(stat.)}   \phantom{}^{+9}_{-8} ~\text{(exp.)}~\phantom{}^{+4}_{-3}~\text{(theo.)}~\fb
\end{align}
for the ATLAS result which is based on the large luminosity of $\sim
80~\fb^{-1}$. The region selected by this analysis, $|y_h|<2.5$, has
an expected cross section times branching ratio of $102\pm
\phantom{}^{+5}_{-7}~\fb$, and the experimental result therefore
agrees well with the prediction, corresponding to
\begin{align}
\mu^{\text{ATLAS}}_{ggh,\gamma\gamma} = 0.97 \pm 0.11~\text{stat.}\phantom{}^{+0.10}_{-0.08}~\text{syst.}
\end{align}
The latest CMS result is based on $\sim 36~\fb^{-1}$ and measures
\begin{align}
\text{CMS:}~\quad\sigma_{ggh} \text{BR}(h \to \gamma\gamma)  =84 \pm 11~\text{(stat.)} \pm 7 ~\text{(syst.)}~\fb
\end{align}
for a theoretically expected fiducial cross section of $73 \pm 4$
fb. Again this is in good agreement with the theory.  Turning to less
inclusive measurements, we show the jet-$p_T$ spectrum in gluon fusion
events in the right-hand side of Fig.~\ref{fig:hzzplot}. As can be
seen, the high-precision predictions detailed above are required to
model the transverse momentum spectrum satisfactorily given that gluon
fusion provides the largest contribution to the Higgs $p_T$
spectrum. The sensitivity of gluon fusion to the top quark threshold
which effectively embeds top quark scattering for center-of-mass
energies larger than $2 m_t$ has motivated a range of interesting and
BSM-relevant analyses. Most of our understanding of the Higgs boson
that was summarized in Sec.~\ref{sec:basic_char_mass} is based on
gluon fusion production, and we will discuss BSM-specific implications
of gluon fusion kinematics in the next
section~\ref{sec:exp_gf_kin}. There we will comment on how the
measurements of the Higgs boson's kinematic properties in the gluon
fusion process, which are already under good control (see
Fig.~\ref{fig:hzzplot} right), can be used to obtain a detailed
picture of the Higgs boson interaction modifications in non-resonant
BSM extensions.

The gluon fusion channel can be used to obtain limits on the new
resonant physics, including heavy pseudoscalars present in the 2HDM,
MSSM, and NMSSM, through the process,
\begin{align}
gg\rightarrow A\rightarrow H_i Z\, ,
\end{align}
where $H_i$ can be the SM Higgs Boson or one of the heavier neutral
scalars present in these models.  Searches have been made for
$Z\rightarrow l^+l^-$ and $H_i\rightarrow b {\overline b}$ and also
$H_i\rightarrow \tau^+\tau^-$.  The exclusions are model
dependent. Given the well-understood phenomenology of these scenarios,
searches often draw from the specific correlations of different
production and decay modes. We will therefore focus on these scenarios
in more detail in Sec.~\ref{sec:exp_global_full}.

\subsubsection{Power of kinematics}
\label{sec:exp_gf_kin}

If we take the definition of the gluon fusion process in
Eq.\eqref{eq:def_gf} seriously, it represents much more than a total
rate measurement. Two obvious signatures are off-shell Higgs
production, as discussed in Sec.~\ref{sec:basic_char}, and boosted
Higgs production recoiling against one or more hard jets illustrated
in Fig.~\ref{fig:higgspt}. Both of these processes allow us to move
beyond the heavy-top limit in the gluon-Higgs coupling discussed in
Sec.~\ref{sec:basis_decs} and probe the structure of the corresponding
loop(s). A simple theory hypothesis based on the dimension-6 SMEFT
Lagrangian combines the top Yukawa coupling with the effective Higgs
coupling to
gluons~\cite{Banfi:2013yoa,Azatov:2013xha,Grojean:2013nya,Buschmann:2014sia},
\begin{alignat}{5}
\lag 
&= \lag_\text{SM} + 
{\alpha_s\over 12 \pi} \left[ \Delta_t {F_{1/2}(\tau_t)\over F_{1/2}(\tau_\infty)} + \Delta_g 
 \right] \; 
\frac{h}{v} \; 
G_{\mu\nu}^aG^{a,\mu\nu} \; ,
\label{eq:lagrangian}
\end{alignat}
where $F_{1/2}$ is defined In Eq.\eqref{eq:etadef} and
$F_{1/2}(\tau_\infty) =-4/3$ is the $m_t\rightarrow \infty$ limit.  It
is obvious from Eq.\eqref{eq:lagrangian} that gluon fusion alone
cannot distinguish between $\Delta_g$ and $\Delta_t$.  Note that for
these terms the modified Higgs couplings from
Sec.~\ref{sec:basis_couplings} and the linearly realized SMEFT from
Sec.~\ref{sec:basis_eft_linear} are equivalent. In terms of physics
beyond the Standard Model we assume that a new particle generates a
finite $\Delta_g$, but without altering the kinematic structure of the
coupling. The appearance of the modified top Yukawa coupling $\Delta_t$ 
indicates that an alternative way to search
for the same scenario includes $t \bar{t}h$ production, discussed in
Sec.~\ref{sec:exp_tth}.

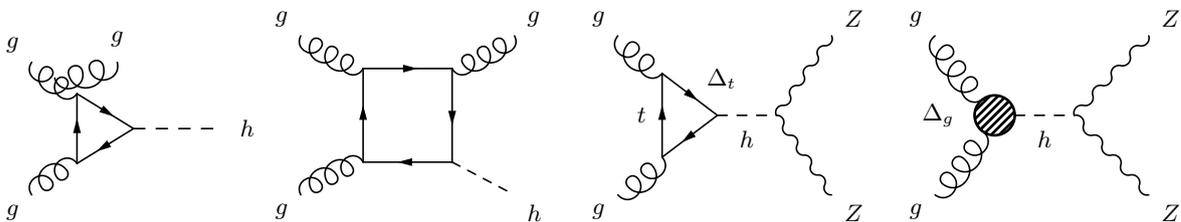
\begin{figure}[b!]
\begin{fmfgraph*}(80,50)
\fmfset{arrow_len}{2mm}
\fmfleft{g1,g2}
\fmfright{h1}
\fmftop{g3}
\fmflabel{$g$}{g1}
\fmflabel{$g$}{g2}
\fmflabel{$g$}{g3}
\fmflabel{$h$}{h1}
\fmf{gluon,width=0.6}{v1,g1}
\fmf{gluon,width=0.6}{g2,v2}
\fmf{fermion,tension=.05,width=0.6}{v1,v2}
\fmf{fermion,tension=.8,width=0.6}{v2,v3}
\fmf{fermion,tension=.8,width=0.6}{v3,v1}
\fmf{dashes,width=0.6}{v3,h1}
\fmffreeze
\fmf{phantom,tension=100,width=0.6}{g2,v4}
\fmf{phantom,tension=100,width=0.6}{v4,v2}
\fmf{gluon,width=0.6}{v4,g3}
\end{fmfgraph*}
\hspace*{5mm}
\begin{fmfgraph*}(100,60)
\fmfset{arrow_len}{2mm}
\fmfleft{g1,g2}
\fmfright{Z1,Z2}
\fmflabel{$g$}{g1}
\fmflabel{$g$}{g2}
\fmflabel{$h$}{Z1}
\fmflabel{$g$}{Z2}
\fmf{gluon,width=0.6}{v1,g1}
\fmf{gluon,width=0.6}{g2,v2}
\fmf{fermion,tension=0.35,width=0.6}{v1,v2}
\fmf{fermion,tension=0.7,width=0.6}{v2,v3}
\fmf{fermion,tension=0.7,width=0.6}{v4,v1}
\fmf{fermion,tension=0.35,width=0.6}{v3,v4}
\fmf{gluon,width=0.6}{v3,Z2}
\fmf{dashes,width=0.6}{v4,Z1}
\end{fmfgraph*}
\hspace*{5mm}
\begin{fmfgraph*}(100,60)
\fmfset{arrow_len}{2mm}
\fmfleft{g1,g2}
\fmfright{Z1,Z2}
\fmflabel{$g$}{g1}
\fmflabel{$g$}{g2}
\fmflabel{$Z$}{Z1}
\fmflabel{$Z$}{Z2}
\fmf{gluon,width=0.6,label.side=left}{g1,v1}
\fmf{gluon,width=0.6}{g2,v2}
\fmf{fermion,tension=.05,label={$t$},label.side=left,width=0.6}{v1,v2}
\fmf{fermion,tension=.8,width=0.6}{v2,v3}
\fmf{fermion,tension=.8,width=0.6}{v3,v1}
\fmf{dashes,tension=1.5,width=0.6,label={$h$},label.side=right}{v3,v4}
\fmf{photon,tension=0.8,width=0.6}{v4,Z1}
\fmf{photon,tension=0.8,width=0.6}{v4,Z2}
\fmffreeze
\fmf{phantom,tension=0.8,width=0.6,label={$\Delta_t$},label.side=right,label.dist=2.5}{v4,v2}
\end{fmfgraph*}
\hspace*{5mm}
\begin{fmfgraph*}(100,60)
\fmfset{arrow_len}{2mm}
\fmfleft{g1,g2}
\fmfright{Z1,Z2}
\fmflabel{$g$}{g1}
\fmflabel{$g$}{g2}
\fmflabel{$Z$}{Z1}
\fmflabel{$Z$}{Z2}
\fmf{gluon,width=0.6,label.side=left}{g1,v1}
\fmf{gluon,width=0.6}{g2,v1}
\fmfv{label.side=right,label.dist=15,label={$\Delta_g$}}{v1}
\fmfv{decor.shape=circle,decor.filled=shaded,decor.size=15}{v1}
\fmf{dashes,tension=1.5,width=0.6,label={$h$}}{v1,v4}
\fmf{photon,tension=0.8,width=0.6}{v4,Z1}
\fmf{photon,tension=0.8,width=0.6}{v4,Z2}
\end{fmfgraph*}
\caption{Sample Feynman diagrams contributing for Higgs production
  with up to two jets (left diagrams) and off-shell Higgs production
  (right diagrams).}
\label{fig:feyn_gf2} 
\end{figure}

As a starting point, we note that we could search for absorptive
threshold contributions in the kinematics of Higgs plus jets
production shown in Fig.~\ref{fig:feyn_gf2}. However, those effects
are too small to give us access to $\Delta_t$ and
$\Delta_g$~\cite{Buschmann:2014twa}. Alternatively, we search for top
mass effects in the $p_T$ spectra of the outgoing Higgs and
one~\cite{Baur:1989cm,Banfi:2013yoa,Grojean:2013nya,Schlaffer:2014osa} or more~\cite{Buschmann:2014twa,Greiner:2016awe} jets
\begin{align}
|\mat_{Hj(j)}|^2 
\propto 
m_t^4 \; \log^4 \frac{p_{T,h}^2}{m_t^2} \; .
\label{eq:pt_log}
\end{align}
In the left panel of Fig.~\ref{fig:top_gluon} we show the
corresponding change in the $p_{T,h}$ slope above the top mass. For a
benchmark point we need to ensure that the inclusive Higgs production
rate is not significantly changed, as for example for
$(\Delta_t,\Delta_g) = (-0.3,+0.3)$. Based on an optimistic
2-dimensional likelihood analysis we can estimate the impact of the
$p_{T,h}$-distribution as ruling out this benchmark point at 95\%
C.L. with an integrated luminosity of $800~\ifb$ of 13~TeV LHC data.

\begin{figure}[t]
\includegraphics[width=0.43\textwidth]{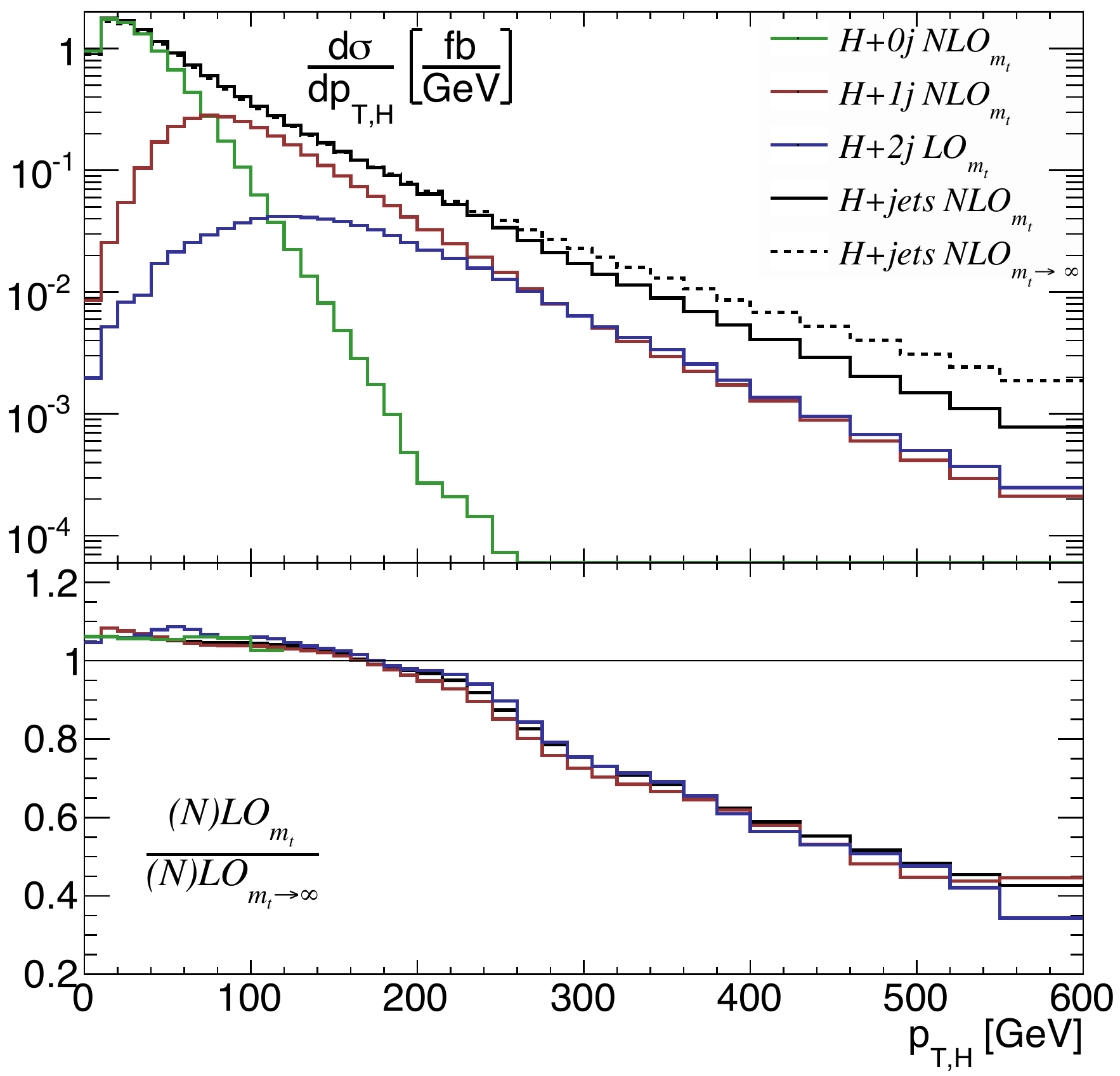}
\hspace*{0.1\textwidth}
\includegraphics[width=0.40\textwidth]{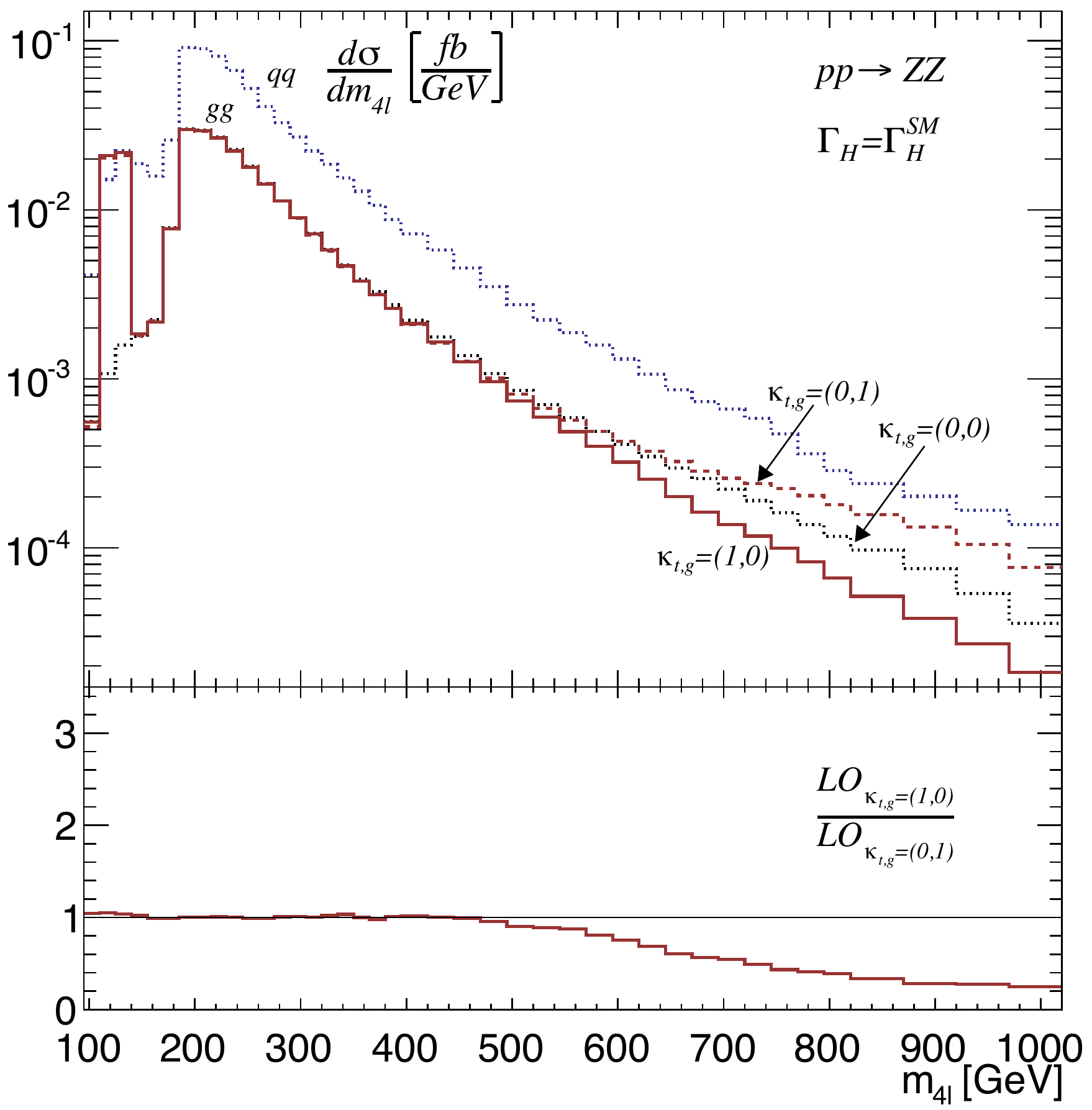}
\caption{Left: transverse momentum distribution $p_{T,h}$ for
  $h\rightarrow WW$+jets production. We use exclusive and merged jet
  samples with finite top mass effects and in the
  low-energy approximation. Right: $m_{4l}$ distributions in
  $q\bar{q}(gg)\rightarrow ZZ$ for the different signal hypothesis and
  the dominant background. Figures from Ref.~\cite{Buschmann:2014sia}.}
\label{fig:top_gluon} 
\end{figure}

A second way to test exactly the same physics hypothesis is off-shell
Higgs production. In this case, the change in the momentum dependence
of the Higgs-gluon coupling will appear through the interplay of the
signal Feynman diagrams shown in Fig.~\ref{fig:feyn_gf2}.  At large
$m_{ZZ}$ we can approximate the external gauge bosons with their
Goldstone modes and we can understand the signal amplitude in terms of
its longitudinal
components~\cite{Glover:1988rg,Azatov:2014jga,Cacciapaglia:2014rla}
\begin{align}
\mat_t^{++00}
=
- 2\; \frac{m_{4l}^2-2M_Z^2}{M_Z^2} \; 
\frac{m_t^2}{m_{4l}^2-M_h^2+i\Gamma_hM_h} \; 
\left[ 1 + \left(1-\frac{4m_t^2}{m_{4l}^2} \right) 
       f\left(\frac{4m_t^2}{m_{4l}^2}\right) \right] \; ,
\label{eq:m4l_full}
\end{align}
with $f$ defined in Eq.\eqref{fundef} and we consider the purely leptonic decay channel, $m_{ZZ}=m_{4l}$. 
 In the low-energy limit $m_t
\gg m_{4l} \gg M_h,M_Z$, it reproduces the usual effective
Higgs--gluon coupling. However for this measurement we are interested
in the low-energy regime $m_{4l} \gg m_t \gtrsim M_h,M_Z$ with
\begin{alignat}{5}
\mat_g^{++00}
&\approx
-\frac{m_{4l}^2}{2M_Z^2} \qqqquad 
&&\text{for} \; m_t \gg m_{4l} \gg m_H,M_Z \notag \\
\mat_t^{++00}
&\approx
+\frac{m_t^2}{2M_Z^2}
\log^2 \frac{m_{4l}^2}{m_t^2}\qqqquad 
&&\text{for} \; m_{4l} \gg m_t \gtrsim m_H,M_Z \notag \\
\mat_c^{++00}
&\approx
-\frac{m_t^2}
{2M_Z^2}\log^2 \frac{m_{4l}^2}{m_t^2} 
&&\text{for} \; m_{4l} \gg m_t \gtrsim M_Z \; .
\label{eq:m4l_tgc}
\end{alignat}
The logarithmic dependence on $m_{4l}/m_t$ is similar to the boosted
Higgs case.  Eventually, the ultraviolet divergence cancels between
the Higgs amplitude and the continuum. The relative sign between the
full top mass dependence and the low-energy limit leads to a
distinctive interference pattern with the $gg\rightarrow ZZ$
continuum, where the top mass dependence predicts a destructive
interference. We show this effect in the right panel of
Fig.~\ref{fig:top_gluon}. Again, an optimistic 2-dimensional
likelihood analysis in terms of $m_{4l}$ and the leading angular
correlation indicates that we can rule out the $(-0.3,0.3)$ benchmark
point at 95\% C.L. with an integrated luminosity of $2000~\ifb$ of
13~TeV LHC data, significantly later than using the boosted Higgs
signature.

A third, indirect way to test yet the same hypothesis is Higgs pair
production~\cite{Azatov:2016xik}, but due to the small rate we do not
expect such a measurement to contribute significantly.

Finally, the elephant in the room is the direct $t\bar{t}h$ rate
measurement. As we will see in Secs.~\ref{sec:exp_tth}
and~\ref{sec:exp_global},  $t\bar{t}h$ production, together with the measurement of the Higgs
production rate in gluon fusion, will likely lead to the best reach. On
the other hand, while these specific kinematic regions appear less
attractive to test a simple and well-defined hypothesis, they offer
important consistency tests of our underlying hypotheses in a
situation where we have to rely on appropriate measurements to
identify BSM physics in the Higgs sector.

\subsection{Weak boson fusion}
\label{sec:exp_wbf}

Weak boson fusion (WBF),
\begin{align}
qq \to qqh + X 
\end{align}
is an especially attractive relevant Higgs production process for a
multitude of theoretical and experimental reasons. As illustrated in
Fig.~\ref{fig:WBFdiag} it directly probes gauge boson scattering and
is therefore a sensitive probe of any physics that relates to
perturbative unitarity conservation~\cite{Lee:1977eg,Lee:1977yc}.

\begin{figure}[b!]
\includegraphics[width=7cm]{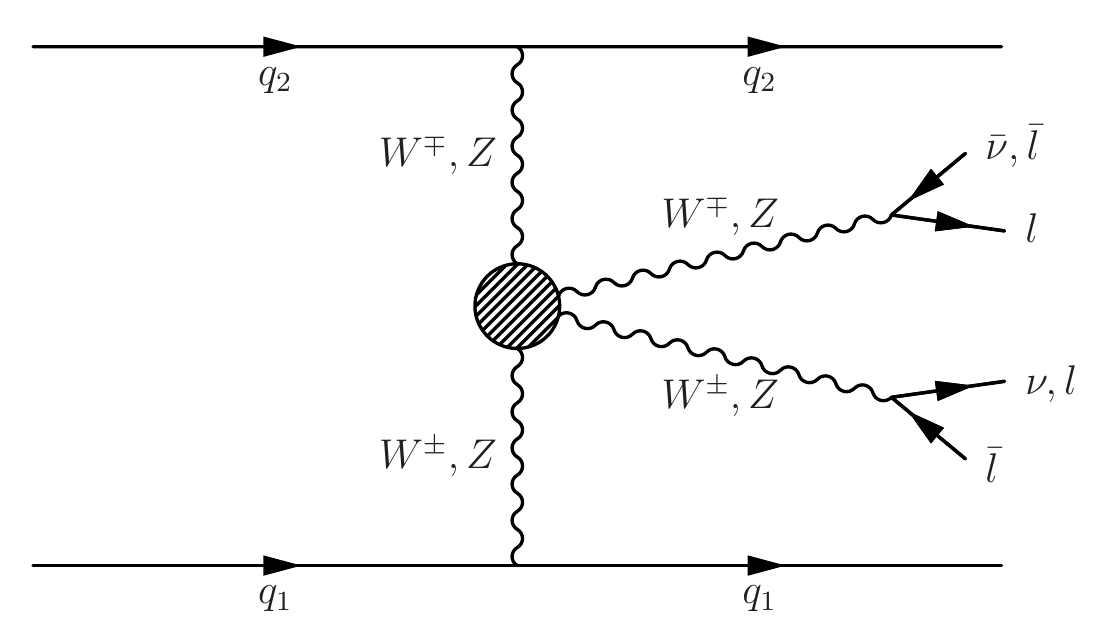}
\caption{Representative Feynman diagrams for weak boson fusion Higgs
  production. The blob represents all possible scatterings, including Higgs exchange diagrams. }
\label{fig:WBFdiag} 
\end{figure}

\subsubsection{Motivation and signature}
\label{sec:exp_wbf_mot}

A key WBF signature from a theory perspective is based on the decay $h
\to WW$~\cite{Cahn:1983ip}.  In the high energy limit, gauge boson scattering $VV \to VV$
($V_i=W^\pm,Z$) is described in terms of the longitudinal degrees of
freedom, and the scalar matrix element has the simple form
\begin{align}
i{\cal{M}}= \alpha^{(2)} s^2 + \alpha^{(1)} s  + \alpha^{(0)} \; ,
\end{align}
where $\sqrt{s}$ is the energy of the sub-process. 
Divergent scattering probabilities are only avoided through a
cancellation of the $\alpha^{(2,1)}$, which are dimensionful constants
including coupling constants and masses~\cite{Lee:1977eg}. In the Standard Model
the relevant couplings are
\begin{alignat}{5}
 \alpha^{(2)}&\qqquad & g_{WWWW}&= g_{WW\gamma}^2+g_{WWZ}^2 \notag \\
 \alpha^{(1)}&\qqquad & 4m_W^2g_{WWWW} &= 3 M_Z^2 g_{WWZ}^2
      + g_{WWh}^2 \; .
\end{alignat}
The contributing quartic and trilinear couplings are $g_{WWWW}=g^2$,
$g_{WW\gamma}=g s_W$, and $g_{WWZ}=g c_W$. This way $\alpha^{(2)}=0$
is protected by gauge invariance.  The second sum rule is sensitive
to the custodial isospin relation $M_Z^2=M_W^2/c_W^2$ and the Higgs
coupling $g_{WWh}=g M_W$~\cite{Birkedal:2004au,Englert:2015oga}.
These cancellations are indeed tantamount to spontaneous symmetry
breaking~\cite{Cornwall:1973tb,Cornwall:1974km}, and although models
with non-SM symmetry breaking realize unitarity cancellations in often
non-trivial ways (for example in the case of Higgs triplets), in generic
perturbative extensions of the Standard Model they remain intact.  This motivates
searches for new resonances linked to the Higgs sector in WBF
production~\cite{Birkedal:2004au,Frank:2012wh,Englert:2015oga}.

One could therefore imagine that physics which modifies the unitarity
sum rules, \eg through the value of $g_{WWh}$, will also lead to large
enhancements of the WBF production rate.  The problem is that all four
gauge bosons couple to light fermions, which means that longitudinal
contributions are mass-suppressed and
the unitarity sum rules
are not directly accessible without high statistics. However, the
tendency of (non-)perturbative models to enforce unitarity at least up
to some intermediate scale with additional resonances still makes weak
boson fusion a well-suited process to measure or constrain the
presence of (non)-resonant new physics as thresholds are also visible
for transverse polarizations (for which the high energy-limit is
trivial)~\cite{Englert:2015oga}.

\begin{figure}[t]
\includegraphics[width=0.42\textwidth]{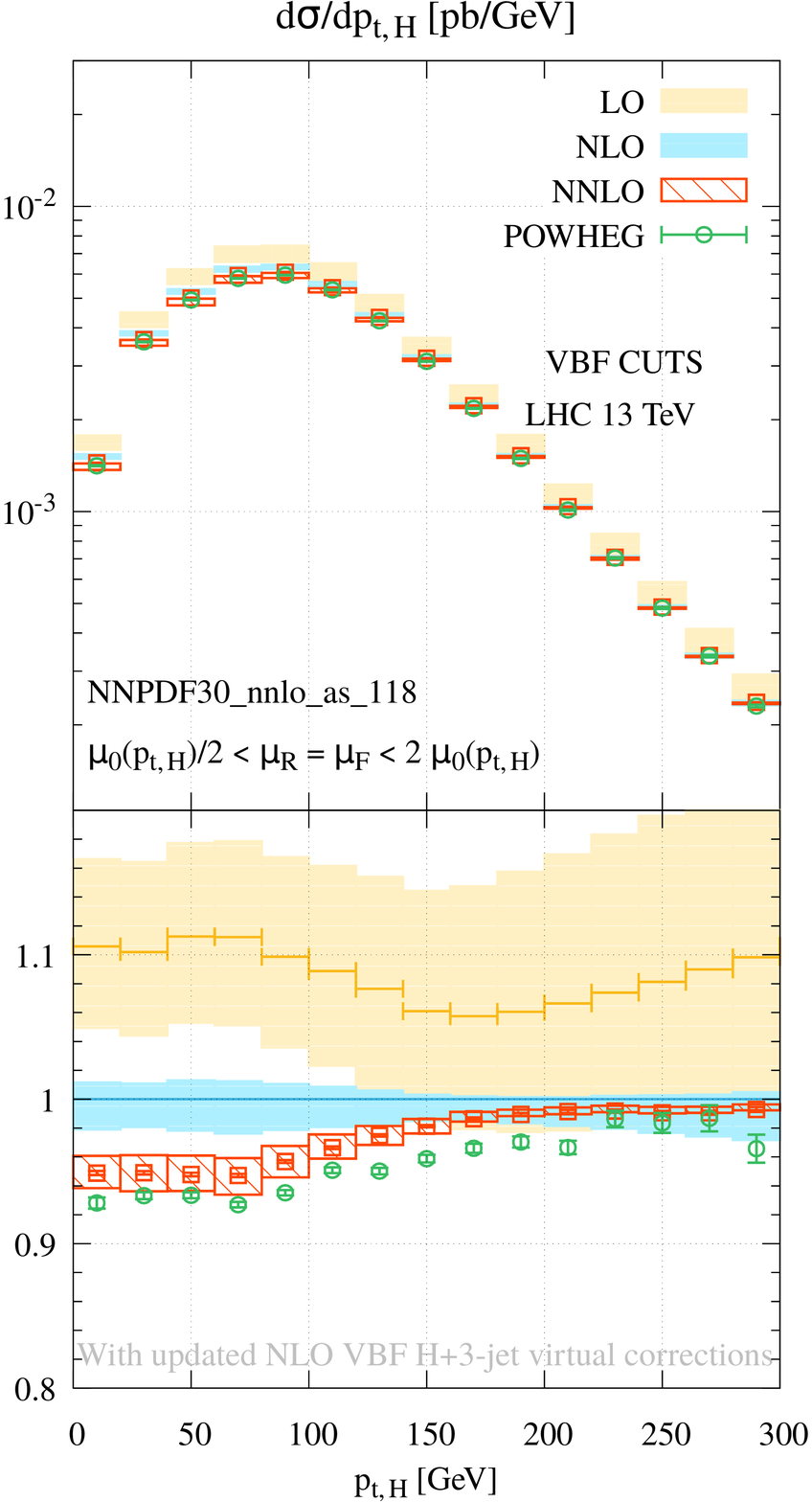}
\parbox{0.42\textwidth}{\vspace{-11cm}\includegraphics[width=0.38\textwidth]{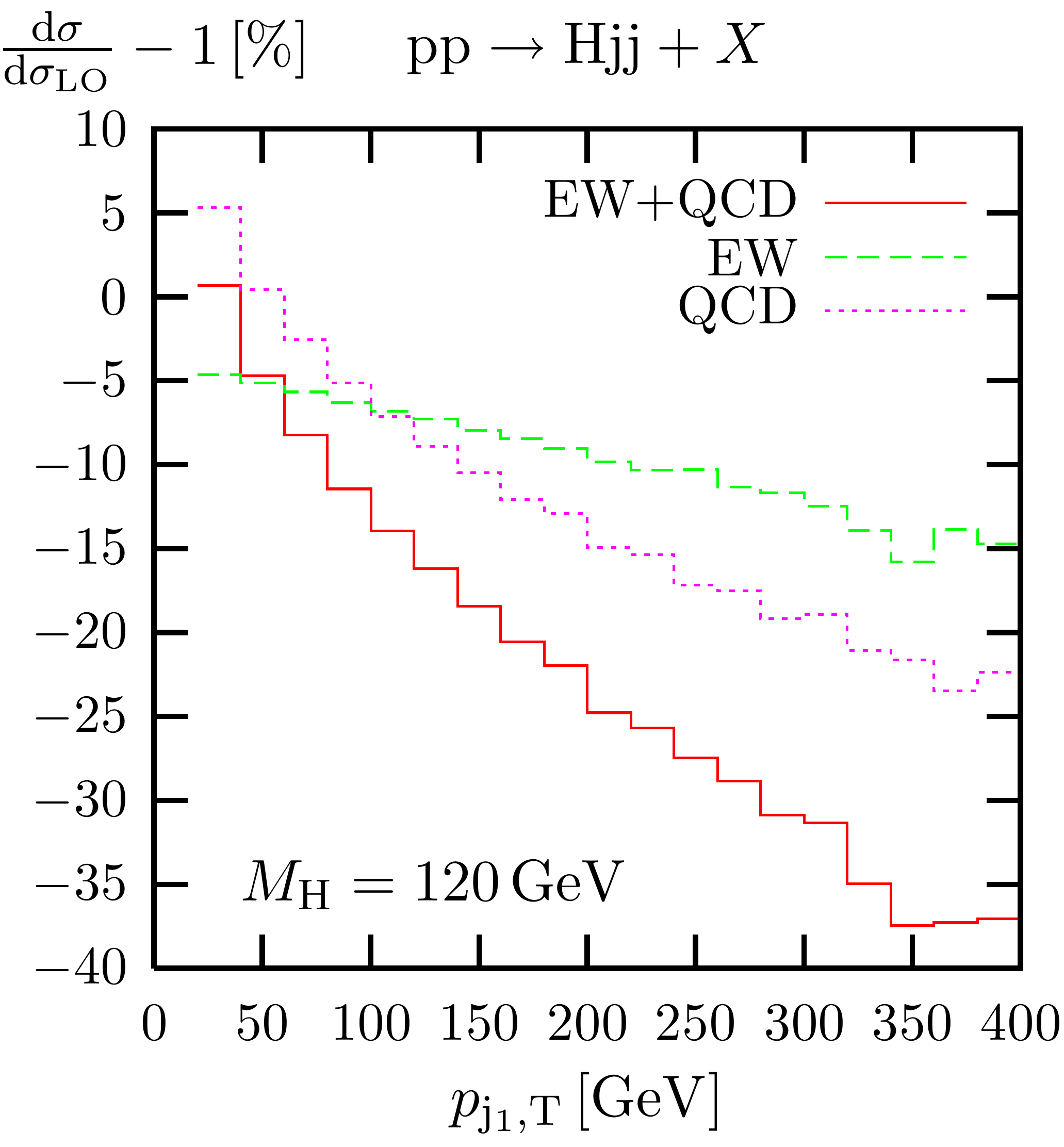}}
\hskip 0.5cm
\caption{Left: Differential NNLO QCD corrections to the transverse
  momentum distribution of the Higgs boson. Figure from
  Ref.~\cite{Cacciari:2015jma}. Right: Impact of the combined
  QCD+electroweak corrections on the leading jet transverse momentum
  distribution in WBF. Figure from Ref.~\cite{Ciccolini:2007ec}.}
\label{fig:WBFpred}
\end{figure}

Experimentally, weak boson fusion processes have a distinct
signature~\cite{Kleiss:1987cj,Baur:1990xe,Barger:1991ib,Rainwater:1996ud,Rainwater:1999sd,Cox:2010ug,Barger:1988vh}. To
produce the Higgs, the incoming quarks are probed at large momentum
fractions while the radiated weak bosons lead to a comparably small
transverse momentum~\cite{Dawson:1984gx}.  This leads to two energetic
forward jets at large invariant mass, separated by a large rapidity
gap. The Higgs is produced centrally and with sizeable transverse
momentum, allowing us to search for the Higgs
decays~\cite{Rainwater:1998kj,Plehn:1999xi,Rainwater:1999sd,Kauer:2000hi,Eboli:2000ze}
\begin{align}
pp \to qqh \to 
\begin{cases} 
 qq \; \tau^+ \tau^- \\ 
 qq \; W^+ W^- \\
 qq \; \text{invisible}  \; . 
\end{cases}
\end{align}
The absence of QCD radiation into the central region follows from the
vanishing color factor when we exchange a gluon between the incoming
quarks, and the tiny interference between the WBF $t$-channel and
$u$-channel amplitudes.  It allows us to use information on the
central jet activity, to first order central jet vetos, as formidable
tools to enhance the Higgs signal over the
backgrounds~\cite{Barger:1991ar,Barger:1994zq,Bernaciak:2013dwa,Biekotter:2017gyu}.
This includes the irreducible gluon-fusion Higgs production that
contributes around 15\% to the Higgs production for typical WBF cuts
that can be reduced to a few percent level as these processes are
characterized by a central QCD emission
profile~\cite{Andersen:2008gc,Andersen:2010zx}.

Finally, an advantage of the WBF process  over Higgs
production in gluon fusion, is that the 3-particle final state allows
us to study many aspects of the $VVh$ interaction. For example, the
sensitivity to the $CP$ properties of this vertex are discussed in
Sec.~\ref{sec:basic_char_cp}. The WBF version of the off-shell Higgs
measurement introduced in Sec.~\ref{sec:basic_char_mass} is actually
the more robust channel to constrain the Higgs
width~\cite{Englert:2014ffa} when we fully correlate the production
and $h\to VV$ decay and in  this way minimize the vulnerability to unknown
BSM loop contributions~\cite{Englert:2014aca,Campbell:2015vwa}.

\subsubsection{Precision predictions}
\label{sec:exp_wbf_pred}

The particular structure of the WBF topology from a QCD point of view
as ``double-deep inelastic scattering'' also renders this process
quite non-standard in terms of expected perturbative QCD
corrections. QCD radiation into the central part of the detector is
highly suppressed.  Interference of the weak boson fusion contribution
with associated Higgs production~\cite{Ciccolini:2007ec} or gluon
fusion is negligible~\cite{Bredenstein:2008tm} for acceptance cuts
that select the WBF signal region.

\begin{figure}[t]
\includegraphics[width=0.55\textwidth]{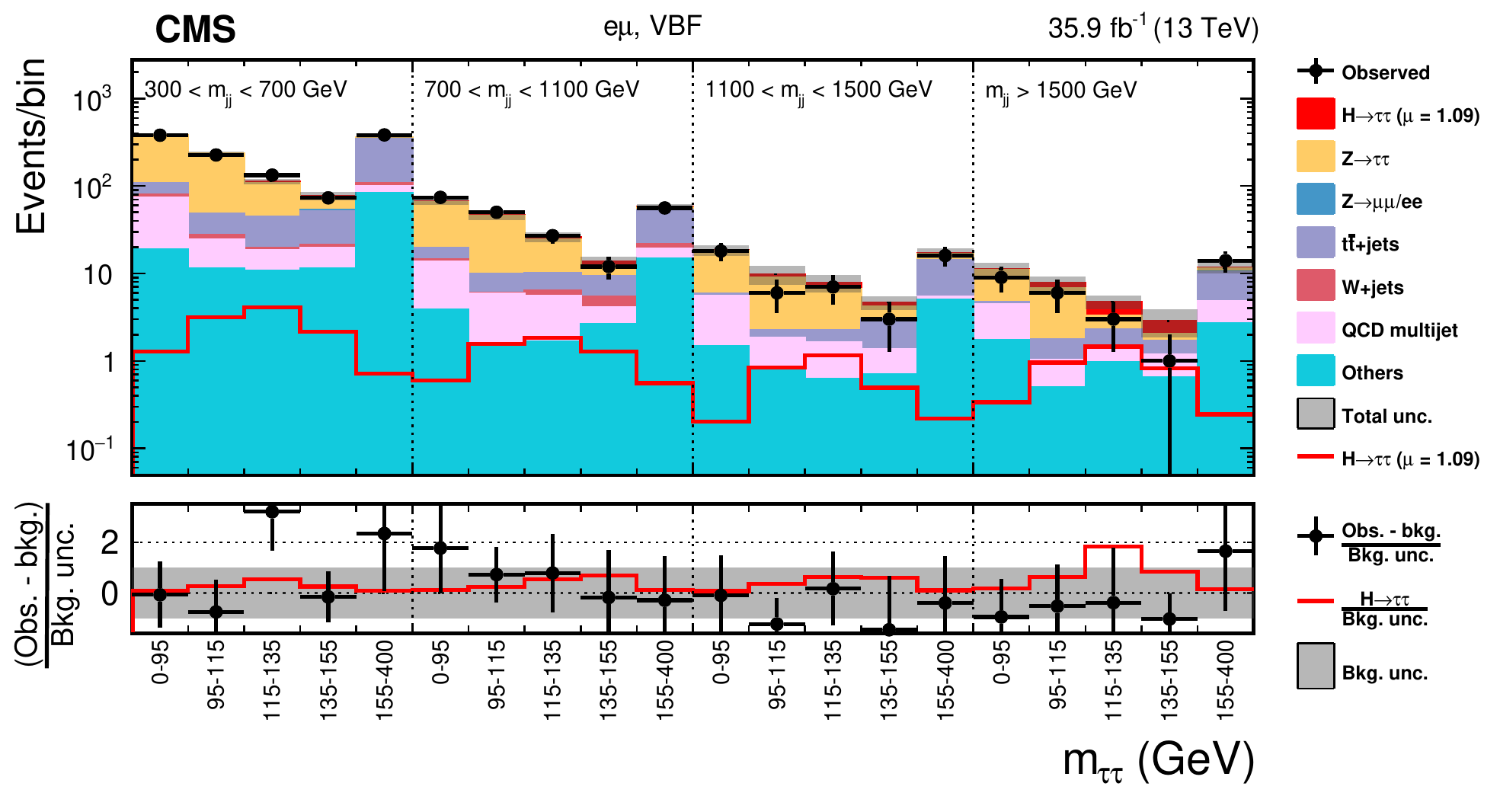}
\hspace*{0\textwidth}
\includegraphics[width=0.40\textwidth]{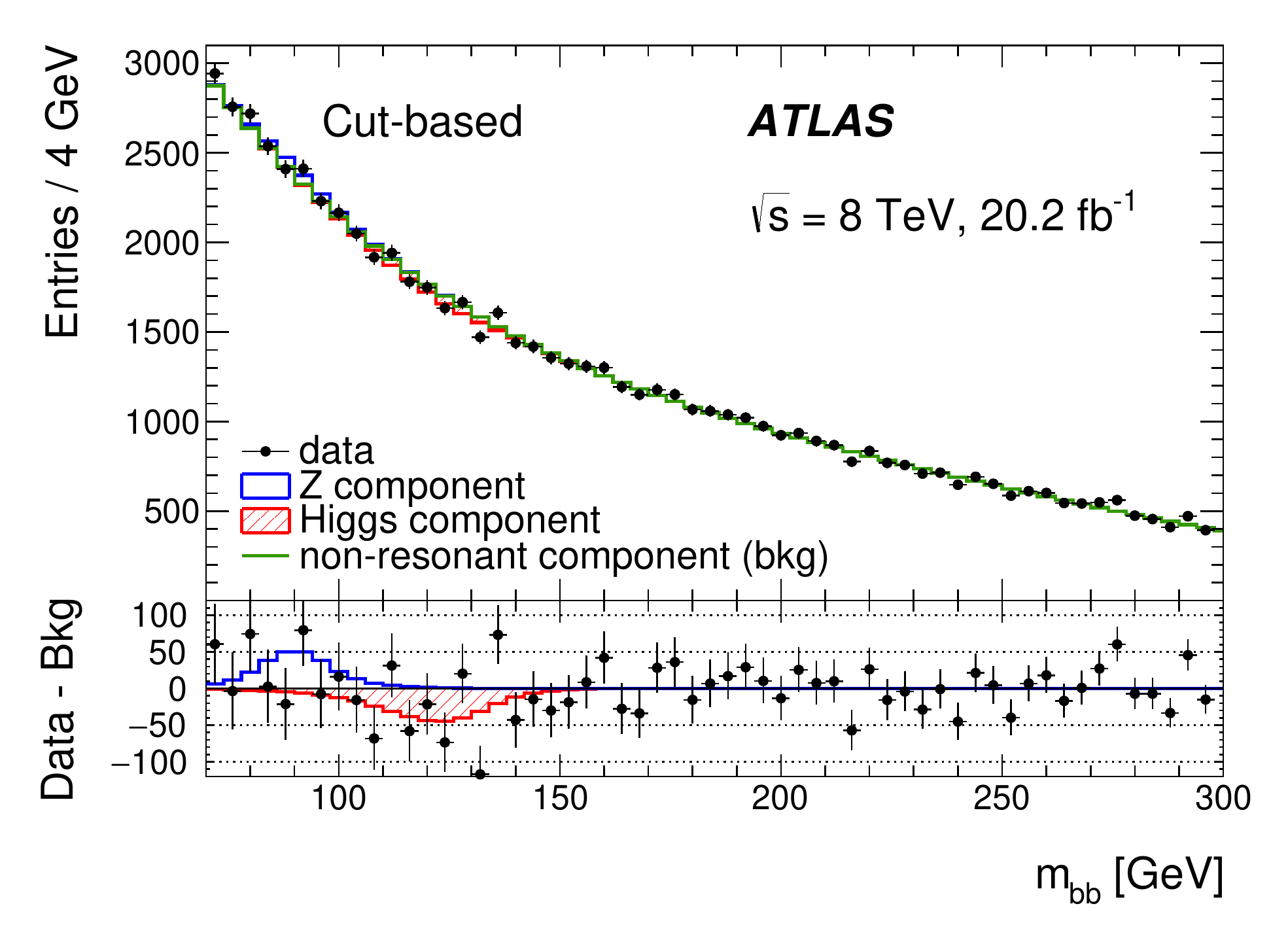}
\caption{Left: 2-dimensional distributions in $m_{\tau \tau}$ for
  slices of the invariant mass of the tagging jets. Figure from
  Ref.~\cite{Sirunyan:2017khh}.  Right: search for $h\to b\bar b$
  decays in the WBF channel.  Figure from Ref.~\cite{Aaboud:2016cns}.}
\label{fig:WBFexpbb}
\end{figure}

The dominant QCD emission effects are therefore related to bremsstrahlung in the
direction of the quark legs. Due to the color singlet $t$-channel
exchange, there are no one-loop contributions that color-connect the
external quark legs. The small NLO QCD corrections can be almost
entirely absorbed into parton densities through choosing the
factorization scales for the incoming partons as the $t$-channel
momentum transfer of  the radiated weak boson, all in agreement with the
deep inelastic scattering
paradigm~\cite{Jager:2006zc,Jager:2006cp,Bozzi:2007ur}.  Considering
the next order in perturbation theory, the NNLO QCD corrections change
the total cross section by around
5\%~\cite{Bolzoni:2011cu,Cacciari:2015jma,Cruz-Martinez:2018rod},
computed again in the structure function approximation that assumes no correlation between the incoming protons.  In 
Fig.~\ref{fig:WBFpred} we show the effects of the NNLO QCD corrections
on the transverse momentum of the Higgs. The radiation of a second jet
renders the transverse momentum spectrum slightly softer, especially
around the maximum $p_{T,h} \approx M_W$. As also shown in this plot,
this kind of effect is expected from the parton shower treatment of
the second jet radiation. Predictions for WBF production in association up to three jets have been discussed 
in Refs.~\cite{Campanario:2013fsa,Campanario:2018ppz}.

Given the small QCD corrections, WBF Higgs production is one of the
LHC processes where we also need to control the electroweak
corrections.  Similar to the NLO QCD corrections, the NLO electroweak
corrections to the total WBF rate range around 5\%. For the total
cross section, the QCD and electroweak corrections largely cancel each
other~\cite{Ciccolini:2007jr,Ciccolini:2007ec,Figy:2010ct}. In the
right panel of Fig.~\ref{fig:WBFpred} we show the effect of the
electroweak corrections on the transverse momentum of the leading
jet. While the QCD corrections switch sign around the maximum of the
$p_{T,j}$ distribution, the electroweak corrections are always
negative and increase towards larger momentum as expected for
electroweak corrections leading to large Sudakov logarithms.

\subsubsection{LHC Analyses}
\label{sec:exp_wbf_ana}

One of the Higgs signatures which relies on the WBF production channel
is the decay to tau leptons~\cite{Sirunyan:2017khh}. The approximate
reconstruction of the invariant mass $m_{\tau\tau}$ only works when
the decaying Higgs has sizeable transverse momentum. In the left panel
of Fig.~\ref{fig:WBFexpbb} we show the signal and backgrounds for
leptonic decays of the $\tau$-pair. The main background is $Z$+jets
production, which can actually be divided into a QCD process with
$\sigma_{Zjj} \propto \alpha_s^2 \alpha$ and an electroweak process
with $\sigma_{Zjj} \propto \alpha^3$. The difference between them is
that the latter has a color structure like the signal, so the analysis
of the central jet activity only suppresses this background based on
small kinematic differences. Typically, the combined $Z$+jets
background is dominated by the QCD process before we take into account
the central jet activity, while after including this information the
two processes are of similar size. Increasing the cut on the tagging
jets through $m_{jj}$ reduces the size of the signal, but it leads to
an even bigger improvement of $S/B$. This shows how we can use the
properties of the hard $2\to 3$ signal process to significantly reduce
the backgrounds.

In the right panel of Fig.~\ref{fig:WBFexpbb} we show an attempt to
extract the Higgs decay $h \to b\bar{b}$ in WBF. Given the pure QCD
nature of the signal process, it is difficult to extract this process 
from the backgrounds. Therefore, available constraints are too weak to obtain limits on
SM-like production, see Fig.~\ref{fig:WBFexpbb}. Proof-of-principle investigations,
however, suggest that relying on advanced analysis strategies this channel
will allow us to set constraints on the associated Higgs interactions~\cite{Englert:2015dlp}.
The $Vh$ channel discussed in Sec.~\ref{sec:exp_vh} provides another avenue to 
test these couplings.

\begin{figure}[t]
\includegraphics[width=0.41\textwidth]{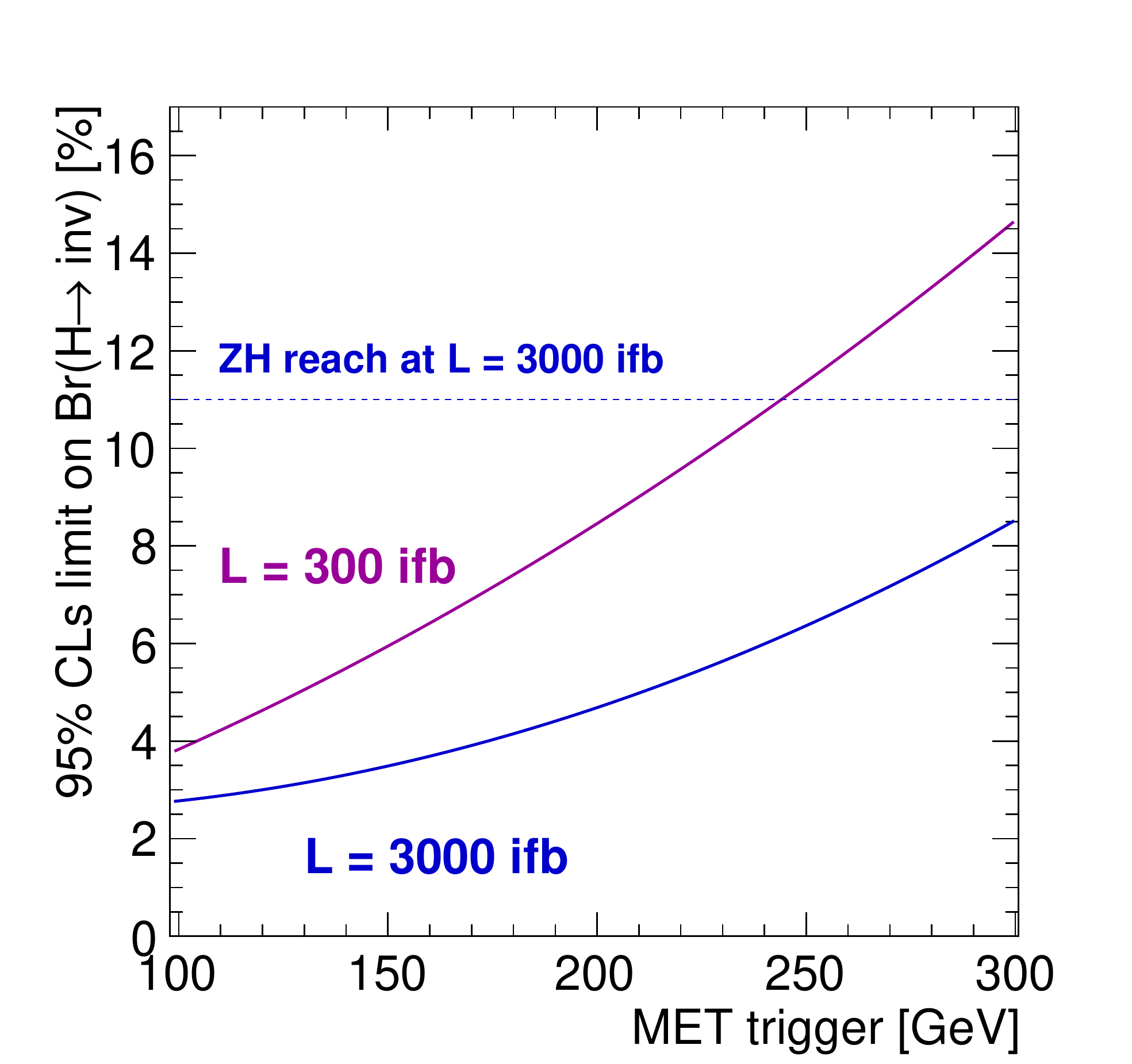}
\hspace*{0.1\textwidth}
\includegraphics[width=0.40\textwidth]{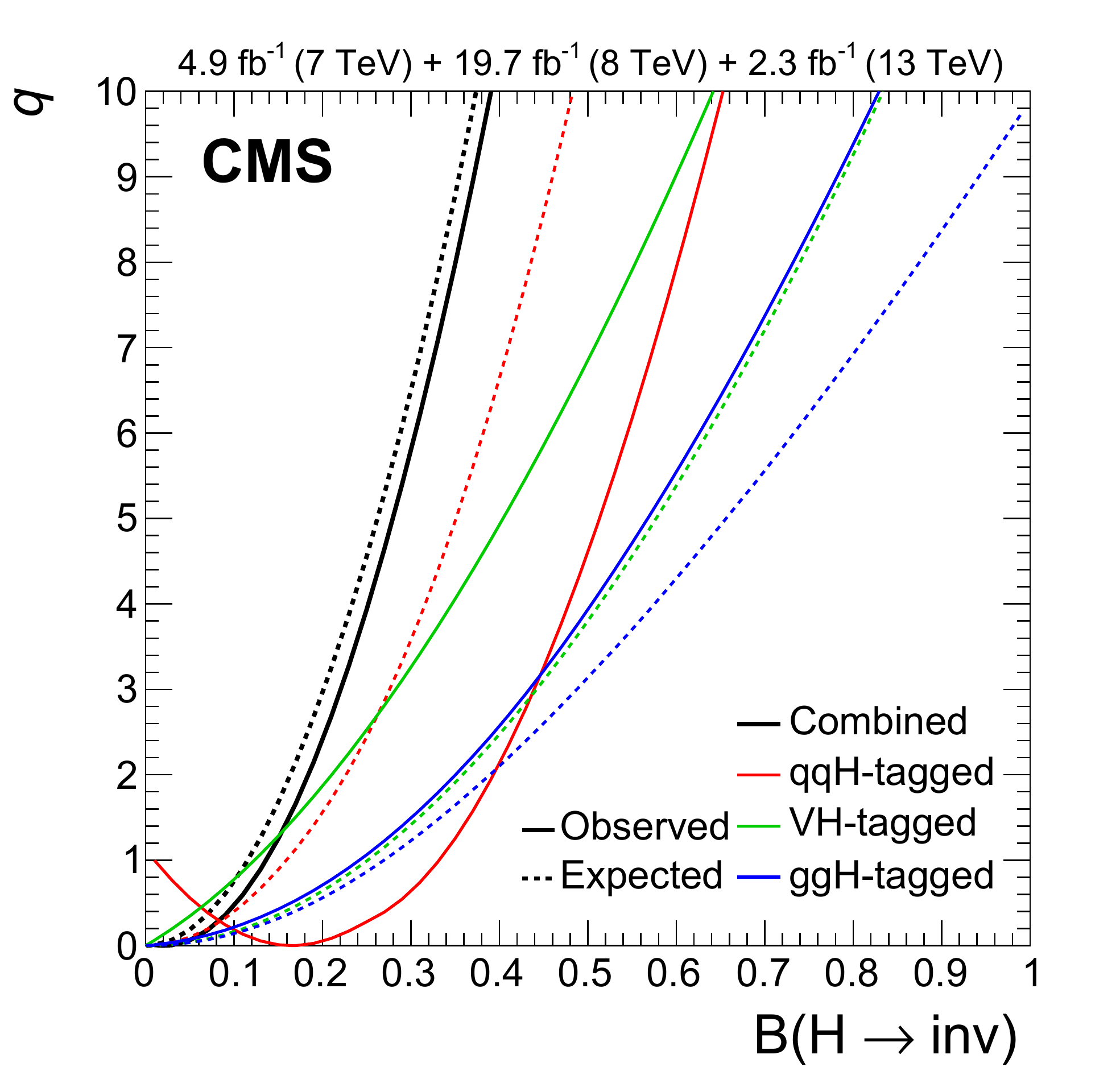}
\caption{Left: Comparison of the capabilities of weak boson fusion
  (solid curves) and $Zh$ production (dotted curve) to observe
  invisible Higgs decays, as a function of the $\met$ threshold in the
  WBF process . Figure from Ref.~\cite{Biekotter:2017gyu}.  Right:
  Profile-Likelihood as a function of the invisible Higgs branching
  ratio for the CMS analysis of Ref.~\cite{Khachatryan:2016whc}. Shown are
  individual channels as well as the combined result. Solid curves are
  computed from data, dashed curves are the SM expectation of no
  invisible decay. The upwards fluctuation in the WBF category is
  consistent with the Standard Model at 95\% confidence level. Figure from
  Ref.~\cite{Khachatryan:2016whc}.}
\label{fig:WBFdm}
\end{figure}

The WBF production process is the best way to search for
invisible SM-like Higgs decays at the
LHC~\cite{Eboli:2000ze,Bernaciak:2014pna,Biekotter:2017gyu}. In the
left panel of Fig.~\ref{fig:WBFdm} we compare the reach of this
channel with the second-best strategy, based on $Zh$ production. While
WBF has an advantage with its sizeable transverse momentum of the
Higgs boson with a maximum around $p_{T,h} \approx M_W$, triggering is
a problem. We have to rely on a combination of missing transverse
energy with two forward tagging jets. This problem can be solved by
$Zh$ production combined with a leptonic $Z$-decay, as will be
discussed in Sec.~\ref{sec:exp_vh}. Towards high LHC luminosities we
see that the WBF analysis always wins over the $Zh$ channel, provided
we can trigger on missing transverse momentum below around
300~GeV~\cite{Biekotter:2017gyu}. In the right panel of
Fig.~\ref{fig:WBFdm} we show the results of a CMS search for invisible
Higgs decays using WBF production, $Zh$ production, and gluon fusion
Higgs production including a hard jet. While the expected hierarchy of
the channels is WBF ahead of $Zh$ ahead of gluon fusion, the observed
limit is driven by $Zh$ production. Combining all analyses gives a
95\% C.L. limit of $\br(h \to \text{inv}) < 0.24$, assuming all
Higgs couplings to be SM-like. A more complete analysis, including
variable Higgs couplings, will be the topic of
Sec.~\ref{sec:exp_global}.  Such invisible Higgs decays can be
interpreted in direct relation with dark matter, in particular through
the Higgs potential role in portaling to a hidden sector through the
SM singlet operator $|\phi|^2$ as outlined in
Sec.~\ref{sec:basic_weak_singlet}. If additional dark fields are
sufficiently light, the Higgs can obtain a new invisible decay channel
that would manifest itself as an invisible branching
ratio~\cite{Butter:2015fqa}.

Finally, WBF production is also an important channel for new physics
searches in visible final states.  $H^\pm\to W\pm Z$ interactions can
be large in the context of the Georgi-Machacek model where a hierarchy
in triplet vs double-induced electroweak symmetry breaking can be
relaxed~Fig.~\ref{fig:WBFH+}. The presence of these vertices links to
how unitarity is restored in such models as discussed above. Searches
for doubly charged Higgs bosons proceed in a similar fashion. As the
branching ratio of $H^{\pm\pm}\to W^\pm W^\pm$ is enhanced for heavy
doubly charged Higgs masses (completely analogous to the case of the
SM decay $h\to W^+W^-$), clean same-sign lepton$+\slashed {E}_T+2$~jet
searches can be used to obtain more stringent constraints on triplet
electroweak symmetry breaking, Fig.~\ref{fig:WBFH+}.

\begin{figure}[t]
\includegraphics[width=0.46\textwidth]{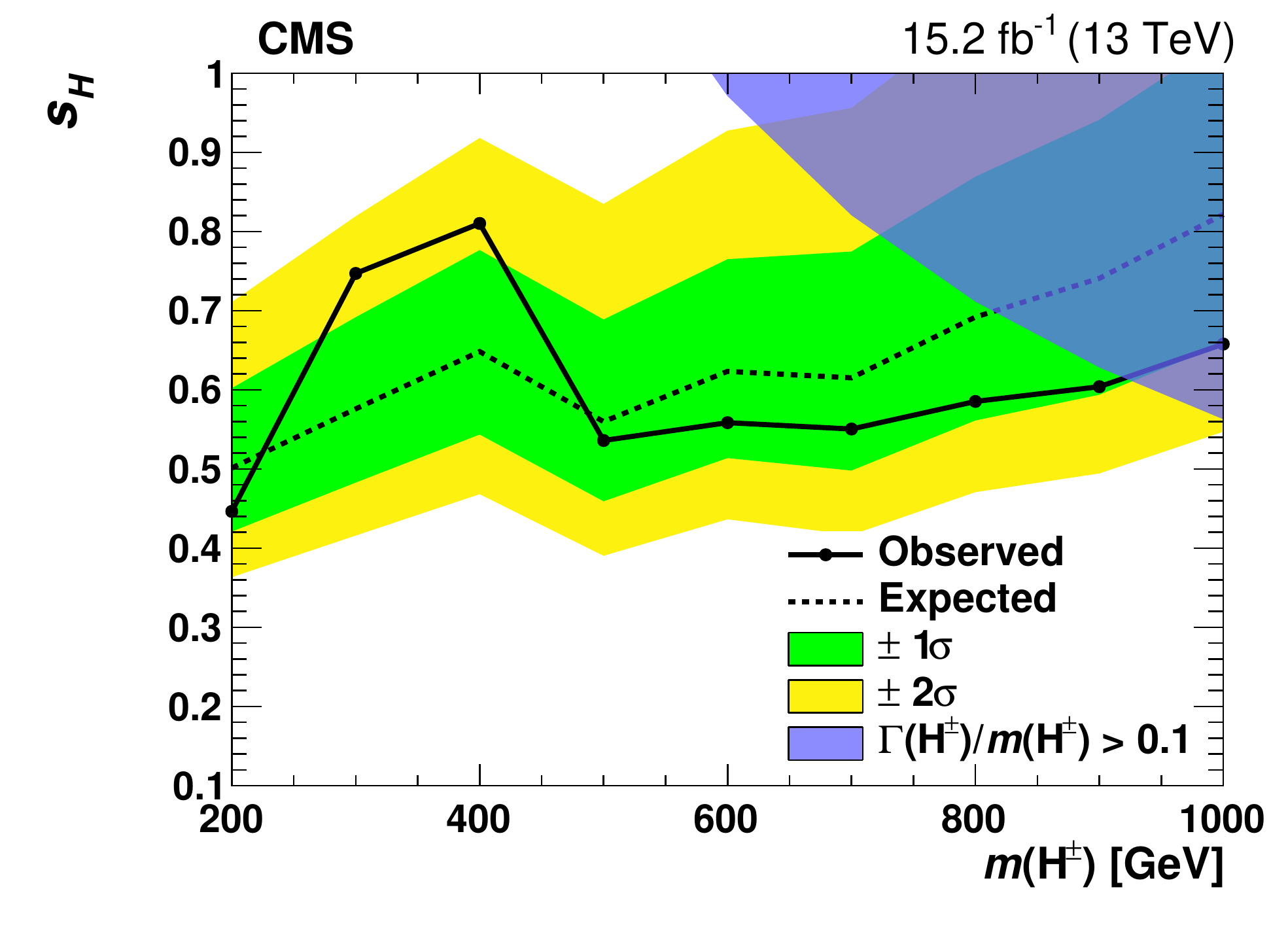}
\hspace*{0.1\textwidth}
\includegraphics[width=0.41\textwidth]{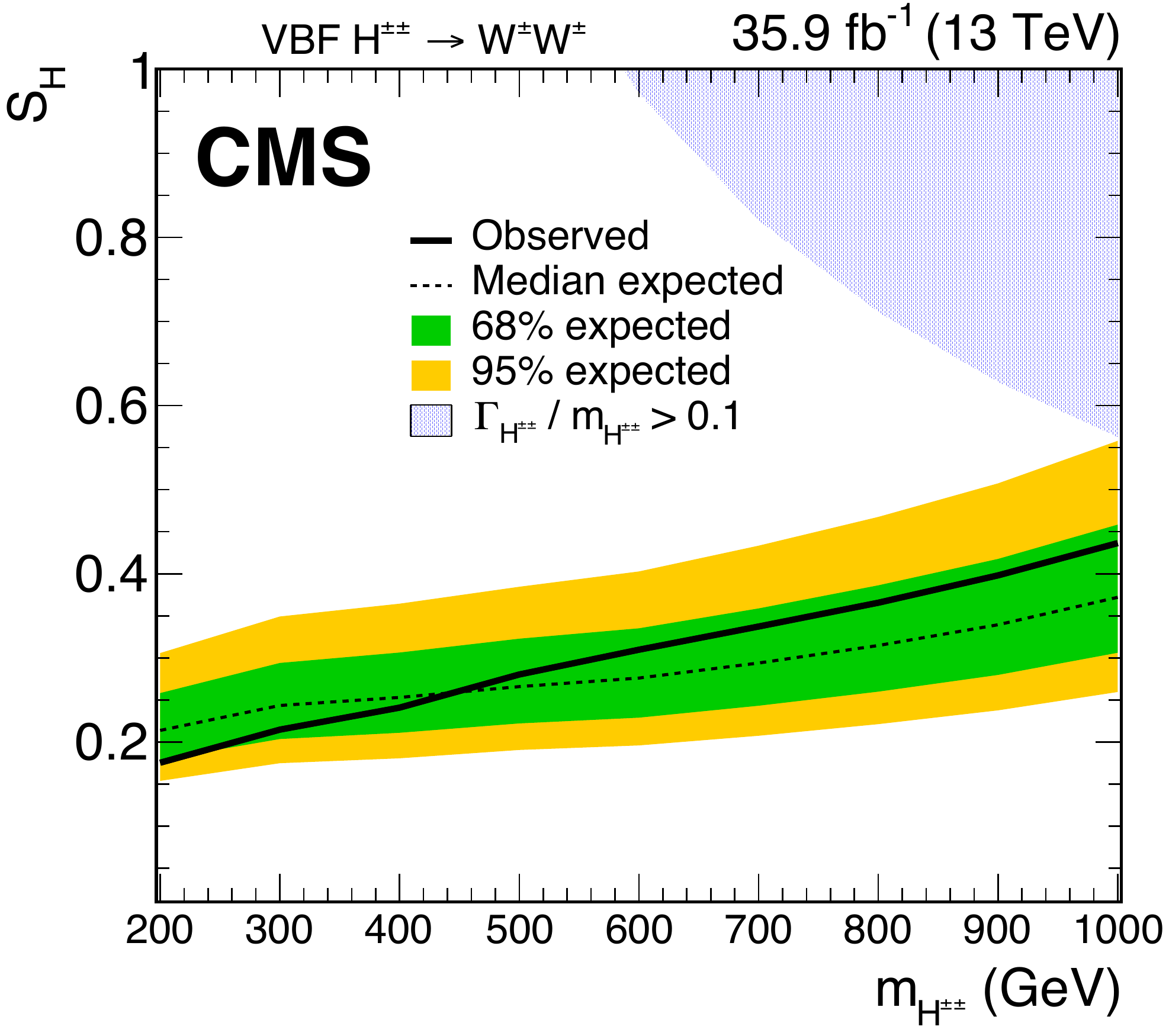}
\caption{Left: Constraints on the production of a charged Higgs in the
  WBF channel, mapped onto the triplet character of the electroweak
  vacuum, with $s_H=0,1$ referring to pure doublet and triplet
  EWSB. Figure taken from Ref.~\cite{Sirunyan:2017sbn}. Right:
  Constraints on the production of a doubly charged Higgs decaying
  into same-sign leptons in the WBF channel, expressed in terms of the
  same mixing parameter. Figure taken from
  Ref.~\cite{Sirunyan:2017ret}.}
\label{fig:WBFH+}
\end{figure}

\subsection{Associated $Vh$ production}
\label{sec:exp_vh}

Associated Higgs production with a weak boson, also called Higgs-strahlung,
\begin{align}
pp \to Vh + X 
\qquad \text{with} \; V=W,Z
\end{align}
corresponds to the working horse channel $e^+ e^- \to Zh$ at LEP or at
future electron-positron colliders.  Unlike at lepton colliders, it is
not the production channel with the leading cross section. As a matter
of fact, it was clear from the very beginning that this channel would
hardly contribute to a Higgs discovery.

On the other hand, $Vh$ production is the most experimentally clean
Higgs production channel, when we take into account triggering, background
suppression, controlled perturbation theory, final state
reconstruction, etc. As we will see, these aspects make it extremely
useful when we want to search for specific Higgs decays, provide
crucial input for a global Higgs analysis, or study the detailed
kinematics of the hard Higgs production process. 

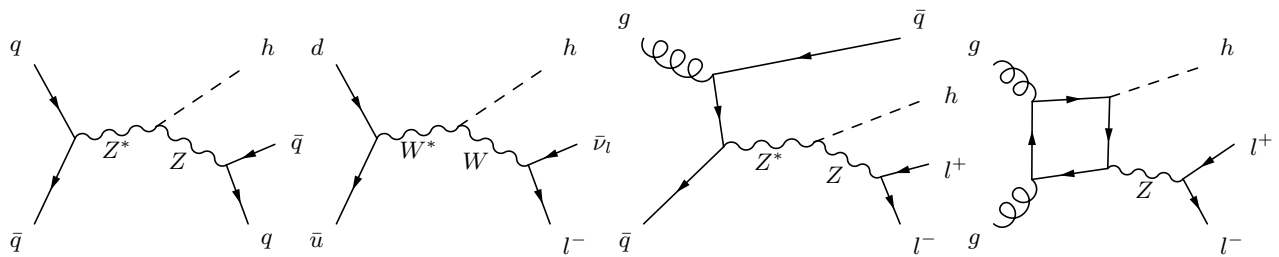
\begin{figure}[b!]
  \begin{center}
    \begin{fmfgraph*}(100,60)
      \fmfset{arrow_len}{2mm}
      \fmfleft{i2,i1}
      \fmfright{o3,o2,o1}
      \fmf{fermion,width=0.6,lab.side=left,tension=2}{i1,v1}
      \fmf{fermion,width=0.6,lab.side=left,tension=2}{v1,i2}
      \fmf{photon,width=0.6,lab.side=right,label=$Z^*$,label.dist=3,tension=2}{v1,v2}
      \fmf{dashes,width=0.6,lab.side=left,tension=1}{v2,o1}
      \fmf{photon,width=0.6,lab.side=right,label=$Z$,label.dist=3,tension=1}{v2,v3}
      \fmf{fermion,width=0.6,lab.side=left,tension=1}{o2,v3}
      \fmf{fermion,width=0.6,lab.side=left,tension=1}{v3,o3}
      \fmflabel{$q$}{i1}
      \fmflabel{$\bar{q}$}{i2}
      \fmflabel{$\bar{q}$}{o2}
      \fmflabel{$q$}{o3}
      \fmflabel{$h$}{o1}
    \end{fmfgraph*}
    \hspace*{3mm}
    \begin{fmfgraph*}(100,60)
      \fmfset{arrow_len}{2mm}
      \fmfleft{i2,i1}
      \fmfright{o3,o2,o1}
      \fmf{fermion,width=0.6,lab.side=left,tension=2}{i1,v1}
      \fmf{fermion,width=0.6,lab.side=left,tension=2}{v1,i2}
      \fmf{photon,width=0.6,lab.side=right,label=$W^*$,label.dist=3,tension=2}{v1,v2}
      \fmf{dashes,width=0.6,lab.side=left,tension=1}{v2,o1}
      \fmf{photon,width=0.6,lab.side=right,label=$W$,label.dist=3,tension=1}{v2,v3}
      \fmf{fermion,width=0.6,lab.side=left,tension=1}{o2,v3}
      \fmf{fermion,width=0.6,lab.side=left,tension=1}{v3,o3}
      \fmflabel{$d$}{i1}
      \fmflabel{$\bar{u}$}{i2}
      \fmflabel{$\bar{\nu}_l$}{o2}
      \fmflabel{$l^-$}{o3}
      \fmflabel{$h$}{o1}
    \end{fmfgraph*}
    \hspace*{3mm}
    \begin{fmfgraph*}(120,70)
      \fmfset{arrow_len}{2mm}
      \fmfleft{i2,i1}
      \fmfright{o3,o2,o1,o0}
      \fmf{gluon,width=0.6,lab.side=left,tension=3}{i1,v0}
      \fmf{fermion,width=0.6,lab.side=left,tension=2}{v0,v1}
      \fmf{fermion,width=0.6,lab.side=left,tension=2}{v1,i2}
      \fmf{photon,width=0.6,lab.side=right,label=$Z^*$,label.dist=3,tension=2}{v1,v2}
      \fmf{dashes,width=0.6,lab.side=left,tension=1}{v2,o1}
      \fmf{photon,width=0.6,lab.side=right,label=$Z$,label.dist=3,tension=1}{v2,v3}
      \fmf{fermion,width=0.6,lab.side=left,tension=1}{o2,v3}
      \fmf{fermion,width=0.6,lab.side=left,tension=1}{v3,o3}
      \fmf{fermion,width=0.6,lab.side=left,tension=1}{o0,v0}
      \fmflabel{$g$}{i1}
      \fmflabel{$\bar{q}$}{i2}
      \fmflabel{$l^+$}{o2}
      \fmflabel{$l^-$}{o3}
      \fmflabel{$h$}{o1}
      \fmflabel{$\bar{q}$}{o0}
    \end{fmfgraph*}
    \hspace*{3mm}
    \begin{fmfgraph*}(100,60)
      \fmfset{arrow_len}{2mm}
      \fmfleft{i2,i1}
      \fmfright{o3,o2,o1}
      \fmf{gluon,width=0.6,lab.side=left,tension=2}{v1,i1}
      \fmf{gluon,width=0.6,lab.side=left,tension=2}{i2,v0}
      \fmf{fermion,width=0.6,lab.side=left,tension=1}{v0,v1}
      \fmf{fermion,width=0.6,lab.side=left,tension=1}{v1,v2}
      \fmf{fermion,width=0.6,lab.side=left,tension=0.3}{v2,v3}
      \fmf{fermion,width=0.6,lab.side=left,tension=1}{v3,v0}
      \fmf{dashes,width=0.6,lab.side=left,tension=0.8}{v2,o1}
      \fmf{photon,width=0.6,lab.side=right,label=$Z$,label.dist=3,tension=1}{v3,v4}
      \fmf{fermion,width=0.6,lab.side=left,tension=1}{o2,v4}
      \fmf{fermion,width=0.6,lab.side=left,tension=1}{v4,o3}
      \fmflabel{$g$}{i1}
      \fmflabel{$g$}{i2}
      \fmflabel{$l^+$}{o2}
      \fmflabel{$l^-$}{o3}
      \fmflabel{$h$}{o1}
    \end{fmfgraph*}
  \end{center}
  \caption{Feynman diagrams describing associated $Vh$ production,
    including a real emission diagram contributing to the NLO
    corrections and the gluon-fusion process contributing at the NNLO
    level.}
  \label{fig:feyn_vh}
\end{figure}

\subsubsection{Motivation and signature}
\label{sec:exp_vh_mot}

Following the logic of Sec.~\ref{sec:basic_char} and especially
Fig.~\ref{fig:feyn_char}, the physics of the $Vh$ production channel
is closely related to the WBF signature discussed in the previous
Sec.~\ref{sec:exp_wbf}. In the left panel of Fig.~\ref{fig:feyn_vh} we
show the leading order Feynman diagram for $Zh$ production combined
with the decay $Z \to q\bar{q}$. Counting the external particles and
the couplings, it is of the same order in perturbation theory as the
WBF process.  Strictly speaking, the two sets of Feynman diagrams are
even connected by gauge invariance, and we can only consider the $Zh$
process independently when we keep the second $Z$ on its mass
shell. This observation is crucial for the interpretation of
measurements in the two channels: given a theoretically solid
hypothesis, the two channels will always test the same kind of
physics. Differences will only occur when they probe different
kinematic regimes or phase space configurations, but even then the two
processes can easily be combined for example by the underlying BSM
physics hypothesis.

A key advantage of the $Zh$ channel is that, combined with a leptonic
$Z$-decay ($Z_l$), triggering is guaranteed independent of the Higgs
decay. This makes this channel especially attractive when we look for
experimentally challenging Higgs decays, like invisible Higgs
decays~\cite{Choudhury:1993hv,Godbole:2003it,Davoudiasl:2004aj,Okawa:2013hda,Goncalves:2016bkl},
Higgs decays to bottom
quarks~\cite{Butterworth:2008iy,Butterworth:2015bya,Tian:2017oza,Goncalves:2018fvn},
or generally Higgs decays to hadrons~\cite{Perez:2015aoa,Carpenter:2016mwd},
\begin{align}
pp \to Z_l h \to 
\begin{cases} 
 Z_l \; \text{invisible} \\
 Z_l \; b\bar{b} \\ 
 Z_l \; \text{jets} \; . 
\end{cases}
\end{align}
Many of the experimental advantages of the leptonically decaying $Z$
carry over to a leptonically decaying $W$ produced in association with
a SM-like Higgs. Furthermore, while it is harder to experimentally
extract a leptonic $W$-decay from the backgrounds, the corresponding
branching ratio is significantly bigger.  For some Higgs decay
channels, the invisible $Z$-decay to neutrinos can also be useful,
defining the experimental 0-lepton, 1-lepton, and 2-lepton analyses of
the $Vh$ production process.

A second advantage of the $Zh$ process is the control it offers over
the backgrounds. For the $Zh$ signal we require that two same-flavor
opposite-sign leptons reconstruct the $Z$ mass within the excellent
experimental resolution at the percent level. This implies that all
background processes without an on-shell $Z$ are automatically
suppressed, leaving us with either hard $Z+$jets or di-boson $ZV$
backgrounds. The latter typically lead to a signal-to-background ratio
$S/B \gtrsim 1$. Whenever the targeted Higgs decay can be related to a
similar $Z$-decay, the di-boson channels also offer an excellent way
to establish a search or to  measure efficiencies. In the left panel of
Fig.~\ref{fig:vh_pheno} we show the, arguably, most interesting $Vh$
search, namely $Vh$ production with a decay $h \to b\bar{b}$. This
specific study~\cite{Butterworth:2008iy} showed that the Higgs signal
and the continuum background can be separated relatively easily in
phase space regions with a small geometric separation between the two
$b$-quarks combined with the requirement $m_{bb} \approx M_h$. It
started the field of sub-jet
physics~\cite{Abdesselam:2010pt,Altheimer:2012mn,Altheimer:2013yza} at
the LHC, with a vast number of applications in many different LHC
searches. Unfortunately, this specific $Vh$ analysis never attracted
the experimental attention it deserves, because the Higgs mass
resolution does not significantly improve over a resolved pair of
$b$-jets. A noteworthy aspect of this analysis is the second peak at
$m_{bb} = M_Z$, which would have helped to establish the new analysis
strategy.

\begin{figure}[t]
\includegraphics[width=0.32\textwidth]{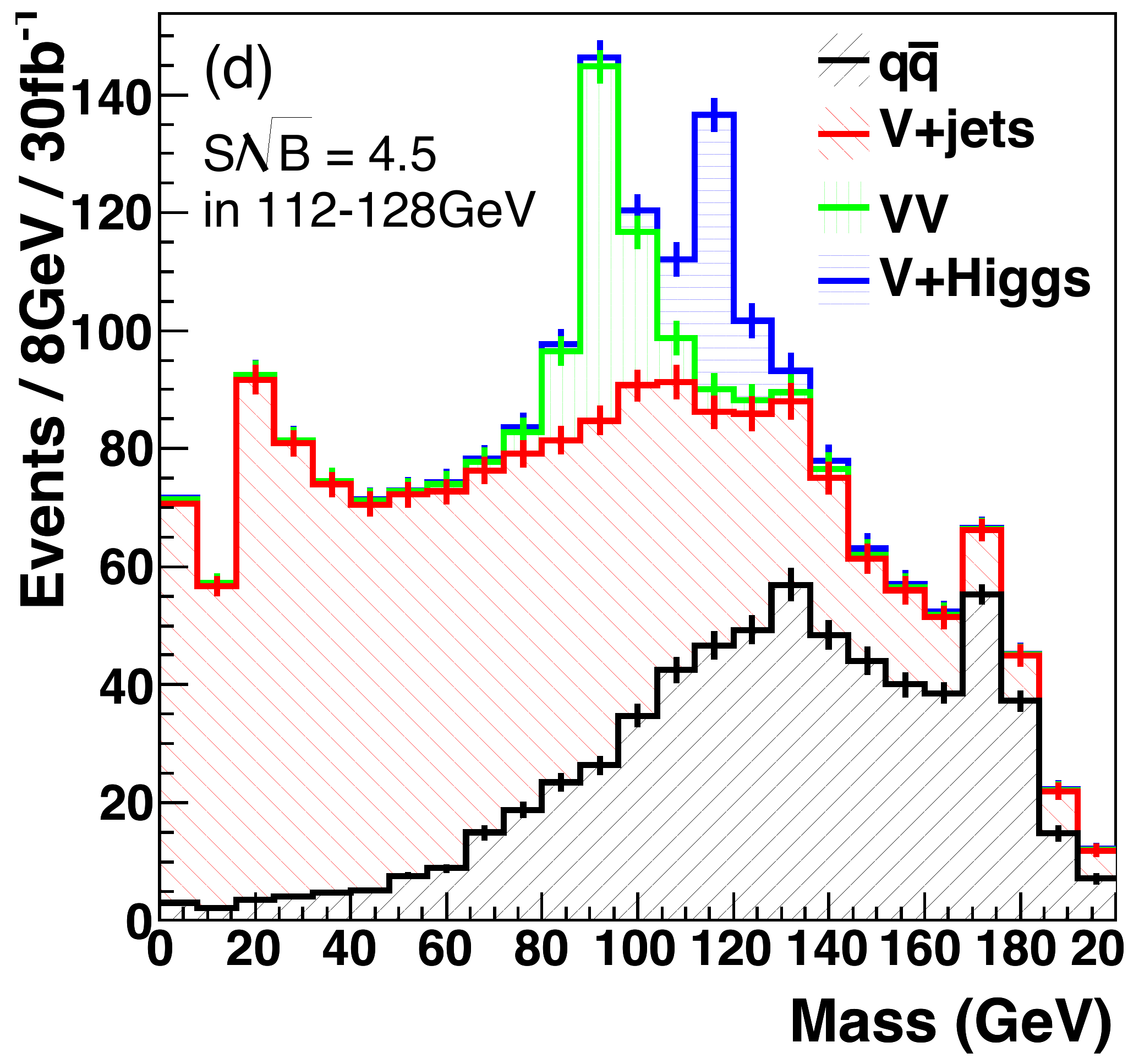}
\hspace*{0.1\textwidth}
\includegraphics[width=0.44\textwidth]{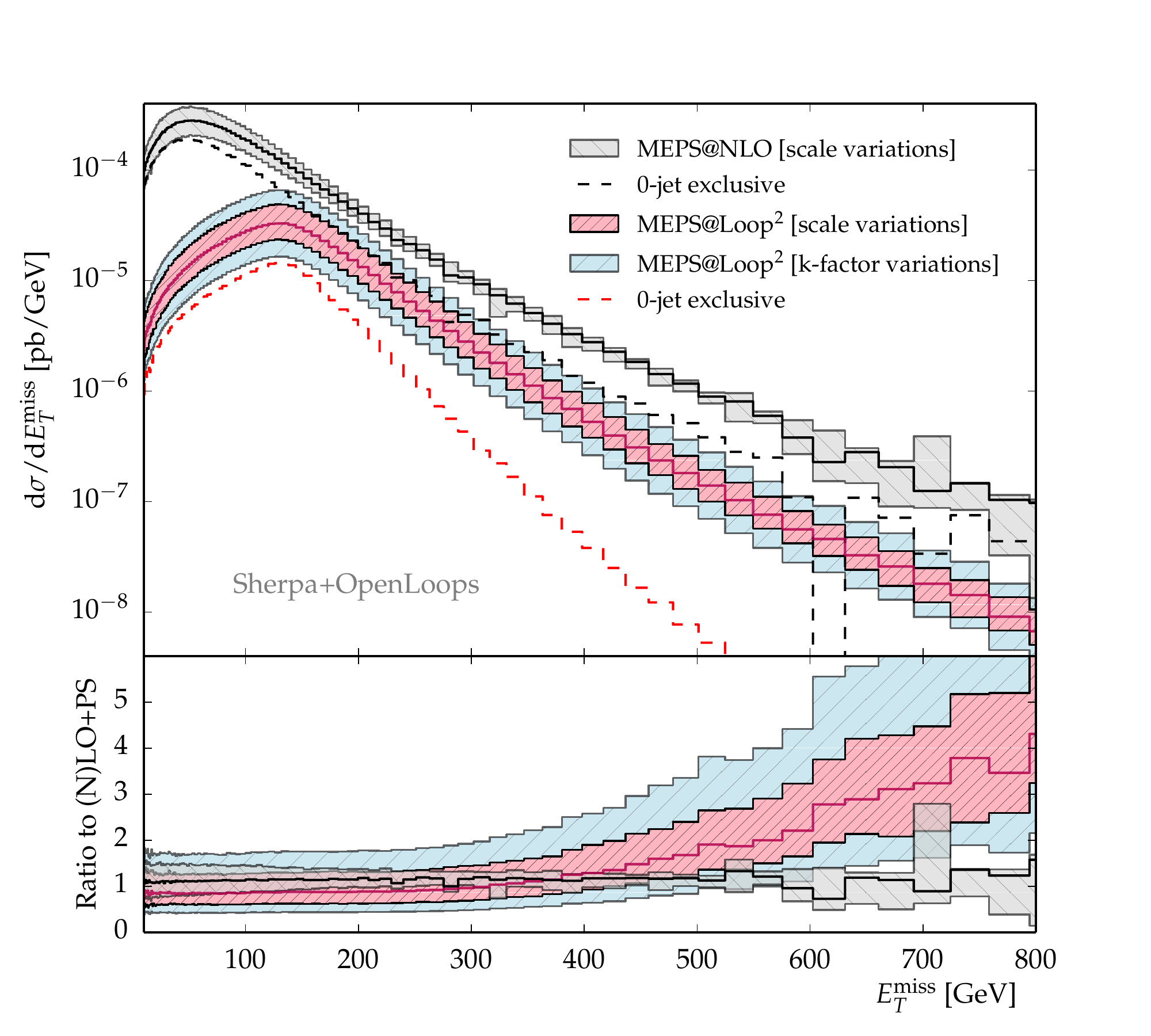}
\caption{Left: Signal and background for an assumed 115~GeV SM Higgs
  decaying to a boosted pair of $b$-jets and identified as a fat Higgs
  jet.  The figure combines 0-lepton, 1-lepton, and 2-lepton channels.
  Figure from Ref.~\cite{Butterworth:2008iy}. Right: Missing
  transverse energy distributions after basic selection cuts for the
  Drell-Yan and the gluon fusion contributions. The theoretical
  uncertainty is extracted from scale and $K$-factor variations.  The
  bottom panel presents the ratios MEPS\@@Loop$^2$ to the
  \textsc{Loop}$^2$+PS prediction and the MEPS\@@NLO to MC\@@NLO
  prediction. Figure from Ref.~\cite{Goncalves:2015mfa}.}
\label{fig:vh_pheno}
\end{figure} 

A third advantage of the $Vh$ production process is the perturbative
control of the rate predictions.  At Run~II theoretical uncertainties
are becoming the leading uncertainties in many Higgs measurements,
especially when analyses utilize specific parts of phase space. In the
right panel of Fig.~\ref{fig:vh_pheno} we illustrate the example of
invisible Higgs searches, which experimentally relies on Higgs
production at large $p_T$. Above the threshold $m_{Zh} \approx 2 m_t$
the one-loop diagram shown to the right in Fig.~\ref{fig:feyn_vh}
becomes relevant, as we will discuss later.  In the right panel of
Fig.~\ref{fig:vh_pheno} we show the rates from the Drell-Yan or
quark-initiated process and from the gluon fusion one-loop process.
From the upper panel  on the right, we see that even in this extreme phase space
region with $p_{T,h} \gg M_h$ all partonic configurations are fully
under control in perturbative QCD.

Finally, the $Z_l$-associated channel makes it especially easy to
fully reconstruct the final state of the hard process. For example, we
could search for alternative Lorentz structures of the $ZZh$
interaction by looking at the reconstructed invariant mass
$m_{Zh}$. This is an advantage over the gluon fusion process discussed
in Sec.~\ref{sec:exp_gf}, where the hard process includes the Higgs
and possible hard jets, and the weak boson fusion process discussed in
Sec.~\ref{sec:exp_wbf} with its experimentally challenging tagging
jets.  

One caveat is that not all four advantages  listed above for the $Zh$ channel immediately apply to
$Wh$ production. Missing transverse momentum is a major experimental
challenge, and the experimental resolution on a transverse mass is
always low. Nevertheless, the $Wh$ channel still allows for easy
triggering through the hard lepton, the perturbative expansion is
under control, and the hard process can be reconstructed reasonably
well. The only problem is new backgrounds, for example single top and
top pair production.

\subsubsection{Precision predictions}
\label{sec:exp_vh_pred}

\begin{figure}[t]
\includegraphics[width=0.42\textwidth]{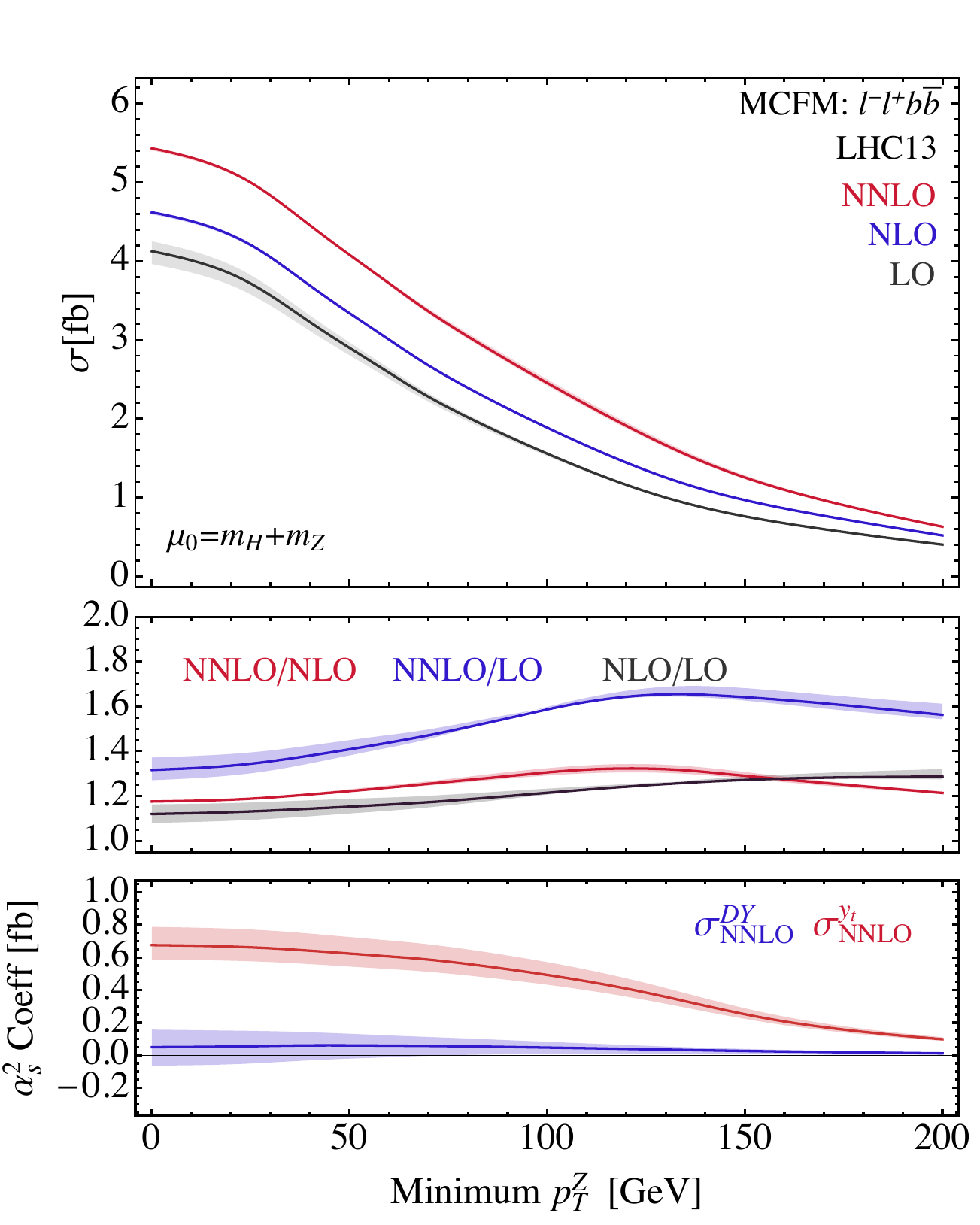}
\hspace*{0.05\textwidth}
\includegraphics[width=0.45\textwidth]{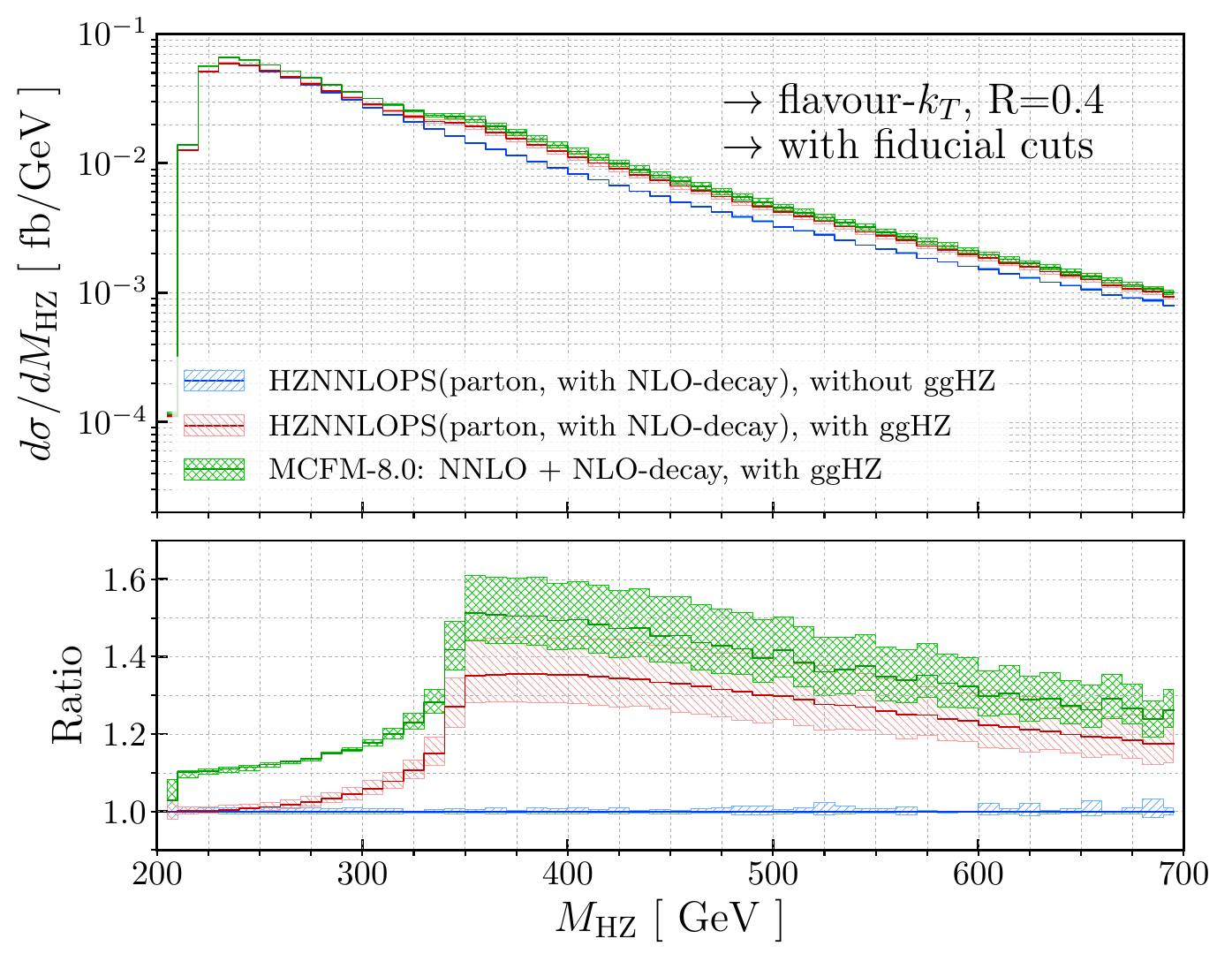}
\caption{Left: cross-section as a function of the minimum transverse
  momentum of the $Z$-boson for the process $pp\rightarrow
  Zh\rightarrow l^+l^-b {\overline b}$. The upper panel presents the
  total cross-section, the middle panel presents the impact of the
  higher order corrections, the lower plot presents the total
  $\alpha_s^2$ coefficient. Figure from
  Ref.~\cite{Campbell:2016jau}. Right: $m_{Zh}$ distributions with and
  without the top-loop contribution. Figure from
  Ref.~\cite{Astill:2018ivh}.}
\label{fig:vh_prec}
\end{figure} 

Given the Higgs signatures we target with its help, the $Vh$ process
is a challenge for precision predictions. It is not sufficient to
predict a total cross section, because the $h \to
b\bar{b}$~\cite{Caola:2017xuq}, the invisible Higgs, and the effective
theory analyses all rely on boosted phase space configurations, which
make up only a small fraction of the total cross
section. Nevertheless, we start by quoting the NNLO cross sections for
the $Vh$ processes at a 14~TeV LHC, including the branching ratios of
the $W$ and $Z$ bosons to one lepton generation
are~\cite{Campbell:2016jau}
\begin{align}
\sigma_{Zh} = 0.74~\fb
\qqqquad 
\sigma_{W^+h} = 34.0~\fb
\qqqquad
\sigma_{W^-h} = 23.9~\fb\, .
\end{align}
A structural complication of the perturbative calculations arises from
terms which do not scale with the leading-order Drell-Yan
process. Normally, one would expect the higher-order production rates
to scale like
\begin{align}
\sigma_{Vh} \propto g^4 \alpha_s^n 
\qquad \text{with $n=0,1,2$ for LO, NLO, NNLO.}
\end{align}
However, for example the process $gg \to Zh$ shown in
Fig.~\ref{fig:feyn_vh} proceeds through a heavy quark loop to which
the heavy bosons couple.  Similarly, the process $q\bar{q} \to Zhg$
includes Feynman diagrams where the Higgs couples to a closed top loop
rather than the gauge boson.  If we square these diagrams we find the
scaling
\begin{align}
\sigma_{gg \to Zh} &\propto g^2 \alpha_s^2 \frac{y_t^2}{(4 \pi^2)^2} \notag \\
\sigma_{qq \to Vhg} &\propto g^2 \alpha_s^3 \frac{y_t^2}{(4 \pi^2)^2} \; ,
\end{align}
including the top Yukawa coupling and a factor $1/(4 \pi^2)^2$ from
the loop in the amplitude. The suppression of the new $2 \to 2$
topology relative to the Drell-Yan process is small, namely
$\alpha_s^2 y_t^2/(4 \pi^2 g)^2$, usually combined with the NNLO
corrections because of the number of loops and the powers of
$\alpha_s$, and only partly compensated by the larger parton densities
for the gluon. However, while their contribution to the total cross
section remains at the $1~...~3\%$ level, the relative importance of the different
contributions changes for kinematic
distributions.

In the left panel of Fig.~\ref{fig:vh_prec} we show the rate above a
minimum cut in $p_{T,Z}$ computed to LO, NLO, and NNLO, including the
top-loop contributions~\cite{Campbell:2016jau}. In the upper two
panels we see that the NNLO corrections are larger than the NLO
corrections, which is an effect of the top-loop contributions and
unique to the $Zh$ process. In the lower panel we see how the
subprocess $gg \to Zh$ completely dominates the NNLO contribution. The
drop towards larger minimum $p_{T,Z}$ values is related to the softer
gluon distributions in the proton.  In the right panel we show a
similar effect, now including the parton shower combined with a NNLO
re-weighting method~\cite{Astill:2018ivh}. Clearly, the top-loop
effects dramatically increase at the threshold $m_{hZ} = 2 m_t$, where
the closed top loop develops an un-suppressed absorptive part. The
slight difference between the two predictions including the top loop
can be explained by a set of detector acceptance (fiducial) cuts on
the leptons and the tagged $b$-jets.

\subsubsection{LHC Analyses}
\label{sec:exp_vh_ana}

\begin{figure}[t]
\includegraphics[width=0.42\textwidth]{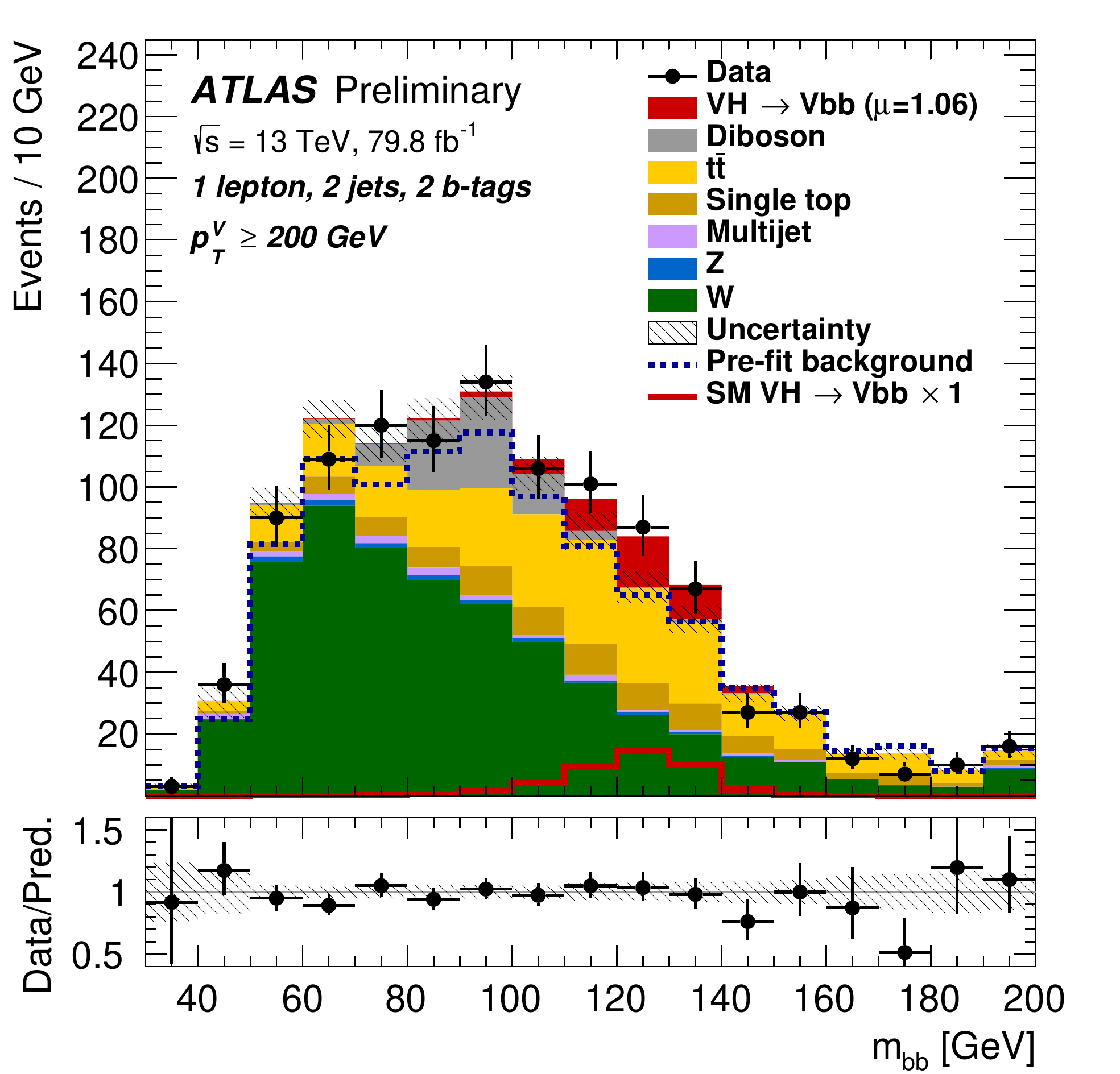}
\hspace*{0.05\textwidth}
\includegraphics[width=0.42\textwidth]{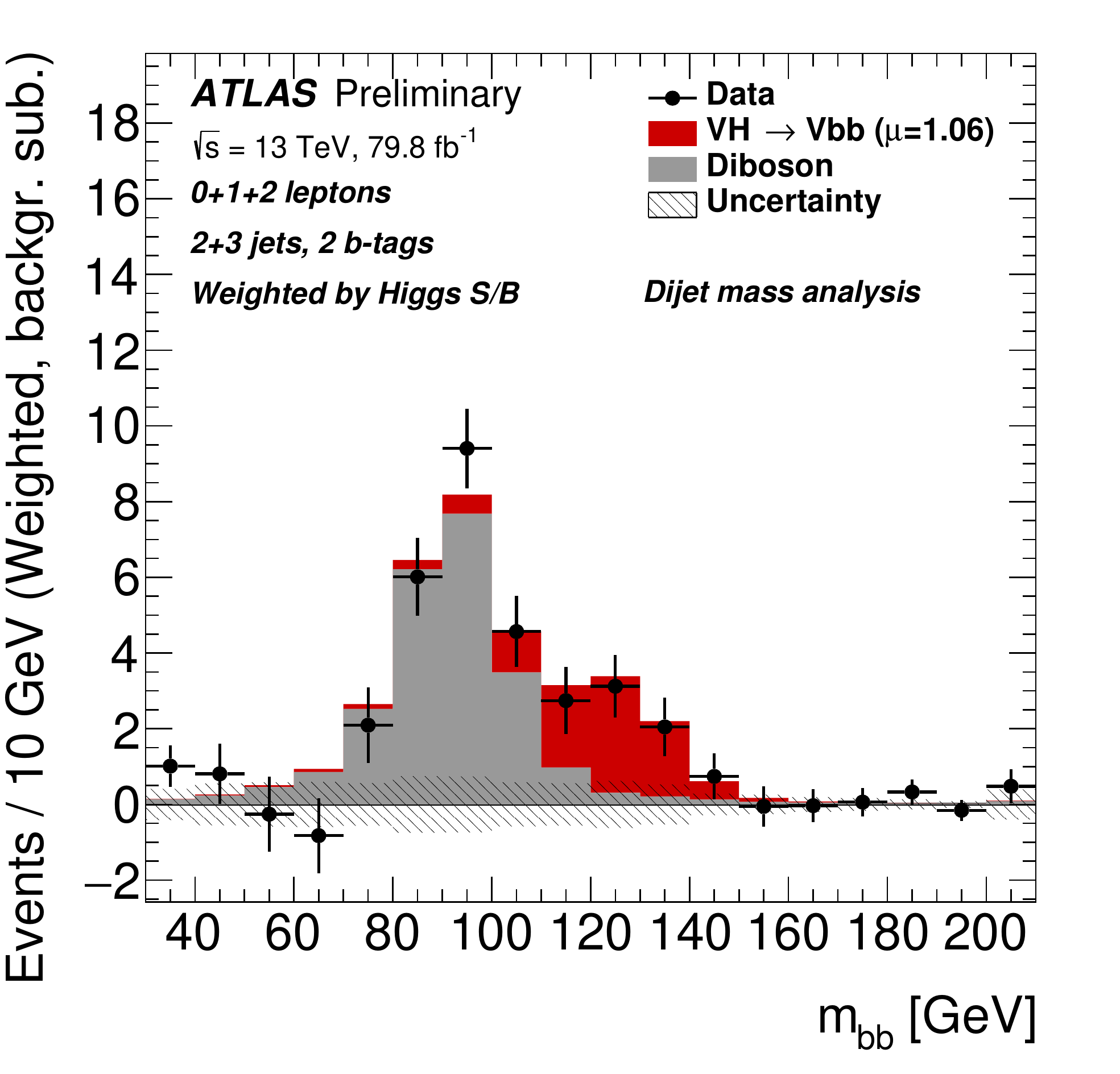}
\caption{Left: $m_{bb}$ distribution in the 1-lepton channel for 2
  $b$-tags and $p_T^V > 200$~GeV. The backgrounds are shown after a
  global likelihood fit, while the signal is normalized to the
  measured signal yield ($\mu=1.06$).  The dashed histogram shows the
  total background from the pre-fit MC simulation.  Figure from
  Ref.~\cite{ATLAS:2018nkp}.  Right: $m_{bb}$ distribution after
  background subtraction indicating the di-boson contribution from $WW$
  and $ZZ$ production. Figures taken from Ref.~\cite{ATLAS:2018nkp}.}
\label{fig:vh_res}
\end{figure} 

Using LHC data at 13~TeV, the focus of ATLAS and CMS in using the $Vh$
channel has been on Higgs decays to bottoms.  ATLAS reports the two
separate and one combined rate measurements from the 0-lepton,
1-lepton, and 2-lepton searches~\cite{Aaboud:2017xsd},
\begin{align}
\sigma_{Wh} \br (h \to b\bar{b}) & = 1.08^{+0.54}_{-0.47}~\pb \notag \\
\sigma_{Zh} \br (h \to b\bar{b}) & = 0.57^{+0.26}_{-0.23}~\pb \notag \\
\sigma_{Vh} \br (h \to b\bar{b}) & =1.58^{+0.55}_{-0.47}~\pb 
\end{align}
Unfortunately, CMS does not report any actual cross section
measurements, neither does ATLAS in their $h\to b\bar b$ observation
at $5.4\sigma$~\cite{ATLAS:2018nkp}.  The combined rate measurement
can be expressed as a single signal strength measurement of $Vh$
production combined with a Higgs decay to bottom quarks. ATLAS and CMS
find~\cite{ATLAS:2018nkp,Sirunyan:2017elk},
\begin{align}
\mu_{Vh,bb}^\text{ATLAS} &= 1.01^{+0.12}_{-0.12} \text{(stat)} ^{+0.16}_{-0.15} \text{(syst)} \notag \\
\mu_{Vh,bb}^\text{CMS} &= 1.19^{+0.21}_{-0.20} \text{(stat)} ^{+0.34}_{-0.32} \text{(syst)}
\; .
\end{align}
The size of the error bars corresponds to the fact that none of these
measurements has the significance required for a discovery of this
signal.  A simple check comes from searching for the corresponding
di-boson signature from $VZ$ production combined with the decay $Z \to
b\bar{b}$~\cite{ATLAS:2018nkp,Sirunyan:2017elk},
\begin{align}
\mu_{VZ,bb}^\text{ATLAS} &= 1.16^{+ 0.16}_{-0.16} \text{(stat)} ^{+ 0.21}_{-0.19} \text{(syst)}  \notag \\
\mu_{VZ,bb}^\text{CMS} &= 1.02 ^{+ 0.22}_{-0.23} 
\; .
\end{align}
In Fig.~\ref{fig:vh_res} we show two kinematic distributions for the
1-lepton channel, to illustrate the kinematic patterns of the
different backgrounds. In the left panel we see that it is indeed
possible to clearly separate the Higgs peak and the $Z$-peak. The
larger Higgs mass helps with the signal extraction, because all
backgrounds are rapidly dropping functions of $m_{bb}$. Because of the
required two $b$-tags and at most one additional jet, the $W$+jets
background with an actual heavy-flavor jet dominates in the
low-$m_{bb}$ range. Under the Higgs peak the top pair and single top
backgrounds become equally important. The global agreement between the
combined backgrounds in the side bands and the data is related to the
post-fit nature of this plot. In this approach the individual
background contributions are floated individually, and a scaling
factor is then applied under the Higgs pole.  In the right panel we
show the Higgs signal after subtracting all backgrounds except the
di-boson contribution, again highlighting the importance of the $VV$
region.

Unlike the above analyses targeted at the decay of the Higgs boson to
bottom quarks, we can also use the same final state to search for new
physics in this channel~\cite{Aaboud:2017cxo}. In the right panel of
Fig.~\ref{fig:vh_res} we show the results for one of the signal
categories, which is especially interesting: for a heavy resonance we
expect that the two $b$-jets coming from the Higgs decay can be
boosted enough to become one observable jet. The invariant mass
$m_{TV}$ in the 1-lepton category is then computed by requiring that
the lepton and the missing energy combine to an on-shell $W$-boson. As
a side effect of this analysis we see that there is no hint of any
unexpected behavior of the $VH$ channel all the way to $m_{VH} \approx
4$~TeV.

Additional searches using the $Vh$ production channel search for
invisible Higgs decays. The corresponding CMS
analysis~\cite{Khachatryan:2016whc} combines this channel with the WBF
and the gluon fusion production mechanism. Because the WBF analysis
was expected to be driving this search, we will discuss it in
Sec.~\ref{sec:exp_wbf}.

\subsection{Associated $t\bar th$ and $th$ production}
\label{sec:exp_tth}

\begin{figure}[b!]
  \begin{center}
    \begin{fmfgraph*}(100,60)
      \fmfset{arrow_len}{2mm}
      \fmfleft{i1,i2}
      \fmfright{t1,h1,t2}
      \fmf{gluon,width=0.6,lab.side=right,tension=2}{i1,v1}
      \fmf{gluon,width=0.6,lab.side=right,tension=2}{v2,i2}
      \fmf{fermion,width=0.6,lab.side=left,tension=1}{t1,v1,vh,v2,t2}
      \fmf{dashes,width=0.6,lab.side=left,tension=1}{vh,h1}
      \fmflabel{$g$}{i1}
      \fmflabel{$g$}{i2}
      \fmflabel{$\bar{t}$}{t1}
      \fmflabel{$t$}{t2}
      \fmflabel{$h$}{h1}
    \end{fmfgraph*}
    \hspace*{6mm}
    \begin{fmfgraph*}(100,60)
      \fmfset{arrow_len}{2mm}
      \fmfleft{i1,i2}
      \fmfright{t1,h1,t2}
      \fmf{fermion,width=0.6,lab.side=right,tension=2}{i1,v1,i2}
      \fmf{gluon,width=0.6,lab.side=right,tension=1.5}{v1,v2}
      \fmf{fermion,width=0.6,lab.side=left,tension=1}{t1,v2,vh,t2}
      \fmf{dashes,width=0.6,lab.side=left,tension=1}{vh,h1}
      \fmflabel{$q$}{i1}
      \fmflabel{$\bar{q}$}{i2}
      \fmflabel{$\bar{t}$}{t1}
      \fmflabel{$t$}{t2}
      \fmflabel{$h$}{h1}
    \end{fmfgraph*}
    \hspace*{12mm}
    \begin{fmfgraph*}(100,60)
      \fmfset{arrow_len}{2mm}
      \fmfleft{i1,i2}
      \fmfright{t1,h1,t2}
      \fmf{fermion,width=0.6,lab.side=right,tension=2}{i1,v1,t1}
      \fmf{photon,width=0.6,lab.side=right,label=$W$,label.dist=3,tension=1.5}{v1,vh}
      \fmf{photon,width=0.6,lab.side=right,label=$W$,label.dist=3,tension=1.5}{vh,v2}
      \fmf{fermion,width=0.6,lab.side=left,tension=1}{i2,v2,t2}
      \fmf{dashes,width=0.6,lab.side=left,tension=1}{vh,h1}
      \fmflabel{$b$}{i1}
      \fmflabel{$u$}{i2}
      \fmflabel{$t$}{t1}
      \fmflabel{$d$}{t2}
      \fmflabel{$h$}{h1}
    \end{fmfgraph*}
    \hspace*{6mm}
    \begin{fmfgraph*}(100,60)
      \fmfset{arrow_len}{2mm}
      \fmfleft{i1,i2}
      \fmfright{t1,h1,t2}
      \fmf{fermion,width=0.6,lab.side=right,tension=2}{i1,v1}
      \fmf{fermion,width=0.6,lab.side=right,label=$t^*$,label.dist=3,tension=2}{v1,vh}
      \fmf{fermion,width=0.6,lab.side=right,tension=2}{vh,t1}
      \fmf{photon,width=0.6,lab.side=right,label=$W$,label.dist=3,tension=1.5}{v1,v2}
      \fmf{fermion,width=0.6,lab.side=left,tension=1}{i2,v2,t2}
      \fmf{dashes,width=0.6,lab.side=left,tension=1}{vh,h1}
      \fmflabel{$b$}{i1}
      \fmflabel{$u$}{i2}
      \fmflabel{$t$}{t1}
      \fmflabel{$d$}{t2}
      \fmflabel{$h$}{h1}
    \end{fmfgraph*}
  \end{center}
  \caption{Feynman diagrams contributing to top
    pair-associated and single top-associated Higgs production.}
  \label{fig:tthfeyn}
\end{figure}
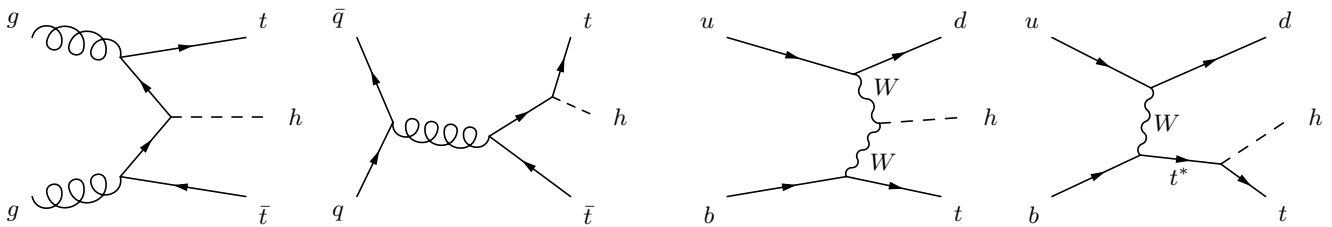

Top pair associated Higgs production 
\begin{align}
pp \to t\bar t h + X
\end{align}
is the dominant channel with direct sensitivity to the top-Yukawa
coupling, which from the fundamental physics perspective of
Sec.~\ref{sec:basis_implications} is one of the two most important
parameters in the Higgs sector (the Higgs self-coupling being the other). In addition, it is crucial in allowing
us to use the effective Higgs-gluon coupling to search for new
particles contributing to the loop.  We show two typical Feynman
diagrams in Fig.~\ref{fig:tthfeyn}. Consequently, $t\bar t h$
production is central to the LHC Higgs phenomenology program. The
$t\bar t h$ measurement program includes all top decay channels as
well as dominant and phenomenologically clean Higgs final states.

The production of a Higgs boson in association with a single top
quark~\cite{DiazCruz:1991cs,Maltoni:2001hu},
\begin{align}
p p \to (t/\bar t) h  + X
\end{align}
features an interference between Higgs emission off a $W$-leg and
Higgs emission of a top-leg, shown in Fig.~\ref{fig:tthfeyn}. This
allows us to constrain the sign of the top Yukawa coupling.

\subsubsection{Motivation and signature}
\label{sec:exp_tth_mot}

Early sensitivity estimates~\cite{ATLAS:1999vwa} motivated top
pair-associated Higgs production as the main discovery channel for a
light SM Higgs boson. A more thorough subsequent
analysis~\cite{Cammin:2003uaa} significantly decreased this estimate,
and isolating $t\bar t h$ production remains one of the experimentally
most challenging tasks of the LHC Higgs program. Relevant channels
are
\begin{align}
pp \to t\bar t h \to 
\begin{cases} 
 t\bar t \; b\bar b \\
 t\bar t\; \gamma\gamma \\ 
 t\bar t \; VV \\
 t\bar t \; \tau \tau \; . 
\end{cases}
\end{align}
where the latter two are similar enough on the detector level to be
combined.  Typically, analyses focus on fully-leptonic or
semi-hadronic top decays, while fully-hadronic $t\bar t h$ production
is extremely challenging due to triggering and multi-jet backgrounds.

\begin{figure}[t]
\includegraphics[width=0.43\textwidth]{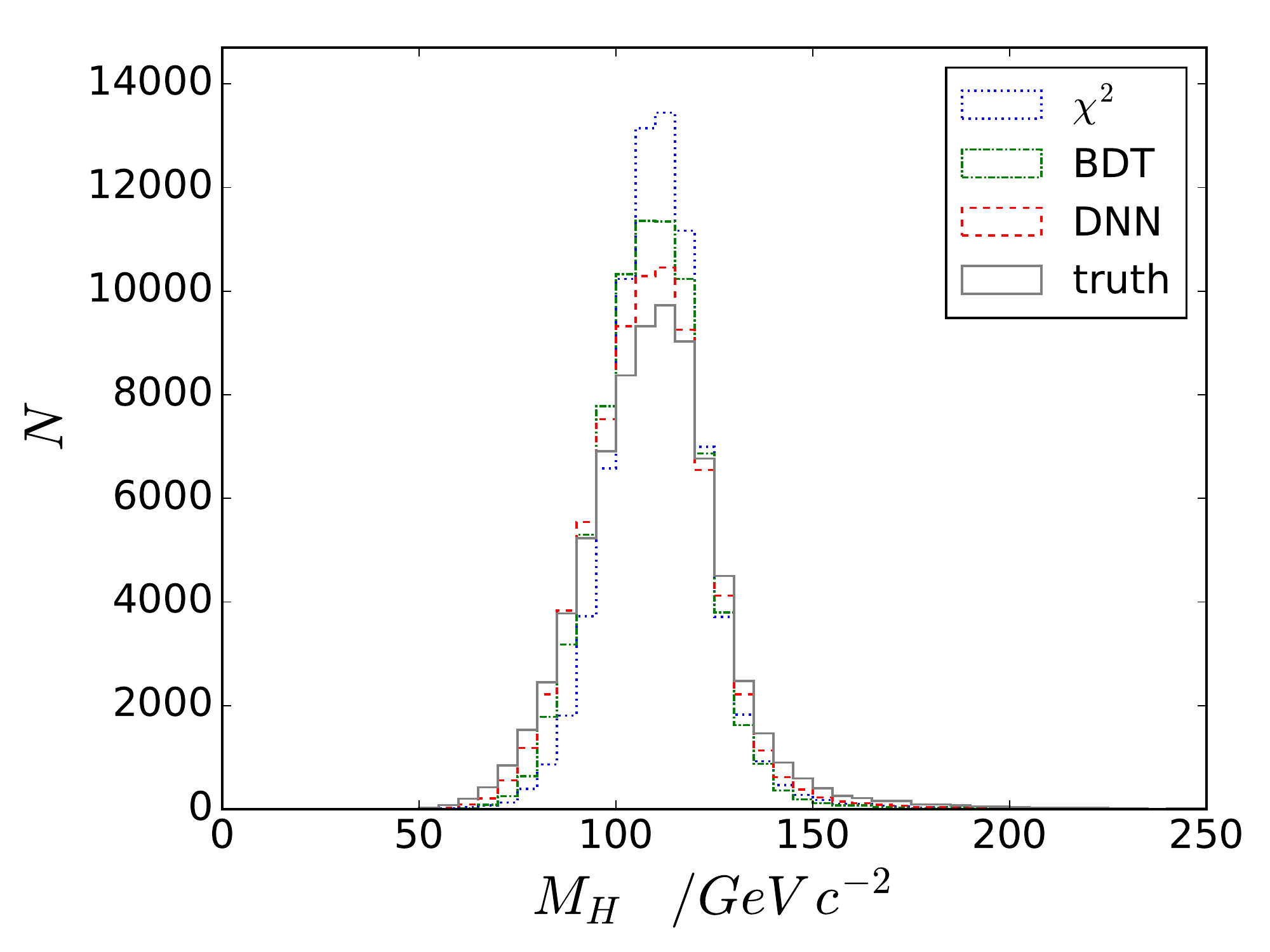}
\hspace{0.1\textwidth}
\includegraphics[width=0.38\textwidth]{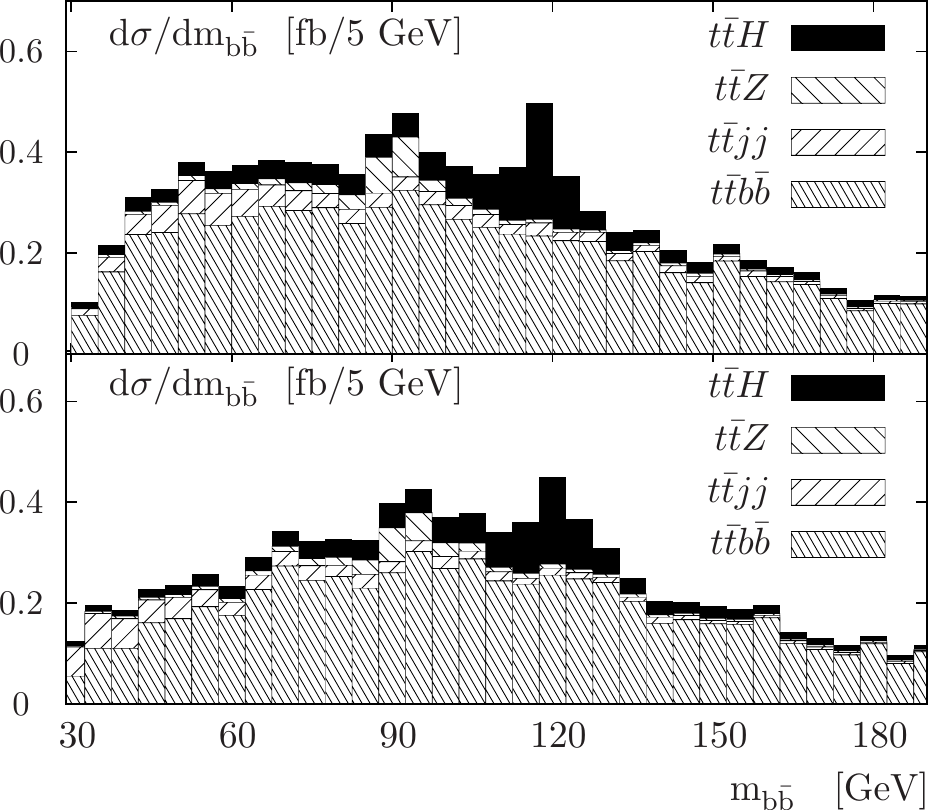}
\caption{Left: reconstructed Higgs mass based on a deep neural
  network, compared with other methods. Figure from
  Ref.~\cite{Erdmann:2017hra}. Right: reconstructed Higgs mass based
  on boosted $t\bar{t}h$ production, including backgrounds and without
  (upper) and including (lower) underlying event. Figure from
  Ref.~\cite{Plehn:2009rk}.}
\label{fig:tth_mbb} 
\end{figure}

While it gives the largest signal rate, the Higgs decay $h\to b\bar b$
suffers from a challenging signal vs background ratio of~15\%. An
improved discrimination of signal and background can be achieved by
combining signal and background matrix elements into a likelihood to
achieve a highly-adapted background rejection performance. For
instance, the matrix-element method described in
Sec.~\ref{sec:exp_data_mem} should improve the
situation~\cite{Artoisenet:2013vfa} and was adopted by CMS and
ATLAS~\cite{Khachatryan:2015ila,Aad:2015gra}.  Given the moderate
signal vs. background ratio and the fundamental interest in the top
Yukawa coupling, $t\bar t h$ production serves as the show-case for
some of the most advanced analyses approaches that are pursued at the
LHC. With increasing luminosity we expect the combination with $h\to
\gamma \gamma$ decays to dominate the $t\bar{t}h$ analysis and provide
the most robust and usable results.

The main challenge for observing  $t\bar{t}h$ in the $h\to b\bar{b}$ channel is
the combinatorics of the four $b$-jets. Without a clear identification
of the correct combination out of $4!/2!=12$ combinations and hence
without information on the Higgs mass it is extremely difficult to
define and control a background rejection~\cite{Cammin:2003uaa}. One
way to address this problem is deep learning. Comparing the
performance of a $\chi^2$ method, a BDT, and a neural network based on
the same observables, the deep neural network clearly gives the best
results~\cite{Erdmann:2017hra}. In the left panel of
Fig.~\ref{fig:tth_mbb}, we show the reconstructed Higgs mass from the
three methods, compared to Monte Carlo truth.  Motivated by the success
of substructure methods in $Vh$ production~\cite{Butterworth:2008iy},
it makes sense to also probe $t\bar t h$ in a phase space region where
the $b$-jet combinatorics is automatically resolved by the boost of
the decaying particles~\cite{Plehn:2009rk,CMS:2016qwm}. The boosted
analysis using subjet methods allows for the usual side-band analysis
in $m_{bb}$ to control the backgrounds, as shown in the right panel of
Fig.~\ref{fig:tth_mbb}.

Single top-associated Higgs production 
\begin{align}
pp \to t/\bar t + h \to 
\begin{cases} 
 t/\bar t \; \tau\tau \\
 t/\bar t\; \gamma\gamma \\ 
 t/\bar t \; VV \; . 
\end{cases}
\end{align}
has the potential to constrain the sign of the top Yukawa coupling
relative to the $W^+W^-h$ coupling. These couplings have different
signs in the Standard Model and current LHC measurements cannot yet 
fully distinguish  between swapped Yukawa
coupling signs~\cite{Corbett:2015ksa,Khachatryan:2016vau} in the
combination of the major production and decay channels, see below.

As a consequence of the destructive interference in the Standard Model, the top
associated Higgs production cross section increases by a factor of
about 15 when the top quark Yukawa interaction sign is swapped
compared to the Standard Model, as shown in Fig.~\ref{fig:thq}. Note that this
would fall outside the assumptions of perturbative deformations of the
Standard Model using dimension-6 SMEFT. While this effect is similar to the the
partial decay width $h\to\gamma\gamma$, which is small due to a
cancellation between the virtual top and $W$ contributions the
sign-degeneracy cannot be fully resolved using Higgs decay observables
like $h\to \gamma\gamma$ alone~\cite{Farina:2012xp}. Including single
top-associated Higgs production, however, breaks the residual
degeneracies as discussed in~\cite{Farina:2012xp,Biswas:2012bd}. This
makes $th(j)$ production an important part of the Higgs phenomenology
program at the LHC.

Finally, single-top-associated Higgs production is particularly
relevant in the search for  for charged Higgs bosons in the
decay $H^+ \rightarrow \tau^+ \nu$~\cite{Roy:1991sf,Assamagan:2002in}.  Charged
Higgs boson from Higgs triplets can be searched for in gauge-induced
pair production or in WBF, while charged Higgs bosons in a 2HDM or in
the MSSM can only be produced through their fermion couplings,
specifically a $t_\beta$-increased bottom Yukawa coupling.  The $\sim
\bar t b H^+ \text{h.c.}$ interaction then gives rise to a top-charged
Higgs final state,
\begin{align}
pp \to \bar{b} t H^+ + X \, .
\end{align}
The main challenge in describing and extracting this signal is the
treatment of the forward bottom
jet~\cite{Plehn:2002vy,Weydert:2009vr}, where a large invariant mass
of the $tH^+$ system induces a reasonably large logarithm, which in
turn can be resummed through the use of bottom parton densities.

\subsubsection{Precision predictions}
\label{sec:exp_tth_pred}

\begin{figure}[t]
\includegraphics[width=0.42\textwidth]{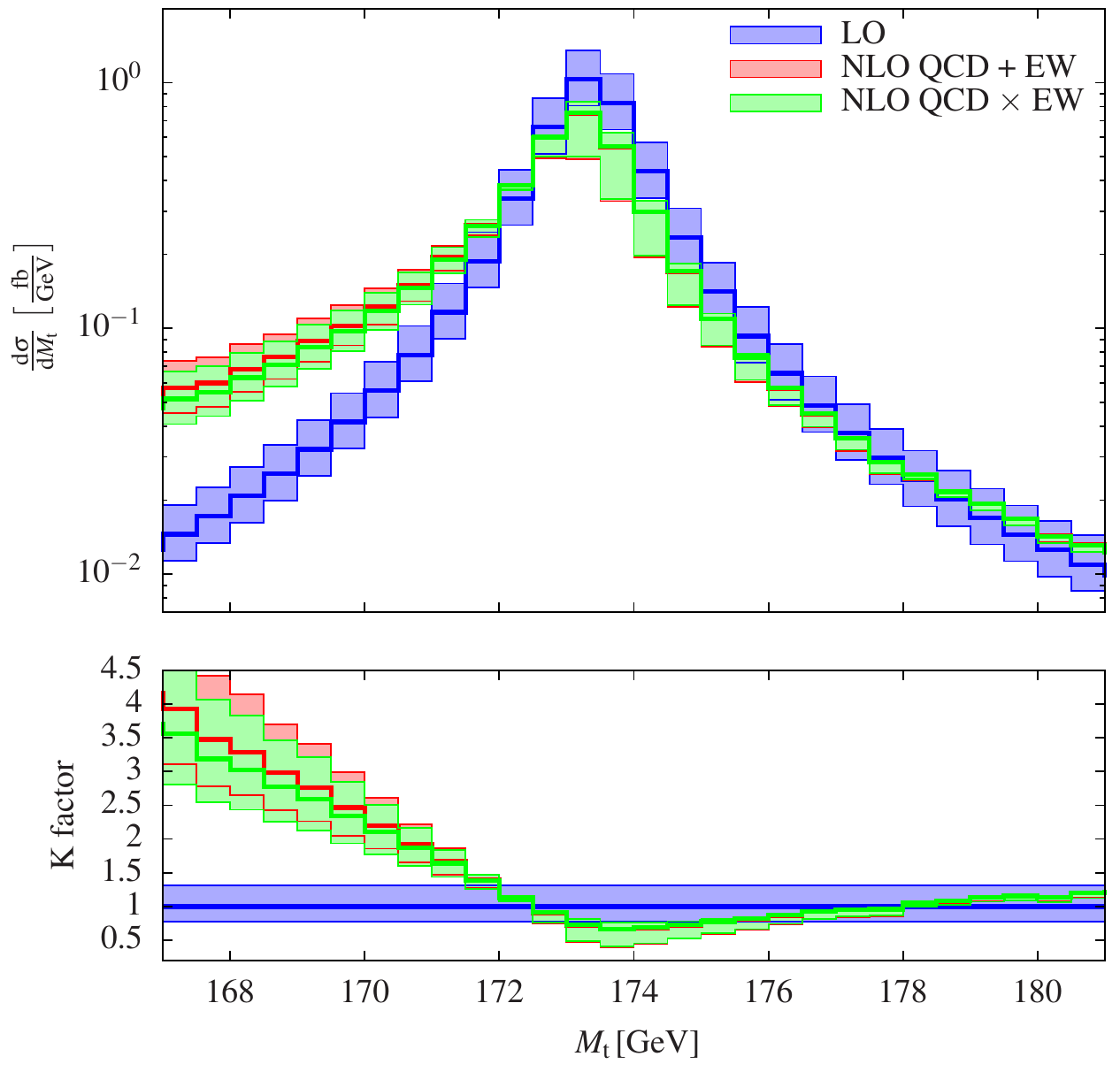}
\hspace{0.1\textwidth}
\includegraphics[width=0.42\textwidth]{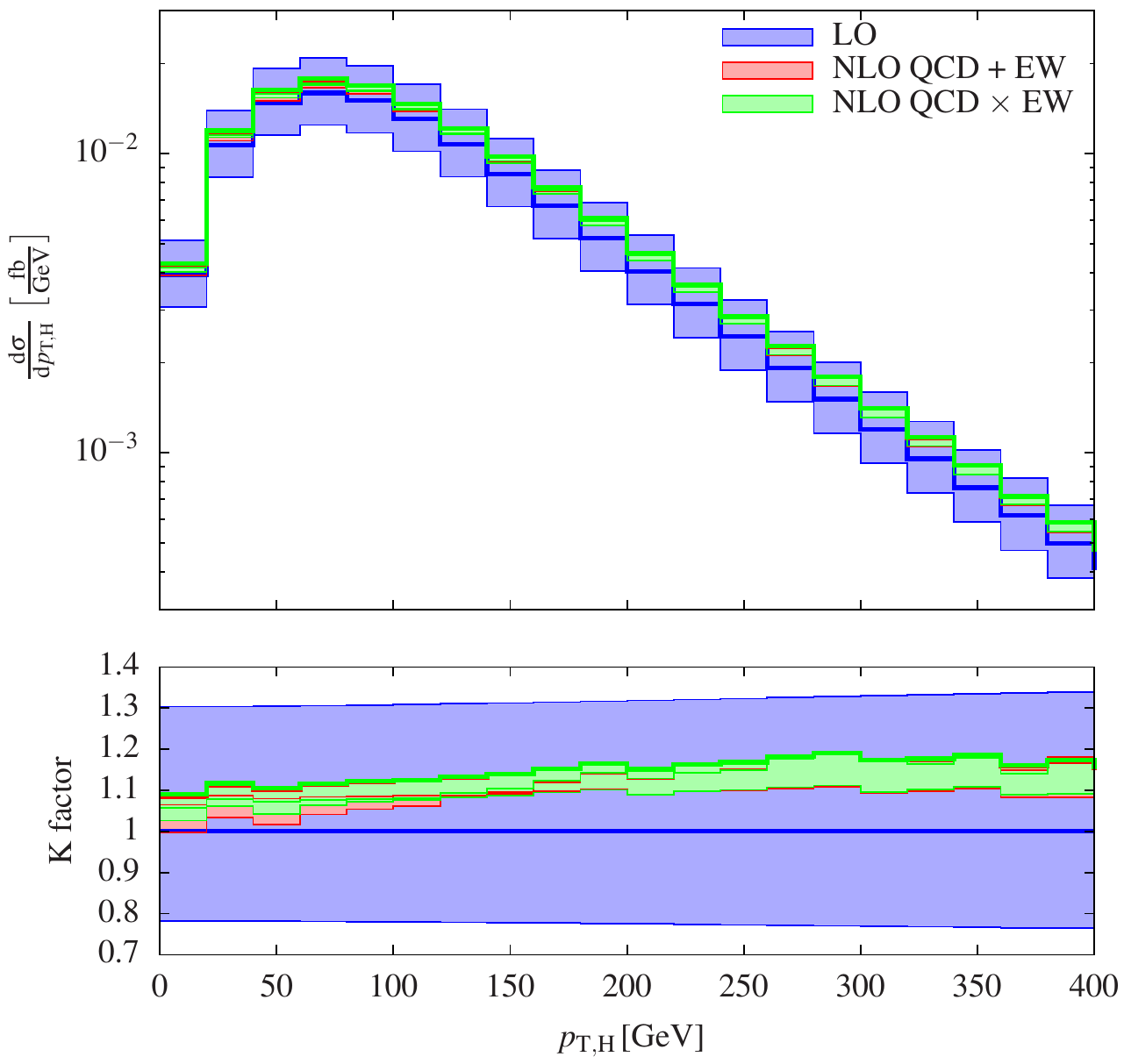}
\caption{Combined QCD and electroweak NLO corrections for the
  reconstructed top mass and the $p_T$ of the Higgs.  Figures from
  Ref.~\cite{Denner:2016wet}.}
\label{fig:tth_nlo} 
\end{figure}

NLO QCD corrections moderately increase $t\bar t h$ production by
$\lesssim
20\%$~\cite{Reina:2001sf,Reina:2001bc,Dawson:2002tg,Dawson:2003zu,Beenakker:2001rj,Beenakker:2002nc}. The
cross section at $13~TeV$ in the Standard Model is~\cite{Dittmaier:2011ti}
\begin{align}
\sigma_{t\bar t h} = 509^{+10\%}_{-15\%}~\text{fb}
\end{align}
However, given the complex analysis techniques, a precision prediction
of the total $t\bar{t}h$ rate falls short of the experimental
requirements. Instead, we need a precise prediction of all kinematic
distributions for the signal as well as for the irreducible
background~~\cite{Bredenstein:2010rs,Denner:2012yc,Gutschow:2018tuk}. This
includes the boosted regime, where we expect any $t\bar{t}$ analysis
to collect most of its significance~\cite{Plehn:2013paa}.
Corresponding results are available to NLO in QCD~\cite{Denner:2012yc}
and including electroweak corrections~\cite{Denner:2016wet}. In
Fig.~\ref{fig:tth_nlo} we show the NLO effects on two key
distributions for the $t\bar{t}h$ signal. The dramatically increased
tail in the reconstructed Higgs mass arises from non-recombined gluons
and photons. While the NLO contributions are large and positive in the
boosted Higgs regime, they lead to a relative drop in the boosted top
regime.

First LHC cross sections for single top-associated Higgs production
were provided in Ref.~\cite{Maltoni:2001hu}; cross sections in the 5
flavor scheme are currently known to NNLO
precision~\cite{Brucherseifer:2014ama} and NLO precision in the 4
flavor scheme~\cite{Campbell:2009ss,Demartin:2015uha}. The flavor
scheme-averaged NLO prediction at $13~TeV$ in the Standard Model is~\cite{Demartin:2015uha}
\begin{align}
 \sigma_{th} = 72.55 \pm 10.1\%~\text{(scale variation and 4/5 flavor scheme)}~^{+3.1\%}_{-2.4\%}\text{(PDF+$\alpha_s$+$m_b$)}\,.
 \end{align}
This amounts to a moderate K factor of $\lesssim 1.2$). The dominant
uncertainty of about 10\% is related to the flavor scheme and scale
dependencies.

\begin{figure}[t]
\includegraphics[width=0.47\textwidth]{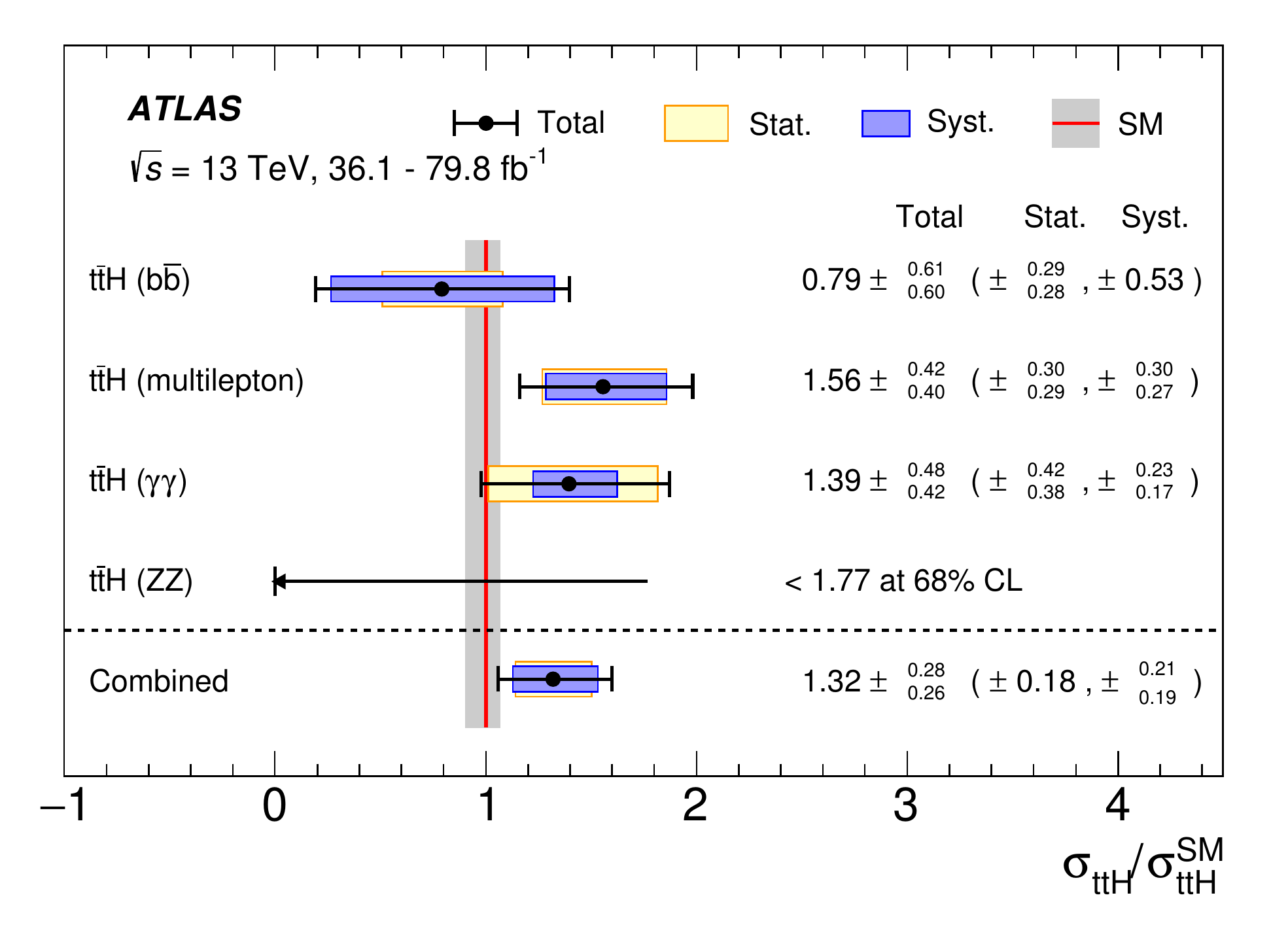}
\hspace{0.05\textwidth}
\includegraphics[width=0.42\textwidth]{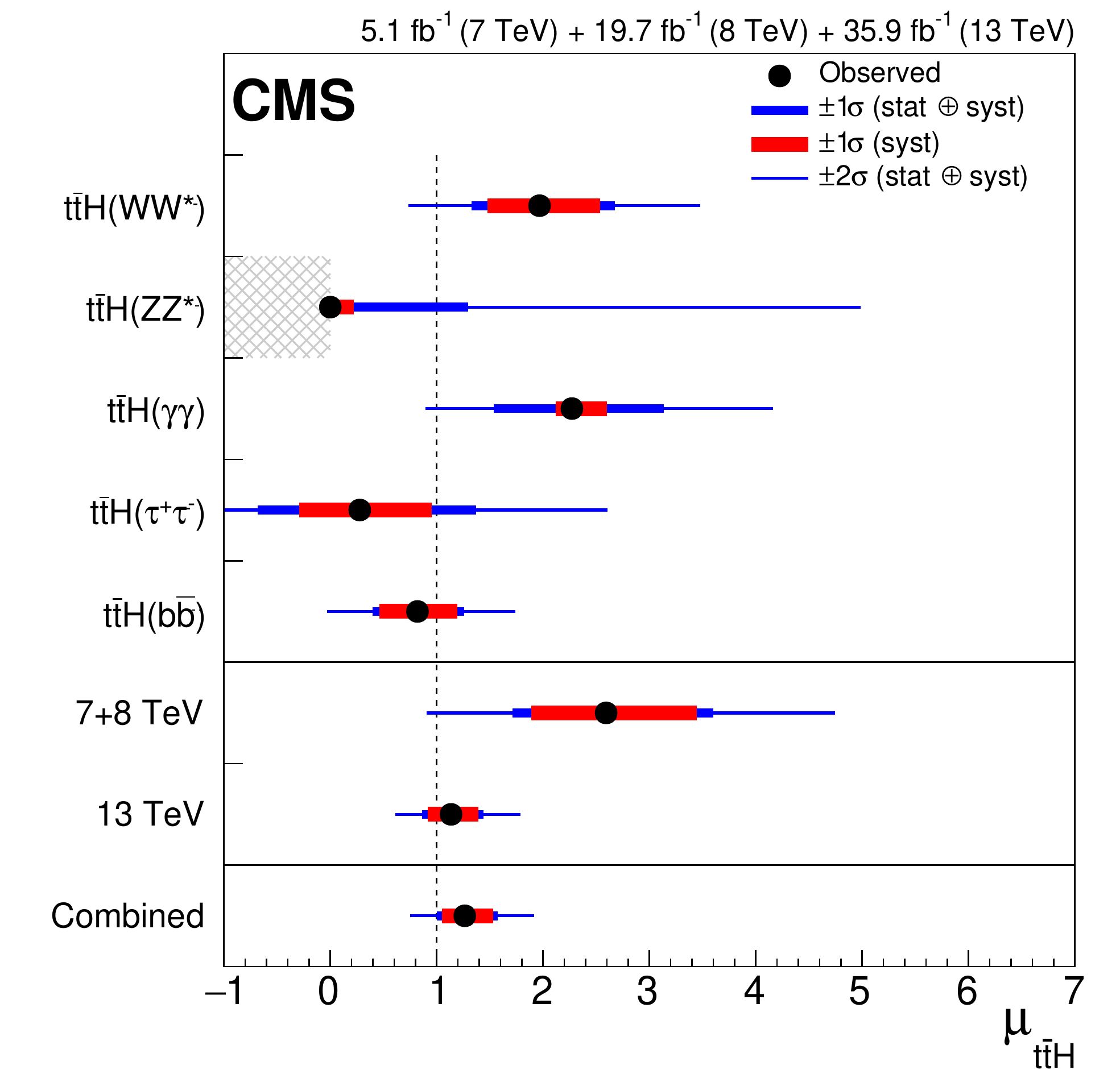}
\caption{Best-fit value of the signal modifier of Standard Model
  $t\bar t h$ production for the different final states, as well as
  their combination. Figure taken from Ref.~\cite{Aaboud:2018urx}
  and~\cite{Sirunyan:2018hoz}. Both measurements translate into $>5$
  sigma Higgs discoveries.}
\label{fig:atlastth} 
\end{figure}

\subsubsection{LHC Analyses}
\label{sec:exp_tth_ana}

CMS has observed evidence for $t\bar t h$ production in leptonic
(including tau) final states~\cite{Sirunyan:2018shy} at the $3\sigma$
level~\cite{Sirunyan:2018shy}; upon combining with run 1 data CMS
reported a discovery at the $5.2\sigma$ level~\cite{Sirunyan:2018hoz}
with a best-fit signal strength of
\begin{align}
\mu^{\text{CMS}}_{t\bar t h}=1.26^{+0.31}_{-0.26}\,.
\end{align}
A comprehensive search for $t\bar t h$ in the context of the Standard Model at 13
TeV most recently by ATLAS~\cite{Aaboud:2018urx} combines results from
$h\to bb$ and $h\to \gamma \gamma$, as well as $h\to ZZ,WW$ and $h\to
\tau \tau$ at a luminosity of 79.8/fb. Combining these final states in
a dedicated boosted decision trees allowed ATLAS to obtain a best fit
value of the $t\bar t h$ signal
\begin{align}
\sigma(t\bar t h) = 670  \pm 90~\text{(stat)}^{+110}_{-100} \text{(syst)}~\fb \; ,
\end{align}
which is consistent with the SM prediction, but shows a slight
preference for overproduction with a signal strength
\begin{align}
\mu^{\text{ATLAS}}_{t\bar t h} =1.32^{+0.28}_{-0.26}\,.
\end{align}
The ATLAS measurement amounts an observed significance of 5.1 standard
deviations, and hence ATLAS provided the observation of top quark pair
associated Higgs production in the Standard Model with the largest overall data
set that has been considered in a Higgs analysis so far.

A fully hadronic analysis of $t\bar t h$ is notoriously challenging
which is apparent in  the rather loose 13~TeV constraint observed by
CMS~\cite{Sirunyan:2018ygk}, which reports a best-fit signal
modified constraint of $\mu=0.9\pm 1.5$. ATLAS constrains
$\mu=1.6\pm 2.6$ at 8~TeV~\cite{Aad:2016zqi}.

CMS has recently looked for single top-associated Higgs production at
13 TeV~\cite{CMS:2017uzk} in the $\tau$ and $W,Z$ Higgs final states,
while the $h\to \gamma \gamma$ analysis is only available at 8
TeV~\cite{Khachatryan:2015ota}. The cross section limit that CMS
obtains under the assumption of a flipped Yukawa coupling compared to
the Standard Model is
\begin{align}
\sigma_{th}  < 0.64~\pb \quad \hbox{for} \quad \frac{1+\Delta_t}{1+\Delta _V}=-1\quad\hbox{at 95}\%~\text{C.L.}\,.
\end{align}
The best-fit signal strength assuming SM couplings is $\mu=1.8 \pm
0.3~\text{(stat)}\pm 0.6~\text{(syst)}$, which amounts to an observed
significance of 2.7$\sigma$. As can be seen from Fig.~\ref{fig:thq},
the LHC has begun to exclude wrong-sign Yukawa couplings (under the
assumptions quoted in Fig.~\ref{fig:thq}). 

\begin{figure}[t]
\includegraphics[width=0.42\textwidth]{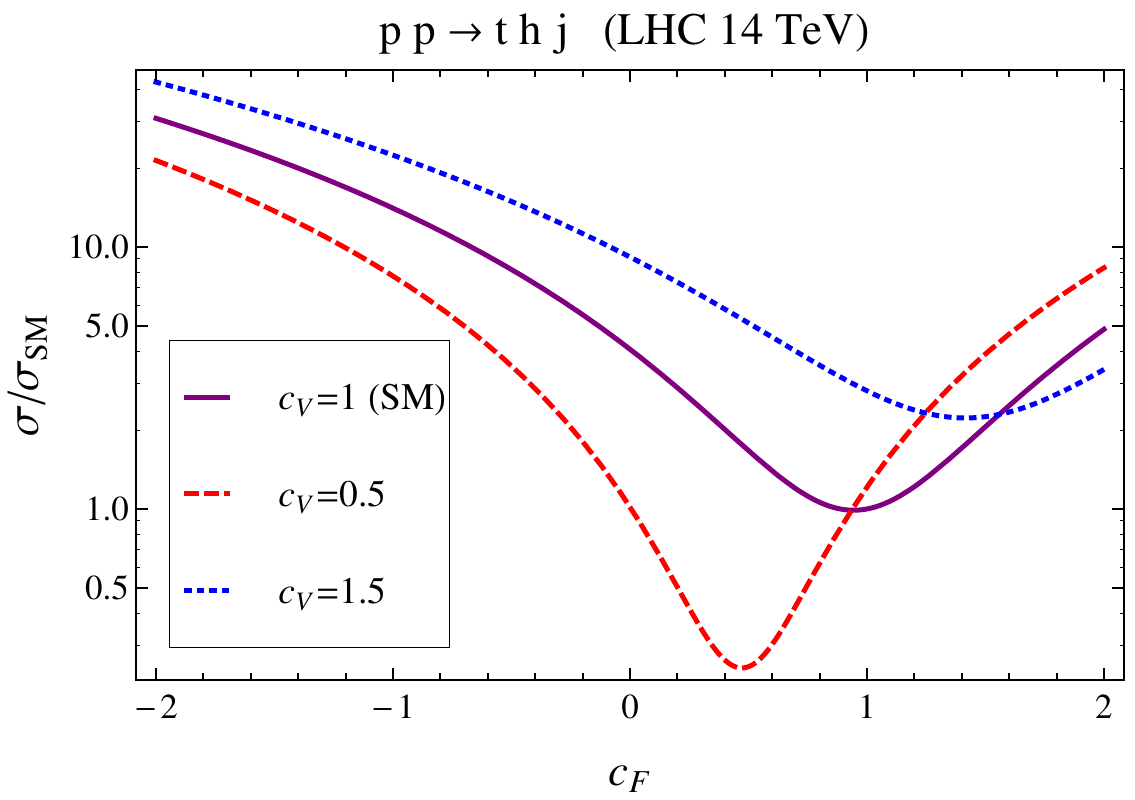}
\hspace{0.1\textwidth}
\includegraphics[width=0.42\textwidth]{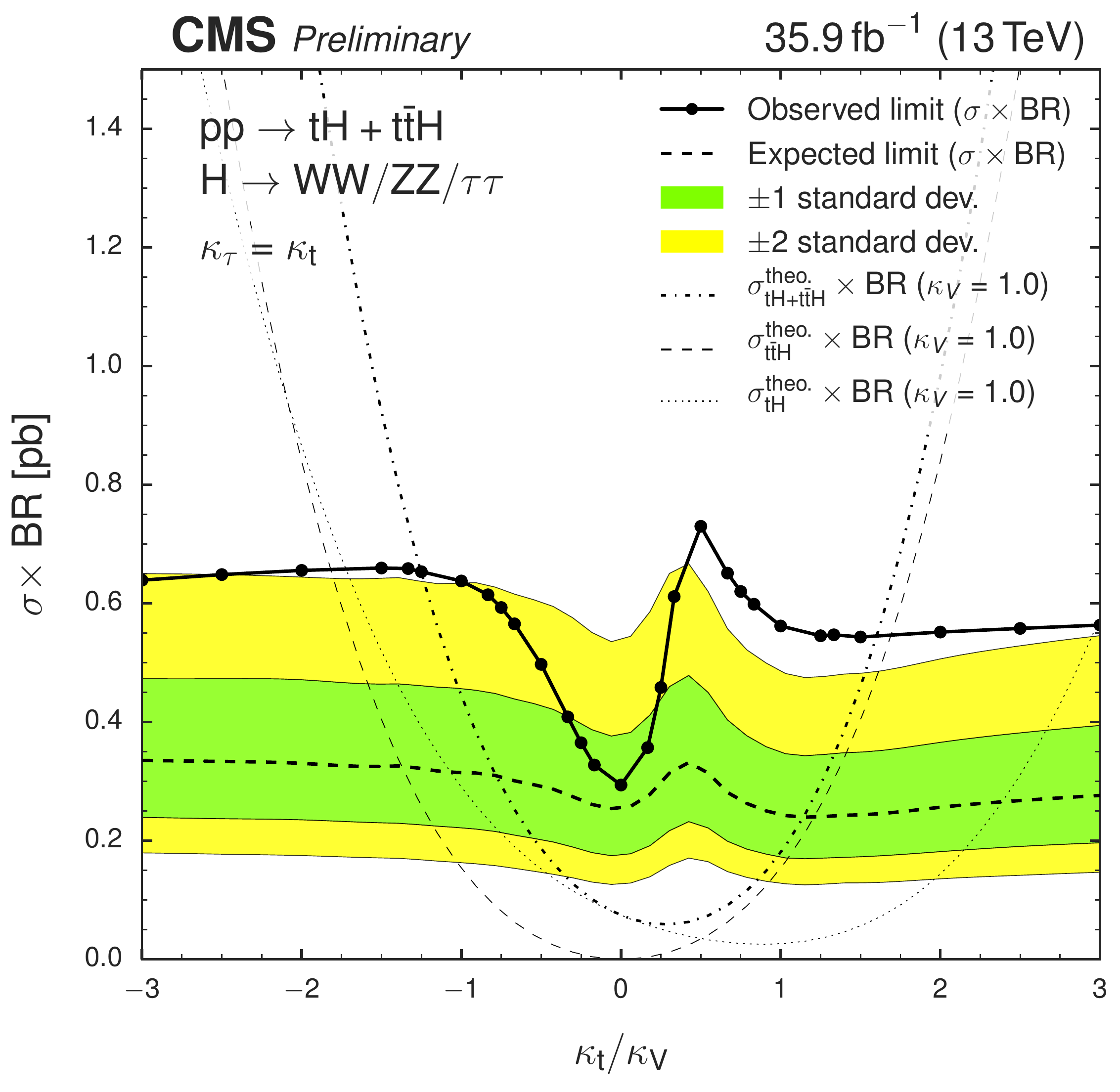}
\caption{Left: Single top-associated Higgs production for different
  values of re-scalings of the Higgs-vector boson $c_V=\kappa_V \equiv
  1+\Delta_V$ and Higgs-fermion couplings $c_F=\kappa_t \equiv
  1+\Delta_t$. Figure taken from Ref.~\cite{Farina:2012xp}. Right:
  95\% confidence level constraints on single top-associated Higgs
  production reported by CMS for an analysis based on 13 TeV data,
  focusing on $h\to WW,ZZ,\tau\tau$. Figure taken from
  Ref.~\cite{CMS:2017uzk}.}
\label{fig:thq} 
\end{figure}

Additionally, charged Higgs production can be searched for in $tb$
final states~\cite{Aad:2015typ,Khachatryan:2015qxa} or
$\tau+\slashed{E}_T$~\cite{Aaboud:2016dig}, as illustrated in
Fig.~\ref{fig:thqcharged}.

\begin{figure}[b!]
\includegraphics[width=0.43\textwidth]{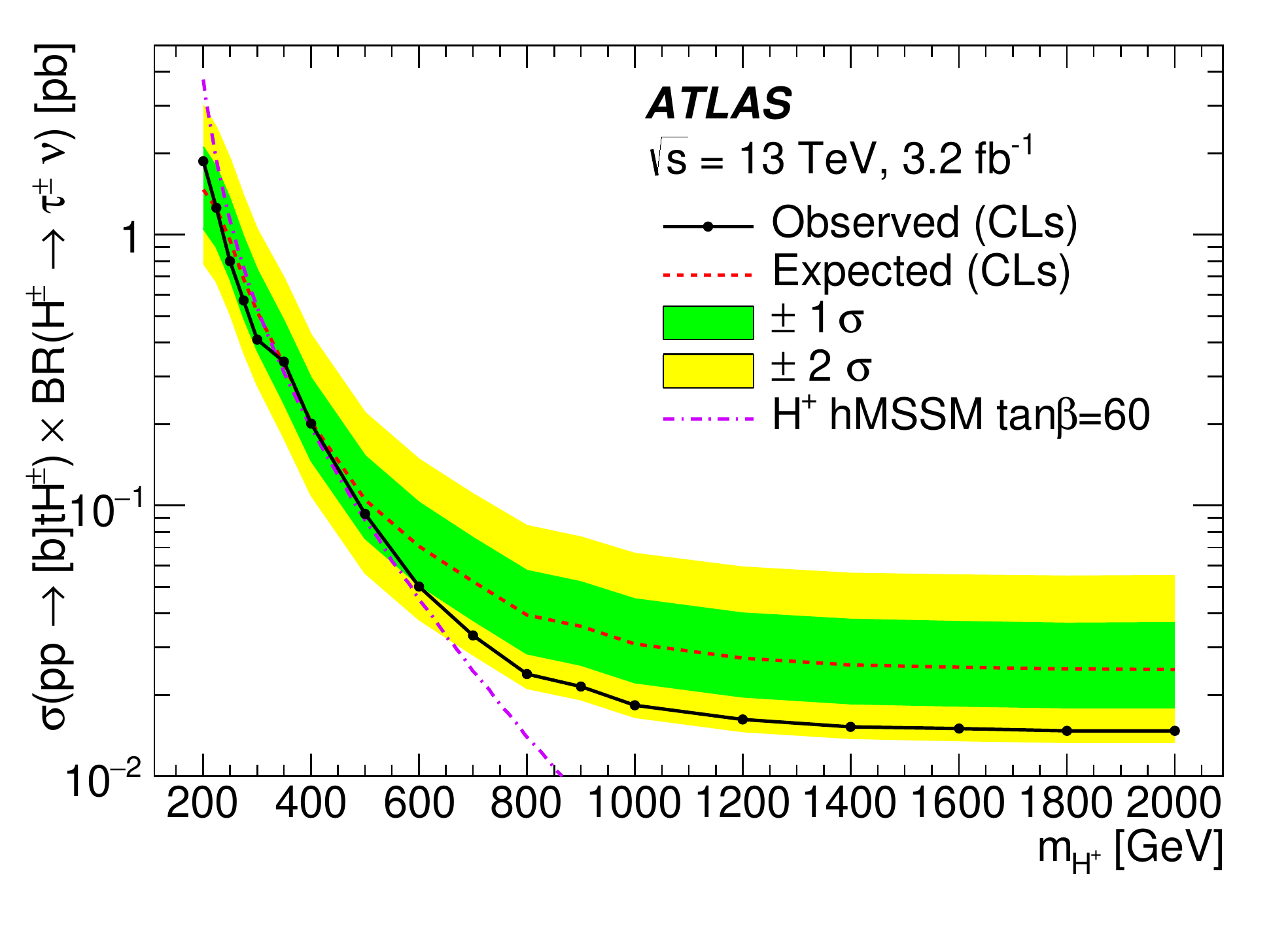}
\hspace*{0.1\textwidth}
\includegraphics[width=0.43\textwidth]{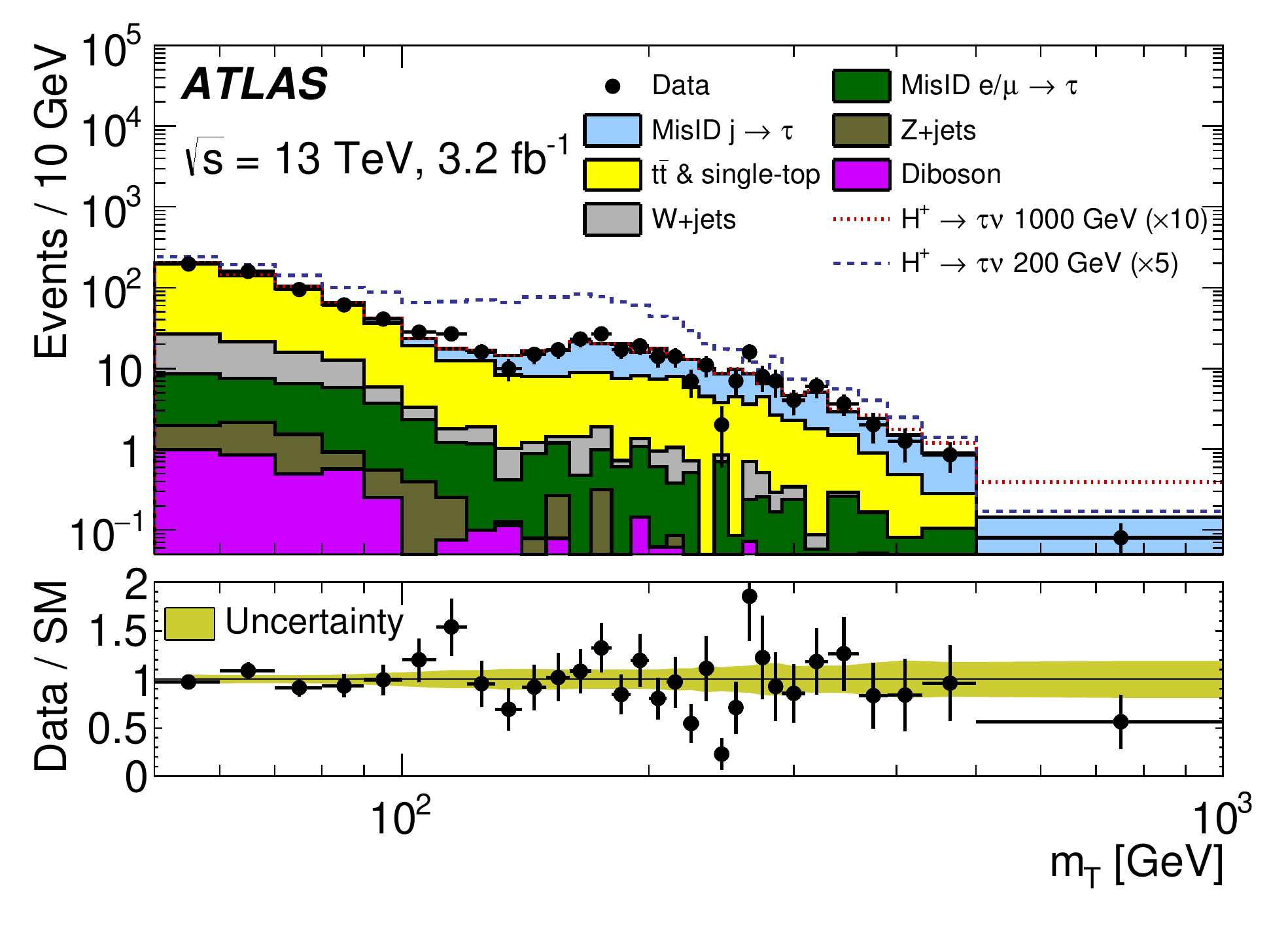}
\caption{95\% confidence level constraints on charged Higgs production
  reported by ATLAS for an analysis based on 13 TeV data, focusing on
  $H^+\to \tau\nu$ (left). Right shows the signal and backgrounds as a
  function of the transverse mass for two benchmark scenarios. Figures
  taken from Ref.~\cite{Aaboud:2016dig}.}
\label{fig:thqcharged} 
\end{figure}

\subsection{Exotic and rare decays}
\label{sec:exp_rare}

\begin{figure}[t]
\includegraphics[width=7cm]{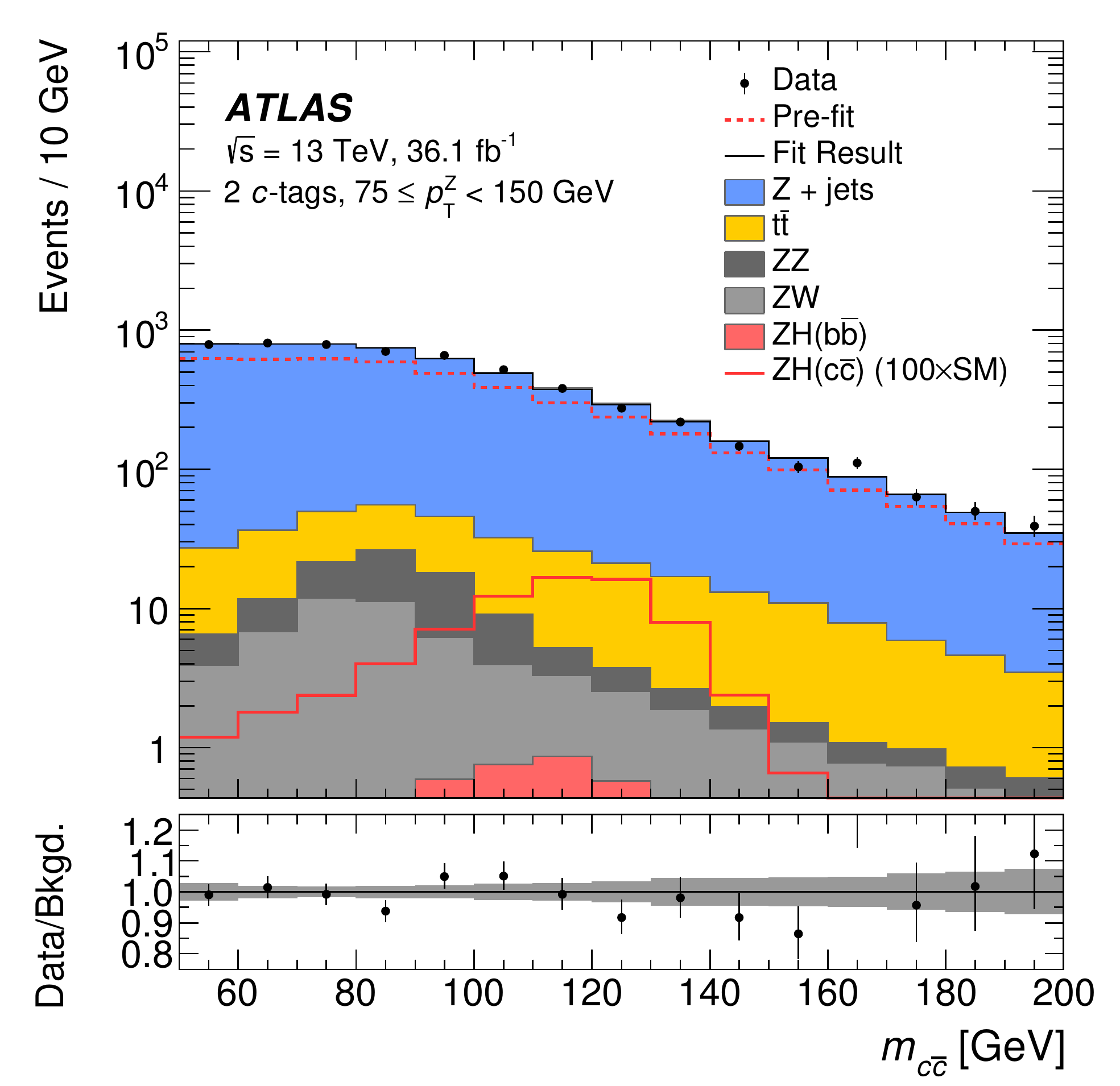}
\hspace{1cm}
\includegraphics[width=7cm]{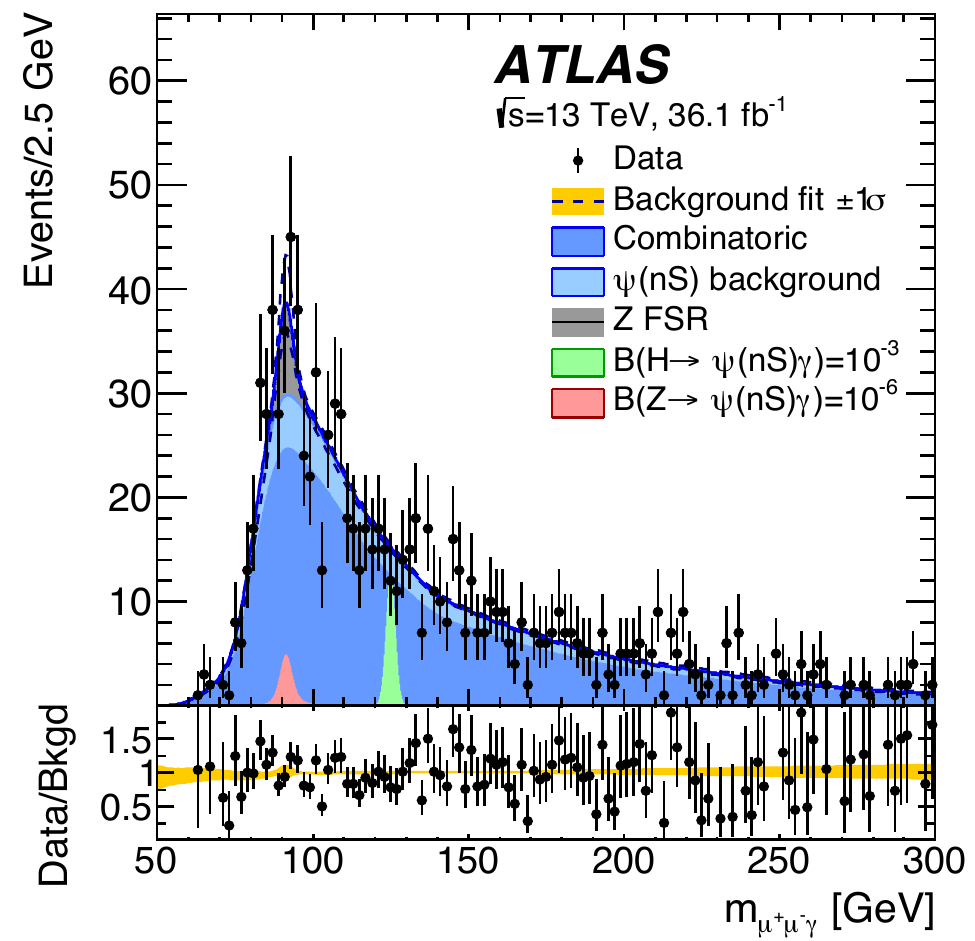}
\caption{Left: Search for $h\to c\bar c$ in associated Higgs
  production, taken from Ref.~\cite{Aaboud:2018fhh}. Right: Search for
  $h\to J/\psi + \gamma$ taken from Ref.~\cite{Aaboud:2018txb}.}
\label{fig:htocc} 
\end{figure}

Constraining Higgs interactions through its decays to rare final
states is a formidable experimental task, in particular because the
expected signal count is tiny. Firstly, trigger criteria can be
satisfied through, \eg, observing the leptonic decay of the $Z$ boson
in associated Higgs production
\begin{align}
pp\to hZ, h\to c\bar c
\end{align}
This process, including the Higgs decay to charm branching ratio, has a cross section 
\begin{align}
\sigma (hZ, h\to c\bar c) = 26\, \fb \qquad \text{excluding $Z$ branching fraction}.
\end{align}
Through adapted jet-tagging techniques the charm content of the Higgs
candidate events can be tagged similar to $b$-tagging,
Fig.~\ref{fig:htocc}. These techniques are experimentally involved and
are based on charmed meson decay
phenomenology~\cite{Aad:2014xca,Chatrchyan:2013uja}).

\begin{figure}[b!]
\includegraphics[height=0.45\textwidth]{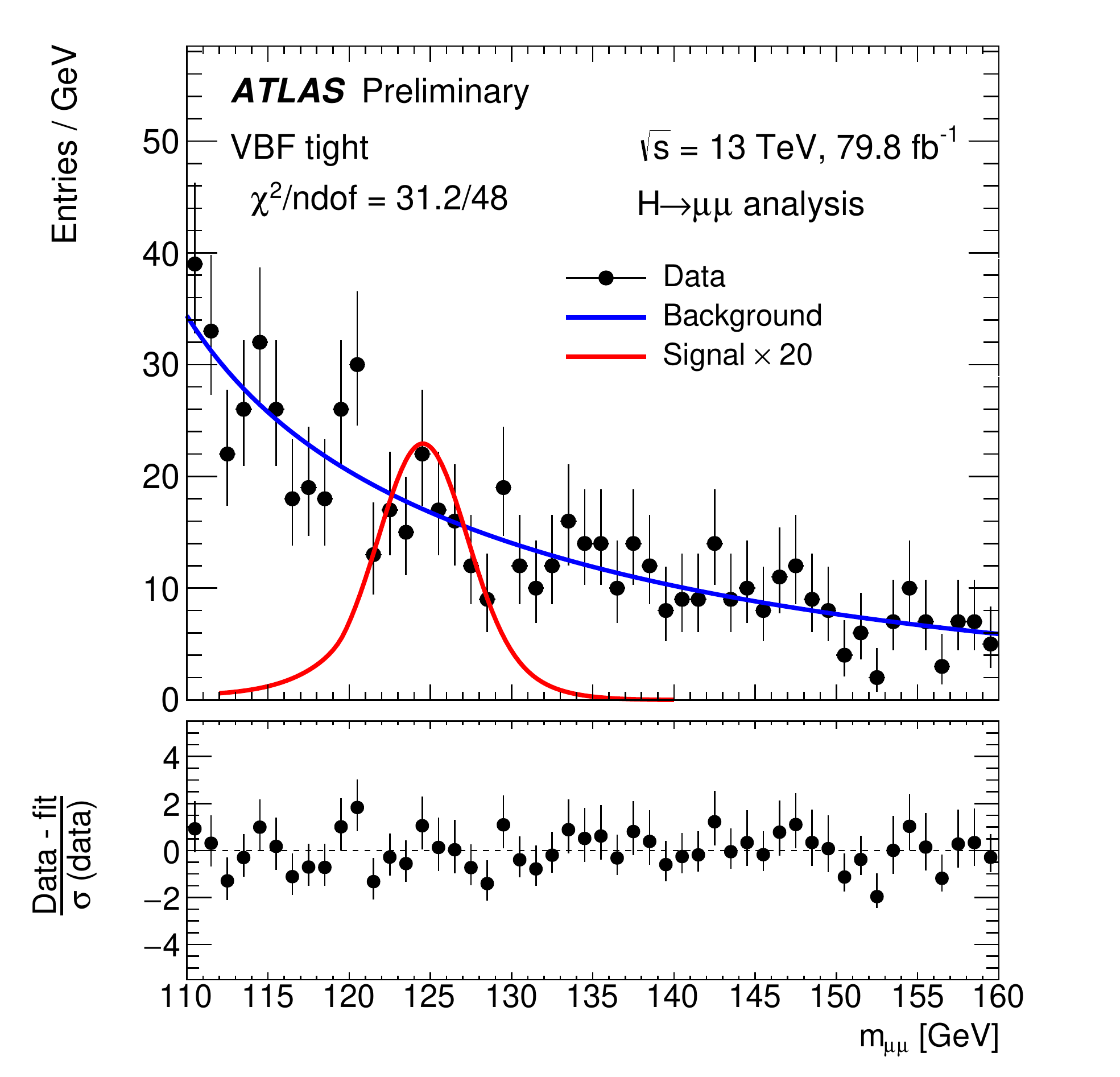}
\hspace*{0.05\textwidth}
\includegraphics[height=0.45\textwidth]{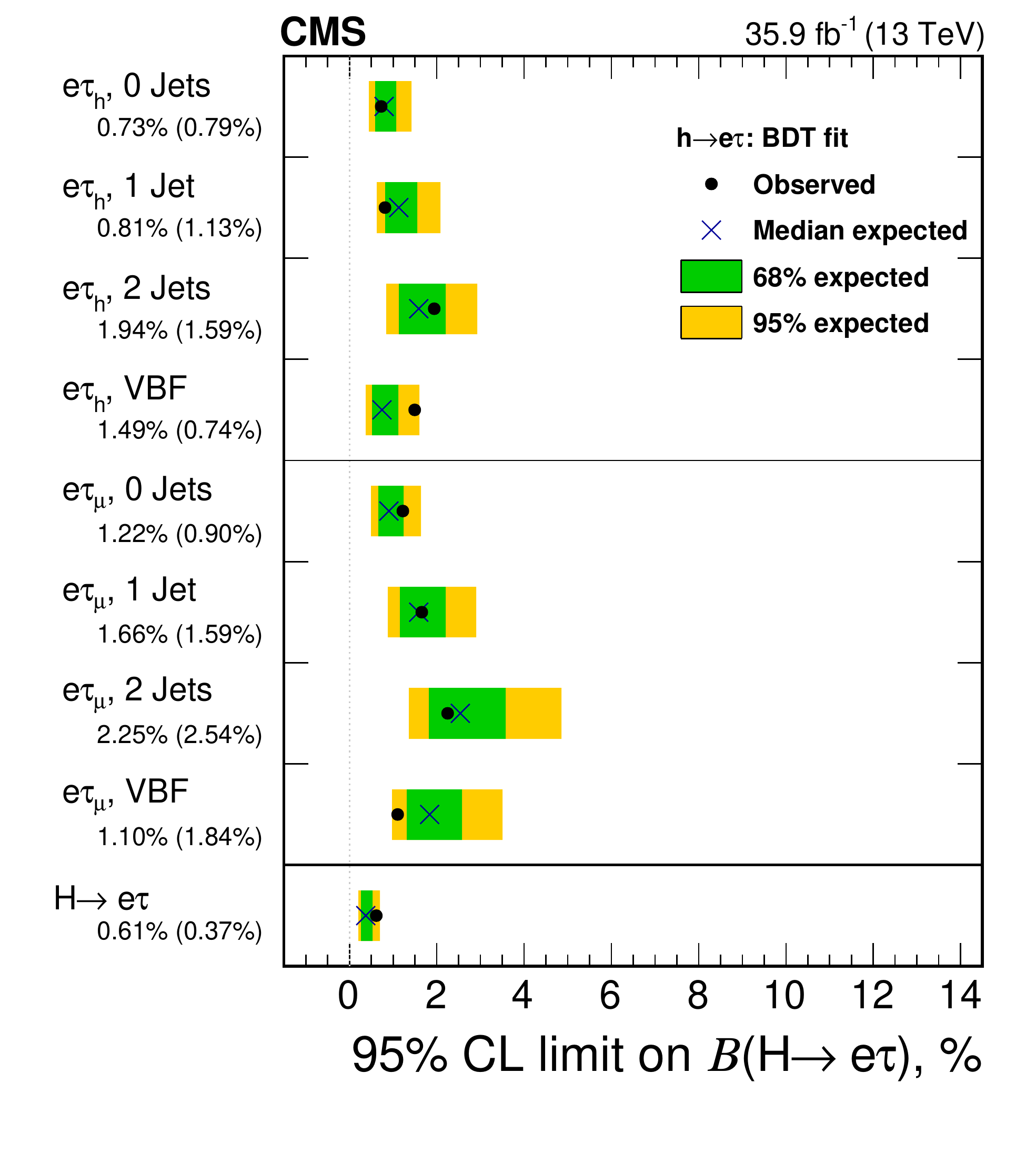}
\caption{Left: Search for $h\to \mu^+\mu^-$ in the weak boson fusion
  category, taken from Ref.~\cite{ATLAS:2018kbw}.  Right: Constraints
  form searches for lepton flavor violating Higgs decays into $e\tau$,
  taken from Ref.~\cite{Sirunyan:2017xzt}.}
\label{fig:htomumu} 
\end{figure}

The second avenue as mentioned in Sec.~\ref{sec:exoticrare} relates to
Higgs decays with an easy to tag final state including photons, $Z$
bosons and $c$- and $b$-flavored mesons (see also~\cite{Aaboud:2017xnb}). Summing over different Higgs
production modes ATLAS has placed the limits~\cite{Aaboud:2018txb}
\begin{align}
\text{BR}(h\to J/\psi \gamma) & < 3.5\times 10^{-4}\,, \notag \\
\text{BR}(h\to \Upsilon(1S) \gamma) & < 4.9\times 10^{-4}\,, \notag \\
\text{BR}(h\to \Upsilon(2S) \gamma) & < 5.9\times 10^{-4}\,, \notag \\
\text{BR}(h\to \Upsilon(3S) \gamma) & < 5.7\times 10^{-4}\,.
\end{align}
These constraints are two to 5 orders of magnitude away from the SM
expectations~\cite{Bodwin:2013gca,Koenig:2015pha}. In the right panel
of Fig.~\ref{fig:htocc} we show how the different Higgs production
modes contribute to the combined analysis. As expected for Run~I, the
gluon fusion process dominates the statistics-limited analysis.

Leptonic decays of the Higgs are Yukawa-suppressed in the Standard Model and the
decay of the Higgs boson into to muons is highly suppressed at an
expected branching ratio value of $2.18\times 10^{-4}$. The signature,
however, is very clean as muons are easy to isolate and the mass
resolution in the di-muon channel is high, Fig.~\ref{fig:htomumu}. The
inclusive search for the Higgs decay $h \to \mu \mu$ is dominated by
gluon-fusion production, the main background is Drell-Yan production
of two muons, and one of the major obstacles is that for the $\mu^+
\mu^-$ final state there are not many handles to separate signal and
background aside from the invariant di-muon
mass~\cite{Han:2002gp}. The situation changes once the integrated
luminosity allows us to combine rate Higgs decays with WBF
production. In this case we can use the central jet activity to reduce
the QCD background to the level of electroweak $Zjj$ production, as
discussed in Sec.~\ref{sec:exp_wbf}. An appropriate multi-variate
analysis of the WBF final state will then significantly enhance the
reach of the LHC~\cite{Plehn:2001qg,Cranmer:2006zs}.  The current
analysis based on inclusive Higgs production~\cite{ATLAS:2018kbw}
constrains the signal strength in the di-muon decay channel to
\begin{align}
\mu(h\to \mu^+\mu^-)= 0.1^{+1.0}_{-1.1}\,,
\end{align}
which corresponds to an observed upper limit on $\mu=3$ at 95\%
confidence level. This improves the 7 and 8 TeV searches by factor of
about two (see \eg the individual CMS and ATLAS analyses of
Refs.~\cite{Khachatryan:2014aep,Aad:2014xva} as well as their
combination~\cite{Khachatryan:2016vau}). The latest CMS 13 TeV
analysis sets a $95\%$ CL constraint of $\mu <
2.64$~\cite{CMS:2017qgo}.

Although lepton-flavor off-diagonal Higgs interactions are absent in
the Standard Model, their observation at the LHC would be clear evidence of new
physics beyond the Standard Model in the Higgs sector~\cite{DiazCruz:1999xe,Han:2000jz,Harnik:2012pb}. This
motivates searches for $h\to \mu \tau$~\cite{Han:2000jz} as performed by the CMS
experiment~\cite{Sirunyan:2017xzt} and ATLAS~\cite{Aad:2015gha}, with
current constraints in the (sub)percent range, Fig.~\ref{fig:htomumu}
(right).

\subsection{Global Higgs analyses}
\label{sec:exp_global}

The ever increasing number of Higgs measurements, based on different
production channels, decay channels, and phase space regions, calls
for global analyses. As introduced in Sec.~\ref{sec:basis_eft}, the
simplest approach starts from a Lagrangian with modified Higgs
couplings to all Standard Model particles. This model can be
incorporated in a consistent quantum field theory as an effective
field theory including the Higgs scalar in the broken phase. An
alternative effective theory approach includes the Higgs field as part
of a doublet with the weak Goldstone modes. We will show results for
this kind of analysis in Sec.~\ref{sec:exp_global_eft}. A general
limitation of all effective field theory approaches is that they
assume that all new particles are decoupled at the LHC. At the same
time, the bases of higher-dimensional operators introduce a cutoff
scale, above which we would have to resort to the full ultraviolet
completion. An alternative set of hypotheses for global Higgs analyses
are consistent ultraviolet completions of the Standard Model. It
includes general extended Higgs sectors as well as their more
predictive supersymmetric counterparts, as introduced in
Sec.~\ref{sec:basis_weak}. We will quote some example results using
these approaches in Sec.~\ref{sec:exp_global_full}.

\subsubsection{Effective theory}
\label{sec:exp_global_eft}

Because of the intermediate Higgs mass of $m_h \approx 125$~GeV, there
exists a multitude of Higgs signatures at the LHC.  Global Higgs
analyses are based on a set of Higgs production channels
\begin{itemize}
\item gluon fusion, loop-induced by heavy quarks in the Standard Model
  (Sec.~\ref{sec:exp_gf})
\item weak boson fusion, combining dominant $WW$-fusion and
  sub-leading $ZZ$-fusion (Sec.~\ref{sec:exp_wbf})
\item associated $Wh$ and $Zh$ production (Sec.~\ref{sec:exp_vh})
\item associated $t\bar{t}h$ and $th$ production (Sec.~\ref{sec:exp_tth})
\end{itemize}
Ideally, each of these categories dominates an experimental signature.
As discussed in Sec.~\ref{sec:exp_gf_kin}, off-shell Higgs production
and boosted Higgs production provide additional information mostly on
the gluon fusion process.  For precision Higgs analyses these
production channels can be combined with SM-like decays
\begin{itemize}
\item decays to fermions $h \to b\bar{b}, \tau \tau, \mu \mu$
  (Sec.~\ref{sec:basis_decs_f})
\item tree-level decays to bosons $h \to WW, ZZ$
  (Sec.~\ref{sec:basis_decs_v})
\item loop-induced decays $h\to \gamma \gamma, Z \gamma$
  (Sec.~\ref{sec:basis_decs_a})
\item invisible Higgs decays $h \to 4 \nu$, negligible in the Standard
  Model
\end{itemize}
These decays can be combined with each of the production processes,
with the exception of gluon fusion Higgs production with a decay $h
\to b\bar{b}$ which is swamped by QCD backgrounds. Moreover, the rare
decays $h \to \mu \mu$ and $h \to Z \gamma$ will only play a
significant role for the upcoming LHC runs.  A straightforward way of
combining the corresponding LHC measurements are the coupling modifiers
defined in
Eq.\eqref{eq:def_delta}~\cite{Zeppenfeld:2000td,Duhrssen:2004cv,Lafaye:2009vr}. As
discussed in Sec.~\ref{sec:basis_couplings}. The notation $\kappa_x =
1 + \Delta_x$ is equivalent, provided we correctly separate the
effective Higgs couplings to gluons and photons into modified
couplings entering the SM loop diagrams and generic new loop
contributions. On the quantum field theory side, the coupling
modifiers can be interpreted as Wilson coefficients in an effective
theory with non-linearly realized electroweak symmetry
breaking~\cite{Buchalla:2016bse}. Because the SM-like coupling
modifiers have very little effect on kinematic distributions, at least
at the Run~I precision level, they can be extracted from measured
total cross sections or so-called signal strengths
\begin{align}
\mu = \frac{\sigma \times \br}{\sigma_\text{SM} \times \br_\text{SM}} \; .
\label{eq:def_sig_strenth}
\end{align} 
%

\begin{figure}[t]
  \center
\includegraphics[width=0.75\textwidth]{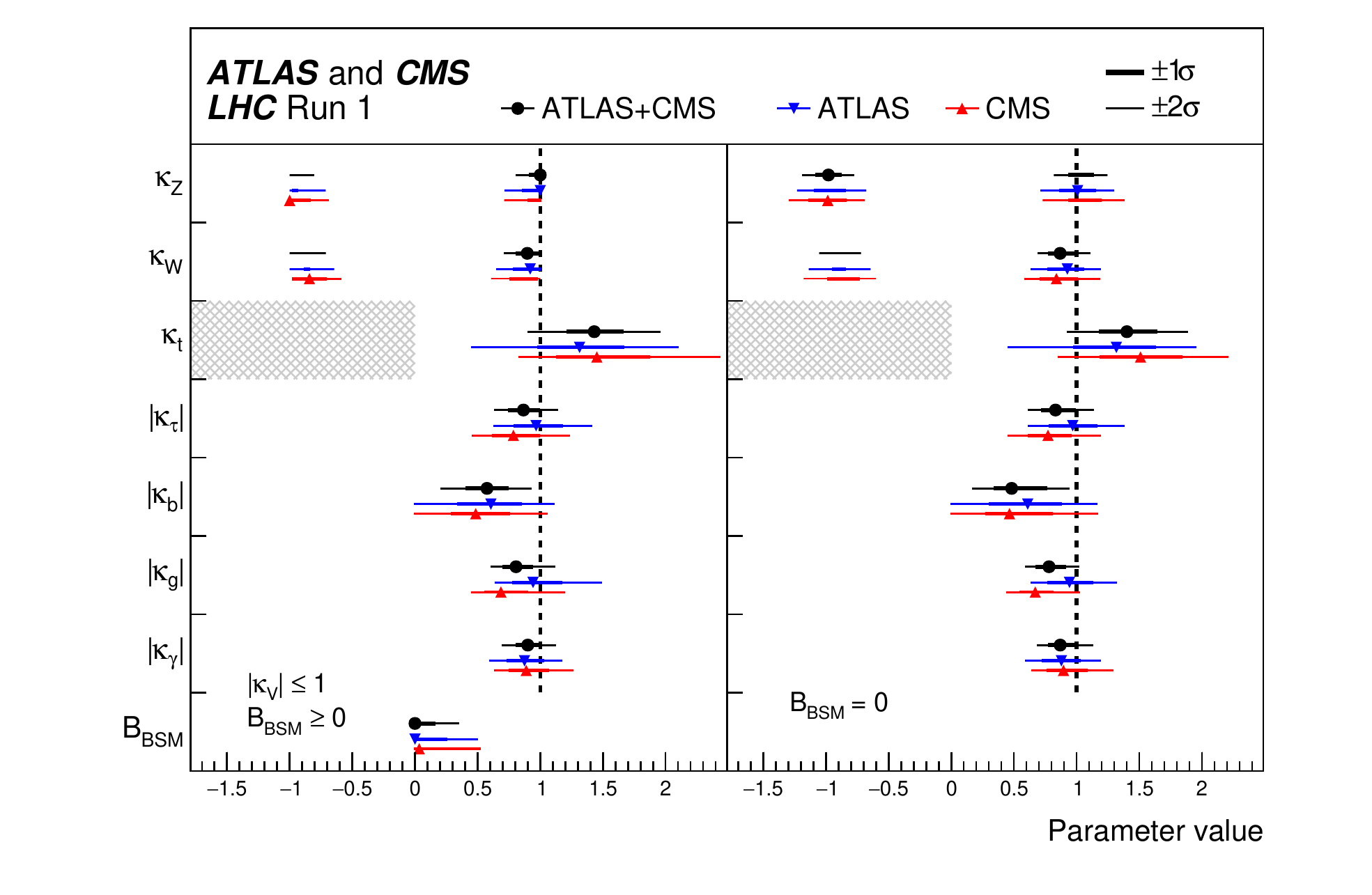}
\caption{Coupling modifiers for two assumptions concerning the total
  Higgs width.  The hatched areas indicate non-allowed regions for the
  top Yukawa. For those parameters with no sensitivity to the sign,
  only the absolute values are shown. Figure from
  Ref.~\cite{Khachatryan:2016vau}.}
\label{fig:kappas}
\end{figure}

The main challenge of the global analysis is not to extract a decent
central value for each coupling modifier based on a set of measured
signal strengths. Given that we have not observed a significant
deviation from the SM predictions, the question is how close the Higgs
couplings are to their SM predictions. In other words, we need to
carefully account for statistical, systematic, and theoretical
uncertainties with full correlations between the observables and the
predictions. ATLAS, CMS, and most theoretical collaborations rely on a
likelihood approach with profile likelihoods for all nuisance
parameters and irrelevant directions in model space.

In Fig.~\ref{fig:kappas} we show the combined ATLAS and CMS limits on
coupling modifiers or Wilson coefficients after Run~I. Each limit is
profiled over all other coupling modifiers, corresponding to the basic
assumption that we do not expect new physics effects to only
contribute to one Wilson coefficient. The two sets of limits shown in
the figure are based on different assumptions about the total Higgs
width, which enters every signal strength. We always start with the
sum of the observed partial width as the lower limit to the total
Higgs width. For the left set of measurements ATLAS and CMS assume
that the Higgs couplings to the weak bosons are at most as large as in
the Standard Model. This can be motivated by unitarity
arguments~\cite{Duhrssen:2004cv} and allows us to include additional
unobserved decays in the total Higgs width. For the right set of
measurements the sum of the observed partial width, including all
coupling modifiers, is identified with the total Higgs width.
Finally, we can switch signs of all Higgs couplings, except for one
global phase in the definition of the Higgs field. ATLAS and CMS fix
the top Yukawa coupling to be positive, as indicated by the hatched
areas. Sensitivity to the relative sign of Higgs couplings arises
through the Higgs decay to photons, where the $W$-loop and top loop
interfere destructively. Looking at actual ultraviolet completions of
the Standard Model, it is not at all obvious how one would switch the
signs of $\kappa_W$ and $\kappa_Z$ without having observed spectacular
signs of new physics at the LHC. With extended Higgs sectors discussed
in Sec.~\ref{sec:basis_weak} it might be possible to switch the sign
of a Yukawa coupling, most easily the bottom Yukawa, but again these
kinds of interpretations should be taken with care. The safe
assumption also based on our estimates of systematic and theoretical
uncertainties would be
\begin{align}
\Delta_x = \kappa_x - 1 \ll 1
\end{align}
for all Higgs couplings analyses.

\begin{figure}[t]
\includegraphics[width=0.40\textwidth]{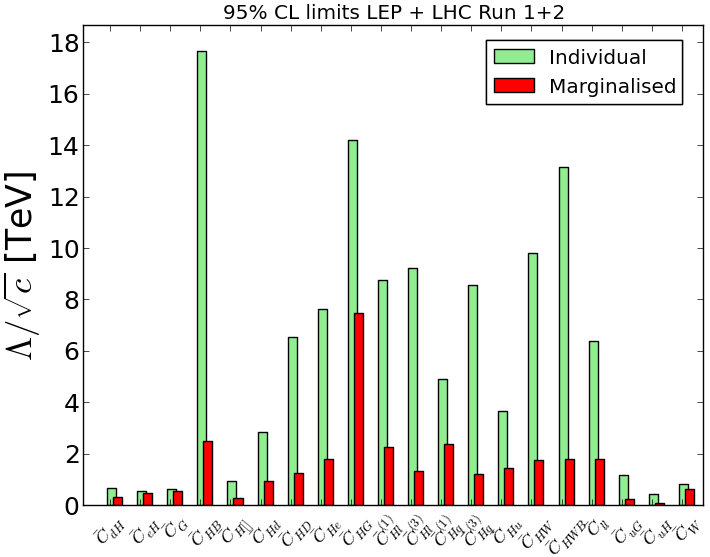}
\includegraphics[width=0.59\textwidth]{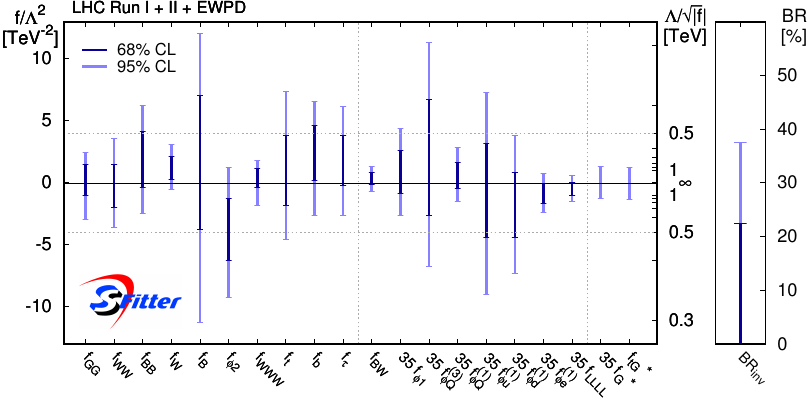}
\caption{Results of a global Higgs and electroweak analysis,
  including direct and
  indirect constraints as well as fermionic operators leading to
  anomalous electroweak gauge couplings.  Figures from
  Ref.~\cite{Ellis:2018gqa} (left) and Ref.~\cite{Biekotter:2018rhp} (right).}
\label{fig:globalres}
\end{figure}

As described in Sec.~\ref{sec:basis_eft_linear}, there are three
shortcomings in the coupling modifier approach: experimentally, it
does not (significantly) affect kinematic distributions and therefore
excludes a wealth of experimental information from a global
fit. Second, it is hard to reconcile with electroweak
renormalizability, which means that we cannot match a percent level
experimental precision with theory predictions. Finally, it does not
allow us to combine information with electroweak measurements, be it
electroweak precision data or di-boson production at the LHC. All
three shortcomings can be cured by extending the model parameters to
Wilson coefficients of an effective theory with linearly realized
symmetry breaking in terms of the Higgs--Goldstone
doublet. Technically, the global analysis is a straightforward
extension of the previous analysis in terms of the coupling
modifiers. It can be run on the same Higgs data set, which means that
we can directly compare the results of the two
approaches~\cite{Corbett:2015ksa}. The main difference is that the
gauge-invariant dimension-6 Lagrangian does not allow for separate
modifications $\Delta_W$ and $\Delta_Z$.

\begin{table}[b!]
\begin{center} 
\begin{tabular}{ll|lccr}
\toprule%
&channel & distribution & \#bins & max [GeV] & \\
\midrule
\multirow{4}{*}{8~TeV}
&$WW\rightarrow \ell^+\ell^{\prime -}+\met \; (0j)$    & leading $p_{T,\ell}$ & 350 & $20.3~\ifb$~\cite{Aad:2016wpd}  \\
&$WW\rightarrow \ell^+\ell^{(\prime) -}+\met \; (0j)$  & $m_{\ell\ell^{(\prime)}}$ & 575 & $19.4~\ifb$~\cite{Khachatryan:2015sga}  \\
&$WZ\rightarrow \ell^+\ell^{-}\ell^{(\prime)\pm}$       & $m_{T,WZ}$           & 450 & $20.3~\ifb$~\cite{Aad:2016ett}  \\
&$WZ\rightarrow \ell^+\ell^{-}\ell^{(\prime)\pm}+\met $ & $p_T^{Z \to \ell\ell}$ & 350 & $19.6~\ifb$~\cite{CMS:2013qea}  \\
\midrule
\multirow{2}{*}{13~TeV}
&$WZ\rightarrow \ell^+\ell^{-}\ell^{(\prime)\pm}$       & $m_{T,WZ}$           & 675 & $36.1~\ifb$~\cite{ATLAS:2018ogj}  \\
&$VH \rightarrow \nu \bar{\nu} \; b\bar{b}$ & $m_{T,Vh}$ & 990  & $36.1~\ifb$\cite{Aaboud:2017cxo} \\
&$VH \rightarrow \ell \nu \; b\bar{b}$ & $m_{Vh}$ &  1210 & $36.1~\ifb$\cite{Aaboud:2017cxo} \\
\bottomrule
\end{tabular} 
\end{center}
\caption{List of kinematic di-boson and Higgs distributions included
  in the global analysis of Ref.~\cite{Biekotter:2018rhp}. The maximum
  value in GeV indicates the lower end of the highest-momentum bin we
  consider.}
\label{tab:vv_data}
\end{table}

Because the effective theory leads to sizeable modifications of
kinematic distributions through operators scaling like $p/\Lambda^2$
with the momentum $p$ flowing through a vertex, rather than
$v/\Lambda^2$. This makes it mandatory to include for instance
transverse momentum distributions in weak boson fusion or associated
$Vh$ production or the invariant mass $m_{Vh}$ in associated $Vh$
production. Technically, this complicates the analysis because we need
to take into account bin-to-bin correlations in all uncertainties. On
the other hand, from the scaling of the effective operators we also
know that most of the relevant information will come from the highest
reliable bins in these kinds of distributions. The corresponding
momentum typically scales with the momentum flow through the effective
vertex and determines the reason in the new physics scale
$\Lambda$. In Fig.~\ref{fig:globalres} we show results from two
global EFT analyses including Higgs and di-boson measurements at the LHC. While it is useful to express the
limits in terms of $f/\Lambda^2$, \ie the prefactor of the effective
operator in the Lagrangian, the relevant physics is typically encoded
in the re-scaled new physics scale $\Lambda/\sqrt{f}$.
The main change
compared to the Run~I legacy results is the quality of the different
fermionic decay measurements, including $t\bar{t}h$, strongly correlated with the effective
Higgs-gluon coupling included as $\ope_{GG}$. The long list of channels
is nicely illustrated in Fig.~\ref{fig:atlastth}. 
The appearance of $\ope_{WWW}$ reflects the
fact that the fit combines Higgs measurements with di-boson production
channels at the LHC. This additional information is crucial to remove
strong non-Gaussian features and correlations from the pure Higgs
analysis. Comparing the LHC limits from di-boson production on
anomalous triple gauge couplings to LEP limits on the same parameters
we see that the LHC has surpassed the LEP precision due to its larger
momentum flow through the corresponding triple gauge vertices. This
effect is not linked to the linear EFT description and also appears in
the not gauge-invariant parametrization of
Eq.\eqref{eq:deltakappalambda}.

Given the accuracy achieved at the LHC Run~II it turns out that 
we cannot simply ignore operators which violate custodial symmetry at tree-level. Instead, we need to include those operators and
fermionic 
operators leading to anomalous gauge couplings~\cite{Baglio:2017bfe,Alves:2018nof,Franceschini:2017xkh} and combine the LHC fit with
electroweak precision data~\cite{Ellis:2018gqa,Almeida:2018cld,Biekotter:2018rhp}. A similar approach is
advertized in the Bayesian analysis of
Ref.~\cite{deBlas:2016ojx}. Focussing on the marginalized, profile
likelihood result, the operator $\ope_{HG}$ looks much
more constrained. This is related to the fact that the
corresponding SM structure only appears at loop level and that unlike
in Eq.\eqref{eq:ourleff} the loop factor is not included in this
definition of the Wilson coefficient. The operators affecting the
fermion couplings of the weak bosons appear in the LHC signals, but
they are largely constrained by electroweak precision data. The fact
that the limits on these Wilson coefficients are similar to the LHC
limits confirms the observation that in the Higgs sector the LHC
really has been turned into a precision experiment.

\begin{figure}[t]
\includegraphics[width=0.43\textwidth]{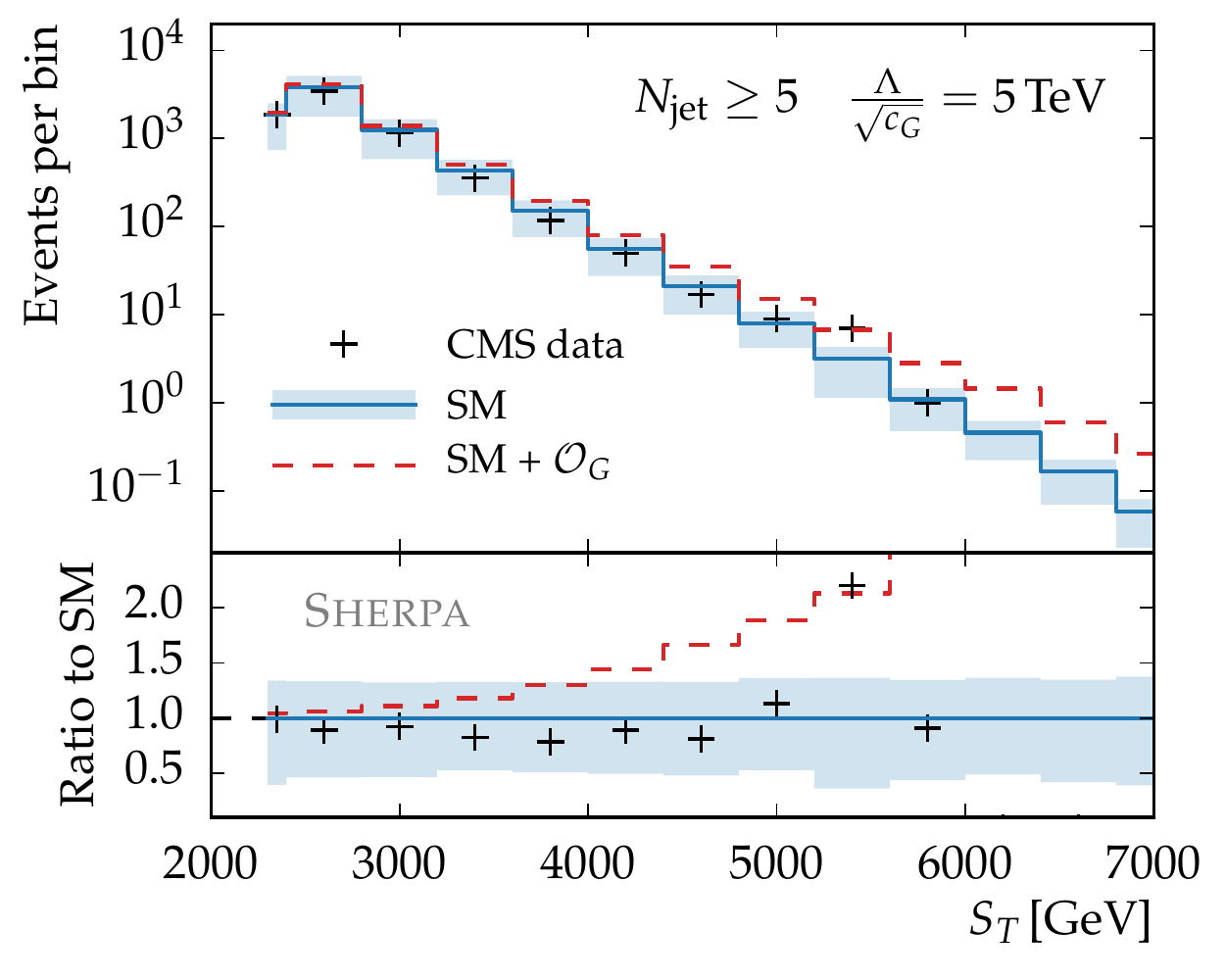}
\hspace*{0.08\textwidth}
\includegraphics[width=0.44\textwidth]{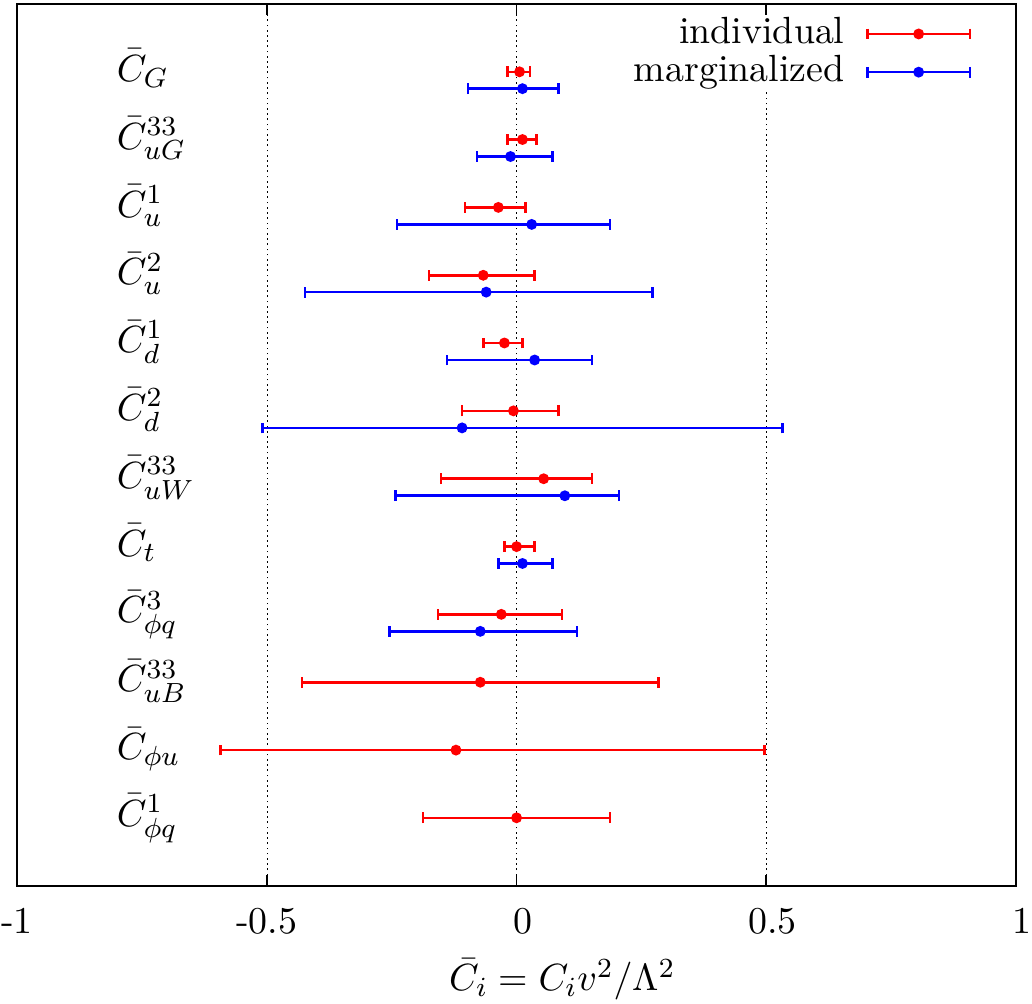}
\caption{Left: limits on the anomalous triple gluon coupling from
  8~TeV multi-jet data. Figure from Ref.~\cite{Krauss:2016ely}. Right:
  results from a fit to top physics data, including early 13 TeV
  results. Figure from Ref.~\cite{Buckley:2015lku}.}
\label{fig:d6else}
\end{figure}

The appearance of the operator $\ope_{G}$ in the Higgs fit shown in
Fig.~\ref{fig:globalres} points towards the need to include
information from entirely different LHC searches. The main reason is
that an anomalous triple gluon interaction
\begin{align}
f_G \ope_G &= \frac{g_s \, f_G}{\Lambda^2} \, f_{abc} 
         G_{a \nu}^\rho G_{b \lambda}^\nu G_{c \rho}^\lambda 
\label{eq:g3}
\end{align}
will affect every LHC process.  This not only includes Higgs
production~\cite{Ghosh:2014wxa,Dawson:2015gka}, but also top pair
production~\cite{Simmons:1990dh,Cho:1994yu,Buckley:2015lku,Buckley:2015nca}
and jet production~\cite{Dixon:1993xd,Krauss:2016ely}. Given the
interference structure with the Standard Model and the available
kinematic regimes, it turns out that this coupling is best probed in
multi-jet production at the LHC. For example the kinematic variable
\begin{align}
S_T = \left( \sum_{j=1}^{N_\text{jets}} E_{T,j} \right) + \left( \met > 50~\gev \right) 
\label{eq:s_t}
\end{align}
combined with high jet multiplicities, usually used to search for
black holes at the LHC, can be used to probe the anomalous triple
gluon coupling to very large scales~\cite{Krauss:2016ely},
\begin{align}
\frac{\Lambda}{\sqrt{f_G}} &> 5.2~\tev 
\end{align}
In Fig.~\ref{fig:d6else} we show the leading distribution entering
this analysis. The reach of the LHC is, as usual, linked to the
maximum momentum flow through the relevant vertex in the hard process.

Finally, the strong link between the Higgs and top sectors call for a
combined analyses of the respective dimension-6 Lagrangians. Global
top sector analyses are already
available~\cite{Bernardo:2014vha,Buckley:2015lku,Buckley:2015nca,AguilarSaavedra:2018nen}
and lead to similar limits on the new physics scale as global Higgs
analyses. As an example we show the results of an EFT fit to abundant
top physics results at the LHC in the right panel of
Fig.~\ref{fig:d6else}. While intermediate Higgs
bosons are phenomenologically not relevant for inclusive top final
states, dimension 6 operators like $\ope_G$ and the chromo-magnetic
moment of the top quark represented by
\begin{align}
\ope^{33}_{uG}=(\bar Q_L \sigma^{\mu\nu} \lambda^a t_R) \tilde \phi G^{a}_{\mu\nu} 
\end{align}
impact inclusive top physics, the latter upon sending the Higgs to its
vacuum expectation value. This also highlights the spirit of EFT
measurements that predict dedicated correlations between different
final states on the basis of gauge symmetry and particle content. The
individual constraints from the top fit, Fig.~\ref{fig:d6else}, show
that the limit on $\ope_G$ from top physics is weaker than the one
obtained from multi-jet production, but the constraint on
$\ope^{33}_{uG}$ amounts to a scale $\Lambda \gtrsim 0.8~\tev$. This
will improve with higher luminosity~\cite{Englert:2016aei}, with
perturbative uncertainties expected to be the limiting factors. As
expected, this scale approximately corresponds to the largest energy
scale probed in the $t\bar t$ analysis included in
Ref.~\cite{Buckley:2015lku}.

\begin{figure}[!t]
\includegraphics[width=0.45\textwidth]{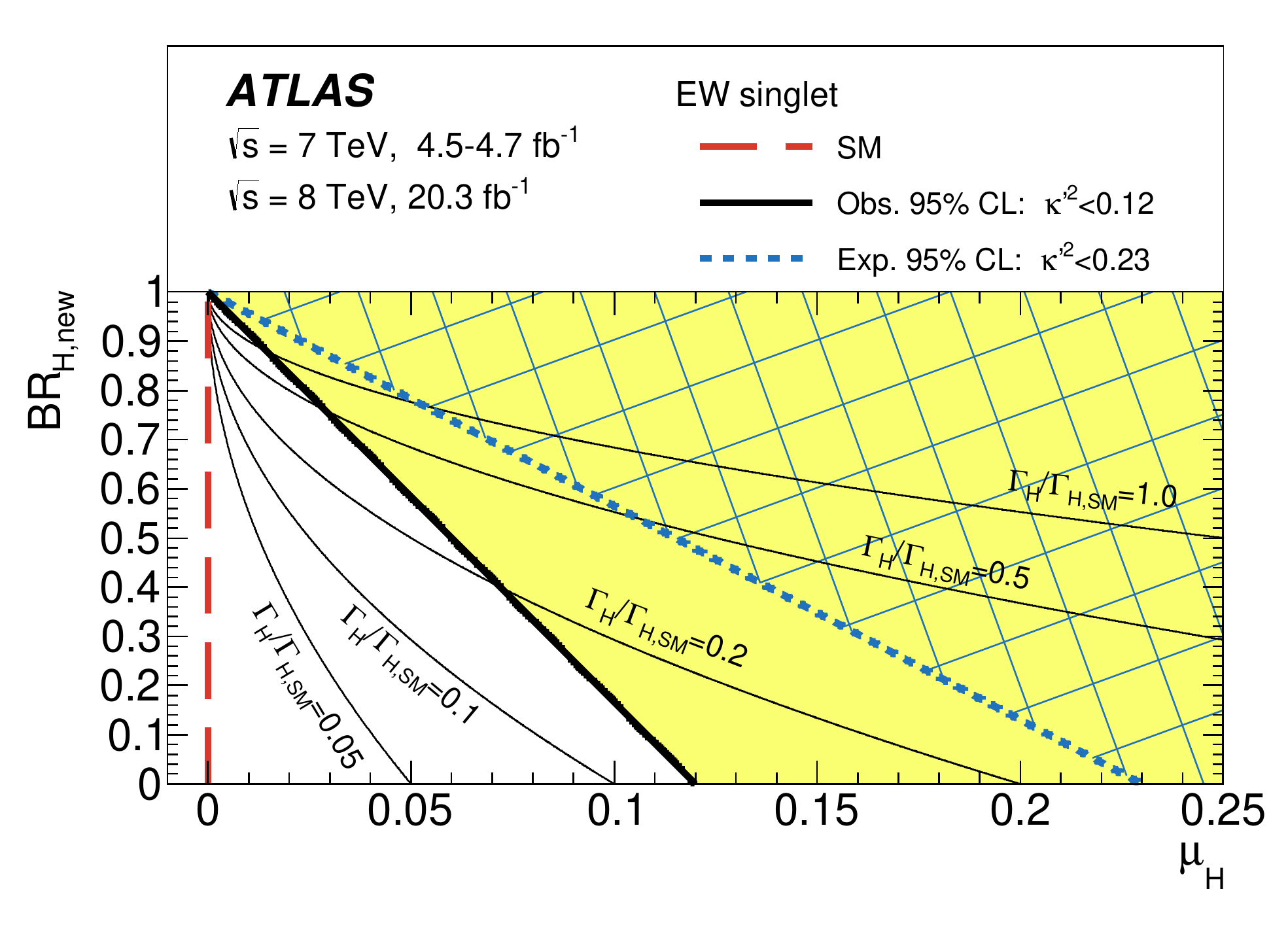}
\hspace*{0.05\textwidth}
\includegraphics[width=0.45\textwidth]{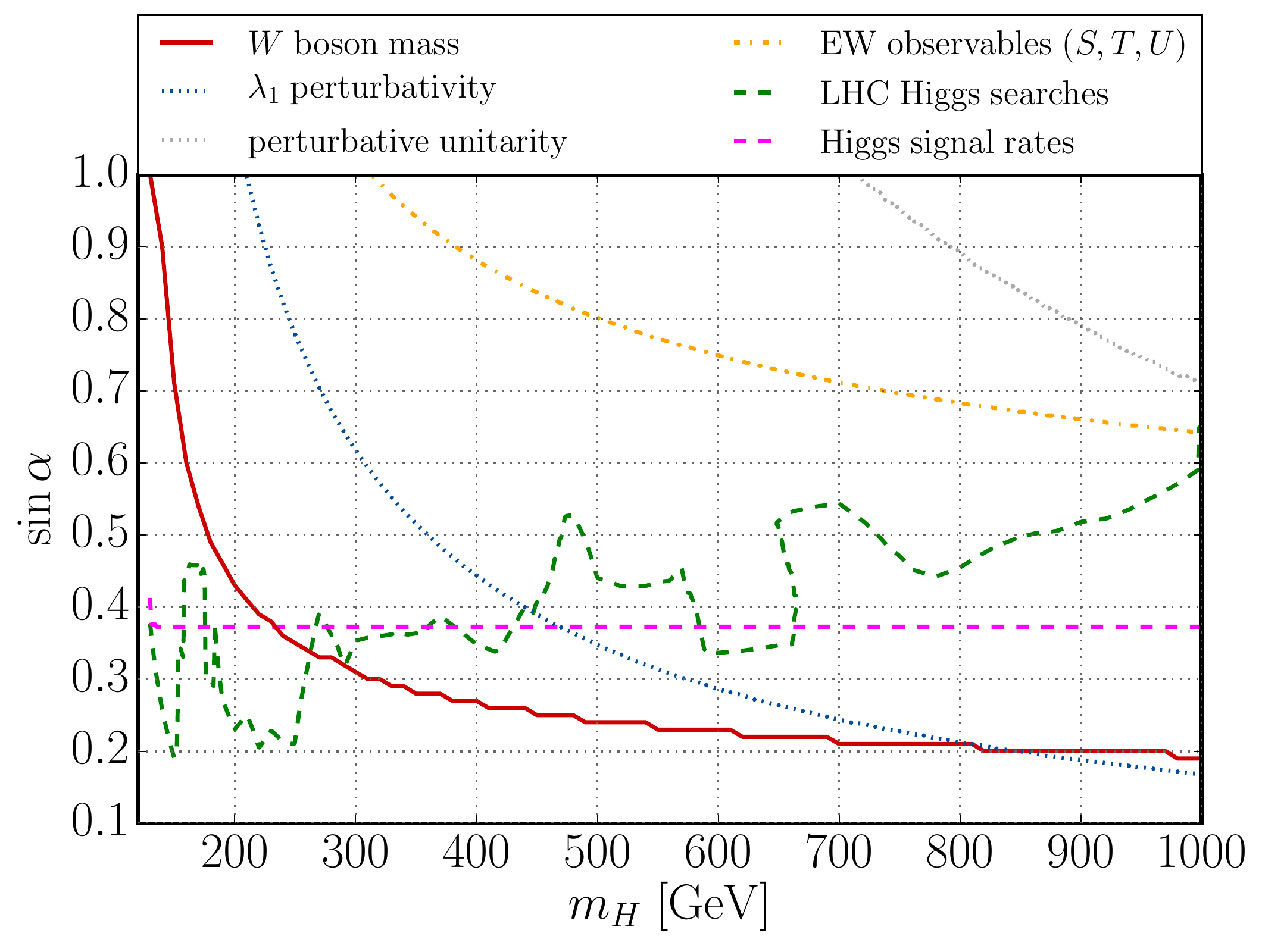}
\caption{Left: Limits on the mixing angle in the singlet model from
  coupling measurements from Ref.~\cite{Aad:2015pla}. Right:
  Comparison of limits on the mixing angle in the singlet model as a
  function of the heavier Higgs mass. Figures from
  Ref.~\cite{Ilnicka:2018def}.}
\label{fg:sing_coups}
\end{figure}

The latter point picks up on a common problem with all effective theory analyses related the
self-consistency of the approach. It always requires that all new
particles can be safely integrated out and do not contribute as
propagating degrees of freedom to a given observable. At the LHC this
can be a problem, first because of the large available partonic energy
for rare events with a parton momentum fraction $x \to 1$, and second
because of the very slow decoupling of $t$-channel processes. On the
phenomenological level, an easy and straightforward way to test the
validity of the effective theory is the appearance of mass peaks. In
other words, we can analyze enhanced tails for example of transverse
momentum distribution, but only to a point where we see a peak. This
requirement is hard to quantify. On the other hand, we can just
consider EFT analyses ways to systematically search for anomalies in
some phase space region of some processes, which we will then try to
explain in terms of a much wider range of models. This approach
largely gets rid of the consistence criterion and instead leaves us
with experimental results on the dimension-6 Lagrangian which some
might find hard to interpret in terms of actual models. Similarly,
theoretical uncertainties derived for example from the truncation of
the dimension-6 Lagrangian do not have to enter the experimental
analysis and can instead be considered as part of the interpretation
of the results in terms of more interesting, full models. This link
can be systematically explored in terms of a proper matching procedure
in the presence of electroweak symmetry
breaking~\cite{Freitas:2016iwx}.

\subsubsection{Ultraviolet completions}
\label{sec:exp_global_full}

For concrete models, correlations between different channels can be
exploited as functions of the new physics parameters to set tight
constraints on the presence of new states. The experiments typically
perform such analyses for well-motivated extensions of the Standard Model such as
two Higgs doublet models or supersymmetric scenarios.

The Higgs singlet model is severely restricted by limits on the Higgs
couplings to SM particles, by direct searches for the second Higgs
particle, $H^0$, and by precision electroweak measurements.  In this
section, we assume $M_{H}>M_{h}$.  If $M_{H}< 2 m_{h}$ (\ie 
it cannot decay to $2 h^0$), then the $h^0$ branching ratios are
identical to those of the Standard Model and the production rates are uniformly
suppressed by $c_\alpha^2$, leading to a limit from a global fit of $|
c_\alpha | < 0.94 $~\cite{Aad:2015pla}.  Once the decay
$H^0\rightarrow h^0 h^0$ is allowed, the total Higgs width and the
branching ratios can be significantly altered.  It is possible to get
branching ratios of ${\cal{O}}(30\%)$ for $H^0\rightarrow h^0
h^0$. The limits from ATLAS are shown in Fig.~\ref{fg:sing_coups}.
Direct limits on the mass of the heavier Higgs boson can be extracted
by re-purposing the SM heavy Higgs search limits and are shown in
Fig.~\ref{fg:sing_coups}.  The strongest limits for a heavy $H^0$ come
from the $W$ boson mass~\cite{Lopez-Val:2014jva}, while direct searches
coupling measurement limits are similar.

\begin{figure}[t]
\includegraphics[width=0.34\textwidth]{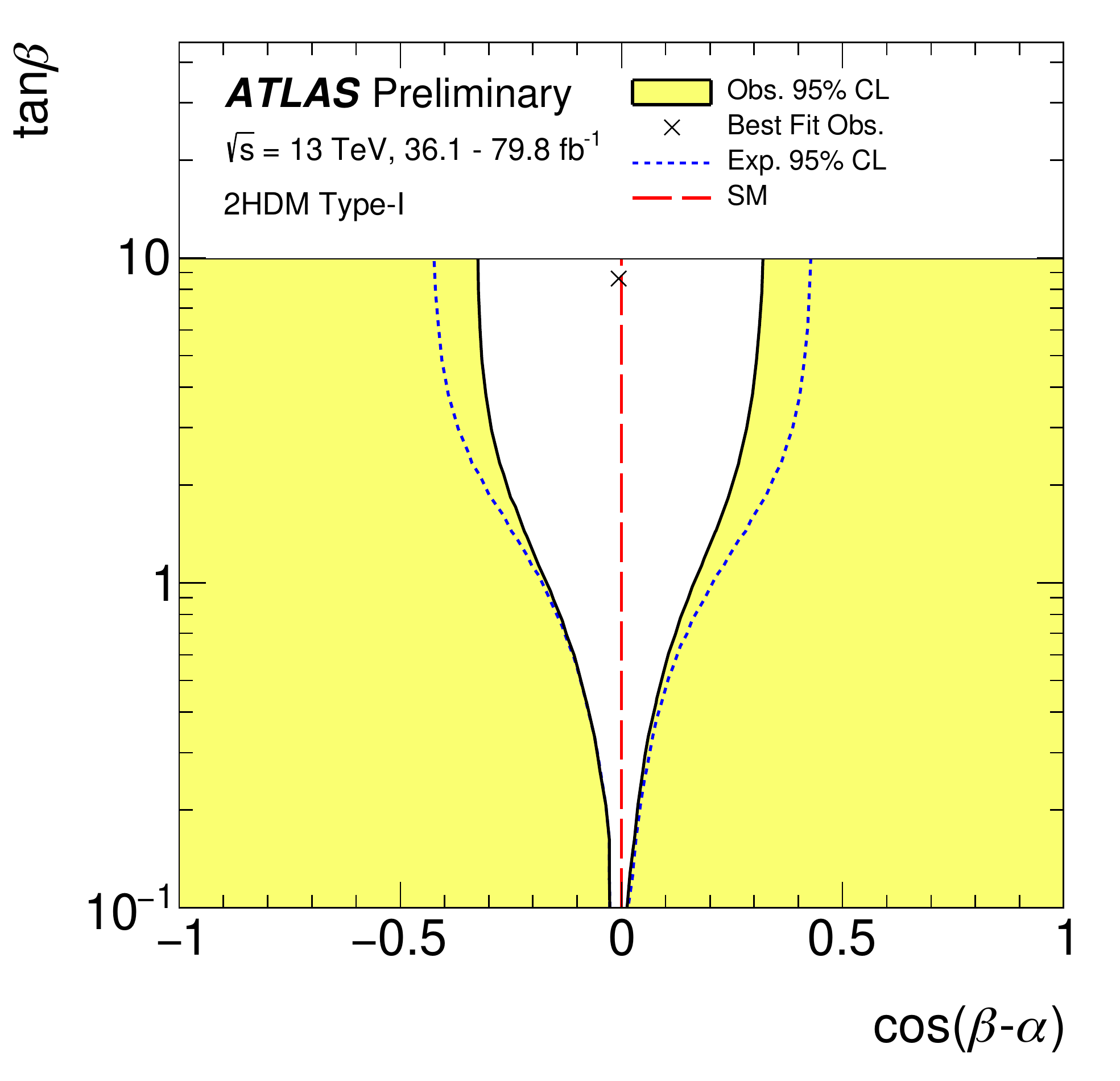}
\hspace*{0.1\textwidth}
\includegraphics[width=0.34\textwidth]{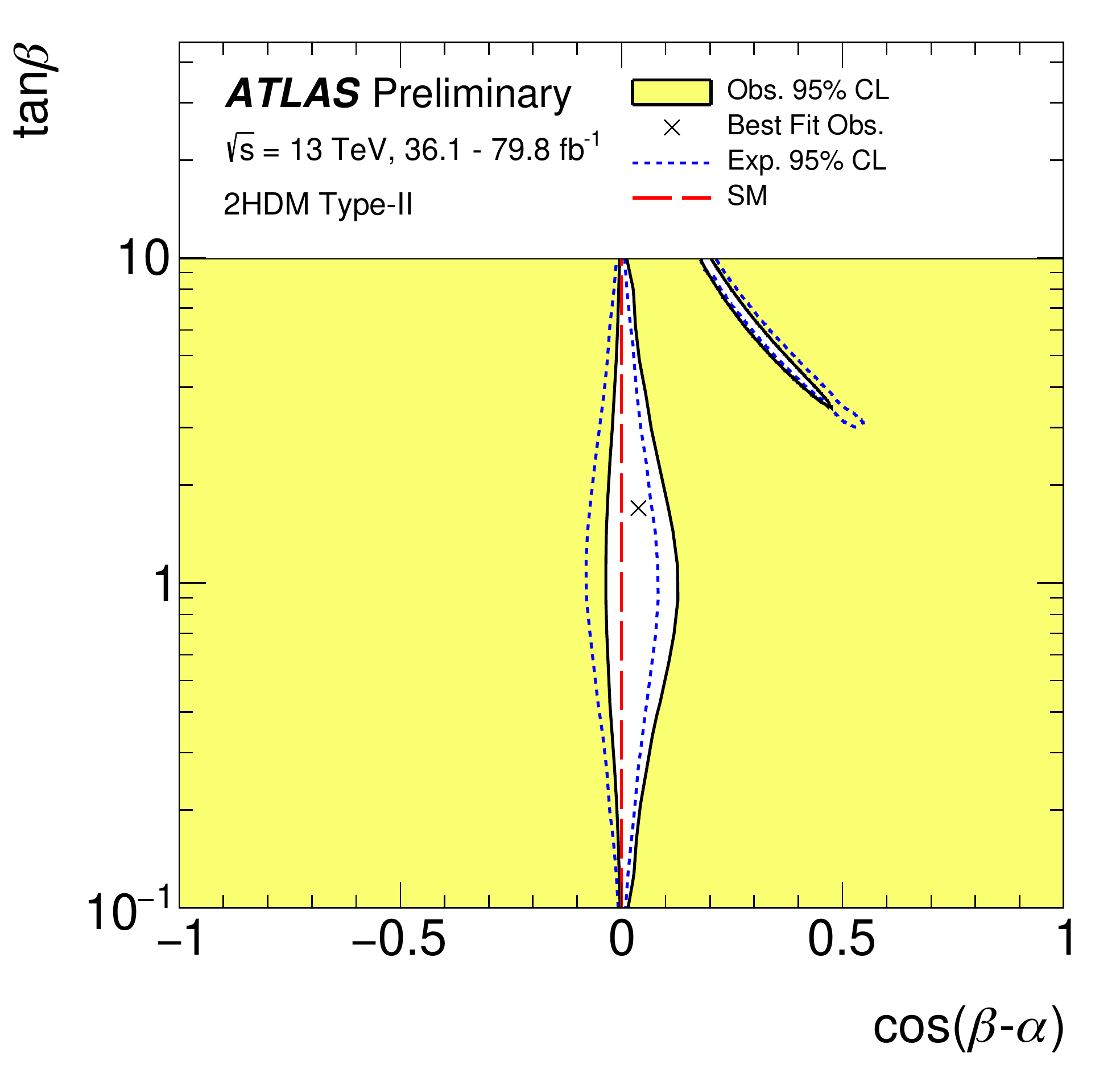}
\caption{Limits on the mixing angle in 2HDM models. The alignment
  limit is given by $\cos(\alpha-\beta)=0$. Taken from
  Ref.~\cite{ATLAS:2018doi}.}
\label{fg:2hdmcoups} 
\end{figure}

The 2HDM can be studied both by analyzing Higgs couplings and by
direct searches for the heavy $H^0,A^0,$ and $H^\pm$.  Coupling fits
from Run-2 in the Type-I and Type-II models are shown in
Fig.~\ref{fg:2hdmcoups} and demonstrate that the couplings are highly
restricted.  Type-I models require $\cos(\beta-\alpha)< 0.3$ from the
coupling fit, with even tighter bounds for $\tan\beta < 1$.  Since the
associated production of $H^0, A^0$ is suppressed by
$\sin(\alpha-\beta)$, these modes are much smaller in the 2HDM than
for a SM Higgs boson of the same mass.  Gluon fusion and $tt H^0$
production rates are $\sim \cot^2\beta$ (in the alignment limit), and
so are promising channels.  In Type-II 2HDMs, the limits from the
coupling fit is quite stringent and requires $\cos(\alpha-\beta)< 1.5$
and so the model must be quite close to the alignment limit.  The
associated productions modes in Type-II 2HDMs are highly suppressed
relative to the Standard Model, while the gluon fusion mode remains
large~\cite{Craig:2015jba}.  The heavier Higgs bosons can be searched
for directly, with the most promising channel being $gg\rightarrow
A^0\rightarrow Zh^0$.

\begin{figure}[!b]
\begin{centering}
\includegraphics[height=6.6cm]{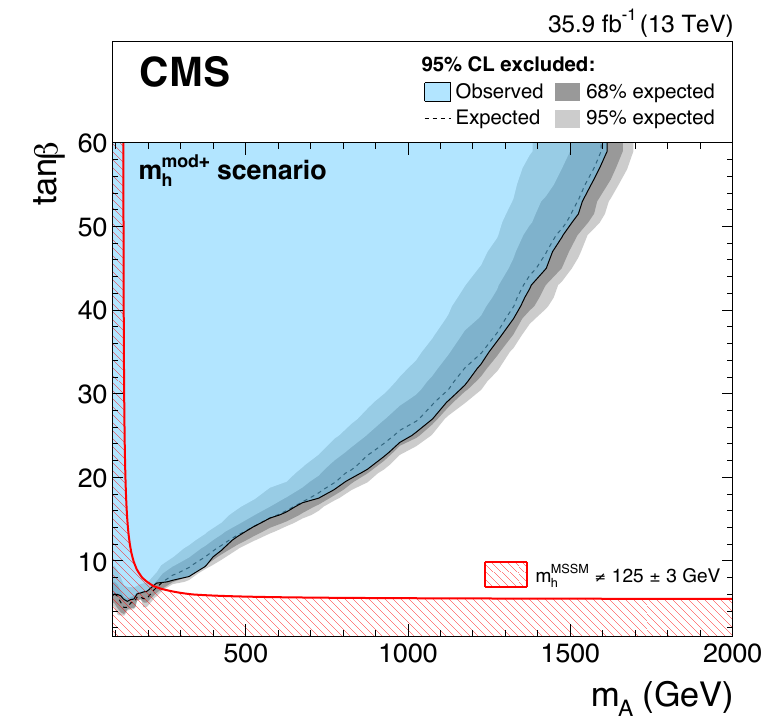}
\hfill
\includegraphics[height=6.6cm]{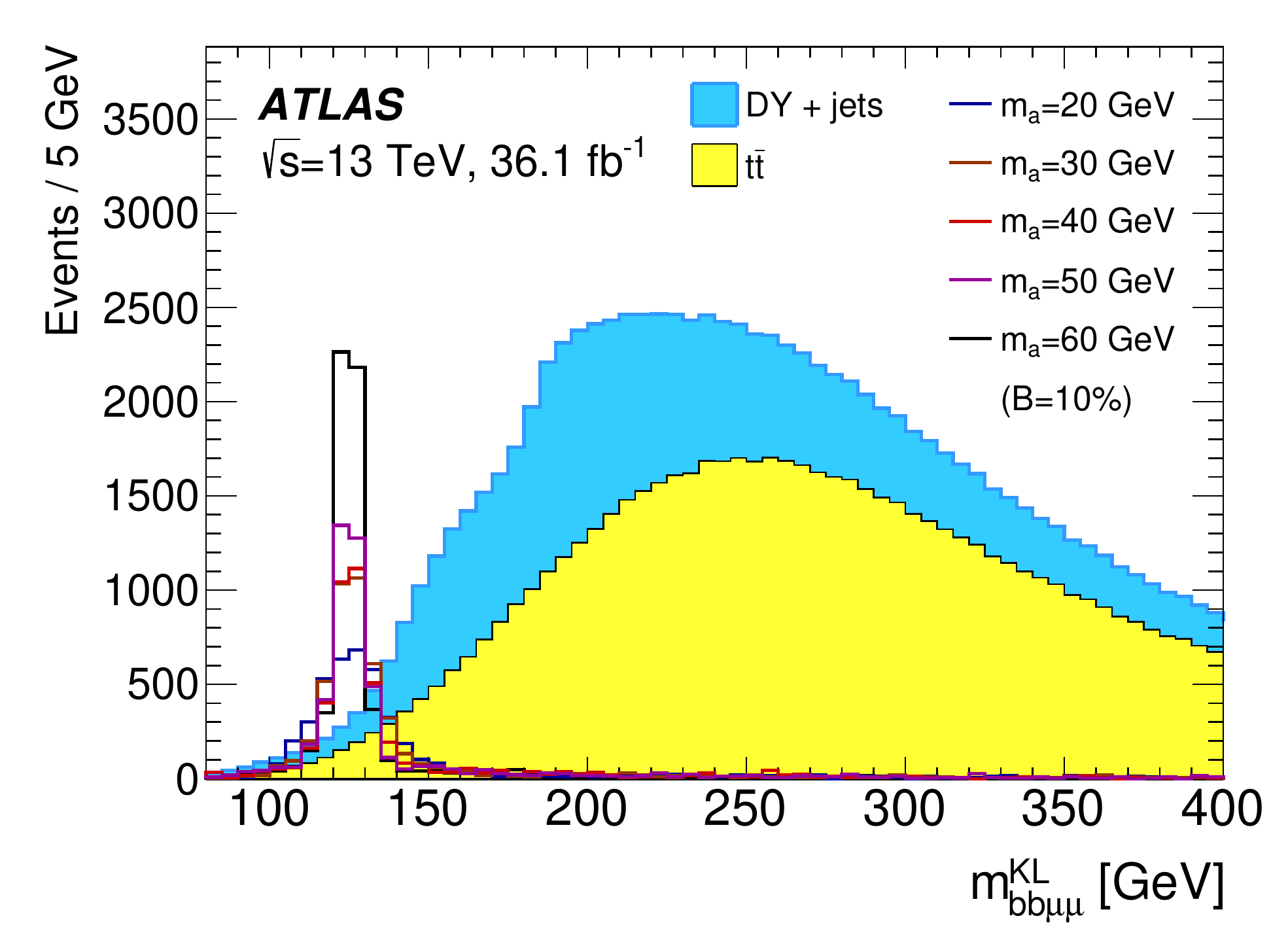}
\end{centering}
\caption{Left: MSSM search limits~\cite{Sirunyan:2018zut}. The
  $m_{h,\text{mod}}^+$ scenario is described in the
  text~\cite{Carena:2013ytb}.  Right: Invariant mass distribution of
  $b\bar b \mu^- \mu^+$ events after a kinematic likelihood (KL) has
  been carried. Also shown is the expected Higgs resonance in the
  presence of a decay $h^0\to a a$ for different pseudoscalar
  masses. Taken from Ref.~\cite{Aaboud:2018esj}.}
\label{fg:mssmgsearch} 
\end{figure}

Production rates for the MSSM Higgs bosons are often compared for
small and large $\tan\beta$ for the specific $m_{h,\text{mod}}^+$
benchmark point
\begin{align}
\begin{split}
m_{h,\text{mod}}^+~\text{scenario}:\quad& M_{\text{SUSY}}=1\tev\,,~\mu=M_2=200~\gev,~A_b=A_\tau=A_t,\\&m_{\tilde{g}}=1.5~\tev,~M_{\tilde{l}_3}=1~\tev,
X_t=1.5~M_{\text{SUSY}}, X_t=1.6M_{\text{SUSY}}
\end{split}
\end{align}
that is part of the more comprehensive benchmarking program carried
out after the Higgs boson discovery~\cite{Carena:2013ytb}. For small
$\tan\beta$, gluon fusion is always the dominant production channel
for $H^0$ and for $A^0$ with $M_A> 350~GeV$.  In the small $m_A$
region, the $bbA$ channel dominates. For large $\tan\beta$, the
$bbA^0$ and $bbH^0$ channels always dominate, due to the enhanced
couplings to $b$ quarks. At large $\tan\beta$, the most stringent
limits come from $A^0/H^0\rightarrow \tau\tau$ shown in
Fig.~\ref{fg:mssmgsearch}.

Another possibility that arises in supersymmetric scenarios is the
exotic decay of the Higgs boson into two pseudo-scalar bosons,
followed by subsequent decays into quarks, photons and leptons. ATLAS
and CMS have studied a plethora of channels at 13 TeV already (see
Fig.~\ref{fg:mssmgsearch} right for an example). Searches exist in the
$2b2\mu$~\cite{Aaboud:2018esj}, $4b$~\cite{Aaboud:2018iil}, $2\gamma
2j$~\cite{Aaboud:2018gmx}, $2\mu
2\tau$~\cite{Sirunyan:2018mbx,Aad:2015oqa}, and $2b
2\tau$~\cite{Sirunyan:2018pzn} final states. So far, no evidence for
such decays has been observed and branching fraction constraints in
percent range (assuming SM production of the Higgs boson) have been
obtained in the $\text{BR}(h^0\to A^0 A^0)\sim 1\%$ region depending
on the model-dependent decay phenomenology and mass of the
pseudoscalar $A^0$.

\clearpage
\section{Perspectives}

Future experiments to further Higgs physics beyond the capabilities of
the LHC divide broadly into two concepts. First, Lepton colliders
that provide a precision investigation of Higgs phenomenology at the
weak scale. The strategy that motivates such machines stand in the
long tradition of a precision spectroscopy that ensues after a
discovery, much like the LEP $Z$ boson precision program that
followed the Super Proton Synchrotron (SpS) which facilitated the
discovery of $W$ and $Z$ bosons. Crucial to the extraction of a
plethora of Higgs couplings is a range of accessible lepton collider
energies (see below).

However, the Standard Model being complete as a renormalizable 
quantum field theory after the Higgs
discovery, a precise understanding of the weak Higgs boson couplings
might not necessarily point in the direction of a more fundamental
theory beyond the weak scale. This situation is qualitatively
different to the early $Z$ boson precision days, when the Higgs boson
and the top quark were the missing pieces to obtain a theoretically
well-defined theory. In this sense, any precision results informed
Higgs physics directly. After the Higgs discovery, only tensions of
the Standard Model can be revealed, but no clear connection to UV physics can be
drawn unless new physics is directly discovered.

This insight motivates a second avenue of future colliders, namely
another (possibly ultimately final) push for a hadron-hadron collider
extrapolated from existing LHC technology. With center of mass
energies hypothesized to push up to 100 TeV, such a machine can not
only access Higgs couplings that are difficult to constrain at the LHC
or lepton colliders, but also opens up the a new range of energy to
observed particle thresholds and resonances, thereby breaking the
status quo of the Standard Model as a complete renormalizable theory.

\begin{figure}[b!]
\includegraphics[width=0.48\columnwidth]{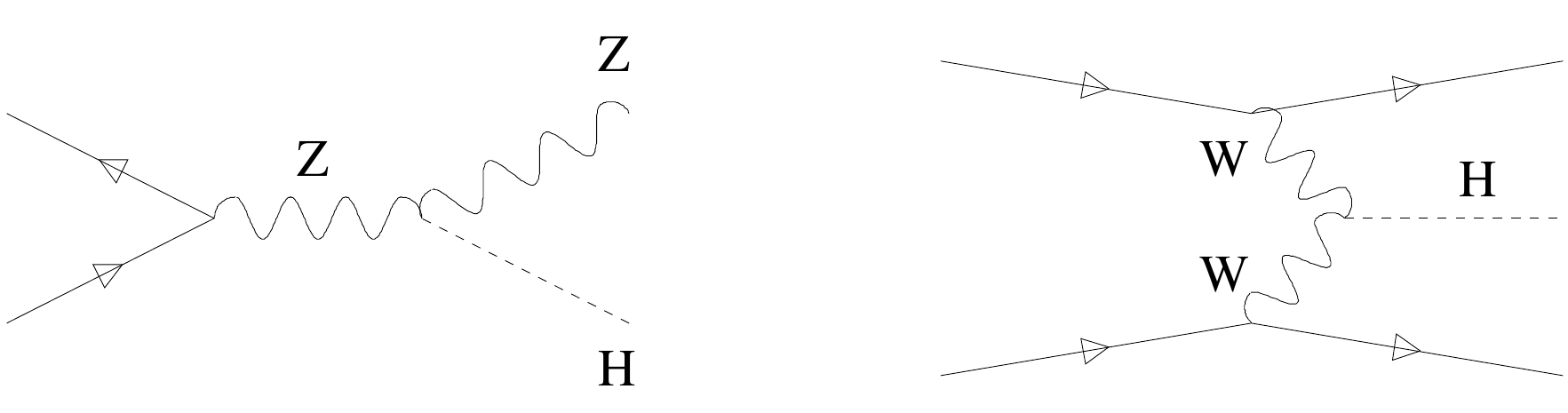}
\caption{Feynman diagrams for Higgs--strahlung and $W$ fusion Higgs production channels at an
  $e^+e^-$ Higgs factory.}
\label{fig:feyn}
\end{figure}

\subsection{Future lepton colliders}
\label{sec:exp_future}

A clear and technologically well-developed avenue for Higgs physics is
coupling measurements at proposed $e^+e^-$ colliders through the
abundant production of Higgs bosons in a clean environment.  There are
two classes of proposals for such Higgs factories: linear and circular
machines.  A well-studied linear collider concept is the International
Linear Collider (ILC) planned with a center-of-mass energy between
250~GeV and 500~GeV, in principle upgradable up to 1~TeV. The technical
design report has been published in
2013~\cite{Behnke:2013xla,Baer:2013cma,Adolphsen:2013jya,Adolphsen:2013kya,Behnke:2013lya},
and can therefore be considered as the most mature proposal of a
future $e^+e^-$ collider.

The Higgs program of such a
machine~\cite{Fujii:2017vwa,Fujii:2015jha,Moortgat-Picka:2015yla} is
based on precision determinations of Higgs couplings.  The level of
uncertainties at an $e^+e^-$ collider are are very different from the
LHC~\cite{Klute:2013cx,Fujii:2017vwa,Han:2012rb}: QCD is less of a nuisance than at hadron
colliders, and electroweak corrections are typically a factor 10 to
100 smaller than QCD corrections. Initial state QED radiation and
Beam-strahlung are well understood~\cite{Boogert:2002jr}, leading to
the possibility of very precise measurements as well as very precise
theoretical interpretations.  The ILC can run on the peak of the
Higgs--strahlung cross section as well as at larger center-of-mass
energy~\cite{Fujii:2017vwa}. The latter option is particularly useful
to extract precise information from $W$-fusion Higgs production as
well as to measure Higgs production associated with top-quark
pairs. The linear design is also used in the Compact Linear Collider
(CLIC) proposal; its conceptual design report was released in
2012~\cite{Aicheler:2012bya} and the Higgs physics program of such a
machine is discussed in Ref.~\cite{Abramowicz:2016zbo}. CLIC aims to
achieve collisions at up to 3~TeV, thus enabling measurements of
various Higgs processes at large momentum transfers which enables a
fine grained picture of Higgs interactions in particular when
considered in the framework of EFT
modifications~\cite{Ge:2016zro,Durieux:2017rsg,Ellis:2017kfi,Barklow:2017suo}.

\begin{figure}[t]
\includegraphics[width=0.48\textwidth]{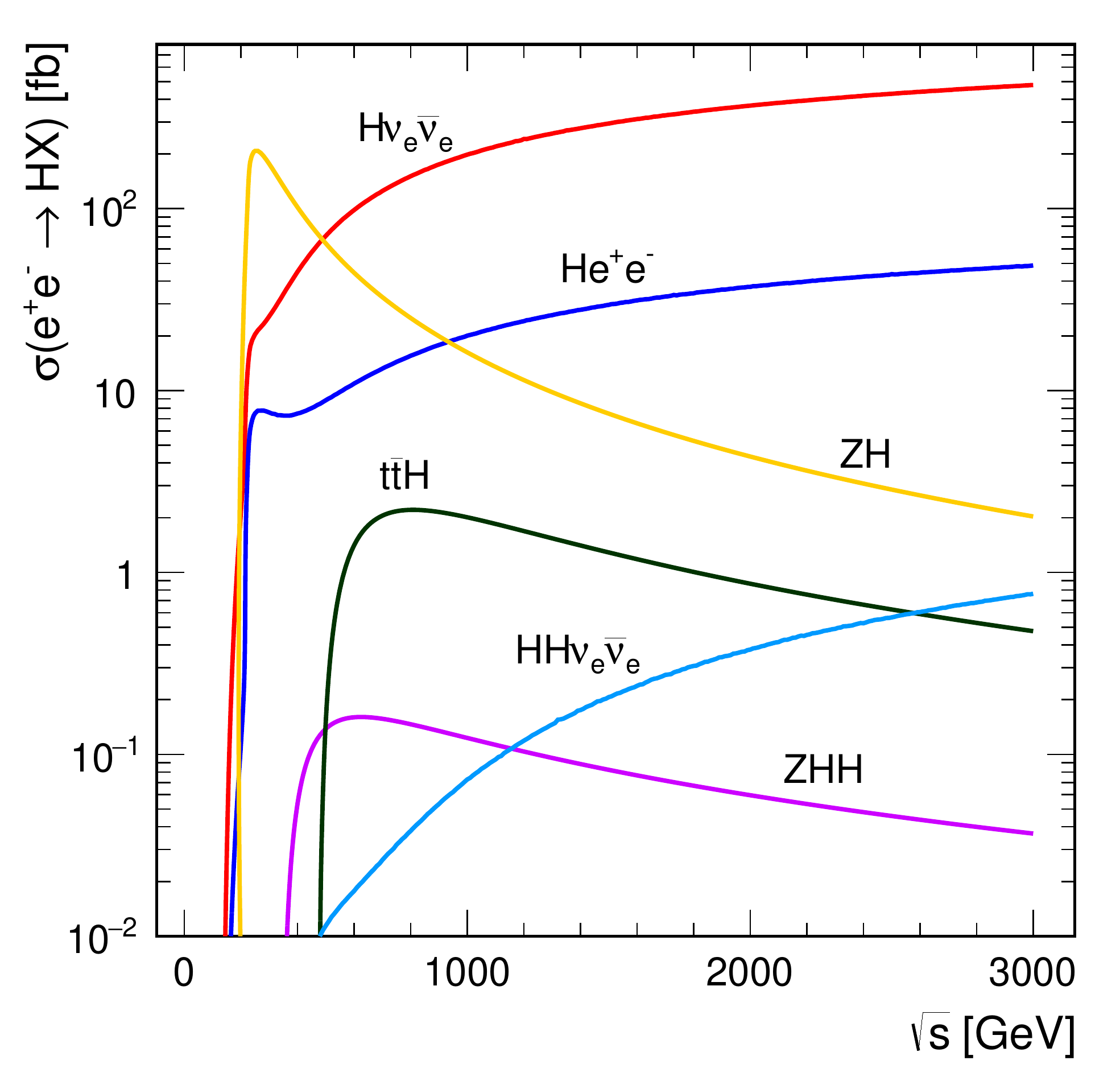}
\caption{Higgs production rates at $e^+e^-$ colliders, taken
  from Ref.~\cite{Abramowicz:2016zbo}.}
\label{fig:lcxsec}
\end{figure}

\begin{table}[b!]
\begin{tabular}{lrr}
\hline
collider         & $\sqrt{s}$ [GeV] & luminosity [$\iab$] \\\hline
HL-LHC           & 14000            & 3                   \\\hline
FCCee/CEPC-base  &   240            & 4                   \\
FCCee/CEPC-350   &   240/350        & 4/1                 \\\hline
ILC-stage        &   250            & 2                   \\
ILC-base         &   250/350/500    & 0.5/0.2/0.5         \\
ILC-lumi         &   250/350/500    & 2/0.2/4             \\\hline
\end{tabular}
\caption{Overview of luminosities of different representative future
  collider experiments. Based on Ref.~\cite{Dawson:2013bba}.}
\label{tab:designs}
\end{table}

The second $e^+e^-$ design possibility is a circular collider
(FCC-ee/TLEP~\cite{Gomez-Ceballos:2013zzn} or
CEPC~\cite{CEPC-SPPCStudyGroup:2015csa,CEPC-SPPCStudyGroup:2015esa})
that could produce very large numbers of $Zh$ events close to
threshold. The strength of circular machines compared to linear
baselines is the large luminosity that can be achieved at the price of
lost control over the beam conditions due to bremsstrahlung. In
particular, polarizations of the $e^+,e^-$ beams cannot be controlled,
which is one of the strengths of the linear collider setups to measure
complementary coupling aspects of Higgs physics~\cite{Fujii:2018mli}. For instance, while
in the Higgs--strahlung process left-chiral and right-chiral fermions
couple to the $Z$--boson with different coupling strengths, only
left-chiral fermions induce $W$-fusion Higgs production. The proposed
integrated luminosity and energies for future $e^+e^-$ colliders are
given in Tab.~\ref{tab:designs} (based on Ref.~\cite{Dawson:2013bba}),
but are therefore not directly comparable.

Both circular and linear $e^+e^-$ concepts feature runs at lower
energies to precisely determine the electroweak input parameters
before measurements at higher energies can be made with high
precision. This includes collecting data at the $Z$ and $W$ pair
thresholds similar to LEP before precision measurements at the $Zh$
and top thresholds are performed. While a $Z$ pair program could be
motivated from a $CP$-violation perspective through constraining
anomalous $ZZZ$ couplings, the expected small
effects~\cite{Corbett:2017ecn} mean that $ZZ$ pair production is
mainly used for data-driven $WW$ analyses.

At an $e^+e^-$ collider the two main Higgs production mechanisms are
Higgs--strahlung and $W$-fusion Higgs production~\cite{Jones:1979bq} as shown in
Fig.~\ref{fig:feyn}. The energy dependence of these processes is given
in Fig.~\ref{fig:lcxsec}. The $t$-channel $W$ fusion process grows
with energy and dominates for energies above $\sim 800~\GeV$,
Higgs-strahlung and $W$ fusion can be complemented by the associated
production of top quark pairs and a Higgs boson and by Higgs pair
production in association with a $Z$-boson. Representative (leading
order, unpolarized) cross sections at 350~(3000)~GeV
are~\cite{Abramowicz:2016zbo}
\begin{align}
\sigma(e^+e^- \to Zh) &=133~(2)~\fb\,,\nonumber\\
\sigma(e^+e^- \to h\nu_e \bar \nu_e) &= 34~(477)~\fb\,,\\
\sigma(e^+e^- \to h e^+ e^-) &= 7~(48)~\fb\nonumber\,.
\end{align}
In the process $e^+e^-\rightarrow Zh$, the observation of the $Z$
boson can be used to tag the process (up to corrections from
radiation) independent of the Higgs boson decay channel, and the
branching ratios of the Higgs boson are observed directly.  The Higgs
mass can be measured precisely with an accuracy of
$14~\text{MeV}$~\cite{Yan:2016xyx} through the recoil of $Z\to
\mu^+\mu^-$ events, Fig.~\ref{fig:epemhmass}. Similarly, the ILC can
constrain an invisible Higgs branching ration as low as
0.4\%~\cite{Asner:2013psa} (such constraints, however, are
model-dependent, as already mentioned).

\begin{figure}[t]
\includegraphics[width=0.4\textwidth]{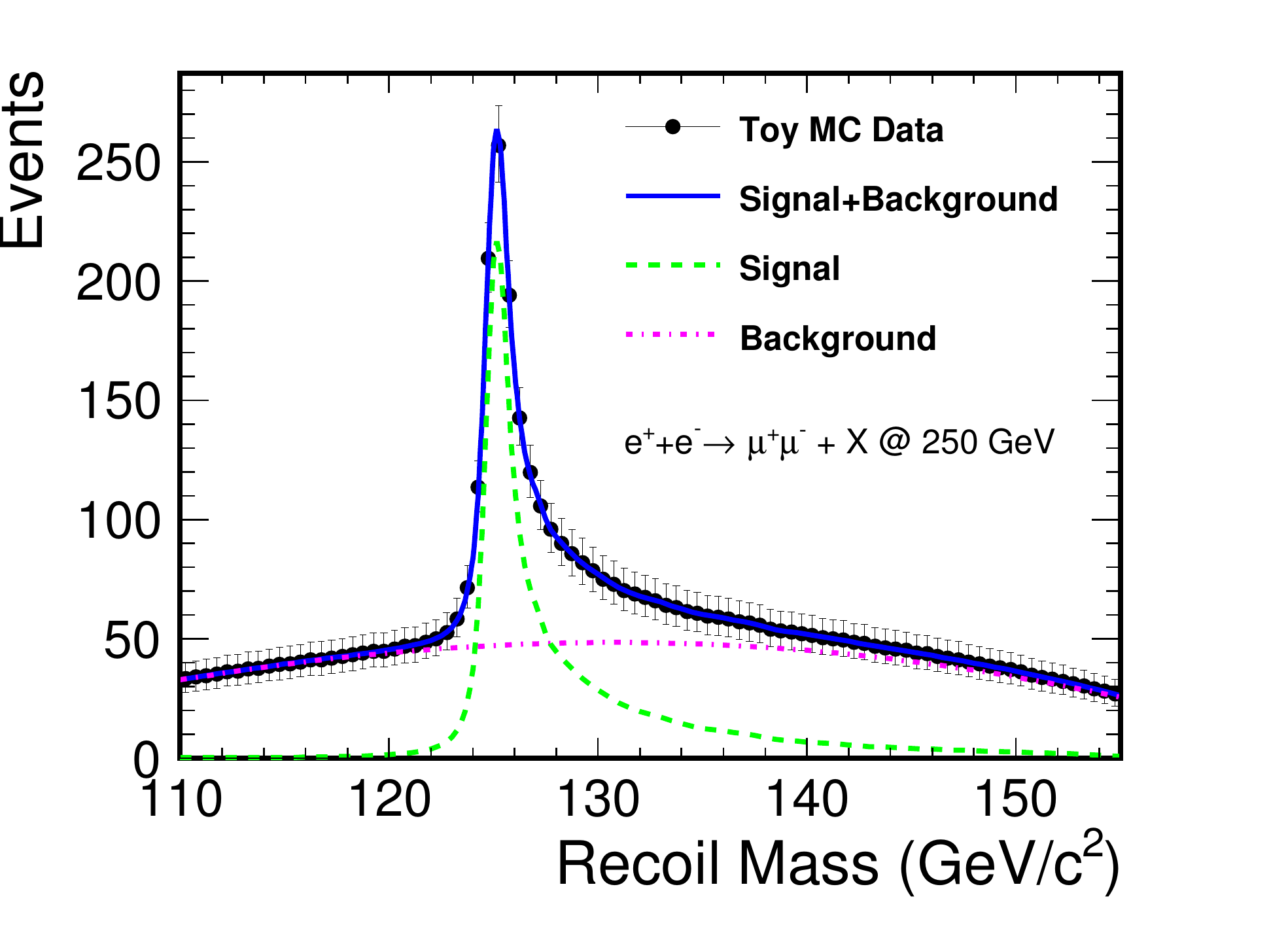}
\caption{Recoil spectrum of $e^+e^-\to hZ$, with $Z\to
  \mu^+\mu^-$. Figure taken from Ref.~\cite{Yan:2016xyx}.}
\label{fig:epemhmass}
\end{figure}

The Higgs coupling strengths are obtained by noting that
\begin{align}
\text{BR}(h\rightarrow XX)=\Gamma(h\rightarrow XX) /\Gamma_h\, .
\end{align} 
Assuming that the Higgs couplings are related to the SM couplings by
the factors $\kappa_X=1+\Delta_X$ described previously, then
\begin{align}
{\sigma(e^+e^-\rightarrow Zh)\over BR(h\rightarrow ZZ)}
={\sigma(e^+e^-\rightarrow Zh)_\text{SM}\over \Gamma(h\rightarrow ZZ)_\text{SM}}\Gamma_h\, .
\end{align}
This yields a measurement of the total Higgs width with the same
assumptions as for the LHC: that is, that there are no non-SM tensor
structures for the interactions of the Higgs boson with the SM
fermions.  As the Higgs branching ratio into $ZZ$ is small, this
requires a significant amount of luminosity.  Another way to determine
the Higgs width is based on four
measurements~\cite{AguilarSaavedra:2001rg,Baer:2013cma} of both
production processes shown in Fig.~\ref{fig:feyn},
\begin{enumerate}
\item Higgs--strahlung inclusive: $\sigma_{Zh}$\,,
\item Higgs--strahlung with a decay to $b\bar{b}$: $\sigma_{Zbb}$\,,
\item Higgs--strahlung with a decay to $WW$: $\sigma_{ZWW}$\,,
\item $W$-fusion with a decay to $b\bar{b}$: $\sigma_{\nu\nu bb}$\,,
\end{enumerate}
involving the four unknown parameters $\kappa_W$, $\kappa_Z$,
$\kappa_b$, and $\Gamma_h$.  Schematically, the total width
can be extracted as
\begin{align}
\Gamma_h\sim
\dfrac{\sigma_{\nu\nu bb}/\sigma_{Zbb}}
      {\sigma_{ZWW}/\sigma_{Zh}}
\times \sigma_{Zh} \; .
\end{align}
This is demonstrated in in Fig.~\ref{fig:base}, where we show the
projected coupling measurements from Ref.~\cite{Lafaye:2017kgf},
assuming that the total width can be computed as the sum of all
observed partial widths plus contributions from second-generation
fermions.  The comparison of FCCee and ILC projections assumes a
combination with the expected HL-LHC results and it is apparent that
the linear and circular designs will both be limited by the
theoretical uncertainty in extracting Higgs couplings from observed
event rates.  The general pattern we see in Fig.~\ref{fig:base} is
that for all couplings with the exception of $\kappa_W$ the FCCee base
design, the ILC base design, and the ILC staging design give
comparable results. As a representative value to be compared with
sensitivities of potential future hadron machines, the fit
of~\cite{Lafaye:2017kgf} gives a best Yukawa coupling precision of
\begin{align}
\kappa_t = 1\pm 1.8\%\,.
\end{align}
The $500$~GeV ILC will probe the $W$-fusion process combined with all
main decay modes, allowing for a higher overall precision.

Improving on studies based on Higgs coupling strength modifiers, the
most generic Higgs coupling modifications from integrated-out UV
states at both the LHC and $e^+e^-$ colliders should be analyzed in
terms of an effective field theory approach. The corresponding
frameworks are discussed in Sec.~\ref{sec:basis_eft}. This introduces
new interactions and thereby breaks the naive correlations between
different production modes at different energies expressed in the
kappa framework. This has profound
consequences~\cite{Ellis:2017kfi,Ellis:2015sca,DiVita:2017vrr,Barklow:2017awn,Durieux:2017rsg,Barklow:2017suo}
for the Higgs coupling extractions as can be seen from
Fig.~\ref{fig:eftilc}.

\begin{figure}[t]
\includegraphics[width=0.8\columnwidth]{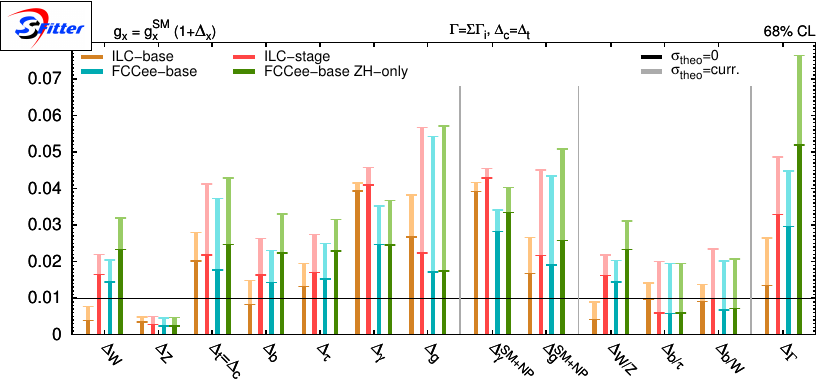}
\caption{Precision of the Higgs couplings extracted in the linear and
  circular baseline scenarios using the current theoretical errors and
  assuming negligible theory errors. We also show results assuming a
  staged low-energy operation of the ILC and the impact of the
  $W$-fusion process by restricting the FCCee measurements to $Zh$
  production. We assume that the total Higgs width is constructed from
  all observed partial widths. Figure taken from
  Ref.~\cite{Lafaye:2017kgf}.}
\label{fig:base}
\end{figure}

\begin{figure}[!b]
\includegraphics[width=15cm]{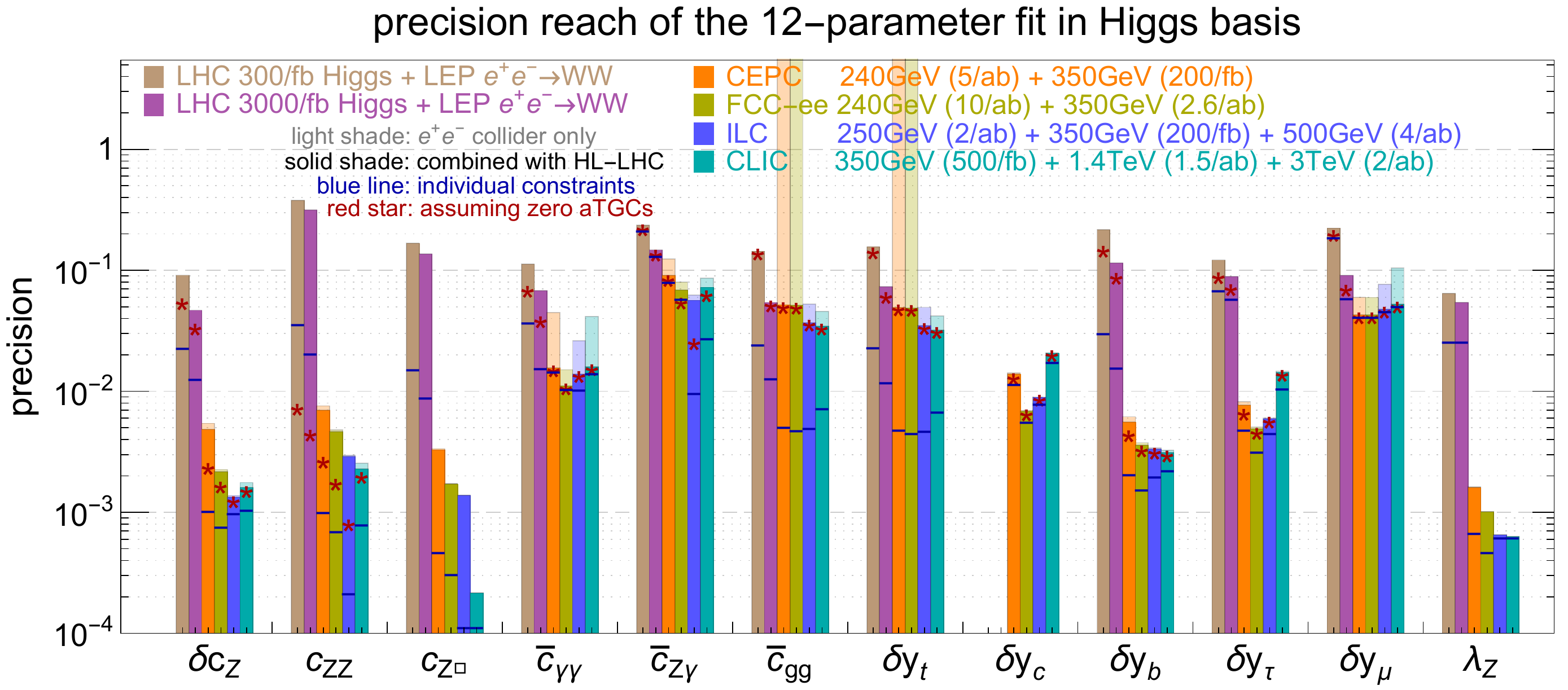}
\caption{Constraints on EFT operators in the Higgs
  basis~\cite{deFlorian:2016spz} at different lepton
  colliders. Taken from Ref.~\cite{Durieux:2017rsg}.}
\label{fig:eftilc} 
\end{figure}

The presence of different Lorentz structures in the EFT framework induces  momentum
dependencies in the Higgs interactions~\cite{Hagiwara:1993qt,GonzalezGarcia:1999fq}. 
This means
that there is a  gain in information when pushing lepton colliders to
larger energy due to  the 
energy-dependent cross section enhancements. This is clearly shown for
the $c_{Z\Box}$ direction $\sim Z_\mu \partial_\nu Z^{\mu\nu}$, which
is not directly related to Higgs physics but demonstrates clearly the
impact of energy coverage.  In addition, the polarization at the ILC can be used
to obtain an increased number of observables (relative to an unpolarized $e^+e^-$
collider), potentially yielding precisions on EFT couplings at the sub-percent level~\cite{Barklow:2017awn}.

One particular coupling that is expected to be only poorly
constrained at the LHC even when considering large luminosities is
the Higgs self coupling.  Lepton colliders close this gap at least
partially.  Direct sensitivity to the Higgs self-interactions requires
the production of (at least) a Higgs pair  through weak boson
fusion and  and/or $Zhh$ production, see Fig.~\ref{fig:lcxsec}. As can be seen
there, energies well above the 240~GeV $hZ$ threshold are
required. CLIC with its target energy of 3 TeV is particularly suited
to achieve such a measurement via weak boson fusion with large
statistics and projections predict a measurement of the trilinear
Higgs coupling at the
\begin{align}
\kappa_\lambda= 1 \pm 40\%~(22\%) 
\end{align}
at 1.4 (3) TeV collisions~\cite{Abramowicz:2016zbo}. Exploiting the
$Zhh$ threshold, similar analyses can be performed in this channel at
the high energy ILC option, and constraints on the Higgs self
interactions can be obtained in a global fit~\cite{DiVita:2017vrr,Barklow:2017awn}.

\subsection{Future hadron colliders}
\label{sec:fcc}

The precision spectroscopy of the Higgs sector that is possible at
lepton colliders is to be compared with that obtained by pushing the LHC's energy frontier
further using proton-proton collisions at up to
100~TeV~\cite{Contino:2016spe,Golling:2016gvc,Mangano:2016jyj}. While
such a machine offers the opportunity of directly observing a more
fundamental scale of physics, a precision Higgs program remains at the
core of these proposals. This is linked to developments at the LHC,
where we have learned how to control experimental systematics to the
level of a few percent, how to provide theory predictions for hard
processes to a similar level, how to precisely simulate events from
first-principles QCD in a multi-jet environment, and how to use jets
in many precision analyses. In the following we focus on a selected 
range of Higgs-relevant processes that demonstrate the improvement 
that can be achieved at such machines compared to the LHC, and how measurements
in these channels will compare against lepton collider concepts.

\subsubsection{Higgs pair production}
\label{sec:fcc_hh}

\begin{figure}[!b]
\includegraphics[width=0.4\textwidth]{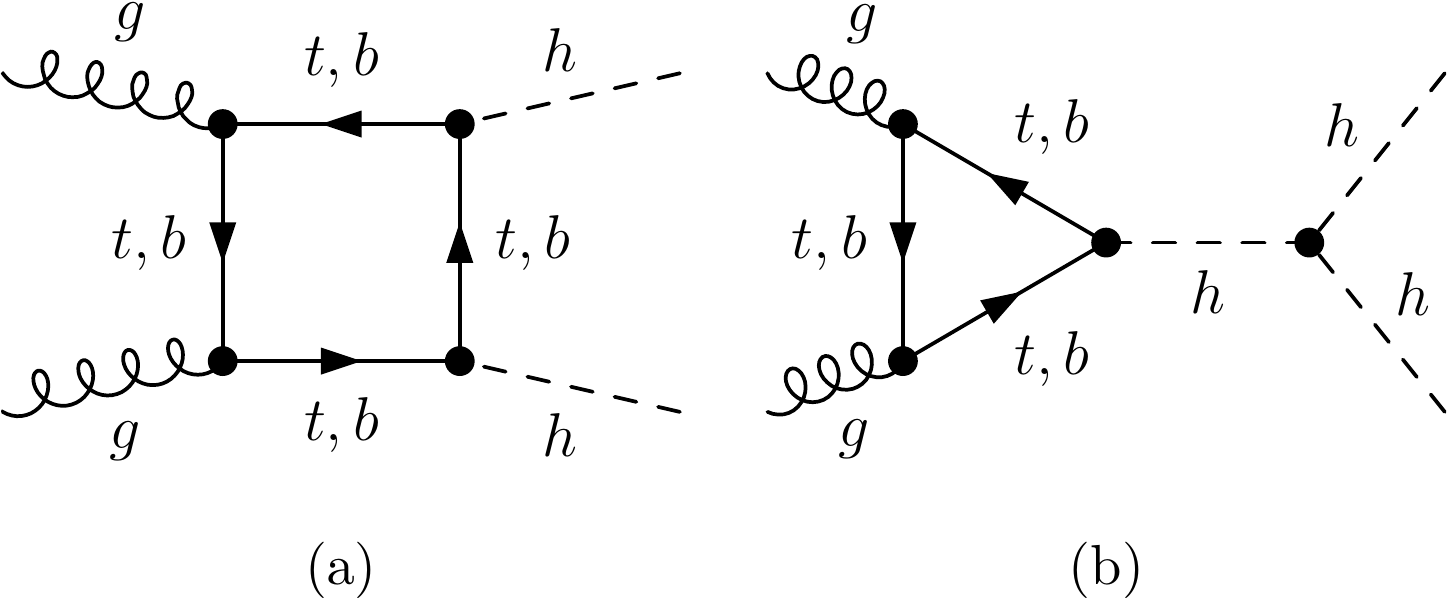}
\caption{Feynman diagrams contributing to $gg\to hh$ production.}
\label{fig:hhfey}
\end{figure}

The last parameter in the Higgs potential of the Standard Model, that
is likely to be left only poorly constrained even after the HL-LHC, is
the Higgs self-coupling. The direct measurement is based on gluon
fusion-induced Higgs pair
production~\cite{Eboli:1987dy,Dicus:1987ic,Glover:1987nx}, which
receives contributions from a triangle loop diagram combined with the
Higgs self-coupling and from a box diagram, as shown in
Fig.~\ref{fig:hhfey}.  Over most of phase space, the box contribution
completely dominates the total rate.  However, there exist three phase
space regions which provide information on the Higgs
self-coupling~\cite{Goncalves:2018qas,Buchalla:2018yce}, which
corresponds to a $(\phi^\dagger \phi)^3$ modification in the effective
field theory language (for enhanced derivative interactions see e.g.~\cite{Goertz:2014qta,Azatov:2015oxa,Gillioz:2012se,Dawson:2012mk,He:2015spf}).  First, there is the threshold
region~\cite{Plehn:1996wb,Djouadi:1999rca,Baur:2002rb,Li:2013rra}
\begin{align}
m_{hh} \approx 2 M_h \; .
\end{align}
The effective Higgs--gluon Lagrangian illustrates the  threshold cancellation between the diagrams at
the amplitude level,
\begin{align}
\frac{\alpha_s}{12 \pi v}
\left( \frac{\kappa_\lambda \lambda_\text{SM}}{s-M_h^2} - \frac{1}{v} \right) 
\to
\frac{\alpha_s}{12 \pi v^2}
\left( \kappa_\lambda -1 \right) \stackrel{\text{SM}}{=} 0 \; ,
\label{eq:higgs_pair}
\end{align}
where $\lambda_\text{SM}={M_h^2\over 2 v^2}$ and $\kappa_\lambda=1$ in
the Standard Model.  While the heavy-top approximation is known to
give a poor description of the signal kinematics as a whole, this
cancellation at threshold remains as an approximate
feature~\cite{Borowka:2016ypz}. Any significant deviation from the SM
value of the self-coupling will spoil this cancellation and lead to an
increased rate at threshold, no matter if the self-coupling is smaller
or larger than in the Standard Model. Second, an enhanced sensitivity
to the self-coupling appears through the top
threshold~\cite{Glover:1987nx}. The triangle contribution shows a
distinctive kink at~\cite{Georgi:1977gs}
\begin{align}
m_{hh} \approx 2 m_t \; .
\end{align}
This is similar to the decay $h \to gg$, as shown in
Eq.\eqref{eq:decgg} with $M_h$ replaced by $m_{hh}$. This is the
leading effect when we search for an anomalous Higgs self-coupling.
The same absorptive imaginary part also leads to a significant dip
around $p_{T,h} \approx 100$~GeV~\cite{Dolan:2012rv,Barr:2014sga} if
we parameterize the $(2 \to 2)$ signal phase space this way.  The
destructive interference effect is shown in the left panel of
Fig.~\ref{fig:hh}, indicating that an increase in the self-coupling
will be clearly visible as a dip relative to the SM distribution.
Finally, the triangular and box amplitudes have a generally different
scaling in the high-energy limit\cite{Plehn:1996wb,Chen:2014xra,Azatov:2015oxa}
\begin{align}
m_{hh} \gg M_h, m_t \; .
\end{align}
Here, the triangle with the intermediate Higgs propagator features an
explicit suppression of either $m_h^2/m_{hh}^2$ or $m_t^2/m_{hh}^2$,
whereas the box diagrams drops more slowly. Whenever we can
reconstruct the di-Higgs final state, an obvious way of extracting the
Higgs self-coupling is through an $m_{hh}$ shape
analysis~\cite{Bauer:2017cov}. We can use a maximum likelihood
analysis~\cite{Cranmer:2006zs} to show that most of the significance
for an anomalous self-coupling comes from around $m_{hh} =
350~...~450$~GeV~\cite{Kling:2016lay,Goncalves:2018qas}.

In Fig.~\ref{fig:pphh} we show the cross sections for different
Higgs pair production channels as a function of the collider
energy. As expected, gluon fusion is the dominant process. The rates
for alternative production mechanisms are at least one order of
magnitude smaller, with no obvious advantages of those channels at the
analysis level. At the 13~TeV LHC the cross section is
\begin{align}
\sigma(gg\to hh, 13~\text{TeV}) = 39.56~\text{fb}\; ,
\end{align}
computed at
NNLO+NNLL~\cite{deFlorian:2013jea,deFlorian:2015moa,deFlorian:2016uhr}
including NLO top-mass
effects~\cite{Borowka:2016ypz,Heinrich:2017kxx,Borowka:2016ehy,Grazzini:2018bsd}. LHC
projections by CMS put the sensitivity in the range of $\lesssim
0.5~\sigma$ at $3~\iab$~\cite{CMS:2017cwx} at the LHC. The rare but
clean $hh\to \gamma\gamma b\bar b$ channel~\cite{Baur:2003gp} will
likely play the dominant role, although $h\to
\tau\tau$~\cite{Baur:2003gpa,Dolan:2012rv} can contribute
significantly given recent level~1 trigger upgrades that have allowed
CMS to reach hadronic di-$\tau$ tagging with a $\sim 70\%$ efficiency
at an acceptable fake rate.

\begin{figure}[t]
\includegraphics[width=0.6\textwidth]{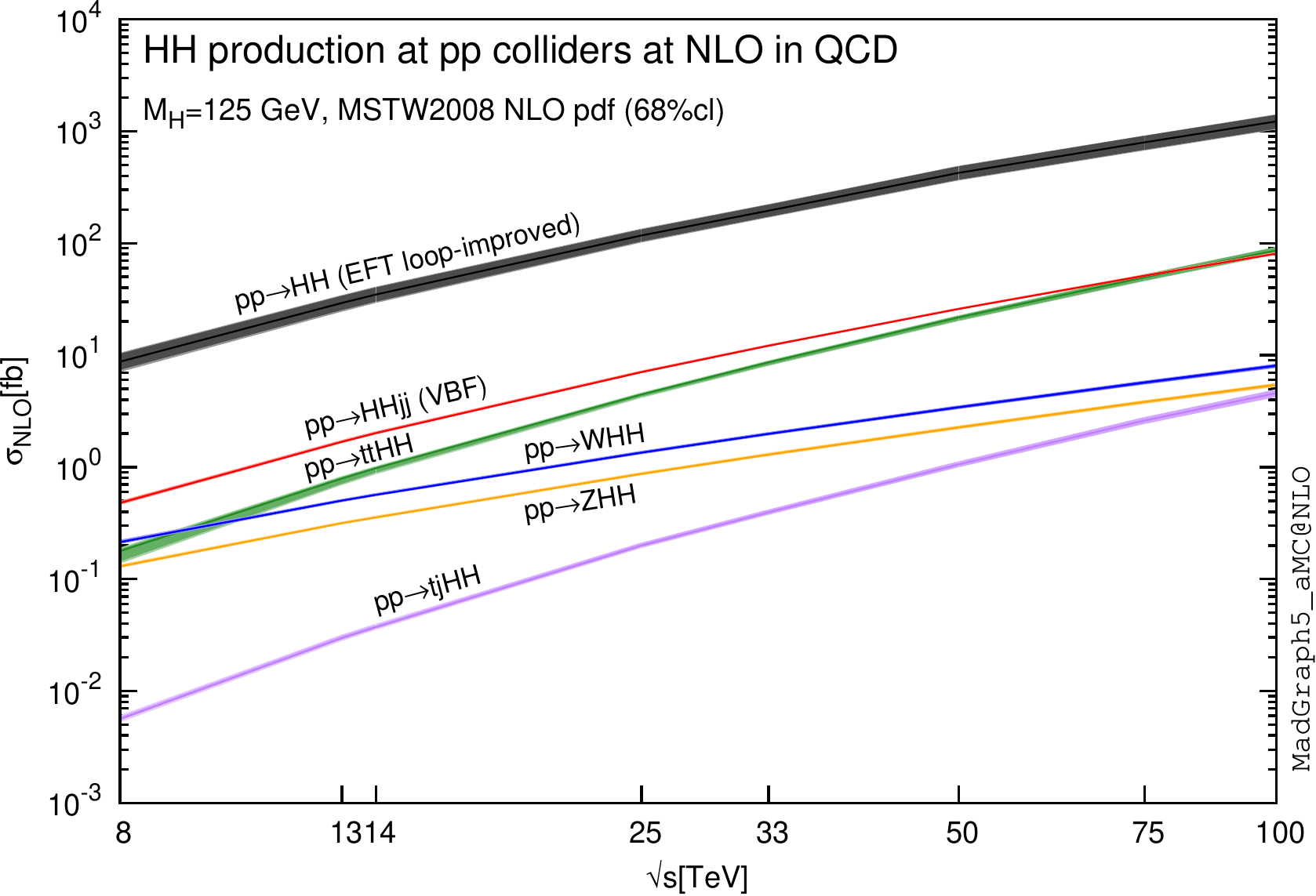}
\caption{Double Higgs production cross sections as a function of the
  center-of-mass energy, taken from Ref.~\cite{Frederix:2014hta}.}
\label{fig:pphh}
\end{figure}

\begin{figure}[t]
\includegraphics[width=0.47\textwidth]{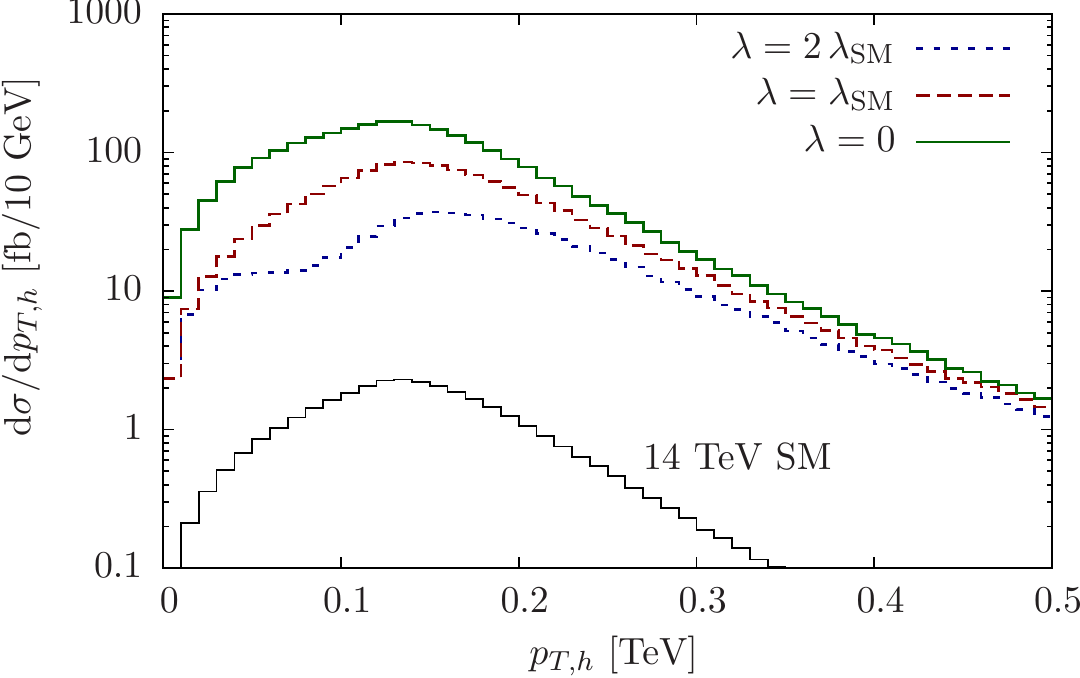}
\hspace{0.1\textwidth}
\includegraphics[width=0.36\textwidth]{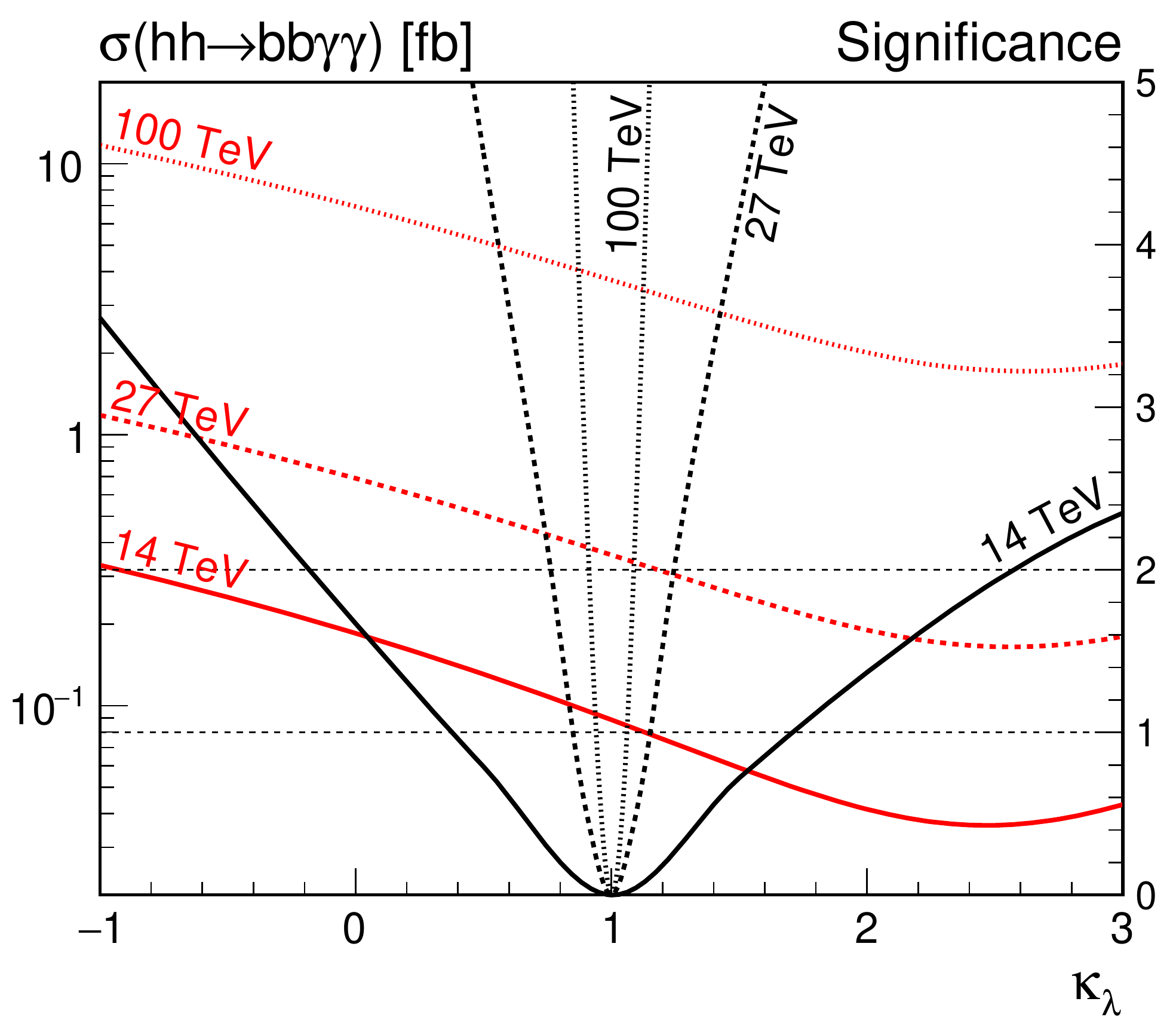}
\caption{Left: Leading order differential di-Higgs production cross
  sections as a function of $p_{T,h}$. Figure from
  Ref.~\cite{Barr:2014sga}. Right: Higgs pair production cross section
  (red lines with left vertical axis) and maximum significance (black
  lines with right vertical axis) for extracting an anomalous Higgs
  self-coupling, as a function of the modified self-coupling. Figure
  from Ref.~\cite{Goncalves:2018qas}.}
\label{fig:hh}
\end{figure}

Going from 13~TeV proton-proton collisions to 27~TeV (HE-LHC) and on
to 100~TeV (FCChh) increases the Higgs pair production rate
significantly, as seen in Fig.~\ref{fig:pphh}. If we combine this
with an expected luminosity of up to $15~\iab$ (HE-LHC) or $30~\iab$
(FCChh), the measurement of the Higgs self-coupling becomes a promising
motivation and benchmark for further pushing the high energy
frontier. In particular, the statistics-limited process $hh\to b\bar b
\gamma \gamma$ at the LHC predicts 5k events already at the HE-LHC.
Using a binned likelihood analysis for $m_{hh}$ including detector
effects and realistic efficiencies, we expect a
reach~\cite{Goncalves:2018qas}
\begin{align}
\kappa_{\lambda} &\approx 1 \pm 15\%
\qquad \text{(HE-LHC, 27~TeV, $15~\iab$)} \notag \\
\kappa_{\lambda} &\approx 1 \pm 5\%
\ \,\qquad \text{(100~TeV, $30~\iab$)\, ,} 
\label{eq:kappa_self_27}
\end{align}
with a very slight margin of error due to the expected detector
effects and $b$-tagging efficiencies~\cite{Contino:2016spe}. Only the Higgs tri-linear
coupling is varied in this analysis.  We can
compare this result to the expected reach using all available phase
space information, shown in the right panel of Fig.~\ref{fig:hh}. The
key feature of these curves is that they do not feature an alternative
minimum at very large anomalous self-couplings, predicting a SM-like
total cross section from inverting the relative size of the triangular
and continuum contributions.

If we go beyond a simple modification of the Higgs potential of the
Standard Model the SMEFT framework introduced in
Sec.~\ref{sec:basis_eft_linear} is the appropriate framework to study
an anomalous Higgs self-coupling. Among the operators listed in
Eq.\eqref{eq:eff} two modify the triple Higgs coupling,
$\ope_{\phi,3}$ through an constant shift and $\ope_{\phi,2}$ through
a momentum-dependent shift. On the other hand we know that
$\ope_{\phi,2}$ also modifies all Higgs couplings through a wave
function renormalization of the physical Higgs field, so we need to
add $\ope_{\phi,3}$ to a global Higgs analysis and include Higgs pair
observables, ideally covering all the distinct kinematic regimes
discussed above. In Fig.~\ref{fig:hheft} we show projections for a
future 27~TeV collider, combining the global Higgs and gauge analyses
extrapolated from Ref.~\cite{Butter:2016cvz}, the invisible Higgs
searches extrapolated from Ref.~\cite{Biekotter:2017gyu}, and the
Higgs pair analysis of Ref.~\cite{Goncalves:2018qas}.  We can then
compare the expected limits on $f_{\phi,3}/\Lambda^2$ from a specific
search for a modified Higgs self coupling with the expected limits
from a global analysis, profiled over all other Higgs and gauge
operators. For the former, we can translate the 15\% measurement from
Eq.\eqref{eq:kappa_self_27} into $|\Lambda/\sqrt{f_{\phi,3}}| >
1$~TeV.  While the effects of $\ope_{\phi,2}$ and $\ope_{\phi,3}$ can
be separated through the different kinematic regimes, the global
correlations nevertheless weaken the corresponding expectations to
$|\Lambda/\sqrt{f_{\phi,3}}| > 430$~GeV, for an optimistic projection
of systematic and theory uncertainties.

\begin{figure}[t]
\centering
\includegraphics[width=0.55\textwidth]{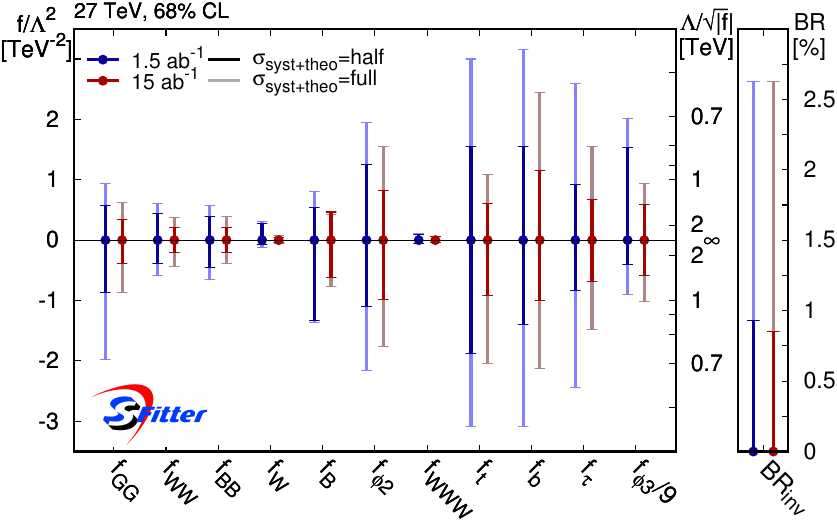}
\caption{Combined global fit of the Higgs-gauge sector including Higgs
  pair production at a future 27~TeV hadron collider. Figure
  from Ref.~\cite{Biekotter:2018jzu}.}
\label{fig:hheft}
\end{figure}

The $2\to 2$ kinematics for $hh$ production links the two sensitive
phase space regions $m_{hh}\sim 2 M_h,~m{t}$ to the suppressed
threshold of the hard production process. Going from the LHC to a
100~TeV collider, the production rate of Higgs pairs with a hard jet
grows by a factor around 80, significantly faster than the total $hh$
production rate~\cite{deLima:2014dta,Goncalves:2018qas}.  This defines
a novel phase space configuration with $m_{hh}\sim 2m_{h,t}$, where a
collimated Higgs pair recoils against a hard jet.  This jet might even
be used to trigger the event.  As a consequence, the phase space
region $m_{hh}\sim 2m_t$ is washed out across $p_T$ and $\Delta
R_{hh}$. The reduced total rate of this process naturally points to
larger Higgs decay rates such as $b\bar b \, \tau \tau$. For this
signature, subjet analyses allow us to to reduce the dominant
backgrounds, especially when the decay $h\to b\bar b$ dominates the
recoil against the hard jet while the softer Higgs is reconstructed in
$\tau\tau$ final states. Relying on this configuration, a precision of
\begin{align}
\kappa_{\lambda} \approx 1 \pm 8\%
\ \,\qquad \text{(100~TeV, $30~\iab$).} 
\end{align}
can be achieved at 68\% C.L.~\cite{Banerjee:2018yxy}.

\subsubsection{Associated $t\bar t h$ production}
\label{sec:fcc_tth}

Following Sec.~\ref{sec:basis_implications} the second most
interesting parameter in the Higgs sector is the top Yukawa coupling,
leading to problems with vacuum stability in the Standard Model. As
discussed in Sec.~\ref{sec:exp_tth}, the measurement of the top Yukawa
coupling is best done in $t\bar{t}h$ production. At the LHC, this
channel is severely limited by experimental systematics and theory
uncertainties. ATLAS and CMS expect a measurement of the top Yukawa
coupling at the 10\% level for the
HL-LHC~\cite{ATL-PHYS-PUB-2014-017,CMS:2017cwx}. This limited
precision leaves a large parameter space of UV models
unconstrained. For example in composite Higgs scenarios it corresponds
to a strong interaction scale around $4\pi f\sim 13$~TeV, barely larger
than the direct LHC reach.

Going to a 100~TeV collider, the production cross section increases
dramatically to
\begin{align}
\sigma (pp\to t\bar t h, 100~\text{TeV}) = 33.9~\text{pb} \; ,
\end{align}
as illustrated in Fig.~\ref{fig:tthscale}. While the increased rate
does not automatically help us control the systematic and theory
uncertainties, it allows us to focus on phase space regions which are
less prone to those limitations~\cite{Plehn:2015cta}. As discussed in
Sec.~\ref{sec:exp_tth} one of those phase space regions is boosted
$t\bar{t}h$ production~\cite{Plehn:2009rk}. It automatically solves
the $b$-jet combinatorics and allows for a side band analysis in
$m_{bb}$. In addition, we can control uncertainties, for example from
parton densities, by evaluating the ratio
\begin{align}
\frac{d \sigma_{t\bar{t}h}}{d \sigma_{t\bar{t}Z}}
\end{align}
over phase space~\cite{Plehn:2015cta}.  The $t\bar{t}Z$ reference
process features the same top decays as the signal, while the hard
process is entirely determined by the gauge structure of the Standard
Model.

\begin{figure}[t]
\includegraphics[width=0.46\textwidth]{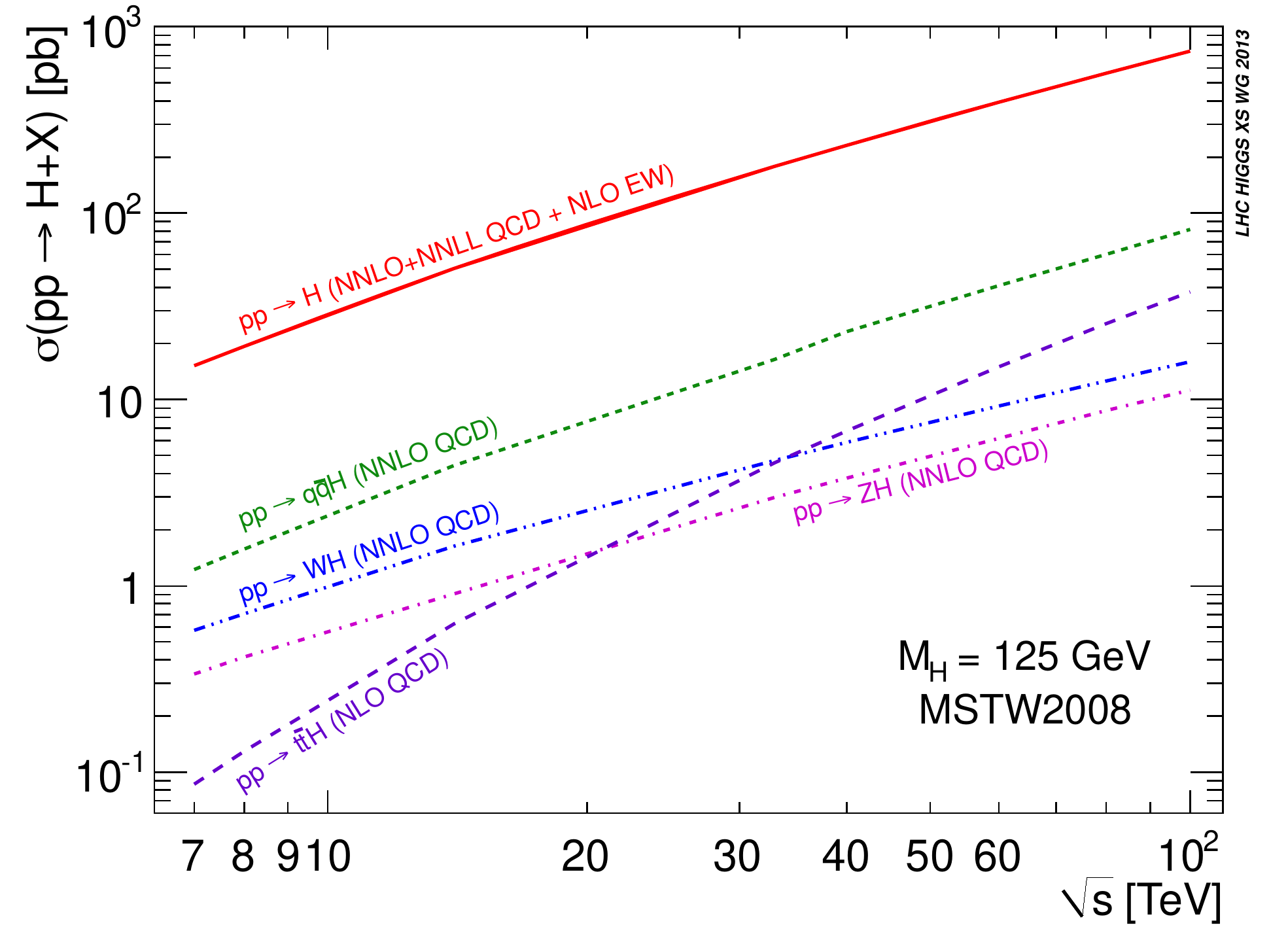}
\hspace{0.1\textwidth}
\includegraphics[width=0.38\textwidth]{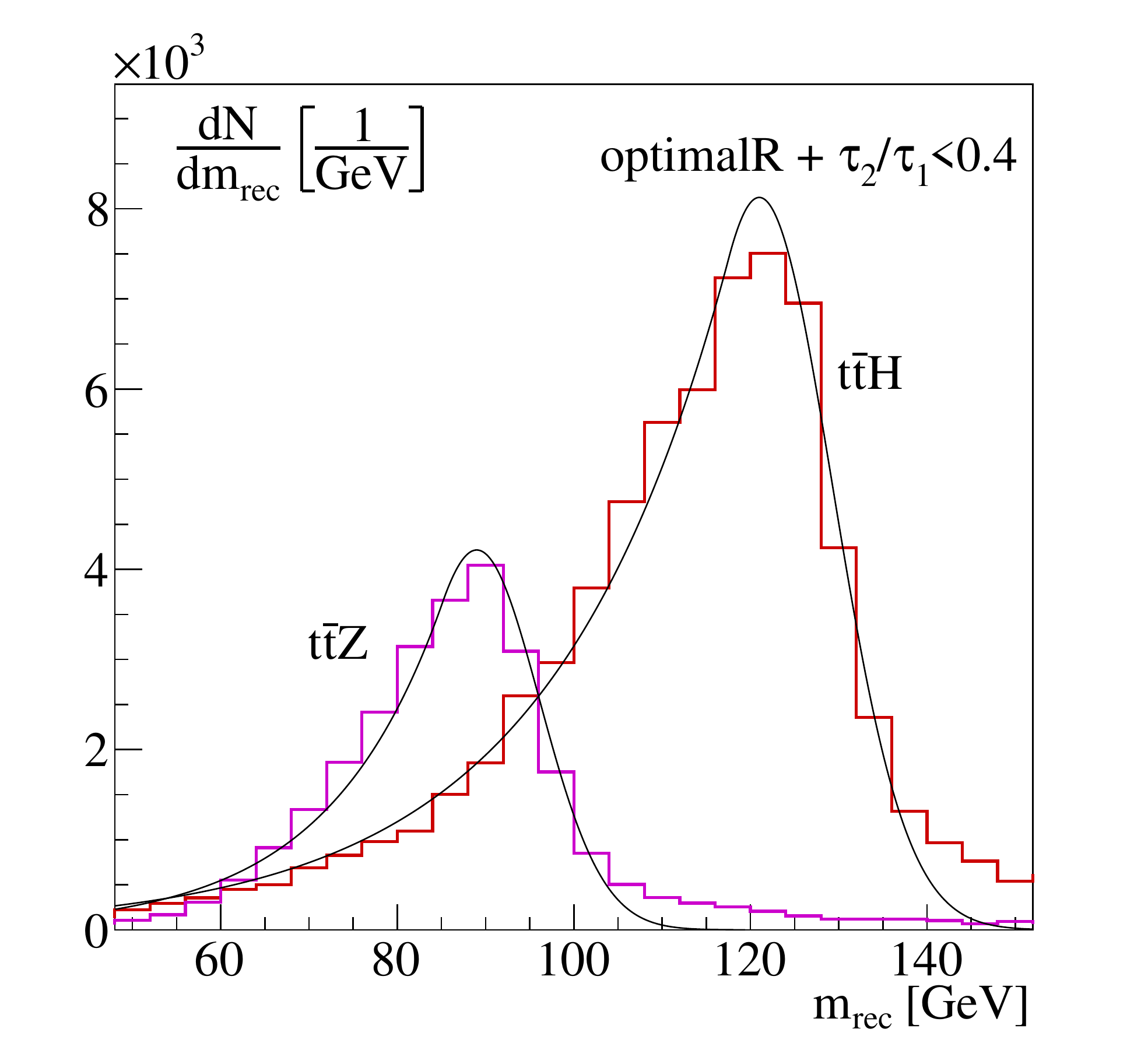}
\caption{Left: Extrapolation of the Higgs production cross sections as
  a function of the center of mass energy. Figure from
  Ref.~\cite{Dittmaier:2011ti}. Right: Two-peak signal structure on
  top of the continuum background, reconstructed with the
  \textsc{HEPTopTagger} and a corresponding Higgs tagger. Figure from
  Ref~\cite{Plehn:2015cta}.}
\label{fig:tthscale}
\end{figure}

Note that we can expect to see the $Z$-peak in the $m_{bb}$
distribution next to the Higgs peak~\cite{Plehn:2009rk}, so the
100~TeV analysis will be based on the two-peak fit on top of the
continuum background shown in the right panel of
Fig.~\ref{fig:tthscale}.  In the signal region of the $m_{bb}$
distribution, the boosted analysis leads to signal-to-background
ratios around $1/3$ and a huge signal significance. It should allow us
to measure the top Yukawa coupling with an
uncertainty~\cite{Plehn:2015cta},
\begin{align}
\kappa_t \approx 1 \pm 1\% \qqquad \text{(100~TeV, $20~\iab$)}  \; .
\end{align}
%

\subsubsection{Invisible Higgs decays}
\label{sec:fcc_inv}

A third benchmark channel for future hadron colliders are invisible
Higgs decays. From the LHC we know that they can be best searched for
in WBF Higgs production~\cite{Biekotter:2017gyu}, as discussed in
Sec.~\ref{sec:exp_wbf}. This electroweak process is much less prone to
theory uncertainties than the usual QCD processes and should therefore
allow us to systematically define precision measurements at a 100~TeV
hadron collider~\cite{Goncalves:2017gzy,Jager:2017owh}. In
Fig.~\ref{fig:wbf100} we show the main kinematic features of the WBF
process at a 100~TeV collider, compared to the LHC. First, we see that
indeed the transverse momentum distributions of the tagging jets are
defined by the $W$-mass, with a peak around 70~GeV. For the 100~TeV
collider we observe a slight change in the spectrum when we limit the
tagging jet rapidities to $\eta <5$, reflecting a preference for a
slightly larger rapidity coverage at higher energies. In the right
panel we show the number of jets in the different signal and
background processes, which reflects the different Poisson and
staircase scaling patterns~\cite{Gerwick:2011tm}. This jet activity
serves as the main handle in separating the WBF and gluon fusion
contributions to $hjj$ production in a precision measurement.

\begin{figure}[t]
\includegraphics[width=0.37\textwidth]{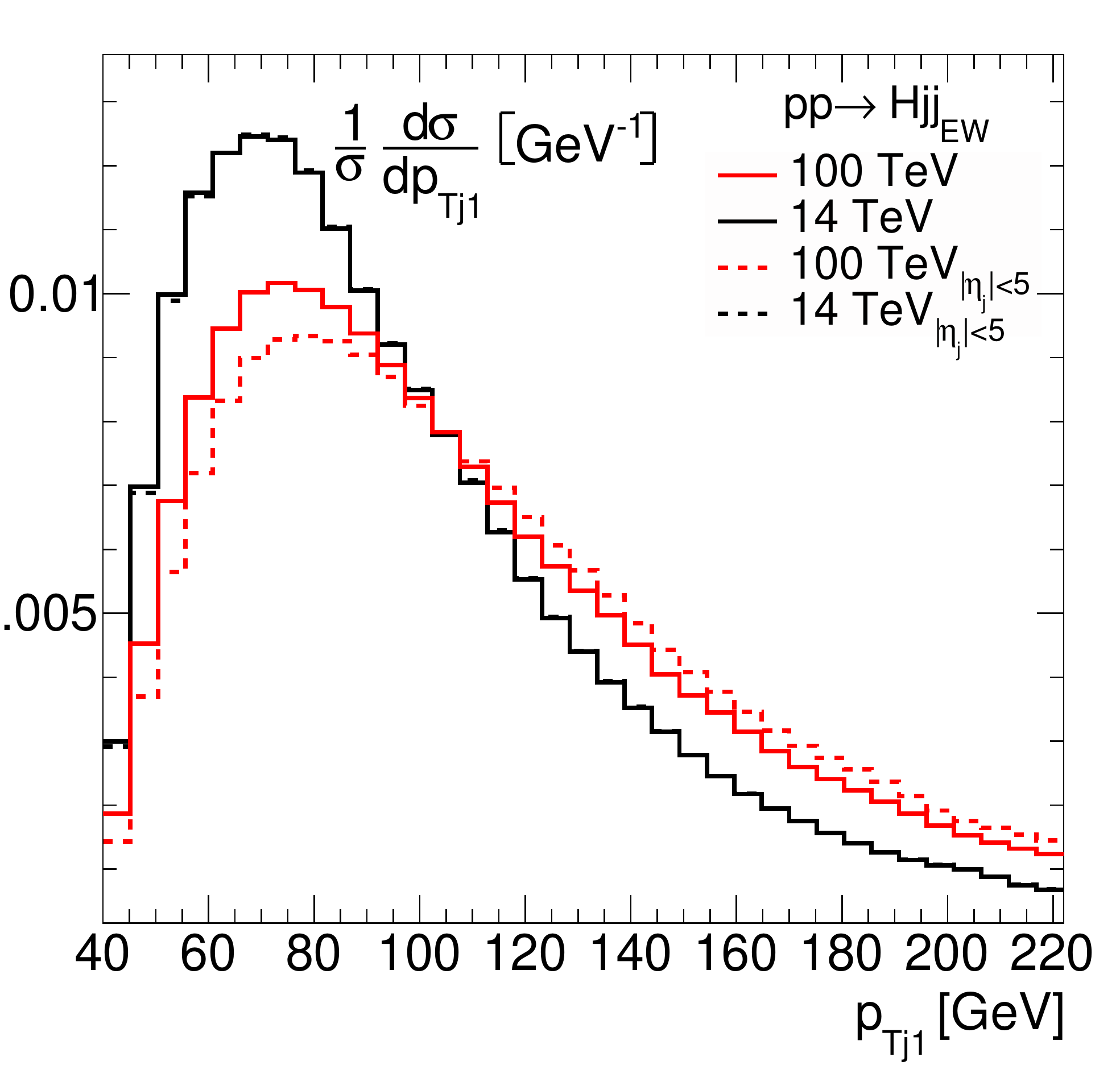}
\hspace{0.15\textwidth}
\includegraphics[width=0.36\textwidth]{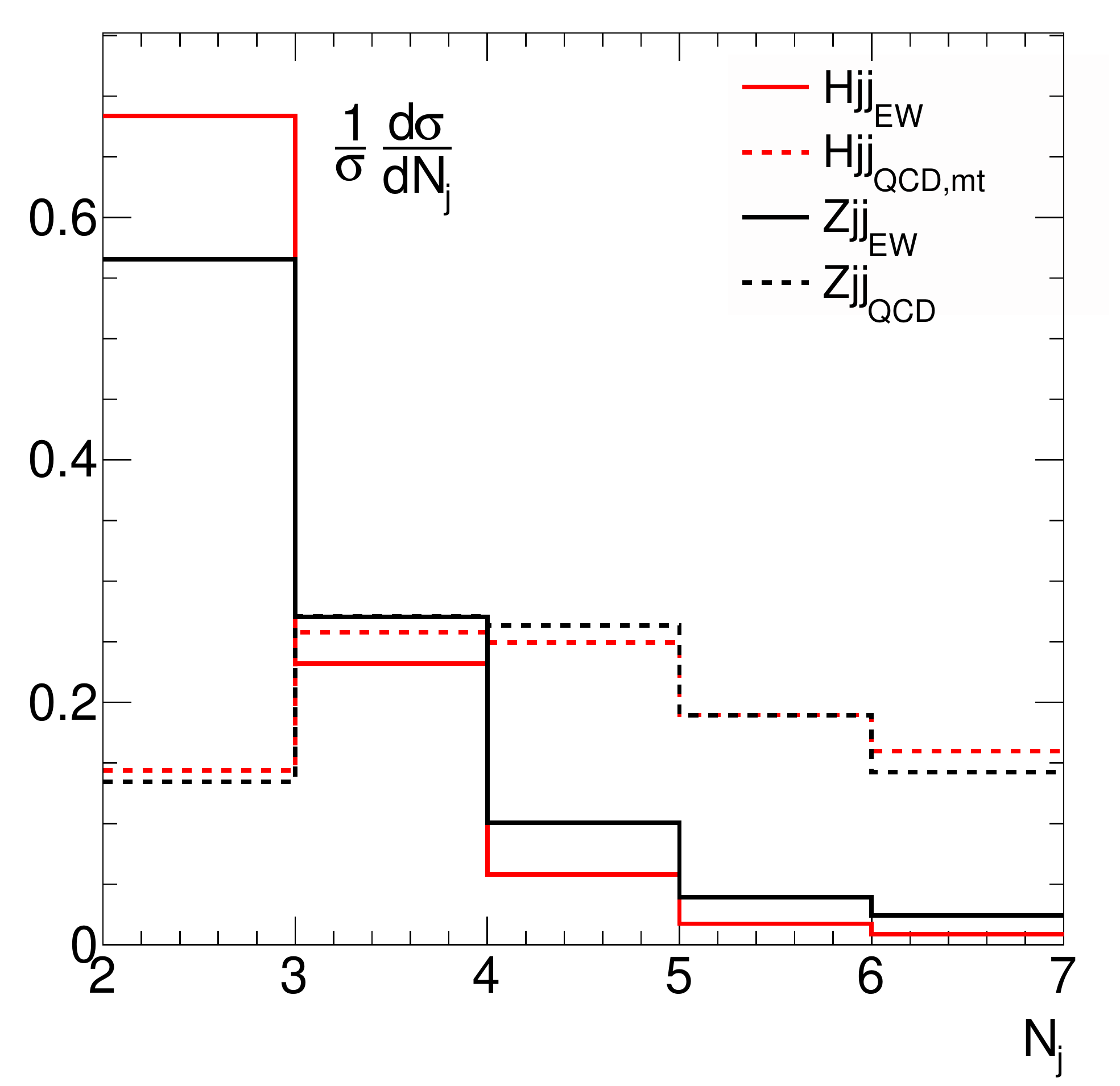}
\caption{Tagging jet kinematics at a 100~TeV hadron collider. We show
  the $p_T$ of the leading tagging jet on the left and the exclusive
  number of jets on the right. Figures from
  Ref.~\cite{Goncalves:2017gzy}.}
\label{fig:wbf100}
\end{figure}

As for the Higgs pair analysis discussed in Sec.~\ref{sec:fcc_hh}, a
WBF analysis at a 27~TeV or 100~TeV collider benefits from including jet
radiation in the signal process. We can for example define 2-jet and
3-jet signal and background samples by (i) vetoing a third jet for
$p_{T,j_3} > p_{T,\text{veto}}$ and (ii) requiring a third jet with
$p_{T,j_3} > p_{T,\text{veto}}$ and vetoing a fourth jet for
$p_{T,j_4} >
p_{T,\text{veto}}$~\cite{Bernaciak:2014pna,Goncalves:2017gzy}. A
typical jet veto scale is $p_{T,\text{veto}} = 20$~GeV. We can then
use the $\met$, $m_{j_1,j_2}$, and $N_j$ distributions to search for
invisible Higgs decays and expect a reach of~\cite{Goncalves:2017gzy}
\begin{align}
\br(h \to \text{inv}) &\lesssim 1\%
\qquad \text{(HE-LHC, 27~TeV, $20~\iab$)} \notag \\
\br(h \to \text{inv}) &\lesssim 0.5\%
\qquad \text{(100~TeV, $20~\iab$)\, ,} 
\end{align}
In both cases the systematic uncertainties become the limiting factor.
It might be possible to use the large statistics of mono-jet
production to search for invisible Higgs decays in the process $pp \to
h$+jet~\cite{Feng:2005gj}. However, there does not exist a reliable signal and
background study, and the sizeable systematic and theory uncertainties
in the underlying QCD process will limit the measurement of this
branching ratio.

\subsection{Theory for modern analyses}
\label{sec:exp_data}

One of the key challenges of contemporary phenomenogy is to keep track
with the experimental analysis strategies, which could allow us to
turn the LHC into a precision experiment. In this section we will
review the ideas behind some of the most relevant advanced analysis
techniques and describe the corresponding challenges and novel ideas
to theoretical LHC physics. We will show how these techniques are
already applied at the LHC and how a coherent effort between
experiment and theory will help us to make the best possible use of
LHC data, in the Higgs sector and elsewhere.

Historically, LHC analyses have started with identifying phase space
regions with a large signal-to-background ratio and extracting them by
applying cuts on well-defined kinematic observables. A simple example
is the mass window in $m_{\gamma \gamma}$ in the Higgs discovery. A
measurement is then based on counting events and comparing them to the
background-only and the signal-plus-background predictions, using
Poisson or Gaussian statistics. Assuming a discovery is quantified in
terms of the statistical significance we typically apply a kinematic
cut if it improves $S/\sqrt{B}$ for a given luminosity.  This kind of
analysis has the fundamental disadvantage that there will always be
kinematic observables and phase space regions which do not contribute
to the analysis.  Multi-variate analyses for example using boosted
decision trees define cuts as non-trivial surfaces in the
multi-dimensional phase space, but still remain cut-and-count
analyses.

One way to improve these analyses is to change the way we organize
events. Instead of a kinematic observable, we can define histograms in
terms of any variable we want. The problem of finding the best test
statistics for a given hypothesis test leads us to the Neyman-Pearson
lemma, optimal observables introduced in Sec.~\ref{sec:exp_data_opt},
and the likelihood ratio discussed in Sec.~\ref{sec:exp_data_mem}.  We
can even classify events by some kind of boosted decision tree output
variable without an obvious physics interpretation.  Just like
kinematic observables, usually defined with the hard process in mind,
we eventually compute these variables from the measured 4-momenta.  In
addition, some test statistics require us to define the underlying
hypothesis to interpret each event. Often, we still apply a
cut-and-count strategy for example on an optimal observable or a
log-likelihood ratio. In general this method automatically identifies
useful events and allows us to include information from all over phase
space and without a clear boundary between useful and not useful
events. This means that well-defined kinematic distributions from
first-principles calculations only serve as cross-checks and as an
illustration of the actual analysis\footnote{The limited relevance of
  first-principles predictions can be measured by the fraction of
  plots with non-physics $x$-axes in the average ATLAS and CMS Higgs
  paper.}. In the analysis we really use physics hypotheses as inputs
to full event simulations and then compare the simulated events with
data with no direct reference to the hard process.

Finally, we can improve any cut-and-count analysis by computing the
probability that a given event corresponds to a given hypothesis and then
multiplying events with their signal and background probabilities, as
done for instance in many LHCb analyses. This probability can be
extracted from a neural network and avoids losing small pieces of
information by cutting away background-like events. Any such
probability will eventually combine information from the hard process
with detector efficiencies and particle identification, so it also breaks
the direct link to first-principles theory predictions. Finally, this approach
begs the question of which objects and observables we use to extract
these probabilities.  Modern machine learning techniques allow us to
rely on low-level observables, like measured 4-momenta, as we will
discuss in Sec.~\ref{sec:exp_data_deep}.

Following this general development, it is clear that theoretical
predictions for the hard process are still at the heart of LHC
physics, but that the relation between the corresponding observables
and the experimental analysis results is at least seriously
blurred. Given the current formats of theoretical and experimental
publications, we cannot track for example the impact of a new
precision prediction for a given signal or background phase space
region on the experimental result. This means that particle theory
needs to develop ways to model and understand modern experimental
analyses. A promising new approach is information geometry, discussed
in Sec.~\ref{sec:exp_data_info}. Moreover, we need to develop tools
which allow us to analyze the impact of first-principles calculations
on likelihood-free machine-learning analyses based on low-level
observables like 4-vectors of the full event. Below, we will
describe some basic concepts and applications which go significantly
beyond the current experimental tools.

\subsubsection{Optimal observables and Fisher information}
\label{sec:exp_data_opt}

Optimal observables are used as an efficient method to optimize an
analysis involving a complicated phase space~\cite{Atwood:1991ka}.
The idea is especially simple to illustrate using an especially
interesting Higgs property like the $CP$ measurements described in
Sec.~\ref{sec:basic_char_cp}. We will use this example to
mathematically argue that there exists a unique, best-suited
observable to extract the available information from a data
set~\cite{Diehl:1993br}.

There are two ways to derive optimal observables. Both start from a
kinematic phase space distribution, which is affected by a vector of
model parameters $g_i$ with the no-signal hypothesis $g_i = 0$. Its
linearized form is
\begin{align}
\frac{d \sigma(\vec{g})}{dx} 
\approx s^{(0)} + \sum_i g_i s^{(1)}_i  \; .
\label{eq:opt_rate}
\end{align}
Under certain assumption the Neyman-Pearson lemma states that the
likelihood ratio is the best discriminator for two hypotheses. If we
simplify our model such that all relevant information is included in
kinematic distributions rather than the total rate, this likelihood
ratio is related to the kinematic information as
\begin{align}
\frac{p(x|\vec{g})}{p(x|\vec{0})} 
= \frac{s^{(0)} + \sum_i g_i s^{(1)}_i}{s^{(0)}}
\label{eq:def_like1}
\end{align}
The probability $p(x|\vec{g})$ for a phase space configuration $x$ to
occur in a model described by the parameters $\vec{g}$ is the basis of
any event generation and at the parton level related to the matrix
element squared or the kinematic distribution given in
Eq.\eqref{eq:opt_rate}. We will discuss ways to compute it in
Sec.~\ref{sec:exp_data_mem}.  For an unbiased estimator, the true
value of $g_i$ maximizes the log-likelihood ratio, defining for
instance the reference point $g_i=0$,
\begin{align}
\frac{\partial}{\partial g_i} \; \log \frac{p(x|\vec{g})}{p(x|\vec{0})} \Bigg|_{\vec{g} = \vec{0}} = 0
\qqquad \Leftrightarrow \qqquad 
\frac{\partial p(x|\vec{g})}{\partial g_i} \Bigg|_{\vec{g} = \vec{0}} = 0 \; .
\end{align}
We can evaluate this condition for our toy model of
Eq.\eqref{eq:opt_rate}
\begin{align}
\frac{\partial}{\partial g_i} \; \log \frac{p(x|\vec{g})}{p(x|\vec{0})}
= \frac{\dfrac{s^{(1)}_i}{s^{(0)}}}{1 + \sum_m g_m \dfrac{s^{(1)}_m}{s^{(0)}}} \; .
\label{eq:opt_obs}
\end{align}
This form indicates that we can extract all information on the
parameter $g_i$ in the vicinity of the reference point just evaluating
one observable. In our model this optimal observable
$\opt_i^\text{opt}(x)$~\cite{Davier:1992nw} is
\begin{align}
\opt_i^\text{opt}(x) = \frac{\partial p(x|\vec{g})}{\partial g_i} = \frac{s_i^{(1)}}{s^{(0)}} \; .
\label{eq:def_opt_obs}
\end{align}
We remind ourselves that this derivation is tied to the Neyman-Person
lemma and the likelihood ratio as the most powerful test statistic. We
will further develop this approach into the matrix element method in
Sec.~\ref{sec:exp_data_mem}.

An alternative derivation of the optimal observable starts with the
general expression for the expectation value of a vector of
observables $\opt_i(x)$~\cite{Diehl:1993br,Nachtmann:2004fy}. We again
use the kinematic distribution in Eq.\eqref{eq:opt_rate} and
expand the expectation value linearly in the $g_i$,
\begin{align}
\text{E}[\opt_i]
&= \frac{\int dx \left( s^{(0)} + \sum_n g_n s^{(1)}_n \right) \opt_i}
        {\int dx \left( s^{(0)} + \sum_n g_n s^{(1)}_n \right)}
= \frac{\int dx s^{(0)} \opt_i}{\int dx  s^{(0)}}
+ \sum_n g_n \left[ \frac{\int dx  s^{(1)}_n\opt_i}{\int dx s^{(0)}}
                  - \frac{\int dx s^{(1)}_n \; \int dx s^{(0)} \opt_i}{\left( \int dx s^{(0)} \right)^2} \right] \; .
\label{eq:expect_obs}
\end{align}
We can also compute the leading term of the the
covariance matrix and find
\begin{align}
\text{Cov}[\opt_i,\opt_j]
= \text{E}[\opt_i \opt_j] - \text{E}[\opt_i] \text{E}(\opt_j] 
= \frac{\int dx s^{(0)} \opt_i \opt_j}{\int dx s^{(0)}}
 -\frac{\int dx s^{(0)} \opt_i \int dx s^{(0)} \opt_i}{(\int dx s^{(0)})^2} \; .
\end{align}
If we apply these results to the optimal observables defined in
Eq.\eqref{eq:def_opt_obs} and use that our toy model implies that
$s^{(0)}$ integrates to one and the $s_n^{(1)}$ integrate to zero, we
find
\begin{align}
\text{E}[\opt_i^\text{opt}]
&= \sum_n g_n \int dx \frac{s_n^{(1)} s_i^{(1)}}{s^{(0)}} \notag \\
\text{Cov}[\opt_i^\text{opt},\opt_j^\text{opt}]
&= \int dx \frac{s_i^{(1)} s_j^{(1)}}{s^{(0)}} \; .
\end{align}
In addition to this covariance matrix in observable space, we can
solve Eq.\eqref{eq:expect_obs} for the coupling vector in order to define a
covariance matrix in model space,
\begin{align}
\text{E}[\opt_i] 
= \text{E}[\opt_i]^{(0)} + \sum_n c_{in} \; g_n 
\qquad \Leftrightarrow \qquad 
g_n 
&= ( c^{-1} )_{ni} \; \left( \text{E}[\opt_i] - \text{E}[\opt_i]^{(0)} \right)
\notag \\
\text{Cov}[g_m, g_n] 
&= (c^{-1})_{m i} \; \text{Cov}[\opt_i, \opt_j] \; (c^{-1}{})_{n j}
\end{align}
with $\text{E}[\opt_i]^{(0)} = \int dx s^{(0)} \opt_i/\int dx
s^{(0)}$. For the optimal observable this simplifies to
\begin{align}
\text{E}[\opt_i^\text{opt}]
&= \sum_n g_n \text{Cov}[\opt_i,\opt_n]
\quad \Rightarrow \qquad
\text{Cov}[g_i, g_j] = \text{Cov}[\opt_i^\text{opt},\opt_j^\text{opt}]^{-1} \; .
\label{eq:optimial_model}
\end{align}
This relation between the covariance matrices can be put in
perspective when we expanding the log-likelihood around the true of
reference point in model space,
\begin{align}
\log p(x|\vec{g}) = \log p(x|\vec{0}) 
+ \frac{1}{2} \sum_{m,n} g_m g_i \frac{\partial^2 \log p(x|\vec{g})}{\partial g_i \partial g_m} \Bigg|_{\vec{g} = \vec{0}}
+ \cdots
\end{align}
The Fisher information matrix is defined as the leading non-trivial
term,
\begin{align}
I_{ij} 
= - \text{E}\left[ \frac{\partial^2 \log p(x|\vec{g})}{\partial g_i \partial g_j} \right] \; .
\label{eq:def_info}
\end{align}
We again compute it for our toy model and find that the Fisher
information is the inverse of the covariance matrix in model space,
\begin{align}
I_{ij} 
&= - \text{E}\left[ \frac{s^{(1)}_i}{s^{(0)}} \frac{\partial}{\partial g_j} 
             \frac{1}{1 + \sum_n g_n \dfrac{s^{(1)}_n}{s^{(0)}}} 
      \right] 
= \int dx \frac{s^{(1)}_i s^{(1)}_j}{s^{(0)}}
= \text{Cov}[g_i, g_j]^{-1}  \; .
\end{align}
This equality is the saturated case of the Cram\'er-Rao bound. The
bound links two tracers of the sensitivity of a measurement in model
space: the (inverse) Fisher information tells us how much information
a given experiment can extract about a set of model parameters; the
covariance matrix gives the uncertainty of the measurements on the
couplings. In general, the minimum value of the covariance matrix must
be larger than the inverse Fisher information,
\begin{align}
\text{Cov}[g_i, g_j] \geq (I^{-1})_{ij} \; .
\end{align}
Using this second derivation we find that the optimal observable
saturates the Cram\'er-Rao bound and hence the amount of information
we can extract from our assumed measurements.

\begin{figure}[t]
\includegraphics[width=0.4\textwidth]{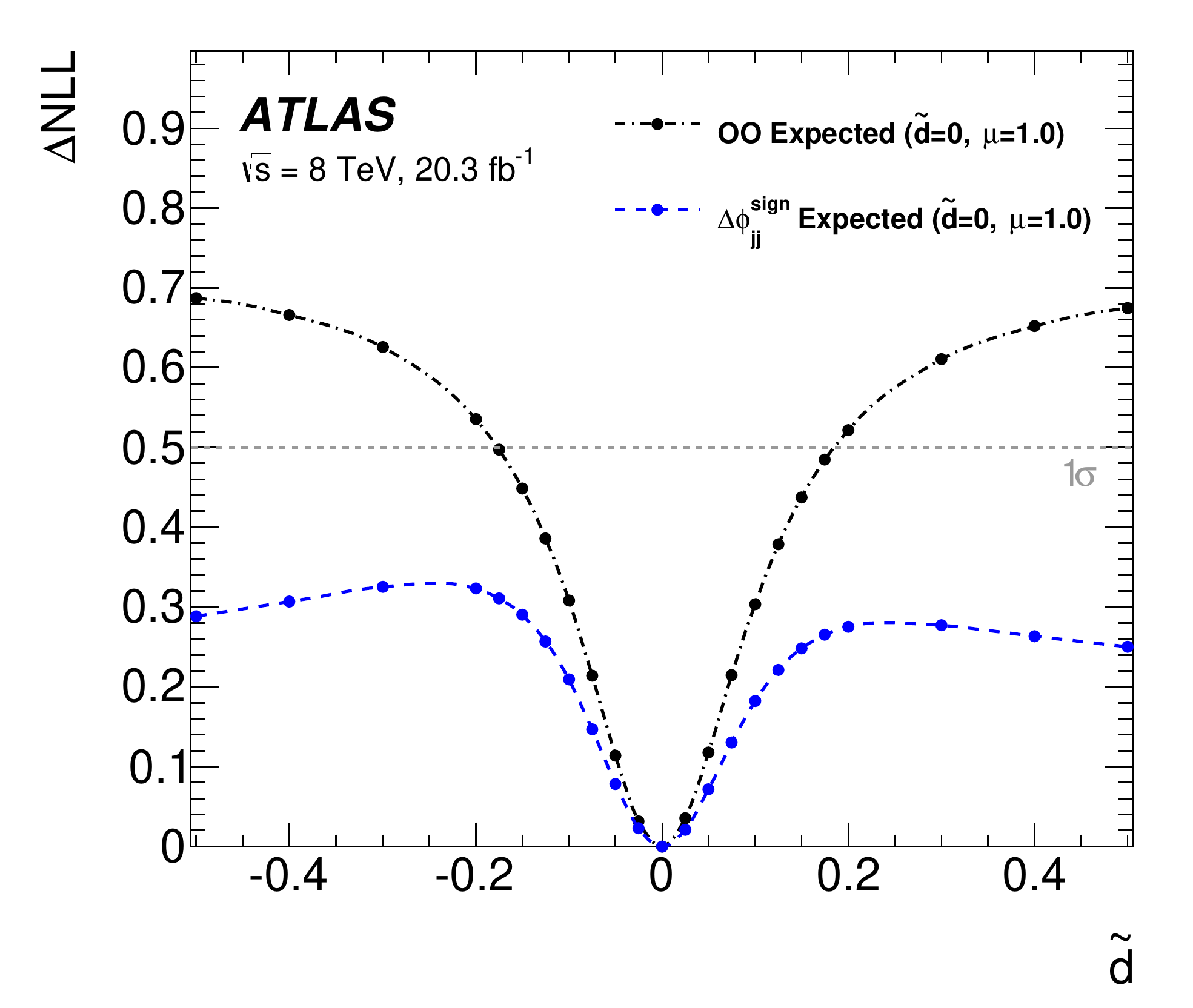}
\hspace*{0.1\textwidth}
\includegraphics[width=0.35\textwidth]{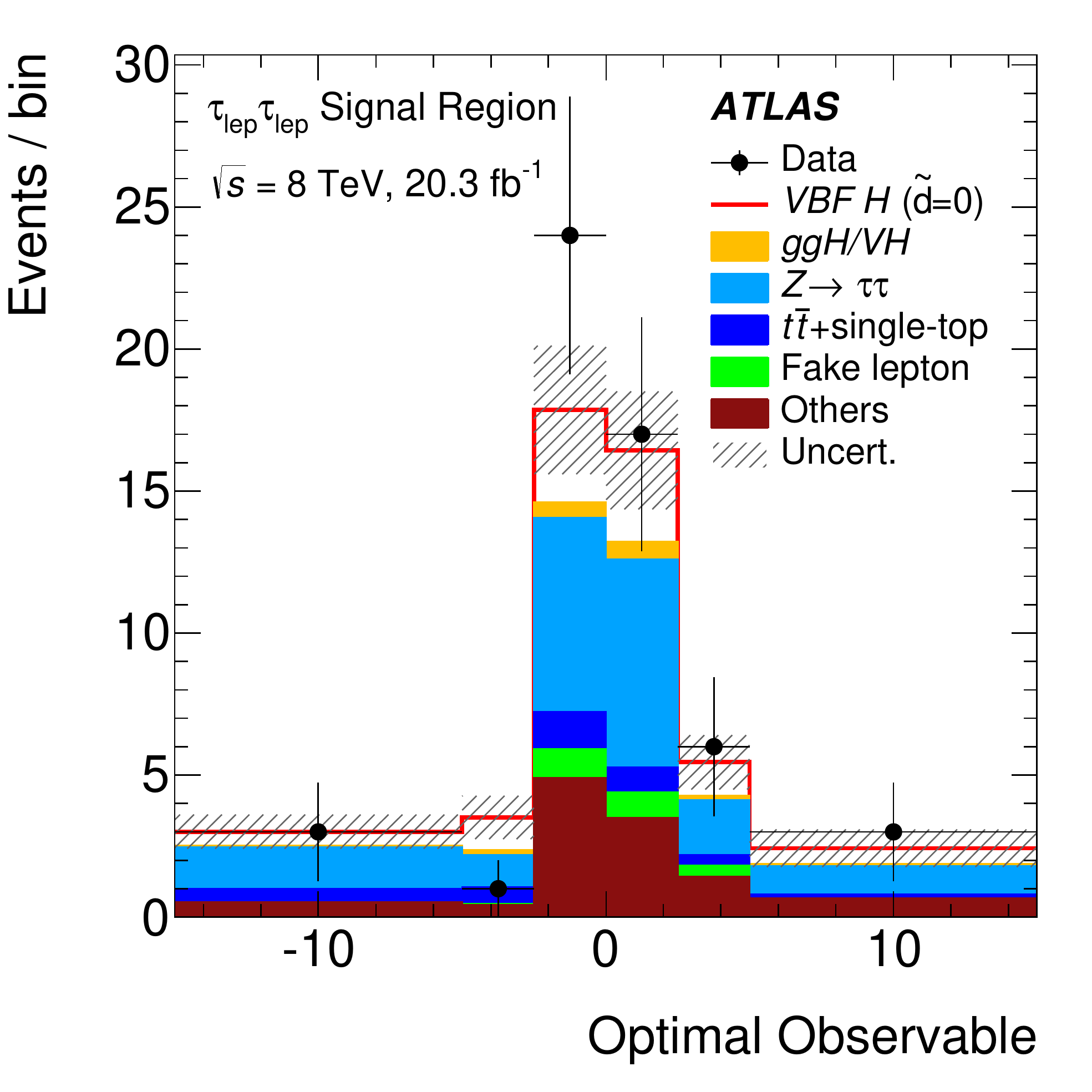}
\caption{Left: comparison of the expected reach of the $CP$-sensitive 
  and the optimal observable in searching for $CP$-violation in
  WBF. Right: distribution of the optimal observable, including the
  background estimates and the signal prediction.  Figures from
  Ref.~\cite{Aad:2016nal}.}
\label{fig:opt_obs}
\end{figure}

When we search for $CP$-violation in WBF Higgs production, the signed
azimuthal angle between the tagging
jets~\cite{Plehn:2001nj,Hankele:2006ma}, introduced in
Sec.~\ref{sec:basic_char_cp}, is the appropriate genuine $CP$-odd
observable.  However, once we write down a model for $CP$-violation in
the Higgs couplings to weak bosons, additional kinematic information
from the form of the corresponding effective operators becomes
available~\cite{Brehmer:2017lrt}.  We can write the optimal observable
from Eq.\eqref{eq:def_opt_obs} in terms of the SM and new
physics matrix elements and find~\cite{Aad:2016nal}
\begin{align}
\opt^\text{opt} 
=\frac{2 \operatorname{Re}(\mathcal{M}_{\text{SM}}^{*}\mathcal{M}_{\text{CP-odd}})}{|\mathcal{M}_{\text{SM}}|^2} \; .
\end{align}
It allows us to extract the pre-factor of a $CP$-violating
contribution to the $WWh$ coupling inducing the new physics amplitude
$\mathcal{M}_{\text{CP-odd}}$. In Fig.~\ref{fig:opt_obs} we show the corresponding
ATLAS results based on lepton-lepton and
lepton-hadron decays of WBF Higgs production with the subsequent
decays $h \to \tau \tau$. In the left panel we start with a comparison
between the expected reach of the $CP$-odd observable $\Delta
\phi_{jj}$ and the optimal observable. Indeed, the latter is more
sensitive on the dimension-6 realization of $CP$-violation. In the
right panel we see the distribution for the signal and background
events after a cut extracting the signal region through a boosted
decision tree. The data peaks at small values of $CP$-odd
contributions, with small tails expected from statistical and
systematic uncertainties, indicating no sign of $CP$-violation in the
weak gauge-Higgs sector.

Is it important to note that our toy model illustration of optimal
observables does not imply that the Neyman-Pearson lemma is equivalent
to the Cram\'er-Rao bound; the two statements are derived in different
formalisms and with completely different assumptions. At parton level
we can use the Neyman-Pearson lemma to systematically study the
potential of a given analysis for separating two hypothesis, like for
example to lead to a discovery of a specific signal in the presence of
known backgrounds~\cite{Cranmer:2006zs}. It is applied to actual
searches through the matrix element method, described in
Sec.~\ref{sec:exp_data_mem}.  The Fisher information with the
Cram\'er-Rao bound describes a continuous dependence on model
parameters, allowing us to estimate the expected precision of any
measurement in a combined signal and background sample. We illustrate
the information geometry approach in Sec.~\ref{sec:basic_char_cp} and
generalize it to the detector-level in Sec.~\ref{sec:exp_data_info}.

\subsubsection{Matrix element method}
\label{sec:exp_data_mem}

Following a similar argument as for optimal observables, but for more
general analysis tasks leads us to the matrix element method, which
allows us to systematically include all relevant information from a
signal phase space based on the Neyman--Pearson lemma. It states that
the likelihood ratio is the most powerful test statistic to tell a
simple null hypothesis --- for example background only --- from an
alternate hypothesis --- for example signal plus background. Here,
maximum power is defined as the minimum probability for false negative
error for a given probability of false positive. The key observation
which relates this statement to LHC analyses is that the likelihood to
populate a given phase space region given a model is nothing but the
exclusive cross section as we use it in an event generator.

We start with a theoretical problem~\cite{Cranmer:2006zs}, where we
assume that we can ignore detector effects and have full control over
the calculation of the differential cross section. In a counting
experiment the likelihood of observing $n$ events in a given phase
space region is given by a Poisson distribution.  To describe phase
space we introduce a observable $x$, where we assume that the
hypothesis $\vec{g}_b$ is described by the normalized distribution $f_b(x)$,
while the alternative hypothesis $\vec{g}_s$ is described by $f_{s+b}(x) =
(sf_s(x)+bf_b(x))/(s+b)$.  The single-event likelihood can then be
factorized into the Poisson likelihood to observe an event and the
normalized event likelihoods,
\begin{align}
q(x|\vec{g}_s,\vec{g}_b) 
= \log \; \frac{p(x|\vec{g}_s)}{p(x|\vec{g}_b)}
= - s +  \log \left( 1 + \frac{sf_s(x)}{bf_b(x)} \right) 
\to -\sigma_{\text{tot},s} \; \mathcal{L} \; + \,
             \log \left( 1 + \frac{d\sigma_s(x)}
                                 {d\sigma_b(x)}  
                 \right) \; .
\label{eq:llr}
\end{align}
This expression replaces Eq.\eqref{eq:def_like1} from our toy
model. Because the log-likelihood ratio is additive when we include
more than one event or phase space configuration, we can integrate it
the same way we integrate a differential cross section to a total or
fiducial cross section.  ${\cal L}$ denotes the integrated
luminosity. In Sec.~\ref{sec:exp_data_info} we will describe some
ideas how to use the likelihood ratio directly in LHC analyses, while
in this section we focus on ways to compute the phase-space dependence
of the resulting statistical significance.

Because the log-likelihood ratio is the most powerful test statistics,
we now generate a $q$-distribution for events where we for example
assume the hypothesis $\vec{g}_b$. We integrate over the entire phase space
with the normalized event weight $d\sigma_0(x)/\sigma_{0, \text{tot}}$
and generate the log-likelihood distributions
\begin{align}
\rho_{0,n=1}(q) 
&= \int dx \; f_b(x) \;
             \delta( q_{n=1}(x) - q) \notag \\
\rho_{0,n}(q) 
&= \rho_{0,n=1} \otimes \rho_{0,n=1} \otimes \cdots \otimes   \rho_{0,n=1} \notag \\
\rho_0(q) 
&= \sum_n \text{Pois}(n|b) \; \rho_{0,n}(q) \; ,
\label{eq:rho_single}
\end{align}
with a convolution in $q$-space. Combining the log-likelihood
distributions for $\vec{g}_b$ and $\vec{g}_s$ we can compute the
maximum significance with which we will, assuming full control, be
able to distinguish between two hypotheses at the LHC. Because the
log-likelihood is additive we can apply the same method after
integrating over part of phase space or after defining slices of phase
space. 

To illustrate this approach and its link to theory problems we show the
maximum significance for the process
\begin{align}
pp \to Zh \to Z \; b\bar{b} \; ,
\end{align}
as discussed in Sec.~\ref{sec:exp_vh}, in Fig.~\ref{fig:mem}. The
individual bins are slices of the reconstructed $p_{T,h}$, indicating
that a slight boost of the decaying final state helps us to separate
the signal from the continuum background. The drop towards large
$p_{T,h}$ reflects the statistical limitation given by first term in
Eq.\eqref{eq:llr}~\cite{Plehn:2013paa}. Similar analyses are available
for Higgs pair production, often used as a benchmark channel for the
high-energy LHC or future hadron
colliders\cite{Kling:2016lay,Goncalves:2018qas}. As a tool on the
Monte Carlo level, it allows us to model and understand the power of
modern analysis methods in testing two theory hypotheses.

\begin{figure}[t]
\includegraphics[width=0.35\textwidth]{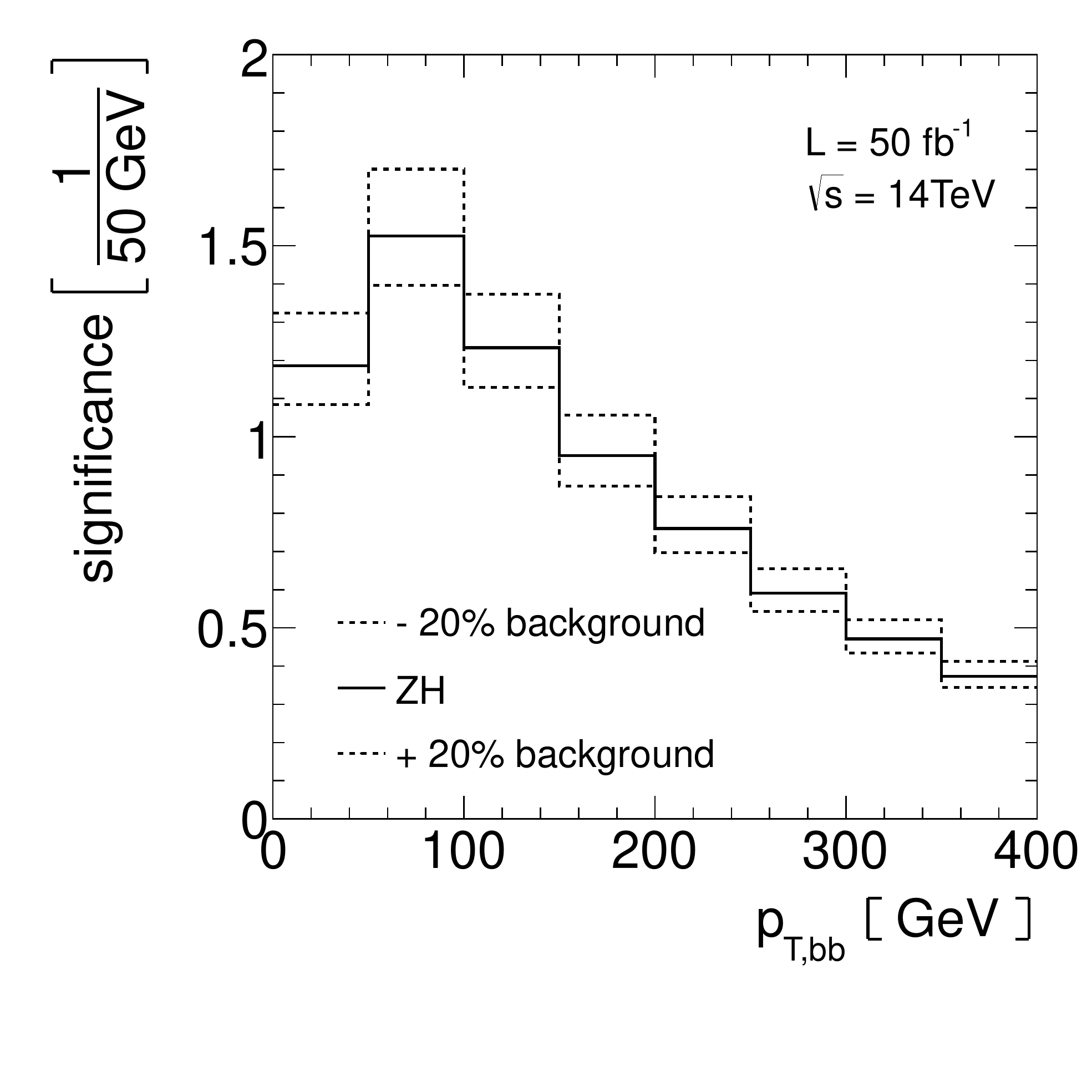}
\hspace*{0.1\textwidth}
\includegraphics[width=0.4\textwidth]{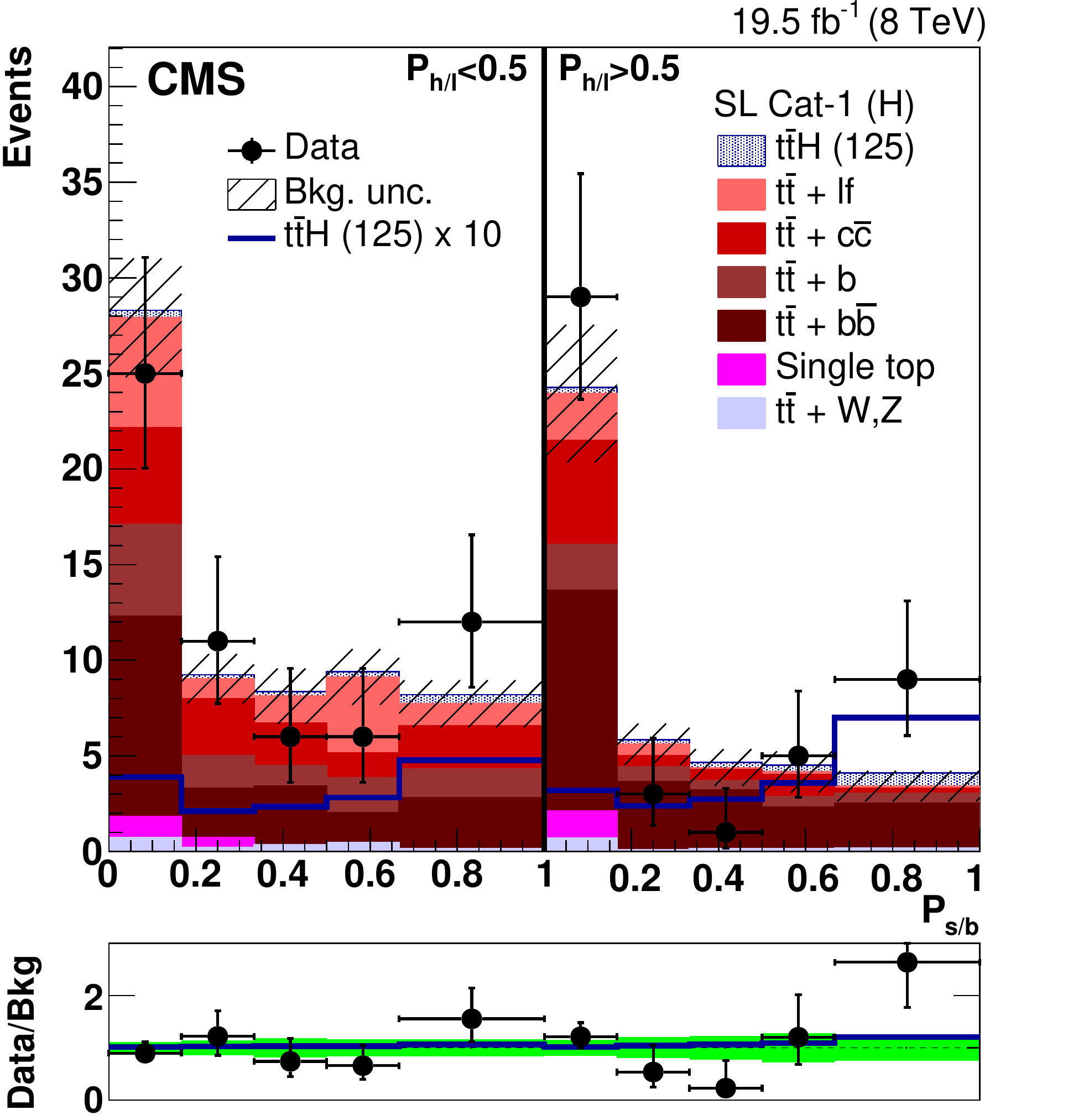}
\caption{Left: bin-wise distribution of the maximum significance in 
  extracting the $Vh$, $h \to b\bar{b}$ signal from the continuum
  background. Figure from Ref.~\cite{Plehn:2013paa}. Right:
  Distribution of the log-likelihood ratio for mostly light-flavor and
  mostly heavy-flavor jets ($P_{\mathrm{h/l}}$) The signal and
  background yields come from a combined fit of all nuisance
  parameters assuming the Standard Model signal rate.  Figure from
  Ref.~\cite{Khachatryan:2015ila}.}
\label{fig:mem}
\end{figure}

Whenever we want to extract a signal from a background at the LHC, we
should be able to use the same test statistics for an actual number of
events. This defines the matrix element
method~\cite{Kondo:1988yd,Abazov:2004cs}.  Following
Eq.\eqref{eq:llr}, we need to compute the differential cross sections
for the signal and background hypotheses for each event. There are three problems with this idea in practice:

A first, obvious problem arises if a background is reducible, so the
signal and background phase spaces cannot be mapped onto each other to
form the ratio $d\sigma_s(x)/d\sigma_b(x)$ for each event's phase
space point $x$. To solve this problem, we have to integrate over the
additional phase space direction given for example by an additional
final state particle missed by the detector.  Next, to compute the
log-likelihood for the matrix element method we need to reconstruct
the 4-momenta of all initial-state and final-state particles. While
for leptons this seems feasible~\cite{Gritsan:2016hjl}, detector
effects will lead to sizeable smearing in the corresponding
4-momentum. Even if by construction of the partonic 4-momenta in the
hard process we manage to not violate energy and momentum
conservation, detector effects turn the ratio
$d\sigma_s(x)/d\sigma_b(x)$ into a ratio of integrals, not an integral
of a ratio.  Finally, when we compute perturbative QCD corrections
involving virtual and real gluons we need to combine phase spaces with
different number of final state particles~\cite{Alwall:2010cq}. This
can be solved by adapting the computation of the real emission
diagrams~\cite{Martini:2015fsa,Martini:2017ydu}.

In principle, all of those problems can be overcome.  One analysis
where the matrix element method is applied in Higgs physics is the
process~\cite{Artoisenet:2013vfa}
\begin{align}
pp \to t_l \bar{t}_{l,h} h \to t_l \bar{t}_{l,h} \; b\bar{b} \; ,
\end{align}
as discussed in Sec.~\ref{sec:exp_tth}. the notation $t_l$
represents a leptonic top decay. CMS has used the matrix element
method in both the single lepton and the double lepton
channel~\cite{Khachatryan:2015ila}, while ATLAS employed it in the
single lepton channel as part of a larger set of input observables to
a neural network~\cite{Aad:2015gra}. A major experimental challenge is
how to include the $b$-tagging in this analysis, so CMS actually uses
two likelihood ratios, one describing the event kinematics and the
other describing the heavy flavor content of the observed jets.  In
the right panel of Fig.~\ref{fig:mem} we show the log-likelihood ratio
distribution measured by CMS, including the expected signal and
background contribution. The left and right sets of curves clearly
indicate the importance of $b$-tagging in extracting the signal.

Just as a side remark, the input to the classic matrix element method
are jets, possibly including tagging information. It is possible to
extend the method to jet constituents, using the fact that Sudakov
factors are nothing but no-splitting probabilities.  This motivates
the so-called shower
deconstruction~\cite{Soper:2011cr,Soper:2012pb,Soper:2014rya}.

\subsubsection{Information geometry at detector level}
\label{sec:exp_data_info}

In the last two section we have seen how likelihoods as a function of
phase space and assuming a theory hypothesis are the central objects
of modern analyses. However, in Higgs physics we want to reduce the
focus on signal vs background analyses or simple observables and apply
these ideas to measuring model parameters using the full
log-likelihood information over the signal phase space. Such a model
parameter dependence leads us to the idea of information geometry,
usually defined in terms of the the Fisher information of
Eq.\eqref{eq:def_info} as the leading Taylor coefficient. It is not
yet used in LHC analyses, but it is a standard approach for example in
cosmological model tests and enjoys an obvious motivation in
theoretical and experimental LHC physics; if we can compute the
information geometry of LHC processes by extending the approach
described in Sec.\ref{sec:exp_data_mem} we gain full control over its
link between LHC measurements and the physics we can extract from
it~\cite{Brehmer:2016nyr,Brehmer:2017lrt}.

One feature common to the optimal observables and matrix element
method is that they are straightforward to implement at the Monte
Carlo level, but that the description of detector effects is
notoriously difficult.  On the other hand, they need to be included
because the likelihood ratio at parton level includes information
which is not present at the detector level.  The technical
complication is that all steps from a parton-level phase space
configuration $x \equiv x_p$ to the measured 4-momenta in data, $x_d$,
are described by Markov processes; those cannot easily be inverted. It
can be solved when we use the fact that these Markov processes are
independent of the underlying theory hypothesis $\vec{g}_s$ and
$\vec{g}_b$. The detector-level equivalent to Eq.\eqref{eq:llr} reads
\begin{align}
q(x_d|\vec{g}_s,\vec{g}_b) 
= \log \; \frac{p(x_d|\vec{g}_s)}{p(x_d|\vec{g}_b)}
= \log \; \frac{\int dx_p \, p(x_d|x_p) \; p(x_p|\vec{g}_s)}{\int dx_p \, p(x_d|x_p) \; p(x_p|\vec{g}_b)} \; .
\label{eq:llr2}
\end{align}
As long as the two hypotheses are irreducible, the full information on
the hypotheses is included in the parton-level information
$p(x_p|\vec{g}_{s,b})$, which can be easily computed with any event
generator. The probability distribution $p(x_d|x_p)$ describes the parton
shower, detector resolution, and analysis requirements. They are
modeled through a large set of random numbers, and they are not
(easily) invertible.

One way to construct the ratio of the two event-level likelihoods is
based on constructing proxies for $(x_d,x_p)$-dependent distributions
as $x_d$-dependent distributions with the test function $p(x_d|x_p)
p(x_p|\vec{g})$. We define this proxy nature by minimizing the
specifically-chosen functional (or metric-like
construction)~\cite{Brehmer:2018eca}
\begin{align}
F(x_d) = \int dx_p \; \left| g(x_d,x_p) - \hat{g}(x_d) \right|^2 \; p(x_d|x_p) \; p(x_p|\vec{g}) \; .
\label{eq:def_distri}
\end{align}
The variational condition $\delta F/\delta \hat{g}=0$ gives us the 
form of the proxy
\begin{align}
\hat{g}_*(x_d) = \frac{\int dx_p \, g(x_d,x_p) \; p(x_d|x_p) \; p(x_p|\vec{g})}{p(x_d|\vec{g})} \; .
\end{align}
One distribution we can approximate this way is the ratio of
parton-level likelihoods from any event generator~\cite{Brehmer:2018eca},
\begin{align}
g(x_d,x_p) 
= \frac{p(x_p|\vec{g}_s)}{p(x_p|\vec{g}_b)}
= \frac{p(x_d|x_p) \; p(x_p|\vec{g}_s)}{p(x_d|x_p) \; p(x_p|\vec{g}_b)}
\qquad \Rightarrow \qquad 
\hat{g}_*(x_d) 
= \frac{\int dx_p \, g(x_d,x_p) \; p(x_d|x_p) \; p(x_p|\vec{g}_b)}{p(x_d|\vec{g}_b)} 
= \frac{p(x_d|\vec{g}_s)}{p(x_d|\vec{g}_b)} \; .
\end{align}
In this case, the input function is the parton-level likelihood ratio
for each phase space point $x_p$ and does not even depend on $x_d$.  By
numerically minimizing the corresponding version of
Eq.\eqref{eq:def_distri} we can construct the observable-level
likelihood ratio in Eq.\eqref{eq:llr2}.

When we compute likelihoods we do not only have access to likelihood
values of individual phase space points, we can also treat the
likelihood as a function over model space. This means that in our
information geometry approach we can define the score as the vector of
first derivatives
\begin{align}
t_i(x_p|\vec{g}) 
= \frac{\partial \log p(x_p|\vec{g})}{\partial g_i}
= \frac{1}{p(x_p|\vec{g})} \;  \frac{\partial p(x_p|\vec{g})}{\partial g_i}
\end{align}
and find that the proxy for the parton-level score is the
detector-level score,
\begin{align}
g(x_d,x_p) 
= t_i(x_p|\vec{g})
= \frac{p(x_d|x_p) \; \partial p(x_p|\vec{g})/\partial g_i}{p(x_d|x_p) \; p(x_p|\vec{g})}
\qquad \Rightarrow \qquad 
\hat{g}_*(x_d) 
= \frac{\int dx_p \, g(x_d,x_p) \; p(x_d|x_p) \; p(x_p|\vec{g})}{p(x_d|\vec{g})} 
= t_i(x_d|\vec{g}) \; .
\end{align}
To use the detector-level information geometry in practice, we need to
minimize the functional $F(x_d)$ first. This is a typical regression
task for machine learning, where the to-be-minimized loss function
will be inspired by the functional $F$ evaluated over all
detector-level phase space points. This training of the
detector-level likelihood ratio might take some time, but the usual
advantage of the machine learning approach is that its evaluation will
then be very fast.

While this construction focuses on the likelihood ratio, it is easily
generalized to the Fisher information matrix defined in
Eq.\eqref{eq:def_info}.  The resulting information geometry, now
including detector
effects~\cite{Brehmer:2016nyr,Brehmer:2017lrt,Brehmer:2018eca}, has
the advantage that it defines a metric, which means that its results
do not depend on the underlying parametrization. It allows us to study
the parameter sensitivity from the entire phase space, parts of phase
space, or slices in phase space independent of the model
details. While this approach is an established way to analyze
potential measurements at the Monte Carlo level, it has not yet been
developed into an analysis strategy for the LHC.

\subsubsection{Data-based machine learning}
\label{sec:exp_data_deep}

\begin{figure}[t]
\includegraphics[width=0.47\textwidth]{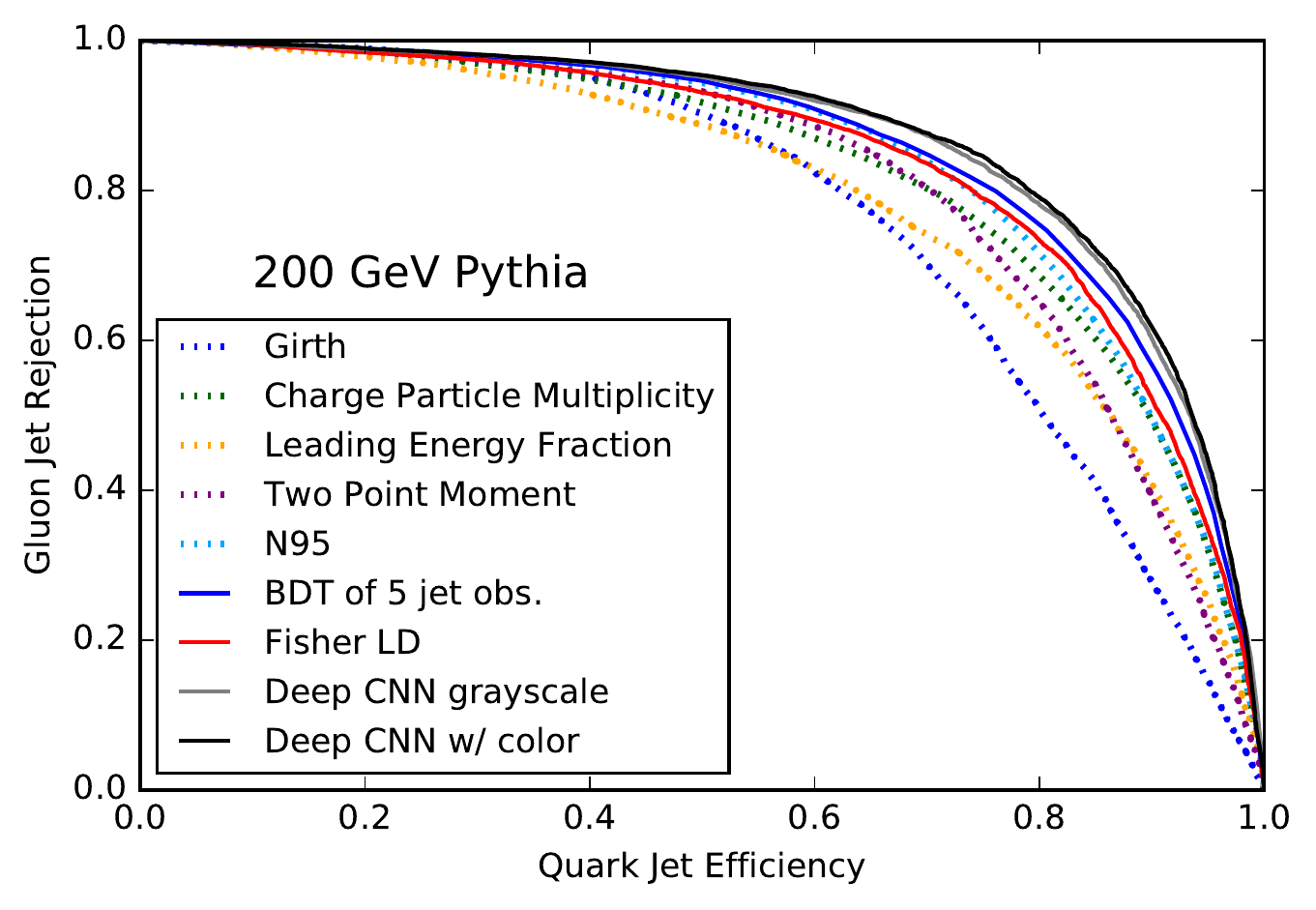}
\hspace*{0.1\textwidth}
\includegraphics[width=0.34\textwidth]{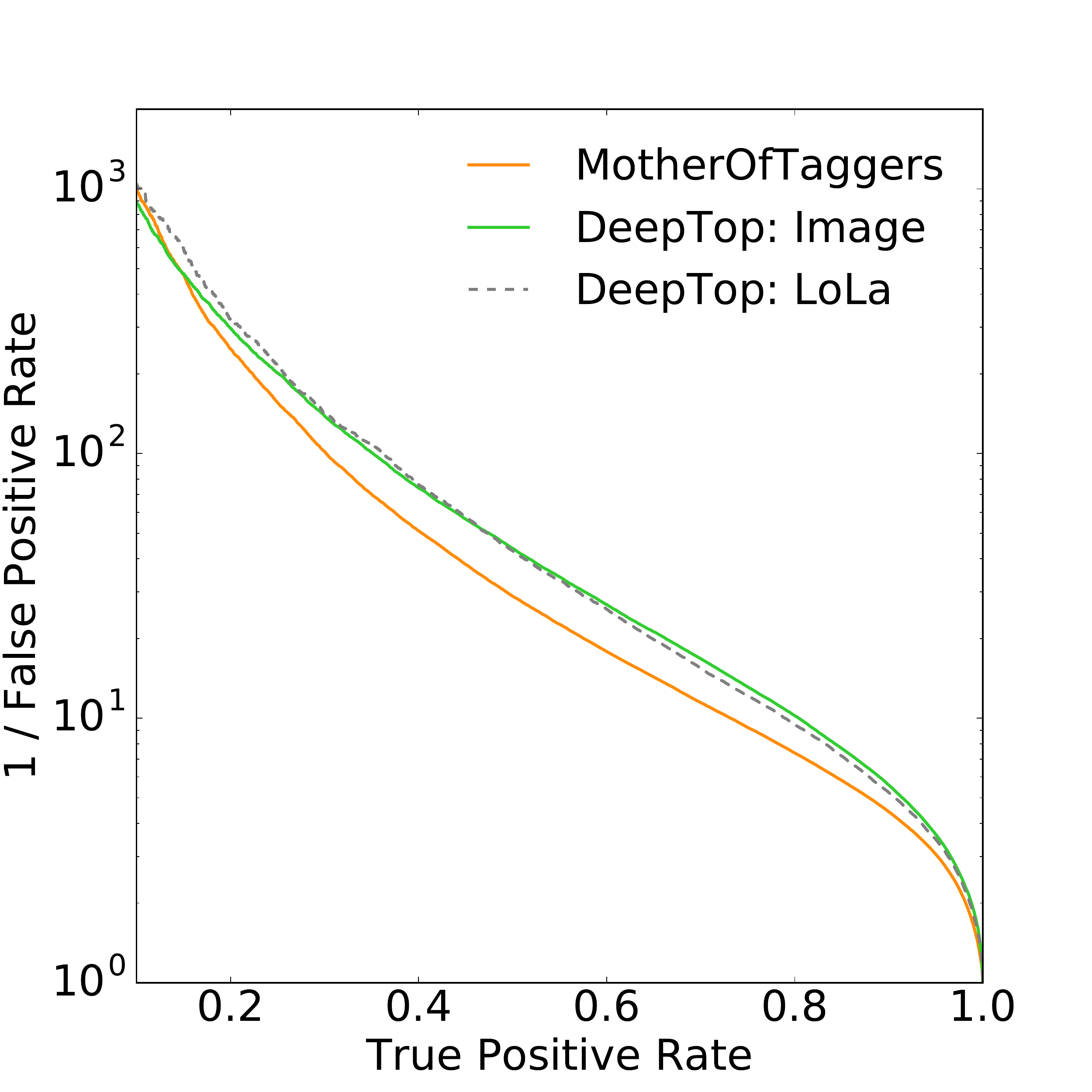}
\caption{Left: ROC curve for an image-based quark-gluon tagger,
  compared with analyses based on high-level observables. Figure from
  Ref.~\cite{Komiske:2016rsd}. Right: Receiver operating
  characteristic (ROC) curves for an image-based top tagger, a
  4-vector-based top tagger, and a set of high-level
  observables. Figure from Ref.~\cite{Butter:2017cot}.}
\label{fig:deep_jets1}
\end{figure}

Machine learning techniques allow us to search for patterns in large
amounts of highly complex data and has long been used by the LHC
experiments.  There, the number of events used in typical analyses is
enormous, especially when we also consider the samples used to measure
the detector performance and the background control regions. In
addition, the LHC has a vast number of output channels for each
event. This information is usually pre-processed into a number of
theoretically well-defined observables.  Multi-variate techniques
working on a number of such high-level observables, including machine
learning tools like neural networks, are standard in all Higgs
analyses. The leading question in many Higgs analyses is why we rely
on high-level observables as input to the neural networks describing
the hard process, while we are happy to directly feed low-level
detector output for example into a $b$-tagging network.  Given that
many state-of-the-art LHC analyses only include a theory hypothesis in
a full event simulation and then compare its output with data, this
choice appears ad-hoc, and we have to ask ourselves if through our use
of theory-motivated observables we lose information. Obviously, this
ongoing paradigm change has profound implications for theoretical
Higgs physics and the way we communicate between theory and
experiment.

An example, where a loss of information from theory-motivated analysis
objects motivates new analysis techniques, is subjet physics, introduced
in Sec.~\ref{sec:exp_tth}. Analyses based on jet-level observables
lose valuable information for example from hadronically decaying top
quarks, and subjet tools made resonance searches in purely hadronic
$t\bar{t}$ final state possible~\cite{Aad:2012ans}. In terms of
analysis objects this means that jets are no longer the main analysis
objects in LHC analyses and merely separate the data into jet-level
and subjet analyses.  While on the theory side we are often struggling
to analytically understand subjet observables beyond the leading-log
approximation~\cite{Dasgupta:2013ihk,Larkoski:2017jix}, this does not
pose a problem for likelihood-free analyses.  As long as we can
simulate distinctive parton splitting patterns for signal and
background processes, we do not have to define a quark or gluon to use
this additional subjet information. As usual, high-level observables
described by Feynman diagrams for the hard process mostly serve as an
illustration.

Rapid progress in machine learning outside particle physics have paved
the way to systematically search for patterns based on a large set of
low-level observables, rather than a limited number of high-level
observables. This development is at the heart of machine learning at
the LHC, where we try to identify analyses where a multi-variate
analysis of high-level observables might miss
information~\cite{Komiske:2017aww}. For such cases, machine learning
tools will extract this information from low-level detector
information. Given the success of subjet physics and the curious
attitude in that field, an obvious question is how much we can improve
the performance of established multi-variate subjet
analyses~\cite{Kasieczka:2015jma,Luo:2017ncs} by using low-level input
and machine learning. To participate in the advances of machine
learning outside particle physics we organize the experimental
information in a standardized manner. We then train a neural network
on some combination of data and simulation, ideally with a cross entropy loss function, such that the
output can be interpreted as a probability for an underlying
hypothesis. This probability can be viewed as the final outcome or as
input to another analysis, for instance going from jets to full
events. The first proposed subjet analyses using machine learning are
calorimeter images evaluated with convolutional
networks~\cite{Cogan:2014oua,deOliveira:2015xxd,Baldi:2016fql,Barnard:2016qma,Kasieczka:2017nvn,Macaluso:2018tck}. Calorimeter
images are 2-dimensional heat maps, encoding the energy deposition in
the rapidity vs azimuthal angle plane. The main problem with the image
approach is the combination of calorimeter and tracking information,
because the two sets of low-level data have a very different
resolution.  In the left panel of Fig.~\ref{fig:deep_jets1} we show a
comparison of a convolutional network using the combined calorimeter
and tracking information with high-level
observables~\cite{Komiske:2016rsd}. A similar picture arises when we
compare low-level machine learning and high-level multi-variate top
taggers~\cite{Kasieczka:2017nvn,Macaluso:2018tck}: machine learning
methods based on low-level inputs can do better than the best high-level
multi-variate analyses.

\begin{figure}[t]
\includegraphics[width=0.34\textwidth]{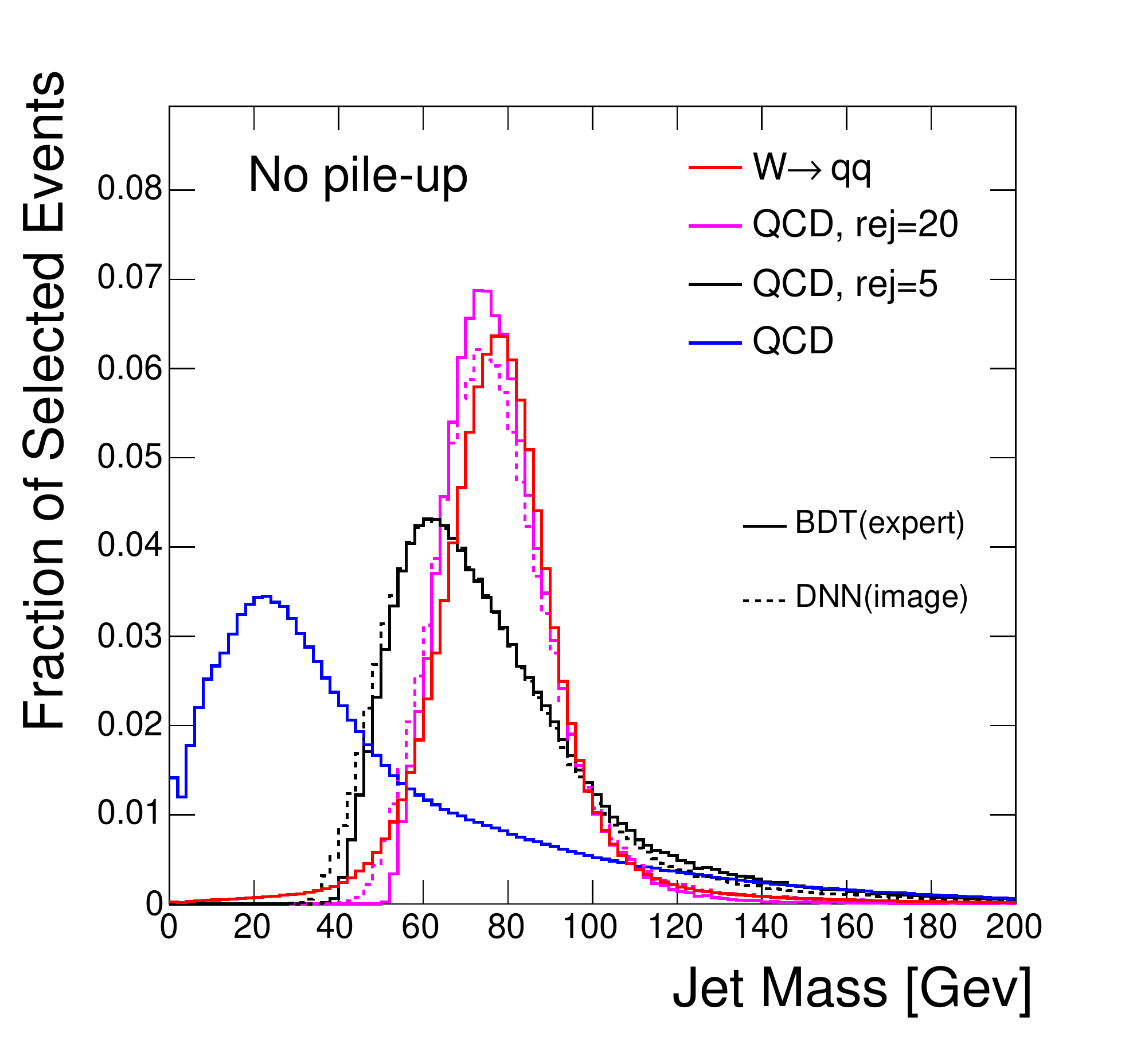}
\hspace*{0.01\textwidth}
\includegraphics[width=0.62\textwidth]{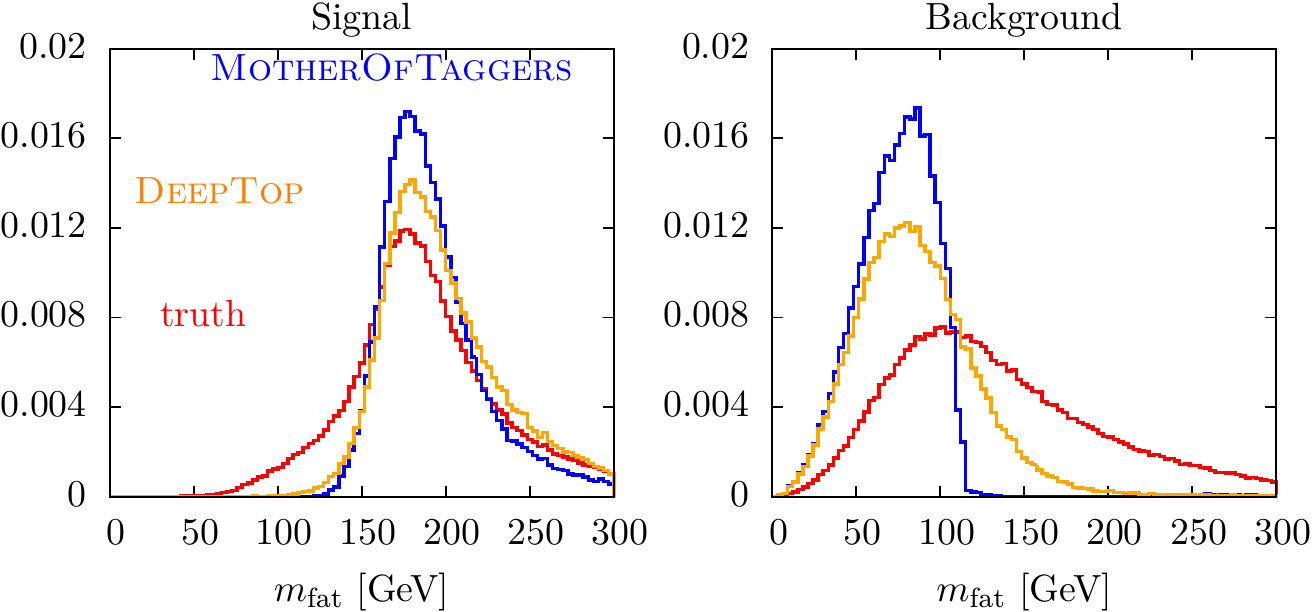}
\caption{Left: jet mass distribution for $W$-tagging at the Monte
  Carlo truth level and after requiring different levels of background
  rejection. Figure from Ref.~\cite{Baldi:2016fql}. Right: jet mass
  for top-tagging for correctly identified signal and background
  events. Again, we show truth level distributions and distributions
  after identification. Figure from Ref.~\cite{Kasieczka:2017nvn}.}
\label{fig:deep_jets2}
\end{figure}

Image recognition is not the only active field in machine
learning. Alternatively, we can buy into new ideas in natural language
recognition and apply them to subjet
analyses~\cite{Louppe:2017ipp,Cheng:2017rdo}. Finally, we can derive
inspiration on the data input format from physics, for instance
organizing the measured energy and momentum of a particle in terms of
4-vectors~\cite{Almeida:2015jua,Pearkes:2017hku} and either using or
learning the Minkowski metric~\cite{Butter:2017cot}.  We compare the
performance of an image-based top tagger and a 4-vector-based top
tagger with identical training samples in the right panel of
Fig.~\ref{fig:deep_jets1}.  The question of which low-level input format
is the most appropriate for subjet analyses is complicated by the fact
that eventually we need to include information on displaced vertices
or leptons, similar to heavy-flavor taggers.

To illustrate how we can understand the performance of machine
learning techniques applied to low-level observables we show the jet
mass distribution for an image-based $W$-tagger~\cite{Baldi:2016fql}
and top tagger~\cite{Kasieczka:2017nvn} in
Fig.~\ref{fig:deep_jets2}. In both cases we compare the Monte Carlo
truth level with a BDT analysis and the convolutional network
result. For the $W$-tagger we can clearly see how a stiffer background
rejection drives the distribution towards the generated signal. For
the top tagger we show correctly identified signal and background
events and compare them with the Monte Carlo truth. The BDT with the
jet mass as one of  its inputs clearly separates the corresponding events,
whereas the neural network also identifies events with a less clear
separation in jet mass. Such events frequently appear in the simulated
data.  These examples illustrate how we can understand the
classification through a neural network in terms of first-principles
observables at least as well as we control a BDT acting on high-level
inputs. The neural network is not a blacker box than a BDT acting on a
large correlated set of high-level observables.

Analyzing whole events using machine learning based on high-level
observables is an established analysis strategy. One problem with this
technique is that the relevant signal regions in phase space might be
plagued with large systematic or theoretical uncertainties, which have
to be implemented by hand. If these uncertainties are part of the even
simulation, we can include them in the spirit of comparing full
simulations to data by using adversarial
networks~\cite{Louppe:2016ylz}. The role of the process
\begin{align}
pp \to h j (j) 
\end{align}
with a moderately boosted Higgs boson and up to two jets has been
discussed in detail in Sec.~\ref{sec:exp_gf_kin}. The final state at
the jet level can be described by a small number of high-level
kinematics observables, like the transverse momentum of the Higgs or
the jets, their rapidities, or their azimuthal angle separations. Of
these variables the transverse momentum of the leading jet carries
information about the nature of the hard interaction. The size of the
corresponding tails of the distribution requires us to understand the
bulk of the distribution, or the total production rate. The latter,
however, is affected by perturbative uncertainties, which can be
described in terms of scale choices in the event generation.  This
allows us to use a set of adversarial networks, where the role of the
adversary is to remove all information that can be covered by the scale
choice~\cite{Englert:2018cfo}. In the left panel of
Fig.~\ref{fig:deep_event} we show the output of the adversarial
network, comparing the signal including a dimension-6 contribution to
the Higgs-gluon interaction with the top-induced Standard Model
prediction. The Standard Model simulations with different scale
choices, and hence significantly different total rates, all score
essentially the same, while the dimension-6 signal shows a clear
preference of a large neural network score.

\begin{figure}[t]
\includegraphics[width=0.43\textwidth]{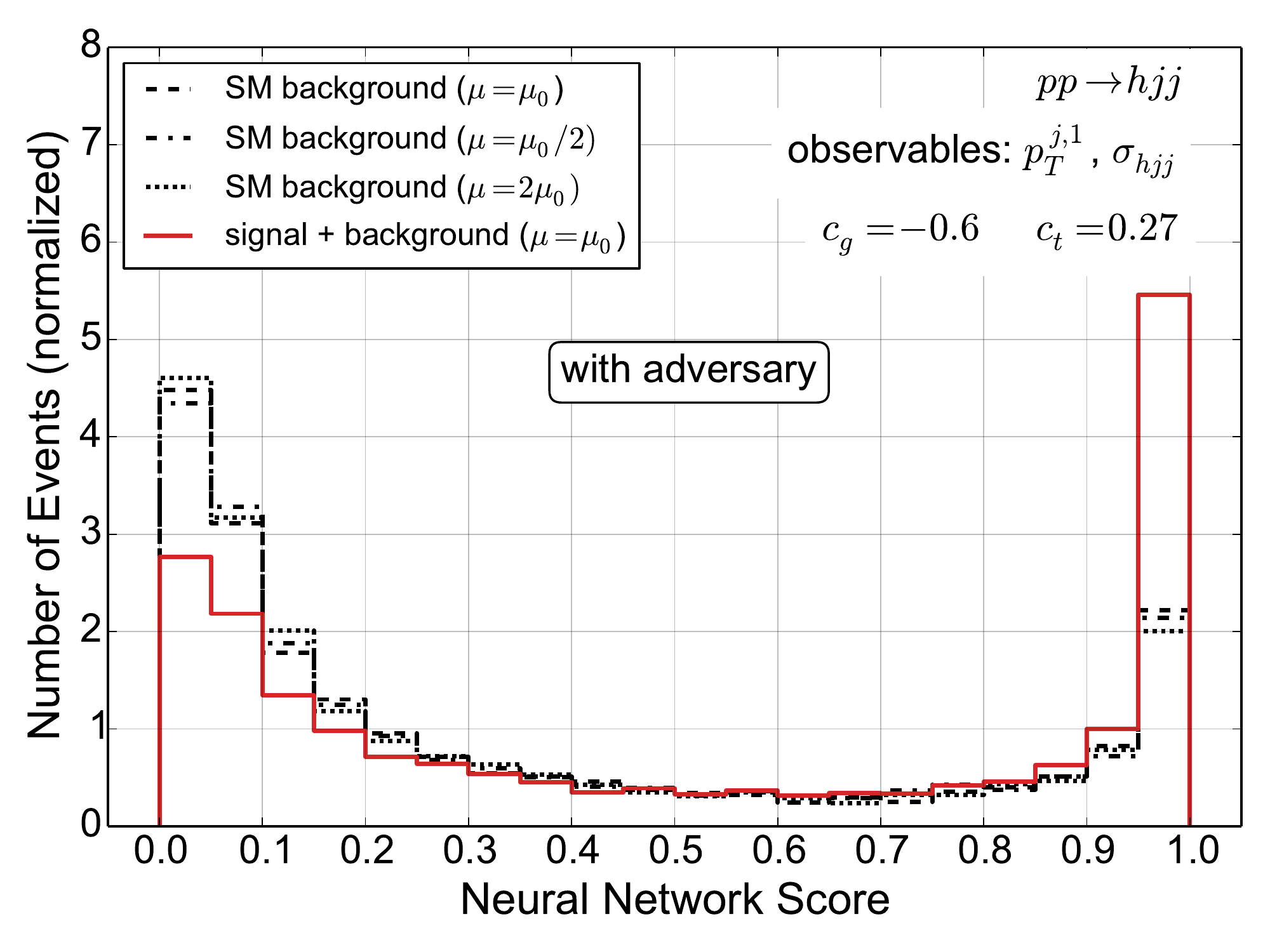}
\hspace*{0.05\textwidth}
\includegraphics[width=0.49\textwidth]{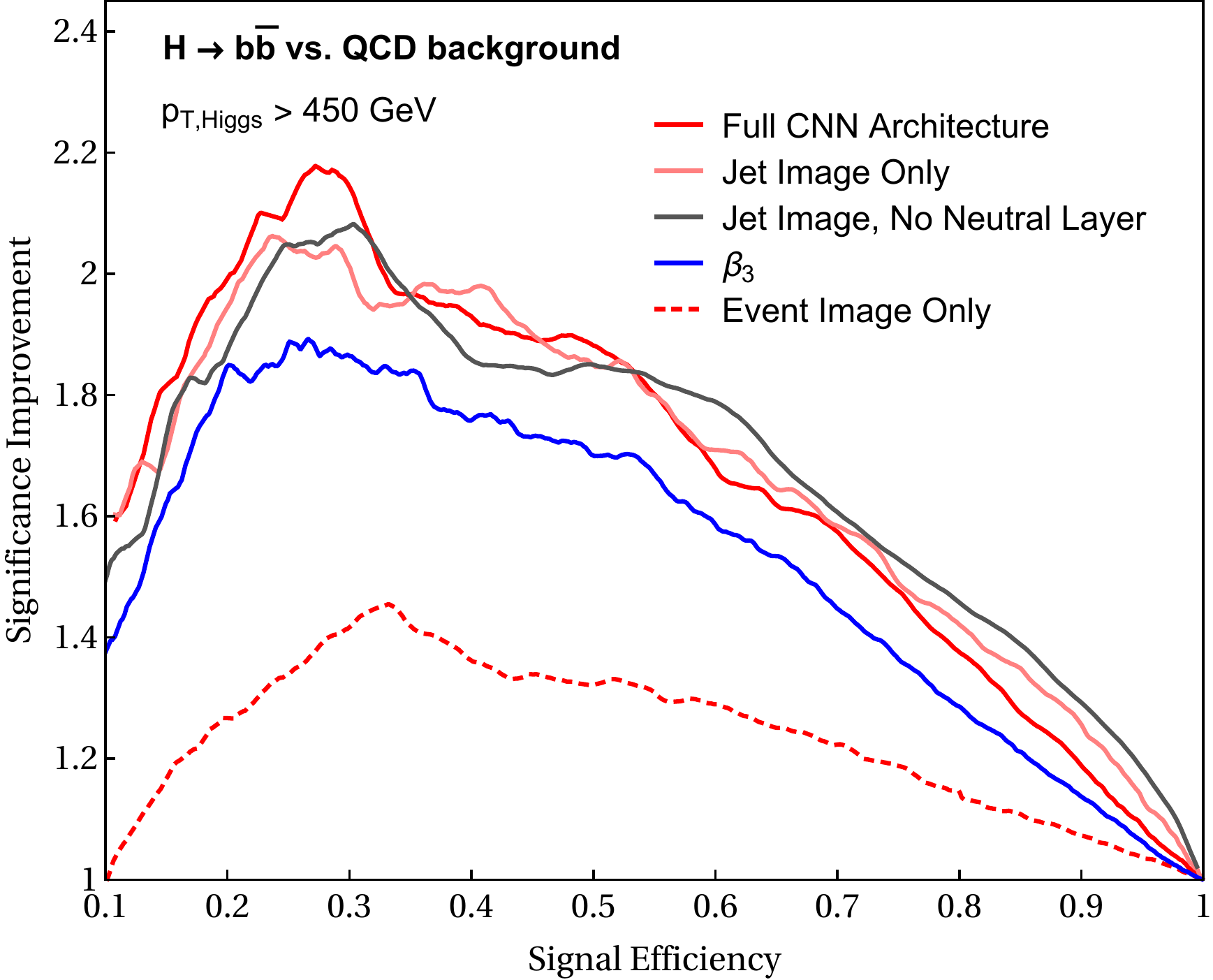}
\caption{Application of neural networks to $h$+jets searches. Left:
  neural network output for an adversarial network taking out the
  perturbative scale dependence. Figure from
  Ref.~\cite{Englert:2018cfo}.  Right: significance improvement from
  two convolutional networks at the jet level and at the
  subjet level. Figure from Ref.~\cite{Lin:2018cin}.}
\label{fig:deep_event}
\end{figure}

Moving beyond the high-level observables altogether, the ultimate goal
in machine learning at the LHC is to analyze entire events based on
low-level observables at the subjet-level and on the jet
level. Instead of generalizing subjet applications to the full
calorimeter and tracker coverage, it is convenient to separate events
into jet-level and subjet information. This can be studied for boosted
Higgs production~\cite{Lin:2018cin},
\begin{align}
pp \to h + \text{jet(s)} \to b\bar{b} \; + \text{jet(s)} \; .
\end{align}
The physics aspects of this process are discussed in
Sec.~\ref{sec:exp_gf_kin}.  Standard resonance search methods have not
been successful for such an all-hadronic channel due to the
overwhelming jet backgrounds. In the boosted configuration, the two
$b$-jets are close to each other and can be identified using subjet
methods, pioneered in Higgs physics by Ref.~\cite{Butterworth:2008iy}.
First, the role of the Higgs tagger can obviously been played by a
convolutional network acting on jet images. It will identify the Higgs
decay through a high-resolution jet image with $40 \times 40$
pixels. For now, a double $b$-tag inside the candidate Higgs jet is
required in addition to the low-level calorimeter information. A
second convolutional network searches for jet-like features in a
coarser $40 \times 40$ pixels image of the entire event. The periodic
boundary conditions are included through pre-processing and the
network setup. This second image does not resolve subjet structures,
but it is of the same order as the typical calorimeter resolution for
jet-level objects. This way, it probes the kinematic correlations
between the Higgs jet and the other even features. For the specific
$h$+jets signal the main physics question then becomes how much
information we can extract from the number of recoil jets and their
kinematics.  In the right panel of Fig.~\ref{fig:deep_event} we show
the improvement of the dual network analysis over a simple kinematic
analysis. We see how the impact of the jet-level network is
significantly exceeded by the subjet-level network, clearly making
the case that a combination of the two analysis tools should be the
ATLAS and CMS default for the coming LHC runs.

\section{Conclusions}

With the discovery of the Higgs boson in 2012 the Standard Model
withstood its ultimate test. Our understanding of perturbative quantum
field theory has been fully validated. The Standard Model is a
sophisticated theory that pulls out all the stops of our field theory
toolbox to make sense of an enormous amount of data. We detailed this
in this review in various contexts, and it would have been easy to
find even more praise for the Standard Model as an incredibly
successful theory.  But this would obscure a more relevant
point. The task of particle physics post Higgs discovery is a
different one. The Standard Model --- been there, done that, wore the
T-shirt. So what lies beyond?

While high energy physicists  are well aware of the philosophical and
practical limitations of the Standard Model, the blessing that is its
renormalizable character which underpins its discovery based on a
precise quantum analysis of its interactions has also become its
curse. Past discoveries like the top quark or the Higgs boson itself
were anticipated. If these fields had not been discovered this would
have challenged our otherwise successful understanding of quantum
field theory.  In this sense, the Higgs discovery was not at all a
coincidence but the pinnacle of a well thought-out plan, which would
have had a positive scientific outcome even in the absence of a
discovery.

However, some six years after the Higgs boson discovery, the sidewalk
has apparently come to an end. On the one hand, all motivated and
established perturbative UV-complete theories that seek to embed the
Higgs into a more fundamental picture of the TeV scale stand 
challenged. On the other hand, strongly-interacting approaches to
the electroweak scale lack experimental motivation and often push our
perturbative QFT toolbox over the edge. This makes concrete
predictions hard to obtain, while first-principle lattice calculations
will require more time to catch up. Adding the so-far negative outcome
of searches for new physics of any kind at the LHC to the mix, one
could develop a rather bleak outlook on particle physics and Higgs
phenomenology in particular.

But Higgs ain't over till it's over. The past years have seen
tremendous progress and success in both experimental and theoretical
particle physics, and most importantly at the intersection of both
communities. While precision and interpretation approaches based on
effective field theory or concrete UV-complete models are continually
pushed forward, novel analyses and search strategies that rely on
the combined expertise of experimentalists and theorists have shaped
the LHC's phenomenology program like no other collider experiment
before. Hadron colliders have been transformed from the historic
discovery machines to a comprehensive precision physics program.

The Higgs boson, as the central part of electroweak symmetry
breaking, is typically considered as a harbinger of an even more
satisfactory UV theory of particle physics that contains the Standard Model and
stands at the center of this development. Novel approaches, which
build on insights ranging from advanced data-analysis all the way to a
better understanding of the theoretical background of
sensitivity-limiting factors will set the scope of a Higgs precision
program that is likely to surpass earlier expectations.  It is
therefore realistic to expect a significant improvement of the current
Higgs property measurements as well as Higgs exotic searches. Together
they provide a coherent approach to understanding the nature of the
electroweak scale and its place in the physics landscape. Theoretical
progress in the precise prediction of new physics phenomena (\eg in
the MSSM or 2HDMs), as well as the interpretation of data in a largely
model-independent way on the grounds of EFT is now met by
experimentally sophisticated techniques to extract information about
the Higgs boson in its rarest final states even at comparably low
statistics. This will (hopefully) enable discoveries of as yet
unanticipated phenomena. The aim of this review is to bridge the basic
ideas that have motivated decades of research to these recent
developments.

\begin{center} \textbf{Acknowledgments} \end{center}

First, we would like to thank all our collaborators on many fun Higgs
projects over the years and everyone who went through this review and 
reminded us of mistakes and shortcomings, explicitly mentioning Michael Spira. 
C.E. and T.P. thank Peter Zerwas for
introducing them to the topic.  We are grateful to the Mainz Institute
for Theoretical Physics for its hospitality. T.P. and C.E.  are
grateful to TASI~2018, where parts of this review were finalized. 
S.D. is supported by the
U.S. Department of Energy under Grant Contract de-sc0012704.  C.E. is
supported by the IPPP Associateship scheme and by the UK Science and
Technology Facilities Council (STFC) under grant ST/P000746/1.

\end{fmffile}

\bibliography{paper}

\end{document}